\documentstyle[sprocl,rotating]{article}

\input{psfig}

\def\be{\begin{equation}}
\def\ee{\end{equation}}
\def\bea{\begin{eqnarray}}
\def\eea{\end{eqnarray}}

\def\bfq{{\bf q}}

\def\bfp{{\bf p}}
\def\bfpp{{\bf p '}}

\def\frac#1#2{{\textstyle{#1\over#2}}}
\def\darr#1{\raise1.5ex\hbox{$\leftrightarrow$}\mkern-16.5mu #1}
\def\){\right)}
\def\({\left( }
\def\]{\right] }
\def\[{\left[ }
\def\si{{}^1\kern-.14em S_0}
\def\siii{{}^3\kern-.14em S_1}
\def\diii{{}^3\kern-.14em D_1}

\def\MeV{{\rm\ MeV}}
\def\CA{{\cal A}}
\def\ket#1{\vert#1\rangle}
\def\bra#1{\langle#1\vert}

\newcommand{\pv}{\vec{\,\!p}\!\:{}}
\newcommand{\qv}{\vec{\,\!q}\!\:{}}
\newcommand{\ii}{\mathrm{i}}

\newcommand{\calI}{\mathcal{I}}

\def\lsc{\Lambda _\chi}
\def\cpt{\chi PT}

\def\vkay{{\vec k}}
\def\vkayprime{{{\vec k}\, '}}
\def\vpee{{\vec p}}
\def\vpeeprime{{{\vec p}\, '}}

\def\si{{}^1\kern-.14em S_0}
\def\siii{{}^3\kern-.14em S_1}
\def\diii{{}^3\kern-.14em D_1}
\newcommand{\gsim}{\raisebox{-0.7ex}{$\stackrel{\textstyle >}{\sim}$ }}
\newcommand{\lsim}{\raisebox{-0.7ex}{$\stackrel{\textstyle <}{\sim}$ }}
\def\pislash{ {\pi\hskip-0.6em /} }
\def\pislashsmall{ {\pi\hskip-0.375em /} }
\def\nopi{ {\rm EFT}(\pislash) }
\def\CzeroS{ {^\pislashsmall \hskip -0.2em C_0^{(\si)} }}
\def\CtwoS{ {^\pislashsmall \hskip -0.2em C_2^{(\si)} }}
\def\CfouroneS{ {^\pislashsmall \hskip -0.2em \tilde C_4^{(\si)} }}
\def\CfourtwoS{ {^\pislashsmall \hskip -0.2em C_4^{(\si)} }}
\def\CzeroT{ {^\pislashsmall \hskip -0.2em C_0^{(\siii)} }}
\def\CtwoT{ {^\pislashsmall \hskip -0.2em C_2^{(\siii)} }}
\def\CfouroneT{ {^\pislashsmall \hskip -0.2em \tilde C_4^{(\siii)} }}
\def\CfourtwoT{ {^\pislashsmall \hskip -0.2em C_4^{(\siii)} }}
\def\CzeroTmone{ {^\pislashsmall \hskip -0.2em C_{0,-1}^{(\siii)} }}
\def\CzeroTzero{ {^\pislashsmall \hskip -0.2em C_{0,0}^{(\siii)} }}
\def\CzeroTone{ {^\pislashsmall \hskip -0.2em C_{0,1}^{(\siii)} }}
\def\CtwoTmtwo{ {^\pislashsmall \hskip -0.2em C_{2,-2}^{(\siii)} }}
\def\CtwoTmone{ {^\pislashsmall \hskip -0.2em C_{2,-1}^{(\siii)} }}
\def\CfouroneTmthree{ {^\pislashsmall \hskip -0.2em \tilde C_{4,-3}^{(\siii)} }}
\def\CfourtwoTmthree{ {^\pislashsmall \hskip -0.2em C_{4,-3}^{(\siii)} }}

\def\CSDzero{ {^\pislashsmall \hskip -0.2em C_0^{(sd)} }}
\def\CSDzeromone{ {^\pislashsmall \hskip -0.2em C_{0,-1}^{(sd)} }}
\def\CSDzerozero{ {^\pislashsmall \hskip -0.2em C_{0,0}^{(sd)} }}

\def\CSDtwoone{ {^\pislashsmall \hskip -0.2em \tilde C_2^{(sd)} }}
\def\CSDtwotwo{ {^\pislashsmall \hskip -0.2em C_2^{(sd)} }}
\def\CSDtwotwotwo{ {^\pislashsmall \hskip -0.2em C_{2,-2}^{(sd)} }}
\def\CSDtwoonetwo{ {^\pislashsmall \hskip -0.2em C_{2,-2}^{(sd)} }}

\def\etasd{\eta_{sd} }

\def\CQuad{ {^\pislashsmall \hskip -0.2em C_{\cal Q} }}
\def\Ltwo{ {^\pislashsmall \hskip -0.2em L_2 }}
\def\Lone{ {^\pislashsmall \hskip -0.2em L_1 }}

\def\pone{{}^3\kern-.14em P_1}
\def\pzero{{}^3\kern-.14em P_0}
\def\ptwo{{}^3\kern-.14em P_2}

\def\CPzero{ {^\pislashsmall \hskip -0.2em C^{(\pzero)}_2  }}
\def\CPone{ {^\pislashsmall \hskip -0.2em C^{(\pone)}_2  }}
\def\CPtwo{ {^\pislashsmall \hskip -0.2em C^{(\ptwo)}_2  }}

\renewcommand{\frac}[2]{{{#1}\over {#2}}}

\begin{document}

\title{From Hadrons to Nuclei: Crossing the Border}

\author{\bf Silas R. Beane$^a$, 
Paulo F. Bedaque$^b$, 
Wick C. Haxton$^b$\\
Daniel R. Phillips$^{a,c}$, and
Martin J. Savage$^{a,d}$   }

\vspace{1cm}

\address{$^a$ Department of Physics,
\\ University of Washington, Seattle, WA 98195-1560}

\vspace{1cm}

\address{$^b$ Institute for Nuclear Theory,
\\ University of Washington, Seattle, WA 98195-1560}

\vspace{1cm}

\address{$^c$ Department of Physics,\\
Ohio University, Athens, OH 45701}

\address{$^d$ Jefferson Laboratory,\\ 12000 Jefferson Avenue, Newport News, 
VA 23606}
\vspace{1cm}
\address{\tt sbeane, bedaque, phillips, savage, haxton@phys.washington.edu}

\maketitle
\abstracts{
The study of nuclei predates by many years the theory of quantum
chromodynamics. More recently, effective field theories have been used
in nuclear physics to ``cross the border'' from QCD to a nuclear
theory. We are now entering the second decade of efforts to develop a
perturbative theory of nuclear interactions using effective field
theory.  This chapter describes the current status of these efforts.
}

%%%%%%%%%%%   INTRODUCTION  %%%%%%%%%%%%%%%%%%%
\section{Introduction}

Nuclei are fascinating objects.  In a nucleus the protons and
neutrons, collectively known as nucleons, are bound together by the
strong nuclear force.  At a fundamental level these interactions are
described by Quantum Chromodynamics (QCD), a theory of quarks and
gluons carrying color charges that are asymptotically-free at
short-distances.  However, the quarks and gluons in a nucleus are very
far from being asymptotically-free.  Instead they comprise individual,
colorless, nucleons, which largely retain their identity in the many-body
system. The color-singlet nucleons are then bound to each other by
what can be thought of as ``residual" QCD strong interactions. This
sketch of nuclear dynamics from the QCD point of view---brief as it
is---makes it clear that from this standpoint the nucleus is an
incredibly complicated, non-perturbative, quantum-field-theoretic,
infinite-body problem.

This makes it truly remarkable that the properties and interactions
of nuclei can be largely understood within the following 
framework, which we shall refer to here as the ``traditional"
picture of a nucleus. The $A$ nucleons in a nucleus are understood to be 
non-relativistic particles interacting via a quantum-mechanical
Hamiltonian consisting of two-body potentials and 
three-body potentials which have considerably smaller, but still
important, effects:
\begin{eqnarray}
H\ =\ 
\sum_{i=1}^N \frac{|\bf{p}_i|^2}{2 M} 
\ +\ 
 \sum_{i<j} V_{NN}({\bf r}_i, {\bf r}_j)
\ +\ 
 \sum_{i<j<k} V_{3N} ({\bf r}_i,{\bf r}_j,{\bf r}_k).
\label{eq:HUrbana}
\end{eqnarray}
This Hamiltonian is then used in the Schr\"odinger equation
\begin{equation}
H|\psi^{(i)} \rangle=E^{(i)} |\psi^{(i)} \rangle,
\end{equation}
to determine the nuclear wave functions $|\psi^{(i)} \rangle$ and
energy-levels $E^{(i)}$.  With modern computing power the
energy-levels of this Hamiltonian can be calculated with $\sim 1 \%$
precision for nuclei as large as $A=8$ and soon calculations for
twelve-body nuclei such as $^{12}C$ should become available.  The
results of this program are in impressive agreement with experiment.
With wave functions in hand, properties of the nuclides such as
magnetic and quadrupole moments, electromagnetic form factors, and the
response of light nuclei to weak probes such as neutrinos can all be
calculated.  For a thorough review of these successes see
Ref.~\cite{CS98}.

However, the success of a picture described by the Hamiltonian in
Eq.~(\ref{eq:HUrbana}) raises a number of other questions, among which
are:
\begin{itemize}
\item Why are nucleons the appropriate degrees of freedom inside
a nucleus?

\item Why are two-body interactions $V_{NN}$ dominant? 
And how should they be described?

\item How do we construct a three-nucleon interaction $V_{3N}$?

\item Are all contributions to observables equally large, or is 
there some set of small parameters which can be used to organize 
the calculation?
\end{itemize}
All of these issues are part of one overarching question, namely:
\begin{quotation}
Can we both understand how the Hamiltonian in Eq.~(\ref{eq:HUrbana}) emerges from
QCD to dominate the dynamics of multi-nucleon systems, and 
systematically refine it to accurately describe all nuclear 
observables.
\end{quotation}

\begin{figure}[h,t,b,p]
\hskip 1.0in
\psfig{figure=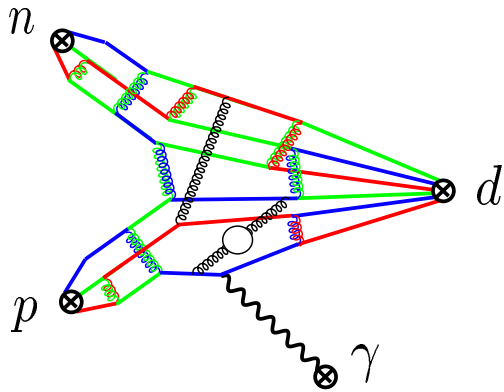,height=2.0in}
\caption{
A ``cartoon'' of a contribution to  $np\rightarrow d\gamma$
in QCD.}
\label{fig:npdgQCD}
\end{figure}

One kind of answer to this question would be provided by computations of
the properties and reactions of nuclei directly from lattice QCD. For
instance, consider what happens when a proton captures a neutron and
forms a deuteron, in the process emitting a photon: the
radiative capture process
$np\rightarrow d\gamma$.  A lattice calculation of $np \rightarrow d
\gamma$ would provide an unambiguous cross-section in terms of the
quark masses and $\Lambda_{\rm QCD}$.  However, such a calculation would
be very difficult to perform.  For instance, a cartoon of one
contribution to $np\rightarrow d\gamma$ in terms of perturbative
quarks and gluons is shown in Fig.~\ref{fig:npdgQCD}.  Sources of
quarks and gluons that have non-zero overlap with the proton, neutron,
and deuteron as well as a source for the photon would be used
to generate the reaction amplitude.  One key problem with using
lattice QCD to calculate this reaction is that a significant amount of
effort goes into forming the asymptotic states themselves---before one
even begins computing their interactions. Indeed, since the deuteron
has such a small binding energy and so is quite extended compared to
the nucleon, considerable effort will be required to generate the
deuteron on the lattice.  The deuteron is yet to be formed in any
lattice calculation, and so, to put it mildly, a comprehensive lattice
study of this simple hadronic process is not imminent.
%
%%%%%%%% np-> d gamma QCD figure %%%%%%%%%%%%%%%
Therefore, given the current state of affairs in lattice QCD, serious efforts
to determine properties of multi-nucleon systems must start with
neutrons and protons as the basic degrees of freedom. This has the
considerable advantage that one is working in terms of the asymptotic
states which are those detected in the laboratory.

Clearly an important first step in describing the dynamics
of a multi-nucleon system is to describe the interaction
between just two-nucleons, i.e. the two-nucleon scattering
amplitude.
This has been accomplished very successfully by fitting the 
nucleon-nucleon ($NN$) potential $V_{NN}$.  There are
many $NN$ potentials on the market, but the most accurate models are
now tuned to fit the over 3000 data which constitute the database for
$NN$ scattering below laboratory energies of 350 MeV~\footnote{In
Ref.~\cite{St93} this was chosen as the uppermost energy for which
data would be included in the $NN$ database because the
pion-production threshold, at $T_{\rm lab}= 280$ MeV, begins to play a
role there, and so the $NN$ collisions are no longer elastic.}. In all
of these models the long-distance part of the interaction,
corresponding to inter-nucleon distances of order $2~{\rm fm} \sim
1/m_\pi$, is described by the model-independent exchange of one
pion---a sixty-five year old idea, due to Yukawa~\cite{Yu35}.  At
shorter distances $\sim 1/(2 m_\pi)$ some potentials include the
exchange of two pions.  This too can be derived in a model-independent
way from chiral perturbation theory ($\chi$PT).  In contrast, once one
looks at the force between nucleons separated by $\sim 1~{\rm fm}$,
there are a variety of short-range components present in the different
$NN$ potentials currently in use.  Some assert that the short-range
interaction arises from the exchange of vector and scalar mesons,
while others look to the naive quark model for inspiration.  More
popular, and more {\it effective}, than either of these models are
potentials which fit the data essentially to within experimental
accuracy~\cite{St94,Wi95,Ma00}, using a short-distance part of
$V_{NN}$ that is essentially a best fit, with no dynamical assumptions other
than continuity, differentiability, and the desire to include all
possible operator structures which can arise in the $NN$ system at low
energies. This part of $V_{NN}$ occurs in the region where the
nucleons themselves have substantial overlap, and
$r~\sim~\Lambda_\chi^{-1}$ with $\Lambda_\chi$ is the chiral-symmetry
breaking scale. 
It is clear that without input from lattice QCD,
assertions about the underlying dynamics in this region are nothing
more than guess-work. Thus the use of these ``best-fit"
short-distance $NN$ interactions is very much in the spirit of
effective field theory.
However, once we move beyond the $NN$ interaction and start considering
three-nucleon forces and the construction of current operators for use
in multi-nucleon systems the picture becomes significantly more
complex.  
As nuclear  systems are inherently non-perturbative
it is
difficult to see which parts of a
computation are essential to the description of a particular
observable, and which parts are not.
Effective field theory (EFT) 
allows one to both systematize calculations in light nuclei, and
simplify the picture of these nuclei down to its essential
ingredients\footnote{There is an old saying:
{\it Those who do not know history are doomed to repeat it.}
However, there is a slightly more modern version that may be 
more relevant:
{\it Those who do not understand history 
will not recognize progress.} 
}.

In this chapter of the {\it Handbook of QCD} we will 
describe the progress that has been made in constructing and 
applying EFT's
to multi-nucleon systems, and the interactions of these systems
with ``external" probes such as $\gamma$'s, $W^\pm$'s, $Z^0$'s,
and $\pi$. Since Weinberg's pioneering efforts~\cite{We90,We91,We92}
in the early 1990s to set up a systematic description of multi-nucleon
interactions using EFT much effort has been gone into realizing this
goal. At present these nuclear EFTs are far from being able to
determine, for instance,
the $B(E2)$ values for transitions in $^{25}$Mg, but
important progress has been made in understanding systems of
two
and three nucleons
as well as in discussion of probes of the
two-nucleons system.
We will also
discuss the progress and problems facing EFT's attempts to discuss
larger nuclei.

An EFT is defined by the most general Lagrange
density consistent with the global and local symmetries of the
underlying theory, a declaration of the regularization and
renormalization scheme, and a well-defined power counting in terms of
small expansion parameters. This issue is somewhat controversial in
the presence of bound- or quasi-bound states because there are
differing criteria for what constitutes a rigorous EFT. 
On the one hand there is cutoff EFT. 
In cutoff EFT the power
counting is essentially dimensional analysis at the level of the
potential: operators are ordered
according to inverse powers of the
momentum space cutoff. 
Since the quantum mechanical potential is not
an observable quantity, the justification for the power counting is
necessarily {\it a posteriori}. 
One computes to a given order in the
expansion and then one performs an error analysis for observables such
as phase shifts in order to verify that cutoff dependence is removed
order-by-order in the expansion of the potential.  This method has
much in common with lattice gauge theory with improved lattice
actions~\cite{Le97}. 
Lattice errors are removed order-by-order in the
lattice spacing by including operators in the action with increasing
powers of the lattice spacing.  On the other hand, there is 
EFT as it is traditionally practiced in a purely perturbative
setting, like chiral perturbation theory ($\cpt$). The power counting
is in the scattering amplitude itself and therefore applies directly
to observable quantities. In this case one can explicitly prove
renormalization scale independence at each order in the EFT 
expansion of observables. Moreover, one can make an {\it
a priori} estimate of the theoretical errors which arise from working
to a given order in the perturbative expansion.  The distinction
between these two criteria becomes an issue only when there is a
bound- or quasi-bound state near threshold. In these cases at least
one local (as in a contact operator) or non-local operator (as in a
particle exchange) must be summed to all orders and so the EFT is
intrinsically nonperturbative. The nonperturbative nature of the $NN$
interaction leads to several unresolved formal issues in a cutoff EFT
of $NN$ interactions. These issues will be discussed below.

The underlying motivation for using EFT to describe systems with
well-separated lengths scales can be found in the Chapters by
Leutwyler, by Mei\ss ner and by Manohar, and we will not repeat it
here. Instead, we begin our tour of EFTs in nuclear physics by looking
at the simplest nuclear system, $A=2$, at momenta well below the pion
mass.  At these very low energies the only relevant degrees of freedom
are the nucleons themselves.  We will show how to systematically
describe both $NN$ scattering and the coupling of electroweak probes
to the $NN$ system.  These calculations of very-low-energy properties
provide analytic results for all observables. They must be compared,
process-by-process, against the results of the ``traditional"
approach.  Subsequently, we move away from the $A=2$ system and
discuss EFT for the $A=3$ system---also at very low energies. The EFT
treatment of the three-nucleon problem provides some truly novel
features, including the emergence of a type of renormalization-group
(RG) behavior not seen before in a single-parameter system.

We then turn to higher energies, where the pion must be included as an
explicit degree of freedom in the Lagrangian.  Attempts to include
pions in perturbation theory are presented and their current status
discussed. We then discuss Weinberg's original suggestion for dealing
with the pions non-perturbatively. This is to apply chiral
perturbation theory ($\chi$PT), as formulated in the Chapters by
Leutwyler and Mei\ss ner, to the $NN$ potential and other two-nucleon
irreducible operators.  We display some successes of this approach,
and discuss some unresolved formal issues. Finally, we discuss recent
progress in incorporating the ideas of EFT into more traditional
methods of calculating nuclear properties.

%%%%  Two Nucleons  %%%%%%%%%%%%%%

\section{Two-Nucleon Systems at Very Low Energies}

Ideas akin to EFT appeared in nuclear physics as early as the 1940s,
when Bethe introduced Effective Range Theory (ERT) as a description of
$NN$ scattering~\cite{Be49,BL50}.  Bethe showed that at low energy the
experimental data did not depend on dynamical details, but only on a
few numbers which summarized the impact of strong interactions on
observables. 
ERT is an attempt to describe both elastic and inelastic processes
involving two nucleons, provided that all momenta involved are less
than the pion mass. 
By construction, strong-interaction data is
reproduced in ERT to arbitrary precision in this low-momentum
regime. However, ERT fails to converge to the measured amplitudes for
processes involving photons or other ``external'' probes. As the
energy of experiments increased, ERT become less relevant, and more
complicated descriptions in the form of potential models, such as the
Hamiltonian (\ref{eq:HUrbana}), became popular. In this section we will
show how to derive an EFT which will reproduce ERT, and hence the $NN$
data, to arbitrary precision, provided that $p < m_\pi/2$. 
EFT goes beyond ERT, as it systematically includes
operators which enter when one
calculates the response of the $NN$ system to external probes.  
We
will give explicit examples of analytic calculations for low-energy
reactions which can be carried out to $\sim 1 \%$ accuracy.

%%%%%%%%%%%%%%  Naive QCD  %%%%%%
\subsection{Naive Dimensional Analysis Really is Naive}

What scales might be reasonably expected in the two-nucleon sector if
one was simply presented with the QCD Lagrangian?  Neglecting quark
masses, QCD is a theory with only one length scale, $\Lambda_{\rm
QCD}$. In the hadronic sector this scale appears as the scale of
chiral symmetry breaking, $\Lambda_\chi\sim 1~{\rm GeV}$, and the
nucleon mass $M$.  At low energies, the effective theory of QCD will
be one whose matter content includes only non-relativistic nucleons and
massless pions. Neglecting factors of $4\pi$ the only mass scale in
the theory is 
$\Lambda_\chi$~\footnote{Strictly speaking this is not true, 
since the mass difference between the nucleon and the
$\Delta$ is $M_\Delta - M = 232~{\rm MeV}$.}.

The inclusion of the quark masses in the QCD Lagrangian explicitly
breaks chiral symmetry. The pion acquires a mass $m_\pi \approx
140~{\rm MeV}$, and so introduces another ``small'' length scale. What
then, is a reasonable scale to expect for the binding energy of
nuclei?  Naively one might expect that the nuclear size will be set
by the pion mass. However, one must remember that the pion is
derivatively coupled, and so it is not clear that the pionic dynamics
are strong enough to be responsible for bound-state formation. Even
accepting that the pion is the particle whose exchange generates the
non-perturbative interaction which leads to the formation of the
deuteron, one would not then expect it to be bound by only
$B=2.224644\pm 0.000034~{\rm MeV}$. This corresponds to a pole in the
$NN$ scattering amplitude at momentum $|{\bf k}| = i \gamma$---with
$\gamma \equiv \sqrt{MB}=45.7025~{\rm MeV}$---in the ${}^3S_1$
channel~\footnote{Here we use the nuclear spectroscopic notation
${}^{2S+1} L_J$, where $S$ is the total spin of the $NN$ system, $J$
denotes its total angular-momentum, and $L$ denotes its orbital
angular momentum.}: significantly less than our prior expectation.  In
the ${}^1S_0$ channel the situation is even more striking.  There is
no bound-state in the $\si$ channel, but a pole on the second
energy-sheet represents a quasi-bound state. In terms of momentum,
this pole sits at $|{\bf k}|\sim (-i)(8~{\rm MeV})$.  
The unnaturally low
energies at which this and the deuteron pole reside are an
experimental fact. Such threshold states can only result from a
fine-tuning between long and short distance physics, and so their
existence can never be understood using naive dimensional 
analysis~(NDA).

In an EFT such fine-tuned bound states result in operators that are
naively irrelevant (in the RG sense), playing a crucial role.
Consequently, power counting based on the NDA we are familiar with in
the zero- and one-nucleon sectors~\cite{MG84} does not apply.  In
fact, as is discussed below, the two-nucleon system is very close to a
non-trivial fixed point in the RG-flow.  Another way to see this is to
remind ourselves of something obvious: the theory is nonperturbative.
Nuclei are bound states of nucleons and so will never arise at any
finite order in perturbation theory.  Unlike the baryon-number zero
and one sectors, the EFT appropriate to nuclei must admit
non-perturbative structure somehow and somewhere.  Thus, an infinite
number of loop-diagrams must enter at the same order in the EFT
expansion.

\subsection{Low-Energy $NN$ Scattering Data and 
Unusually Large Length Scales}

Over the years there have been literally hundreds of experiments
performed to measure $NN$ cross-sections and polarization observables
below the pion-production threshold. In 1993 the Nijmegen group
performed a phase-shift analysis (PSA) of these
experiments~\cite{St93}.  This analysis summarizes all of this data in
terms of the phase shifts and mixing parameters that describe
scattering in the different spin-angular momentum channels.  Clearly
any theory which reproduces these phase shifts will also reproduce the
data. Conversely, although the mapping from data to phase shifts is
not mathematically unique, the Nijmegen group's imposition of
``reasonable" physical constraints apparently makes it quite difficult
to obtain phase shifts which differ radically from those found in
Ref.~\cite{St93}. We will always compare the EFT phase shifts and
mixing parameter(s) with the Nijmegen PSA and not directly with data.

At low energies the $NN$ cross-section is dominated by S-wave
scattering.  Let us recall the behavior of the phase shift in a
strongly-interacting system. Firstly, the S-matrix of two
non-relativistic nucleons scattering in the $\si$ channel 
with center-of-mass momentum $|{\bf k}|=\sqrt{M T_{\rm lab}/2}$ is 
\begin{eqnarray}
  S & = &  e^{i2\delta}
\ =\ 1\ +\  {2 i|{\bf k}|\over |{\bf k}|\cot\delta - i|{\bf k}|}
\ =\ 1\ +\ i {|{\bf k}| M\over 2 \pi  } {\cal A}
,
\label{eq:Smat0} 
\end{eqnarray}
where $\delta$ is the phase shift, and $T_{\rm lab}$ is the 
kinetic energy of the incoming nucleon in the laboratory frame.
We use a normalization such that 
explicit computation of Feynman diagrams, in an arbitrary frame, gives
the non-relativistic amplitude ${\cal A}$.
From elementary scattering theory, $|{\bf k}|\cot\delta$ is an
analytic function of external kinetic energy  with a radius of convergence
bounded by the nearest t-channel singularity. In the case of $NN$ 
scattering this is the pion cut, which occurs at $|{\bf k}|=m_\pi/2$.
Within this radius of convergence we may use the expansion:
\begin{eqnarray}
|{\bf k}|\cot\delta & = & -{1\over a}\ +\ {1\over 2} r_0  |{\bf k}|^2
\ +\  r_1 |{\bf k}|^4\ + \ldots .
\label{eq:pcot0}
\end{eqnarray}
While Eq.~(\ref{eq:pcot0}) corresponds to expanding about $|{\bf k}|=0$,
$|{\bf k}|\cot\delta$ can, of course, be expanded about any
point within the region of convergence. In general, the size of
the coefficients $r_0$ (the effective range) and $r_1$ (the shape
parameter) is determined by the range of the interaction between the
nucleons, which is $\sim 1/m_\pi$~\cite{Wi55,PC97}.  One might also
expect the scattering length $a$ to be ``natural" in this
sense. However, $a$ is not constrained by the range of the underlying
interaction, and, as we have already discussed, the scattering lengths
for $NN$ scattering in both the $\si$ and $\siii$ channel are
significantly larger than $1/m_\pi$. In the ${}^1S_0$ channel the
empirical values of the parameters $a$, $r_0$, and $r_1$
are~\footnote{Throughout, we quote only the
strong-interaction part of the phase shift and give the parameters in
the $np$ channel.  There are corrections for charge-independence
breaking and charge-symmetry breaking which shift the value of both
$a$ and $r_0$ if one considers $nn$ or $pp$ scattering.}:
\begin{eqnarray}
a = -23.714~{\rm fm},\quad r_0 = 2.73~{\rm fm}, \quad
r_1 = -0.48~{\rm fm}^3.
\label{eq:singparams}
\end{eqnarray}

We could also expand about $|{\bf k}|=0$ in the spin-triplet S-wave channel,
the ${}^3S_1$. However, since the deuteron is present in this
partial wave it is convenient to instead make the
expansion of $|{\bf k}|\cot\delta$ about the bound-state
pole position~\cite{Be49,BL50}. In fact, if one truncates the expansion
in Eq.~(\ref{eq:pcot0}) 
at order $|{\bf k}|^2$, it is 
straightforward to relate the location of the pole,
$|{\bf k}| = i\gamma$, to the 
${}^3S_1$ scattering length and effective range,
\begin{eqnarray}
\gamma & = & {1\over r_0}
\left[ 1 - \sqrt{ 1 - \frac{2 r_0}{a}}\right]
\ \ \ .
\label{eq:poleplace}
\end{eqnarray}
Neglecting relativistic effects, and
expanding around the deuteron pole, $|{\bf k}| = i\gamma$ one obtains:
\begin{eqnarray}
  |{\bf k}|\cot\delta  = -\gamma \ +\
  {1\over 2}\rho_d\  (|{\bf k}|^2+\gamma^2)\ +\
  w_2\ (|{\bf k}|^2+\gamma^2)^2\ +\ \ldots,
\label{eq:kcot}
\end{eqnarray}
with~\cite{deS95}:
\begin{eqnarray}
&& \gamma = 45.7025~{\rm MeV}=\left(4.318946~{\rm fm}\right)^{-1},
\nonumber\\
&& \quad\rho_d =  1.764~{\rm fm}, \quad
w_2 = 0.389~{\rm fm^3}.
\label{eq:tripparam}
\end{eqnarray}

Finally, we should mention that since the $NN$ interaction is not
spherically symmetric it mixes states of different orbital angular 
momentum. In particular, the deuteron is not
solely a bound state in the ${}^3S_1$ channel, but instead represents
a bound state in the coupled ${}^3S_1-{}^3D_1$ channels. Below we shall
show that this mixing effect is sub-leading in the low-energy $NN$ EFT,
but still numerically significant.
A more detailed discussion of this mixing is presented below.

%%%%%%%%%   Scattering  %%%%%%%%
\subsection{You Can't Choose Your Neighbors: Living near a Fixed Point} 

We now begin our attempts to build an EFT which reproduces this
low-energy $NN$ data.  The location of the deuteron pole and other
low-energy $NN$ observables are {\it input} for the EFT, and so it
might seem that such an EFT can accomplish very little. However, as
explained above, a low-energy pole in the scattering amplitude
necessarily involves short-distance physics, and so the EFT {\it must} have
this information as input. It simply cannot be expected to yield it as
output. (See Ref.~\cite{Ka97} for an early and explicit implementation of
this idea.) Putting the bound state in ``by hand'' in this way does
not vitiate the ability to make predictions with the EFT.  Once the
physics of the low-energy pole is included in the theory we can use
symmetries to predict other observables, such as the cross-section for
$np \rightarrow d\gamma$.

The extreme situation where $\gamma=0$ corresponds to a bound state at
threshold. If, in addition, the mass scale of the potential is much
greater than the incident center-of-mass momentum the theory is
approximately scale invariant. This leads to a somewhat peculiar EFT,
which exhibits a non-trivial fixed point.

Our goal is to recover such an
$NN$ scattering amplitude from a Lagrange density with
local four-nucleon momentum-independent operators alone:
\begin{eqnarray}
{\cal L}& = &  N^{\dagger}\left[ iD_0\ +\  {{\bf D} ^2\over 2 M} \right] N
\ - \ C_0 \left(N^T P N\right)^\dagger\left(N^T P N\right),
\label{eq:invar}
\end{eqnarray}
where $P$ is a spin and isospin projector.  Note that in the normal
$\chi$PT counting the contribution of the ${\bf D}^2$ term is
suppressed. However, in a non-relativistic two-body bound state its
contribution is of the same order as the contribution from the $i D_0$
term in Eq.~({\ref{eq:invar}). In other words, the kinetic energy of
the $NN$ bound state is the same order as both its potential energy
and its total energy.   Further, the size of
the interaction term can be estimated by NDA,

\begin{eqnarray}
C_0 \sim \frac{1}{M \Lambda},
\label{eq:NDA}
\end{eqnarray}
where $\Lambda$ is the ``high" scale in this theory, which 
is $\sim m_\pi$.

The Lagrange density in Eq.~(\ref{eq:invar}) is sufficiently simple
that the sum of all Feynman diagrams contributing to $NN$ scattering,
shown in Fig.~\ref{fig:bubbles}, can be found exactly.  Note that in
constructing the $NN$ amplitude in this way, contributions from an
irrelevant operator are being resummed.  
If the size of $C_0$ were in fact given by NDA, as
in Eq.~(\ref{eq:NDA}), each additional loop in the bubble sum would be
suppressed by one power of $k/\Lambda$.  Thus, without any further
work we see that the NDA scaling of Eq.~(\ref{eq:NDA}) cannot
produce a low-energy $NN$ bound state.

%%%%%%%% bubble chain figure %%%%%%%%%%%%%%%
\begin{figure}[!ht]
\hskip 1.5in\psfig{figure=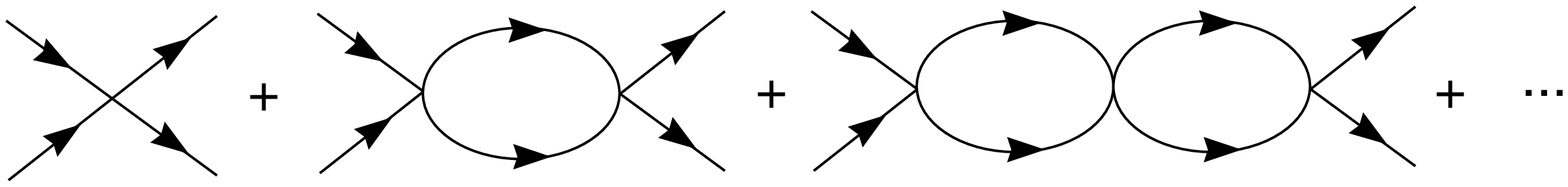,height=1.0in}
\caption{Feynman diagrams contributing to the 
scattering amplitude from the Lagrange density in
Eq.~(\protect\ref{eq:invar}). 
}
\label{fig:bubbles}
\end{figure}

If, in spite of this, we proceed to sum the bubble chain
depicted in Fig.~\ref{fig:bubbles} we obtain an amplitude:
\begin{eqnarray}
{\cal A} & = &  
-\left[\ 
C_0 +  C_0 I_0 C_0 + C_0 I_0 C_0 I_0 C_0 + \ldots\ \right]
\ =\ {-1\over {1\over C_0}-I_0}
\ \ ,
\label{eq:geom}
\end{eqnarray}
where $I_0$ is the momentum-independent single-loop integral, 
which in the $NN$ center-of-mass has the form:
\begin{eqnarray}
I_0 & = & 
-i\left({\mu\over 2}\right)^{4-n} 
\int {{\rm d}^n q\over (2\pi)^n} \ 
\left({i\over E + q_0 -{\bf q}^2/ 2M + i\epsilon}\right)
\left({i\over - q_0 -{\bf q}^2/2M + i\epsilon} \right) 
\nonumber\\
& = & \left({\mu\over 2}\right)^{4-n} 
\int {{{\rm d}}^{(n-1)}{\bf  q}\over (2\pi)^{(n-1)}}\  
\left({1\over E  -{\bf q}^2/M + i\epsilon}\right) 
\nonumber\\
&=& -M\   (-|{\bf k}|^2-i\epsilon)^{\left({n-3\over 2}\right)}\ 
\Gamma\left({3-n\over 2}\right)
\ {\left({\mu\over 2}\right)^{4-n}\over  (4\pi)^{\left({n-1\over 2}\right)}},
\label{eq:loopi}
\end{eqnarray}
with $|{\bf k}|=\sqrt{ME}$ the three-momentum of each nucleon 
in the center-of-mass frame. Note that
this $I_0$ is linearly divergent, and so we have evaluated it in 
$n$ space-time dimensions. 
In order to go further, a subtraction scheme must be specified;
different subtraction schemes amount to a reshuffling between
contributions from the vertices and contributions from the the UV part
of the loop integration. 
The usual $\overline{\rm MS}$ scheme corresponds to taking 
the $n \rightarrow 4$ limit of the expression
in Eq.~(\ref{eq:loopi}), obtaining:
\begin{eqnarray}
I_0 & \stackrel{\overline{\rm MS}}{\longrightarrow} 
& - \left({M\over 4\pi}\right) i|{\bf k}|.
\end{eqnarray}
In this case each successive term in the bubble-sum in
Eq.~(\ref{eq:geom}) involves the product $C_0 M |{\bf k}|$, and so, in
order to produce a bound state of momentum $\gamma$ it is 
necessary to have the coefficient $C_0$ scale as $C_0\sim 1/( M
\gamma)$. This unnatural scaling in $\overline{\rm MS}$, 
and the problems it leads to, 
were first discussed by Kaplan, Savage, and Wise~\cite{Ka96}.  
$C_0$ is simply unnaturally large: in fact it is ill-defined for the case of a 
threshold bound state.

There were many discussions as to how to deal with this problem in
EFT, with some advocating the use of cutoff regularization, instead of
dimensional regularization, so that the unnaturally large value of
$C_0$ would not arise~\cite{Co96,Le97,Pa97,Be97B}.  
A number of papers~\cite{AG97,Ph97,SF98A,Ph98,Gh98,Ge98B}
discussed the peculiar features which arise in this EFT when one
naively implements different regularization
schemes.  Ultimately though,
the issue here was not associated with the use of a specific
regularization scheme.  Instead, the solution to this difficulty lay
in the development of a new power
counting~\cite{Lu95,vK97,Ka98B,Ka98A,vK98}. Subsequently it was shown
how to implement this power counting on a diagram-by-diagram basis,
using a general subtraction procedure called Power Divergence
Subtraction~\cite{Ka98B,Ka98A} (PDS).

In PDS both the $n=4$ and $n=3$ poles are subtracted from each
bubble. The logarithmic divergence in $n=3$---which corresponds to the
power-law divergence in $n=4$---is then included in the expression for
the bubble.  PDS thereby keeps track of linear
divergences (see  Ref.~\cite{Ph99} for extensions of this idea).  
This allows for a fine-tuning between the
coefficient $C_0$ and the linear divergence which produces a shallow
two-body bound state. 
In PDS, the integral $I_0$ in Eq.~(\ref{eq:loopi}) is defined to be
\begin{eqnarray} 
I_0 & \stackrel{\rm PDS}{\longrightarrow} 
& - \left({M\over 4\pi}\right) (\mu + i|{\bf k}| ).
\label{eq:ipdszero} 
\end{eqnarray} 
A similar expression is obtained with a momentum cut-off~\cite{vK97}
or with a momentum-space subtraction~\cite{Ge98A,Ge98B,MS98}.
If PDS is employed the bubble sum 
(\ref{eq:geom}) yields
\begin{eqnarray}
{\cal A} = -{4\pi\over M}\  { 1 \over
{4\pi\over M C_0} + \mu + i |{\bf k}|}.
\label{eq:bubsumLO}
\end{eqnarray}
Matching to the leading-order ERT amplitude
in the ${}^3S_1$ channel, given by Eq.~(\ref{eq:Smat0}) 
with $|{\bf k}|\cot\delta  = -\gamma$, we find
\begin{eqnarray}
{4\pi\over M C_0}\ =\ 
\gamma-\mu.
\end{eqnarray}
In the limit of vanishing binding energy, it is then clear that 
\begin{eqnarray}
\mu {d\over d\mu} \left[ \mu\  C_0(\mu) \right]
& = & 0.
\label{eq:rgzero}
\end{eqnarray}
At the fixed point the theory is conformally invariant (in the
two-body sector)~\cite{Meh99}.  Therefore, in order for the
four-nucleon interaction in Eq.~(\ref{eq:invar}) to generate a system
with a shallow bound state (i.e.  $\gamma \sim 0$), $C_0$ must be near
a non-trivial fixed point, as given by Eq.~(\ref{eq:rgzero}).  The
existence of this non-trivial fixed point~\cite{Ka98A} is independent
of the particular regularization and renormalization scheme
chosen~\cite{Bi99}.

The explicit introduction of the subtraction scale $\mu$ into the
expression for the amplitude results in $C_0$ having its scale set by
$\mu$ and not by $\gamma$. Ultimately all of the couplings associated
with higher-dimensional two-nucleon operators also scale with $\mu$,
as well as with the underlying scale, $\Lambda$.  
A useful
power-counting emerges if $\mu$ is taken to be of order
the external momentum $|{\bf k}|$.  
This leads to a regulator-independent
power-counting~\cite{vK98,Ka98B,BvK97,vK97,Ka98A,Ge98A,Ge98B,MS98,Bi99},
known as Q-counting. In Q-counting $\gamma$ and the external momentum
${\bf k}$~\footnote{Consequently $\mu$ is also denoted by $Q$.}  
are both
denoted by $Q$, and Feynman diagrams are analyzed in terms of the
power of this parameter that they carry.  The result is an expansion
in powers of $Q/\Lambda$, where $\Lambda$ is the underlying scale
which sets the size of the higher-order coefficients in the
effective-range expansion, i.e. $\Lambda \sim 1/r_e \sim m_\pi$.
Q-counting is regulator-independent, but if PDS is used  
then each diagram has a definite order in the Q-expansion
and gauge invariance is explicitly preserved. These advantages
greatly simplify higher-order calculations of processes other than $NN$
scattering~\cite{Ka98B,Ch99}.

This completes the discussion of the Lagrangian in Eq.~(\ref{eq:invar})
in the $NN$ system. 
Now we
generalize this analysis to the most general S-wave $NN$
interaction that is consistent with Galilean invariance. In this case
the sum of all graphs contributing to $NN \rightarrow NN$ can be
determined exactly if PDS is used to define the divergent loops.
Suppose the tree-level amplitude for $NN$ scattering is
\begin{eqnarray}
{\cal A}_{\rm tree}  =
-\left({\mu\over 2}\right)^{4-n} 
\sum_{r=0}^\infty C_{2r}(\mu)\  |{\bf k}|^{2r}.
\label{eq:tree}
\end{eqnarray}
Here the coefficients $C_{2r}(\mu)$ are linear combinations of
couplings in the Lagrangian for operators with $2r$ gradients which
contribute to S-wave scattering.  If {\it this} vertex---instead of the
vertex $C_0$---is used for each interaction in the bubble chain of
Fig.~\ref{fig:bubbles} the resultant amplitude is
\begin{eqnarray}
{1\over {\cal A}}
\ = \ 
-{1\over \sum C_{2r}(\mu)\  |{\bf k}|^{2r}} \ +\  
i {M \over 4\pi} \left( \mu + i |{\bf k}| \right).
\label{eq:answer}
\end{eqnarray}
The $\mu$-independence of this amplitude gives RG
flow equations for each of the coefficients $C_{2r} (\mu)$.  Matching
to the ERT amplitude, Eqs.~(\ref{eq:Smat0}) and (\ref{eq:kcot}), for
$NN$ scattering, requires that for $\mu\sim\gamma\sim Q$ the couplings
$C_{2r}(\mu)$ scale as
\begin{eqnarray}
C_{2r}(\mu) \sim {4\pi \over M \Lambda^r Q^{r+1}}.
\label{eq:cscale}
\end{eqnarray}
Thus the $r$th term in the sum for ${\cal A}_{\rm tree}$, $C_{2r}
k^{2r}$, has a Q-counting order of $Q^{r-1}$.  The counting
rules for this low-energy effective theory of $NN$ scattering are then as
follows:

\begin{itemize}
\item Each nucleon propagator scales as $1/Q^2$;

\item Each loop scales as $Q^5$;

\item Each vertex $C_{2r} k^{2r}$ scales as $Q^{r-1}$.
\end{itemize}

In this way each $NN$ loop contributes one factor of $Q$, but the
vertex $C_0$ scales as $Q^{-1}$, and so, by construction, the
leading-order contribution in this EFT, is the complete bubble-chain
shown in Fig.~\ref{fig:bubbles}, with $C_0$ operators at all the
four-nucleon vertices.  The leading-order amplitude, ${\cal A}^{(-1)}$,
is then given by Eq.~(\ref{eq:bubsumLO}).  Contributions to the $NN$
amplitude which scale as higher powers of $Q$ come from perturbative
insertions of derivative interactions, $C_{2r} k^{2r}$, $r \geq 1$,
but these are always ``dressed" by the leading-order (LO) amplitude
${\cal A}^{(-1)}$, since the product of ${\cal A}^{(-1)}$ with a loop
is of order $Q^0$.

%%%%%%%%%%%%%%  Low Energy, EFT, ERT  %%%%%%

\subsection{Low-Energy Nucleon-Nucleon Interactions: 
${\rm ERT}~\rightarrow \nopi$}

The previous section contains all that is needed in order to set up
the EFT for low-energy $NN$ interactions.  
We will call this theory 
$\nopi$ since it does not contain pions as explicit degrees of
freedom.  $\nopi$ reproduces the ERT $NN$ scattering amplitude ${\cal
A}$ order-by-order in a $Q$-expansion.  If this was all that it did,
then $\nopi$ would represent no more than ERT.  However, $\nopi$'s
power becomes apparent when external probes of the $NN$ system are
considered.  One of the very interesting results obtained in $\nopi$
is that ERT generally gives the LO contribution to observables where
the $NN$ system is coupled to low-momentum external probes, such as
photons and weak gauge bosons~\cite{Ch99,Ph99}.  This justifies the
use of ERT to estimate, for instance, the magnetic moment of the
deuteron.  However, ERT computations of deuteron properties are not
always successful, since ERT only gives the LO contribution to these
observables, and subleading corrections can be numerically
significant.  Operators in $\nopi$ exist which are absent in ERT, and
these operators contribute to electroweak observables.  Including such
operators allows $\nopi$ to produce closed form, analytic expressions
for electroweak processes such as $np \rightarrow d \gamma$ and $\nu d
\rightarrow \nu d$ which are accurate at the $1\%$ level.

%%%%%%%    1S0  %%%%%%%%%%%%%
\subsection{NN Scattering in the $\si$ Channel}

Nucleon-nucleon scattering in the $\si$ channel is described by 
the Lorentz-invariant one-body Lagrange density
 \begin{eqnarray}
{\cal L}_1 & = &  N^{\dagger}\left[ iD_0\ +\  {{\bf D} ^2\over 2 M}\ -\
    {D_0^2\over 2M}\right] N,
\label{eq:lagone}
\end{eqnarray}
and the two-body Lagrange density---up to N$^2$LO:
\begin{eqnarray} 
{\cal L}_2 & = & 
- \CzeroS \left(N^T \overline{P}^a N\right)^\dagger 
\left(N^T \overline{P}^a N\right)
\nonumber\\
& - & 
\CtwoS {1\over 2} 
\left[ \left(N^T \overline{P}^a N\right)^\dagger 
\left(N^T {\cal O}_2^a N\right)
\ +\ {\rm h.c.}\right]
\nonumber\\
& - & \CfourtwoS\ 
 \left(N^T {\cal O}_2^a N\right)^\dagger \left(N^T {\cal O}_2^a N\right)
\nonumber\\
& - & 
\CfouroneS\ 
{1\over 2} 
\left[ \left(N^T  \overline{P}^a N\right)^\dagger 
\left(N^T {\cal O}_4^a N\right)
\ +\ {\rm h.c.}\right],
\label{eq:lagtwo}
\end{eqnarray}
where $ \overline{P}^a$ is the spin-isospin projector for this channel:
$\left[J=0, I=1\right]$
\begin{eqnarray}
 \overline{P}^a & \equiv &  {1\over \sqrt{8}} \sigma_2\  \tau_2\tau^a
\ \ \ , 
\ \ \  {\rm Tr} 
\left[  \overline{P}^{a\dagger}  \overline{P}^b \right]
\  =\ {1\over 2} \delta^{ab},
\label{eq:progS}
\end{eqnarray}
and where the derivative operators are
\begin{eqnarray}
{\cal O}_2^a & = & 
-{1\over 4}
\left[  \overline{P}^a \overrightarrow {\bf D}^2 
+\overleftarrow {\bf D}^2  \overline{P}^a
    - 2 \overleftarrow {\bf D}  \overline{P}^a \overrightarrow {\bf D}\right]
\nonumber\\
{\cal O}_4^a & = & 
+{1\over 16}
\left[  \overline{P}^a \overrightarrow {\bf D}^4 
- 4\overleftarrow {\bf D}  \overline{P}^a \overrightarrow {\bf D}^3
+ 6 \overleftarrow {\bf D}^2  \overline{P}^a \overrightarrow {\bf D}^2
-4\overleftarrow {\bf D}^3  \overline{P}^a \overrightarrow {\bf D}
+\overleftarrow {\bf D}^4  \overline{P}^a
\right].
\label{eq:twoder}
\end{eqnarray}
The subscript on the coefficient denotes the number of derivatives in the 
operator, and 
$\overrightarrow {\bf D}^2 = \overrightarrow {\bf D}^j 
\overrightarrow {\bf D}^j$.
We have employed the covariant derivative, 
$D_\mu = \left(\partial_0+i e Q A_0 , -{\bf \nabla} + i e Q {\bf A}\right)$,
in defining the two and four
derivative operators in Eq.~(\ref{eq:twoder}) 
where $A_\mu$ is the electromagnetic field and $Q$ is the electromagnetic
charge matrix.
Contributions to
the Lagrange density in Eq.~(\ref{eq:lagtwo}) including more than four
derivatives can be found in Refs.~\cite{Ch99B,Ru99}.  Hereafter
relativistic corrections to observables will not be included in
expressions, unless explicitly stated otherwise. It is always
straightforward to include the third term in Eq.~(\ref{eq:lagone}),
which generates these corrections, in perturbation theory~\cite{Ch99}.

At LO the scattering amplitude found from
Eqs.~(\ref{eq:lagone}) and (\ref{eq:lagtwo}) is 
that shown in Eq.~(\ref{eq:bubsumLO}). 
Matching this to the ${}^1S_0$ amplitude from
ERT, Eq.~(\ref{eq:pcot0}), gives
\begin{eqnarray}
\CzeroS (\mu) \ =\  
- {4\pi\over M }\xi^{(\si)} (\mu) 
\ \ \ \ {\rm with}\ \ \ \ 
\xi^{(\si)} (\mu) \  = \  {1\over\mu-1/a^{(\si)}}
\ \ \ .
\label{eq:xidef}
\end{eqnarray}
Note that in Q-counting $r_0 \sim 1/\Lambda$, while $1/a, {\bf k} \sim
Q$.  Using this counting and expanding the ${}^1S_0$ ERT amplitude
defined by (\ref{eq:Smat0}) and (\ref{eq:pcot0}) in powers of $Q/\Lambda$
we can match to the NLO amplitude in $\nopi$. This yields:
\begin{eqnarray}
\CtwoS (\mu)  =   
{4\pi\over M } {{r}_0\over 2}\ \left( \xi^{(\si)} (\mu)\right)^2.
\end{eqnarray}
Carrying out the same procedure at N$^2$LO, the matching gives:
\begin{eqnarray}
\CfourtwoS (\mu)\ +\ \CfouroneS (\mu) 
= - {4\pi\over M }\left(\xi^{(\si)} (\mu)  \right)^3 
\left[{1\over 4}
(r_0^{(\si)} )^2 - {r_1^{(\si)}\over\xi^{(\si)} (\mu)}
\right],
\label{eq:cvalS}
\end{eqnarray}
It is interesting to note that only the sum $\CfourtwoS\ +\
\CfouroneS$ can be constrained by $NN$ scattering.
RG scaling~\cite{Ka98A,Ka98B} of the operators in
Eq.~(\ref{eq:lagtwo}) indicates that while the contribution from
$\CfourtwoS$ is N$^2$LO, the contribution from $\CfouroneS$ is
N$^3$LO.  The time-ordered product of two $\CtwoS$ operators does not
induce the momentum structure of the $\CfouroneS$ operator.  
However, inelastic
processes can be sensitive to different combinations of $\CfourtwoS$ and
$\CfouroneS$.  

The convergence of $\nopi$ can be tested by an 
examination of the behavior of the phase shift $\delta^{(\si)}$ order by 
order in the $Q$-expansion. The results are shown in Fig.~\ref{fig:SphaseS}.
It is clear from Fig.~\ref{fig:SphaseS} that in the $\si$ channel
the expansion must be carried out to N$^4$LO or higher to describe the 
strong interactions for $|{\bf k}|\lsim 100~{\rm MeV}$.
Also, it is clear that $\nopi$ will not be valid for  
$|{\bf k}|\gsim 150~{\rm MeV}$. 
At these momenta the full  $|{\bf k}|\cot\delta$
description of the phase shift starts to deviate from the observed
phase shift. Given the location of the t-channel pion-exchange cut in
the $NN$ amplitude this deviation is  expected.

%%%%%%%%%%   1S0-phase shift   %%%%%%%%%%%%%%%
\begin{figure}[!ht]
\hskip 1.0in \psfig{figure=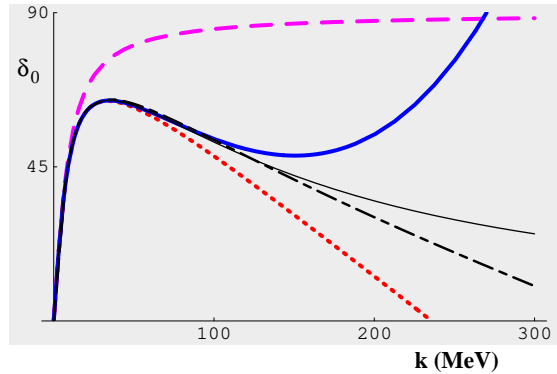,height=2.0in}
\caption{
The phase shift $\delta^{(\si)}$ as a function of the 
center of mass momentum $|{\bf k}|$. 
The dashed curve corresponds to the LO result, 
the dotted curve is the result at N$^2$LO,
the thick solid curve corresponds to N$^4$LO.
The thin solid curve corresponds to the phase shift
from ERT, and the dot-dashed curve are the phases of the Nijmegen PSA. }
\label{fig:SphaseS}
\end{figure}

%%%%%%%%%   3S1--3D1 Scattering  %%%%%%%
\subsection{NN Scattering in the Coupled $\siii-\diii$ Channel}

Scattering in the $J=1$ channel is somewhat more complicated because both
$L=0$ ($\siii$) and $L=2$ ($\diii$) states contribute.  The S-matrix
describing scattering in the coupled channel $J=1$ system is written
as
\begin{eqnarray}
  S & = & \left(
    \matrix{ e^{i2\delta_0}\cos 2\overline{\varepsilon}_1
      & i e^{i(\delta_0+\delta_2)}\sin2\overline{\varepsilon}_1
      \cr
       i e^{i(\delta_0+\delta_2)} \sin2\overline{\varepsilon}_1
      &
       e^{i2\delta_2}\cos 2\overline{\varepsilon}_1
}\right)
\ \ \  ,
\label{eq:Smat}
\end{eqnarray}
where we use the ``barred'' parameterization of Ref.~\cite{St57}, 
which was also used in Ref.~\cite{dS95}.
It will follow naturally from the $\nopi$ that $\overline{\varepsilon}_1$ 
is supressed by
$Q^2$ compared with $\delta_0^{(0)}$, 
and therefore, up to N$^4$LO, the
$\siii-\siii$ element of this matrix may be written as:
\begin{eqnarray}
  S_{00} & = &  e^{i2\delta_0}\ =\ 1\ +\  {2 i\over \cot\delta_0 - i}.
\label{eq:Smat00}
\end{eqnarray}
The phase shift $\delta_0$ has an expansion in powers of $Q$,
$\delta_0=\delta_0^{(0)}+\delta_0^{(1)}+\delta_0^{(2)}+...$, where the
superscript denotes the order in the $Q$ expansion.

Now, taking the expressions in Eqs.~(\ref{eq:Smat0}) and 
(\ref{eq:kcot}), and expanding in powers of $k/\Lambda$ and
$\gamma/\Lambda$, we obtain the LO and NLO contributions to ${\cal A}$,
which are
\begin{eqnarray}
{\cal A}^{(-1)}\ =\ 
{4\pi\over M} \ {1\over \gamma+ i |{\bf k}|}, \qquad
{\cal A}^{(0)}\ =\ 
{4\pi\over M} \ {Z_d - 1 \over \gamma+ i |{\bf k}|},
\label{eq:deutLO}
\end{eqnarray}
where $Z_d$ is the residue of the $NN$ scattering amplitude at
the bound-state pole. In this way,
the position and residue 
of the deuteron pole are correct in the NLO amplitude.
In terms of the parameters $\gamma$ and $\rho_d$,  $Z_d$ is
\begin{eqnarray}
Z_d \ =\ {1\over 1 - \gamma \rho_d}\ .
\label{eq:zed}
\end{eqnarray}
Consequently $Z_d - 1$ is of order $Q$.  $Z_d$ is also related to the
asymptotic S-state normalization of the deuteron wave function by
$A_S^2=2 \gamma Z_d$. Using the Nijmegen PSA value for $A_S$ we get
$Z_d=1.69$.  

Forming the logarithm of both sides of Eq.~(\ref{eq:Smat00}), using
Eqs.~(\ref{eq:Smat0}) and (\ref{eq:deutLO}), and expanding in powers
of $Q$, it is straightforward to obtain the LO and NLO results
\begin{eqnarray}
  \delta_0^{(0)} (|{\bf k}|) & = &
  \pi - \tan^{-1}\left({ |{\bf k}|\over \gamma}\right), \qquad
  \delta_0^{(1)}(|{\bf k}|) \ =\  -{Z_d-1\over 2}  |{\bf k}|,
  \label{eq:phaseexp}
\end{eqnarray}
which are shown in Fig.~\ref{fig:Sphase}.
%%%%%%%%%%   SD-phase shifts   %%%%%%%%%%%%%%%
\begin{figure}[!ht]
\hskip 1.0in \psfig{figure=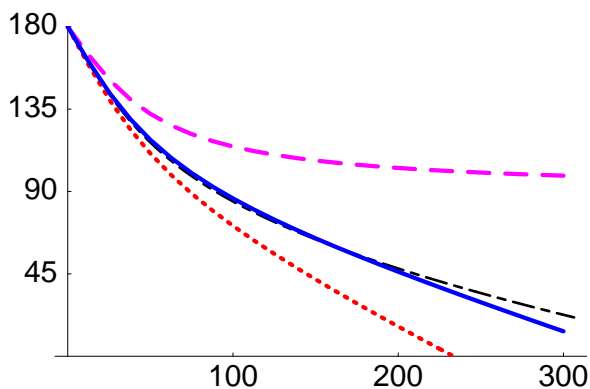,height=2.0in}
\caption{
The phase shift $\delta_0$ as a function of the 
center of mass momentum $|{\bf k}|$. 
The dashed curve corresponds to $\delta_0^{(0)}$, 
the dotted curve corresponds to $\delta_0^{(0)}+\delta_0^{(1)}$,
the solid curve corresponds to 
$\delta_0^{(0)}+\delta_0^{(1)}+\delta_0^{(2)}$,
and the dot-dashed curve is the Nijmegen PSA~\protect\cite{St94}.}
\label{fig:Sphase}
\end{figure}
The N$^2$LO piece of the phase shift, $\delta^{(2)}_0$, also
depends only on $Z_d$ and $\gamma$.
Up to N$^2$LO, the shape parameter term,
$w_2$ does not contribute to the S-wave phase shift.
In fact, $w_2$ is also
numerically small (about a factor of 5 smaller than $\rho_d$), 
but even if it were not, 
it enters the Q-expansion only at high orders.
The convergence of the Q-expansion is clearly 
demonstrated in Fig.~\ref{fig:Sphase}.

The Lagrange density describing scattering in the $\siii$ 
channel is the same 
as that in the $\si$ channel, but the $C$'s 
and the spin-isospin projectors
are different. It is thus the ${\cal L}_1$ of Eq.~(\ref{eq:lagone}) plus
\begin{eqnarray} 
{\cal L}_2 & = & 
- \CzeroT \left(N^T P^i N\right)^\dagger \left(N^T P^i N\right)
\nonumber\\
& - & 
\CtwoT {1\over 2} 
\left[ \left(N^T P^i N\right)^\dagger \left(N^T {\cal O}_2^i N\right)
\ +\ {\rm h.c.}\right]
\nonumber\\
& - & \CfourtwoT\ 
 \left(N^T {\cal O}_2^i N\right)^\dagger \left(N^T {\cal O}_2^i N\right)
\nonumber\\
& - & 
\CfouroneT\ 
{1\over 2} 
\left[ \left(N^T P^i N\right)^\dagger \left(N^T {\cal O}_4^i N\right)
\ +\ {\rm h.c.}\right],
\label{eq:lagtwoT}
\end{eqnarray}
where $P^i$ is now
the spin-isospin projector for the  
$\left[J=1, I=0\right]$ channel,
\begin{eqnarray}
P^i & \equiv &  {1\over \sqrt{8}} \sigma_2\sigma^i\  \tau_2, \qquad
{\rm Tr} \left[ P^{i\dagger} P^j \right]\  =\ {1\over 2} \delta^{ij},
\label{eq:progT}
\end{eqnarray}
and where the derivative operators are
\begin{eqnarray}
{\cal O}_2^i & = & 
-{1\over 4}
\left[ P^i \overrightarrow {\bf D}^2 +\overleftarrow {\bf D}^2 P^i
    - 2 \overleftarrow {\bf D} P^i \overrightarrow {\bf D}\right]
\nonumber\\
{\cal O}_4^i & = & 
+{1\over 16}
\left[ P^i \overrightarrow {\bf D}^4 
- 4\overleftarrow {\bf D} P^i \overrightarrow {\bf D}^3
+ 6 \overleftarrow {\bf D}^2 P^i \overrightarrow {\bf D}^2
-4\overleftarrow {\bf D}^3 P^i \overrightarrow {\bf D}
+\overleftarrow {\bf D}^4 P^i
\right].
\label{eq:twoderT}
\end{eqnarray}
Matching the amplitudes in Eq.~(\ref{eq:deutLO})
to the Q-expanded amplitude generated by the Lagrangian
in Eqs.~(\ref{eq:lagone}) and (\ref{eq:lagtwoT}), gives~\cite{Ph99},
up to NLO
\begin{eqnarray}
  \CzeroTmone & = & -{4\pi\over M}{1\over  (\mu-\gamma)}, \qquad
  \CzeroTzero \ =\  {2\pi\over M}{ \gamma (Z_d-1)\over (\mu-\gamma)^2},
\nonumber\\
  \CtwoTmtwo & = &   {2\pi\over M}{Z_d-1\over \gamma (\mu-\gamma)^2}.
\label{eq:ZCs}
\end{eqnarray}
Here, the coefficients themselves have been expanded as a power series in $Q$,
e.g. $\CzeroT = \CzeroTmone+\CzeroTzero+\CzeroTone+...$,
so that the location of the deuteron pole is unchanged 
order-by-order in the $Q$ expansion.

Turning our attention now to the off-diagonal elements of the S-matrix
in Eq.~(\ref{eq:Smat}), the $\overline{\varepsilon}_1$ parameter 
defines the amount of mixing between the $\siii$ and
$\diii$ channels.  It has a momentum expansion in terms of
the constants $E_1^{(2)}$ and $E_1^{(4)}$,
\begin{eqnarray}
\overline{\varepsilon}_1 & = &
E_1^{(2)} { |{\bf k}|^3\over\sqrt{|{\bf k}|^2+\gamma^2}}
\ +\ 
E_1^{(4)} { |{\bf k}|^5\over\sqrt{|{\bf k}|^2+\gamma^2}}
\ +\ \ldots .
\label{eq:ep1exp}
\end{eqnarray}
$E_1^{(2)}$ and $E_1^{(4)}$ are fit to the Nijmegen PSA~\cite{St94},
and are found to be $E_1^{(2)}= 0.386 \ {\rm fm^2}$ and $E_1^{(4)} \
=\ -2.800 \ {\rm fm^4}$.  
%%%%%%%%%%  ep1 shift   %%%%%%%%%%%%%%%
\begin{figure}[!ht]
\hskip 0.5in \psfig{figure=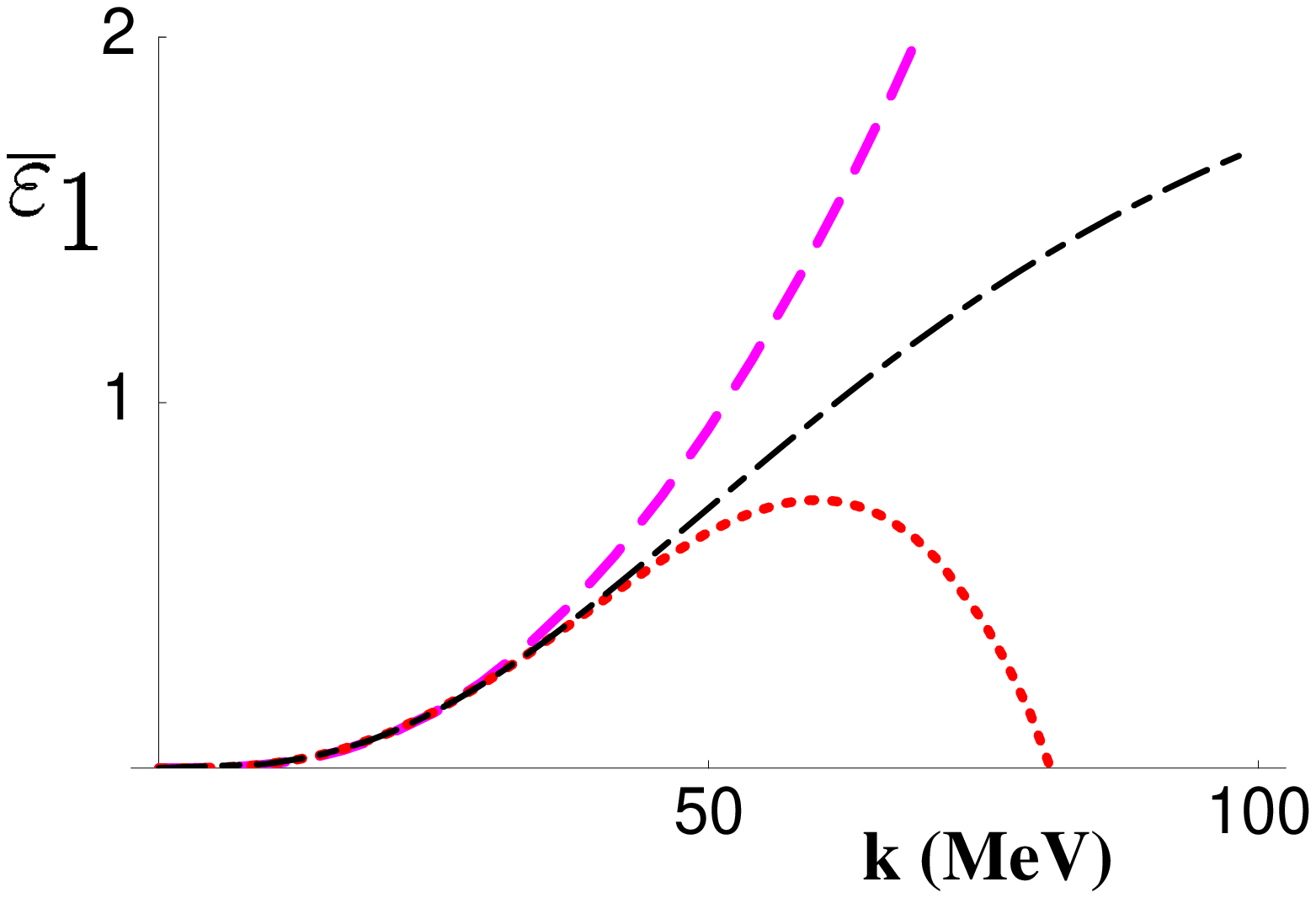,height=2.0in}
\caption{
The mixing parameter $\overline{\varepsilon}_1$ 
as a function of the 
center of mass momentum $|{\bf k}|$. 
The dashed curve corresponds to $\overline{\varepsilon}_1^{(0)}$
and the dotted curve corresponds to 
$\overline{\varepsilon}_1^{(0)}+\overline{\varepsilon}_1^{(1)}
+\overline{\varepsilon}_1^{(2)}$, 
as determined in  Eq.~(\protect\ref{eq:ep1exp}).
The dot-dashed curve is the Nijmegen PSA~\protect\cite{St94}.}
\label{fig:ep1phase}
\end{figure}
As far as the power counting is concerned it is important to note that 
the $E_1^{(Q)}$'s  are set by physics at the  
scale $\Lambda$ and so contributions of order $Q^{-1}$ must
vanish~\cite{Fl99A}. Thus the first term here is $O(Q)$, and the
second $O(Q^3)$. Corrections to Eq.~(\ref{eq:ep1exp}) are suppressed
by two further powers of $Q$. The LO and N$^2$LO contributions to
$\overline{\varepsilon}_1$ are shown in Fig.~\ref{fig:ep1phase}.

In order to calculate $\overline{\varepsilon}_1$ up to N$^3$LO we only
require one insertion of ${\cal A}^{\rm tree}_{sd}$, dressed with the
leading-order amplitude for $\siii-\siii$ scattering
(\ref{eq:deutLO}).  The Lagrange density describing the first two
orders of S-D mixing interactions is~\cite{Ch99B}
\begin{eqnarray}
&&  {\cal L}^{(sd)} = -{\cal T}^{(sd)}_{ij,xy}
\left[
\CSDzero \left( N^T P^i N \right)^\dagger
  \left( N^T {\cal O}^{xy,j}_2 N\right) + 
\right.\nonumber\\
& & \left.
\CSDtwotwo
\left( N^T  {\cal O}^{ll,i}_2  N \right)^\dagger
  \left( N^T {\cal O}^{xy,j}_2 N\right)
+  \CSDtwoone
\left( N^T P^i  N \right)^\dagger
  \left( N^T  {\cal O}^{mm, xy,j}_4 N\right)
\right] \nonumber\\
&& \quad + \ {\rm h.c.}, 
\label{eq:sdlag}
\end{eqnarray}
where
\begin{eqnarray}
{\cal T}^{(sd)}_{ij,xy}
& = &  
\delta_{ix}\delta_{jy} - {1\over n-1} \delta_{ij}\delta_{xy},
\nonumber\\
{\cal O}^{xy,j}_2 & = &
-{1\over 4}\left(
      \overleftarrow {\bf D}^x  \overleftarrow {\bf D}^y P^j
+ P^j \overrightarrow {\bf D}^x  \overrightarrow {\bf D}^y
- \overleftarrow {\bf D}^x P^j\overrightarrow {\bf D}^y
-\overleftarrow {\bf D}^y P^j\overrightarrow {\bf D}^x
\right).
\label{eq:sdop}
\end{eqnarray}
The tree-level amplitude 
for an $\siii\rightarrow\diii$ transition
resulting from this Lagrange density is 
\begin{eqnarray}
{\cal A}^{\rm tree}_{sd} & = & 
-\left( \CSDzero + \left[\CSDtwotwo+\CSDtwoone\right] p^2
+ \ldots \right)\left[p^i p^j-{1\over n-1} p^2\delta^{ij}\right].
\nonumber\\
\label{eq:SDtree}
\end{eqnarray}
Like the S-wave coefficients,
the coefficients that appear in 
Eq.~(\ref{eq:sdlag})
have their own expansion in powers of $Q$,
e.g. $\CSDzero = \CSDzeromone + \CSDzerozero + ...$.

It is convenient to introduce the quantity $\etasd$.
This is the asymptotic D/S ratio of the deuteron,
and may be extracted from the $\siii-\diii$ scattering information
by analytic continuation of the amplitude to the deuteron
pole. In fact,
\begin{eqnarray}
-2 \left({\etasd\over 1-\etasd^2}\right)
\ =\ 
\left. { \tan\left(2 \overline{\varepsilon}_1\right)
\over
\sin\left(\delta_0-\delta_2\right)} 
\right|_{|{\bf k}|=i\gamma}.
\label{eq:asymDS}
\end{eqnarray}
The value 
is then found from
a continuation of the Nijmegen PSA into the unphysical region:
$\etasd=0.02543\pm 0.00007$~\cite{St94}.

Matching the $\siii-\diii$ scattering amplitude 
to the Q-expansion of the off-diagonal elements of the S-matrix 
in Eq.~(\ref{eq:Smat})
evaluated at the deuteron pole, gives the coefficients in
Eq.~(\ref{eq:sdlag}) in terms of  $\{\etasd , E_1^{(4)} \}$, 
\begin{eqnarray}
& & 
\CSDzeromone\ =\  -\etasd\ {6\sqrt{2}\pi \over M \gamma^2 (\mu-\gamma)}
\ ,\ 
\CSDzerozero\ =\ 
\etasd\ (Z_d-1)\ {3\sqrt{2}\pi\over M\gamma(\mu-\gamma)^2}
\nonumber\\
& & \CSDtwotwotwo+\CSDtwoonetwo\ =\ 
\etasd\ (Z_d-1) {3\sqrt{2}\pi\over M\gamma^3 (\mu-\gamma)^2}
\ \ \ .
\label{eq:sdcoeffs}
\end{eqnarray}
In this way one ensures that that {\it all}
of the asymptotic physics of the deuteron state: its binding energy,
and its asymptotic S and D-state normalizations, are reproduced
in $\nopi$ at low orders in the $Q$-expansion.

Finally, the four-nucleon operator that first contributes
to scattering in the $\diii$ channel involves four derivatives. It is
also renormalized by the operators contributing to
$\overline{\varepsilon}_1$. Analysis of the powers of $Q$ which appear
in the $\diii$ phase shift leads us to conclude that it contributes to
the observables we examine in this chapter at a much higher order than
we will be concerned with, and therefore we do not consider it
further.

%%%%%%%  The Deuteron Field %%%%%%%%%%
\subsection{Processes Involving the Deuteron: Formalism}

When considering elastic or inelastic processes involving the deuteron
it is convenient to define an interpolating field for the deuteron
${\cal D}_i (x) = N^T P_i N (x)$, where $P_i$ is the projector defined
in Eq.~(\ref{eq:progT}). The Green's function associated with this
interpolating field is then just that for the ${}^3S_1$
channel:
\begin{eqnarray}
 \int \ d{\bf x}\ e^{-i(Et-{\bf k}\cdot {\bf x})}\, \langle 0|
 {\rm T}\left[{\cal D}_i^{\dagger}(x){\cal D}_j (0)\right]
 |0\rangle 
\equiv G(\overline{E}) \delta_{ij},
\label{eq:fullprop2}
\end{eqnarray}
where, by Lorentz invariance, the propagator only depends on the
energy in the center of mass frame, namely
$\overline E \equiv E - {{\bf k}^2\over 4M} +  \ldots$
with the ellipses referring to relativistic corrections.

The Green's function $G$ is straightforwardly related to the amplitude
${\cal A}$ in the ${}^3S_1$ channel. However, some relationships
between observables are clearer when expressed in terms of this
deuteron-deuteron Green's function. Since there is a
pole in this channel we may always write $G(\overline{E})$ as:
\begin{eqnarray}
G(\overline{E})=i {{\cal Z}(\overline{E}) \over \overline{E} + B},
\label{eq:Gdeut}
\end{eqnarray}
where $B$ is the deuteron binding energy and 
the numerator ${\cal Z}$ is
a function which is smooth near
the deuteron pole. When evaluated for $E \approx B$, ${\cal Z}$ gives
the wavefunction renormalization $Z_\psi$. This is related to the
$Z_d$ defined in Eq.~(\ref{eq:zed})
by simple factors:
\begin{eqnarray}
{\cal Z}(-B) & \equiv &  Z_\psi = -i \left[ {{\rm d}G^{-1}(\overline E)\over
{\rm d} E}\right]_{\overline E=-B}^{-1}
\ =\ -{8\pi\gamma\over M^2} Z_d.
\label{eq:Zpsi}
\end{eqnarray}

It is now useful to define ``$C_0$-irreducible'' Green's functions,
since observables can be computed solely in terms of these
quantities. 
A $C_0$-irreducible Green's function is the sum of all
possible diagrams which do not involve the leading-order interaction
$\CzeroTmone$ and contribute to a given n-point function.
The irreducible deuteron-deuteron function is denoted by
$\Sigma$.  It has the expansion shown in Fig.~\ref{fig:sigma}.

%%%%%%%%%%  Green  function Sigma  %%%%%%%%%%%%%%%
\begin{figure}[!ht]
\hskip 0.5in \psfig{figure=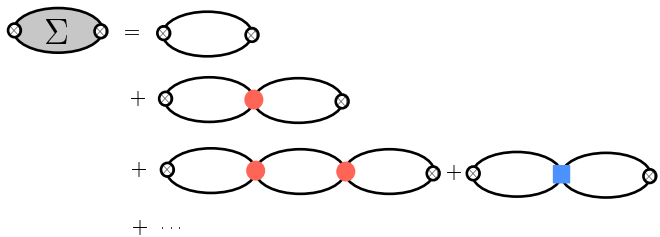,height=1.25in}
\caption{
The perturbative expansion of $\Sigma$. 
The first row has the
LO result, where $\otimes$ 
represents an insertion of the deuteron interpolating
field.  
The second row is the NLO  contribution,
arising from a single insertion of the 
$\CzeroTzero$ and $\CtwoTmtwo$ operators.
The third row is the N$^2$LO  contribution,
arising from two insertions of the
$\CzeroTzero$ and $\CtwoTmtwo$ operators,
and a single insertion of the 
$\CzeroTone$, $\CtwoTmone$, $\CfouroneTmthree$ and 
$\CfourtwoTmthree$.
}
\label{fig:sigma}
\end{figure}

A simple consequence of this definition is that the relationship
between the full $G$ and $\Sigma$ is:
\begin{eqnarray}
G & = & {\Sigma\over 1+i \CzeroTmone\Sigma}.
\label{eq:gsig}
\end{eqnarray}
This can also be seen graphically in Fig.~\ref{fig:gfig}.

%%%%%%%%%%  Green  function G   %%%%%%%%%%%%%%%
\begin{figure}[!ht]
\hskip 0.5in \psfig{figure=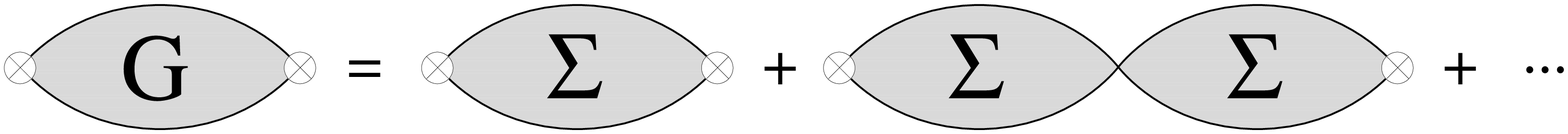,height=0.30in}
\caption{
The expansion of of the full 2-point function $G$ in terms of the
irreducible 2-point function $\Sigma$.}
\label{fig:gfig}
\end{figure}

Recall that we wish to choose the coefficients $\CzeroTmone$, 
$\CzeroTzero$ and $\CtwoTmtwo$ so as to 
reproduce the Green's function (\ref{eq:Gdeut}),
with the function $\cal Z$ having the precise value (\ref{eq:Zpsi}) at the 
deuteron pole. This imposes the following conditions on $\Sigma$:
\begin{eqnarray}
\Sigma (\overline{E}=-B)
& = & {i\over \CzeroTmone}, \qquad
\left[
{1\over \Sigma^2} {{\rm d}\Sigma\over {\rm d} E}
\right]
(\overline{E}=-B) = {i\over  Z_\psi}.
\label{eq:zb}
\end{eqnarray}
These conditions hold at each order of the Q-expansion, and so
restrict the number of independent higher-order couplings
which appear in the deuteron channel.

We are now in position to compute the matrix element of an electroweak
current between two deuteron states. We first define the three-point function
\begin{equation}
G^\mu_{ij}(\overline E,\overline E', {\bf q})
=
\int d{\bf x} d{\bf y}
e^{-i(Ex^0-{\bf k}\cdot {\bf x})}
e^{i(E'y^0-{\bf k}^\prime\cdot {\bf y})}
\ \langle 0|
{\rm T}
\left[ 
{\cal D}_i^{\dagger}(x) J_{ew}^\mu (0) {\cal D}_j(y)
\right]
|0\rangle
\ ,
\end{equation}
where $q^\mu = (E'-E,{\bf k}^\prime-{\bf k})$ 
is the momentum transferred to the deuteron system. 
$G^\mu$ is related to the desired form factor via the LSZ formula
\begin{eqnarray}
\langle {\bf k}^\prime , j|
J^\mu_{ew} 
|{\bf k}, i\rangle
& = & 
Z_\psi
\left[
G^{-1}(\overline E)G^{-1}(\overline E')  
G^\mu_{ij}(\overline E,\overline E', {\bf q})
\right]_{\overline E,\ \overline E'\to -B},
\label{eq:jme}
\end{eqnarray}
where $G(\overline E)$ is defined in Eq.~(\ref{eq:fullprop2}).
As promised, this formula may now be reexpressed in terms of $\Sigma$
and the ``$C_0$-irreducible" 3-point function, $\Gamma^\mu$. Again, it is
then a simple matter of definition that:
\begin{eqnarray}
G^\mu_{ij} (\overline E,\overline E', {\bf q}) 
& = &  {\Gamma^\mu_{ij} 
(\overline E, \overline E', {\bf q}) G(\overline E)
G(\overline E')
\over \Sigma(\overline E)\Sigma(\overline E') }.
\label{eq:Gmureln}
\end{eqnarray}
This relationship is depicted in Fig.~\ref{fig:gmu}.
%%%%%%%%%%  Green  function Gmu   %%%%%%%%%%%%%%%
\begin{figure}[!ht]
\hskip 0.5in \psfig{figure=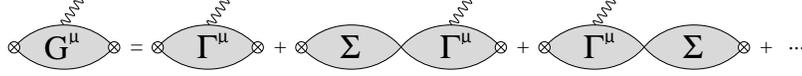,height=0.45in}
\caption{
The expansion of full three-point function $G^\mu$ in terms of the
irreducible two- and three--point functions $\Sigma$, $\Gamma^\mu$.
}
\label{fig:gmu}
\end{figure}
Making use of 
Eqs.~(\ref{eq:Gmureln}), (\ref{eq:gsig}), (\ref{eq:zb}), and (\ref{eq:jme}),
S-matrix elements can then be found in terms of the $C_0$-irreducible vertices
of the theory 
\begin{eqnarray}
\langle{\bf k}^\prime, j | J^\mu_{em} | {\bf k},i\rangle
& = & 
i \left[  {\Gamma^\mu_{ij}(\overline E,\overline E', {\bf q})
\over
{\rm d}\Sigma(\overline E)/{\rm d}E}
\right]_{\overline E,\  \overline E'\to -B}.
\label{eq:current2}
\end{eqnarray}
Since $\Sigma$ and $\Gamma$ both have a simple perturbative expansion
in terms of Feynman graphs the right-hand side of this equation can be
straightforwardly calculated to an arbitrary order in the Q-expansion,
thereby providing a result for the current matrix element appearing on
the left-hand side.

%%%%%%%%%   Phenomenology   %%%%%%%%%%%%%%%%
\section{Phenomenological Applications of $\nopi$}

%%%%%%%%%   EM Form Factors %%%%%%%
\subsection{Electron-Deuteron Scattering}

A deuteron with four-momentum $p^\mu$ and polarization vector
$\epsilon^\mu$ is
described by the state $\ket{\bfp ,\epsilon}$, where the polarization vector
satisfies $p_\mu\epsilon^\mu=0$.  
It is convenient to choose the basis polarization vectors so that in
the deuteron rest frame $\epsilon_i^\mu=\delta^{\mu}_i$.
We then write $\ket{\bfp,i} \equiv \ket{\bfp,\epsilon^\mu_i}$, and these
states have the normalization condition
$\langle\bfpp,j\vert\bfp,i\rangle =(2\pi)^3\delta^3(\bfp-\bfpp)
\delta_{ij}$.
In terms of these states $|\bfp, i \rangle$ the nonrelativistic expansion of 
the matrix element of the electromagnetic current is 
\begin{eqnarray}
\bra{\bfpp,j} J^0_{em}\ket{\bfp,i}\! &=&\! 
e \[  F_C(q^2) \delta_{ij} + {1\over 2
M_d^2}F_{\cal Q}
(q^2)\(\bfq_i\bfq_j-{1\over 3}\bfq^2 \delta_{ij}\)\]
\left({E+E^\prime\over 2 M_d}\right)
\nonumber\\
\bra{\bfpp,j} {\bf J}^k_{em}\ket{\bfp,i}\! &=& \! {e\over 2 M_d} \[ F_C(q^2)
\delta_{ij}(\bfp+\bfpp)^k + F_M(q^2)\(\delta_j^k\bfq_i - \delta_i^k\bfq_j\)
\right.\nonumber\\
&&\qquad\quad \left.+{1\over 2M_d^2} F_{\cal Q}(q^2) \(\bfq_i\bfq_j
-{1\over 3}\bfq^2\delta_{ij}\)(\bfp+\bfpp)^k\] .
\label{eq:emmatdef}
\end{eqnarray}
Relativistic corrections to these formulae enter at N$^2$LO.
In Eq.~(\ref{eq:emmatdef}) $q^2=q_0^2-|{\bf q}|^2$,
is the square of the four-momentum transfer. 
Momentum conservation implies
${\bf q}={\bf p}^{\prime}-{\bf p}$.
The dimensionless form factors defined in Eq.~(\ref{eq:emmatdef})
are normalized such that
\begin{eqnarray}
F_C(0) &=& 1, \quad
{e\over 2 M_d}F_M(0) \ =\  \mu_{M}, \quad
{1 \over M_d^2} F_{\cal Q}(0) \ =\  \mu_{\cal Q},
\label{eq:normalization}
\end{eqnarray}
where $\mu_M= 0.85741 \ {e\over 2M}$ is the
deuteron magnetic moment,
and $\mu_{\cal Q}=0.2859\,{\rm fm}^2$ is the
deuteron quadrupole moment~\cite{BC79,ERC83}.

As an example, the  calculation of the 
electric-quadrupole form-factor of the deuteron
will be outlined. 
In $\nopi$ the physics of this form factor at LO is exactly
the same as in ERT. The photon couples to the D-state component of the
deuteron wave function, which in $\nopi$ appears through a
perturbative insertion of $\CSDzeromone$.  
Some of the NLO corrections, 
those that  serve to give the deuteron its correct
asymptotic S-state normalization,
could  be computed in
ERT. 
However, there is also a contribution
beyond ERT, depicted in diagram (d) of Fig.~\ref{fig:dquad}, where a
quadrupole photon mediates an S-to-S transition in the deuteron.
%%%%%%%%%% Quadrupole Diagrams  %%%%%%%%%%%%%%%
\begin{figure}[!ht]
\hskip 0.75in \psfig{figure=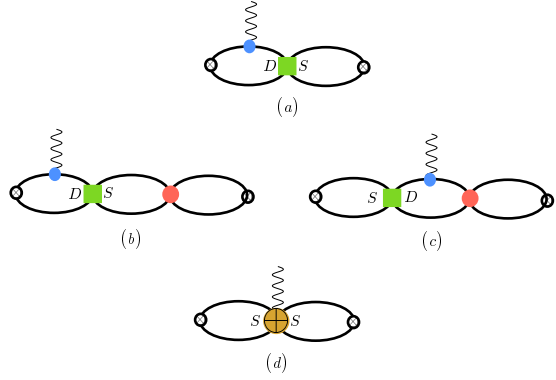,height=2.0in}
\caption{
Some LO and NLO contributions to the deuteron quadrupole form factor.
Diagrams of the form of (a) contribute at LO and higher.
Diagrams of the form of (b) and (c) contribute at NLO and higher.
At NLO there is a contribution from a 
local counterterm, diagram~(d). 
}
\label{fig:dquad}
\end{figure}

At LO, the contribution from diagram (a) in Fig.~\ref{fig:dquad} 
is, after performing the relative-energy-integrations,
\begin{eqnarray}
\Gamma^{(a)}_{\rm LO} & = & 
M^3 \CSDzeromone
\int {d^{n-1}{\bf k}\over (2\pi)^3} {d^{n-1}{\bf l}\over (2\pi)^3}\ 
{({\bf k} + {{\bf q}\over 2})^a ({\bf k} + {{\bf q}\over 2})^b
- {\delta^{ab}\over n-1} |{\bf k} + {{\bf q}\over 2}|^2
\over
\left[ |{\bf l} + {{\bf q}\over 2}|^2+\gamma^2\right]
\left[ |{\bf k} + {{\bf q}\over 2}|^2+\gamma^2\right]
\left[ |{\bf k}|^2+\gamma^2\right]
}
\nonumber\\
& = & 
{1\over |{\bf q}|^2}
\left[ {\bf q}^a  {\bf q}^b - {\delta^{ab}\over n-1} |{\bf q}|^2\right]
M^3 \CSDzeromone
\left({n-1\over n-2}\right) 
\left({\mu-\gamma\over 4\pi}\right)
\nonumber\\
& & 
\int {d^{n-1}{\bf k}\over (2\pi)^3}
\left[
{ {1\over 2}-{1\over n-1}\over  |{\bf k}|^2+\gamma^2}
+
{ {\gamma^2\over n-1} + {|{\bf q}|^2\over 16}
\over \left[ |{\bf k} + {{\bf q}\over 2}|^2+\gamma^2\right]
\left[ |{\bf k}|^2+\gamma^2\right]
}\right]
\nonumber\\
& = & 
{1\over |{\bf q}|^2}
\left[ {\bf q}^a  {\bf q}^b - {\delta^{ab}\over n-1} |{\bf q}|^2\right]
M^3\  \CSDzeromone
\left({\mu-\gamma\over 4\pi}\right)
\nonumber\\
& & 
\quad {3\over 4\pi}
\left[ -{\gamma\over 12}
+\left({\gamma^2\over 3}+{|{\bf q}|^2\over 16}\right){1\over|{\bf q}|}
\tan^{-1}\left({|{\bf q}|\over 4\gamma}\right)\right]
\ \ \ ,
\label{eq:quadcalc}
\end{eqnarray}
where ${\bf q}$ is the three-momentum transfer to the deuteron,
and $n$ is the number of space-time dimensions.
The integrals appearing in Eq.~(\ref{eq:quadcalc}) have been regulated with
PDS.
There are no $n-3$ poles in 
the ${\bf k}$ integration in Eq.~(\ref{eq:quadcalc}) because
the integral is convergent. 
Including the hermitean conjugate of diagram (a) and
wavefunction renormalization
the LO quadrupole form factor is
\begin{eqnarray}
{1\over M_d^2} F_{\cal Q} (|{\bf k}|)
& = & 
{3\etasd Z_d \over 2\sqrt{2} |{\bf k}|^3}
\ 
\left[
- 4\gamma |{\bf k}| + \left( 16\gamma^2 + 3 |{\bf k}|^2\right)
\tan^{-1}\left({|{\bf k}|\over 4\gamma}\right)
\right]
\  .
\label{eq:quadLO}
\end{eqnarray}

At NLO, there are contributions from all of the diagrams in
Fig.~\ref{fig:dquad}.  They are straightfoward to compute, and we will
not present the calculation here.  The contribution from a
four-nucleon-one-quadrupole-photon operator~\cite{Ch99}, diagram (d) in
Fig.~\ref{fig:dquad}, has a coefficient $\CQuad$ defined by the
Lagrange density
\begin{eqnarray}
{\cal L} &=& { e\ \CQuad\over 2} \ \left( N^{T}P_{i}N\right) ^{\dagger }
\left( N^{T}P_{j}N\right)
\left[ \ \nabla^i{\bf E}^j\ +\  \nabla^j{\bf E}^i \ - \
  {2\over n-1}\delta^{ij} \ \nabla\cdot {\bf E}
\ \right]
\nonumber\\
& = &  -e\ \CQuad \ \left( N^{T}P_{i}N\right) ^{\dagger }
\left( N^{T}P_{j}N\right)
\left( \nabla ^{i}\nabla ^{j}-\frac{1}{n-1}\nabla ^{2}\delta ^{ij}\right)
A^{0}
\ +\ ....,
\label{eq:counterQ}
\end{eqnarray}
where ${\bf E}$ is the electric field operator.

At N$^2$LO the only correction comes from a modification to the
one-body charge operator. This modification introduces the finite
charge-radius of the nucleon into the calculation. Thus, the deuteron
quadrupole form factor in $\nopi$ is, up to N$^2$LO:
\begin{eqnarray}
&{1\over M_d^2} & F_{\cal Q} (|{\bf q}|)  \ =\  
\delta\mu_{\cal Q}
\nonumber\\
& + &  
{3 Z_d\ \etasd \over 2\sqrt{2} \gamma |{\bf q}|^3}
\left[ \ 
-4 |{\bf q}| \left( \gamma 
+{1\over 6} |{\bf q}|^2 {Z_d-1\over\gamma Z_d}\right)
+ \left(3 |{\bf q}|^2+16\gamma^2\right)
\tan^{-1}\left({|{\bf q}|\over 4\gamma}\right)
\ \right]
\nonumber\\
& + &  
{\etasd \ \langle r_{N,0}^2\rangle\over 4\sqrt{2}\gamma |{\bf q}|}
\left[ 4 \gamma |{\bf q}| + 
\left(3 |{\bf q}|^2+16\gamma^2\right)
\tan^{-1}\left({|{\bf q}|\over 4\gamma}\right)
\right],
\label{eq:QFF}
\end{eqnarray}
where $r_{N,0}$ is the isoscalar nucleon charge radius.
Note that we have eliminated the bare coupling $\CQuad$ in favor
of the RG-invariant quantity $\delta \mu_{\cal Q}$.
The value of $\delta \mu_{\cal Q}$ is chosen so as 
to reproduce the observed deuteron quadrupole
moment, 
$\mu_Q = F_{\cal Q}(0)/M_d^2 = \delta\mu_{\cal Q} 
+ {\etasd \over \sqrt{2}\gamma^2}$. 
The value that does this is
$\delta\mu_{\cal Q}=-0.0492~{\rm fm}^2$.
%%%%%%%%  Quad FF  %%%%%%%%%%%%%
\begin{figure}[!ht]
\hskip 0.5in \psfig{figure=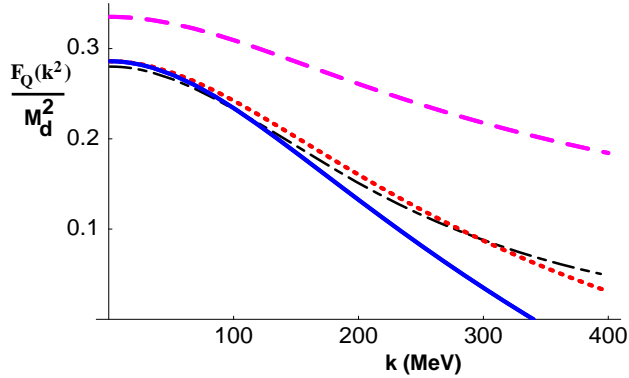,height=2.0in}
\caption{
The deuteron quadrupole form factor.
The dashed, dotted and solid curves correspond to the 
LO, NLO and N$^2$LO predictions.
The dot-dashed curve corresponds to a 
calculation with the Bonn-B potential
in the formulation of  \protect\cite{Ad93}.
}
\label{fig:QFF}
\end{figure}

Once the quadrupole moment is fixed we have a prediction for the $q^2$ 
dependence of $F_Q$. The form factor computed at LO, NLO and N$^2$LO is shown 
in Fig.~\ref{fig:QFF},
along with a potential-model calculation performed with the Bonn-B potential
in the formulation of Ref.~\cite{Ad93}.
The second and third terms in Eq.~(\ref{eq:QFF}) would be reproduced in 
ERT, as long as $\siii-\diii$ mixing and the contribution
from the finite size of the nucleon were included in ERT.
The first term in Eq.~(\ref{eq:QFF}) is a contribution beyond ERT, and
corresponds to an insertion of the lowest deuteron quadrupole counterterm
not constrained by $NN$ scattering alone. 
The presence of 
this counterterm at NLO indicates that $\mu_{\cal Q}$ is sensitive
to short-distance physics---a fact confirmed by results gleaned
from potential-model calculations, and calculations in EFTs with pions
(see below). The absence of the appropriate short-distance physics
in potential model calculations is presumably the reason that the best
of these calculations underpredict $\mu_{\cal Q}$ by about 
5 \%~\cite{CS98,Wi95}.

In contrast, we can define the deuteron's quadrupole ``radius", via:
\begin{equation}
\langle r_Q^2 \rangle = -{1 \over 6 M_d^2} {d \over d|{\bf q}|^2} 
F_Q(|{\bf q}|).
\end{equation}
At N$^2$LO this quantity is given by:
\begin{eqnarray}
\langle r_Q^2 \rangle & = & 
{9\  \etasd\over 80 \sqrt{2} \gamma^4}
\left( Z_d + {80\over 9} \langle r_{N,0}^2\rangle \gamma^2\right)
\ \ =\ \ 1.398~{\rm fm}^4,
\label{eq:rQ}
\end{eqnarray}
independent of the quadrupole counterterm.
$\langle r_Q^2 \rangle$ is, not surprisingly, less sensitive to
short-distance physics than $\mu_{\cal Q}$. 
Therefore, we expect that potential-model calculations of this quantity
will agree substantially better with observation and with each other, 
than potential-model calculations of the quadrupole moment will.

Similar calculations can be performed for the charge and the magnetic form
factors.
At LO, NLO and N$^2$LO the contributions to the
charge form factor are
\begin{eqnarray}
  F_C^{(0)} ( |{\bf q}|) & = & {4\gamma\over |{\bf q}|} \tan^{-1}\left(
    { |{\bf q}|\over 4\gamma}\right)
\ \ ,\ \ 
F_C^{(1)} ( |{\bf q}|)\ =\ 
(Z_d-1) \left( F_C^{(0)} ( |{\bf q}|)\ -\ 1\right)
\nonumber\\
F_C^{(2)} ( |{\bf q}|) & = & 
-{1\over 6} \langle r_{N,0}^2\rangle \  |{\bf q}|^2\ 
F_C^{(0)} ( |{\bf q}|)
\ \ \ .
\label{eq:EFTcffz}
\end{eqnarray}
The sum of the LO, NLO and N$^2$LO contributions 
is exactly what one obtains in ERT.
At the next order, N$^3$LO, there is a contribution 
from a four-nucleon-one-photon interaction that represents
physics beyond ERT.  
The deuteron charge radius that arises from the sum of the form factors 
in Eq.~(\ref{eq:EFTcffz}) is,
including the leading relativistic effect~\cite{Ch99},
\begin{eqnarray}
 \langle r_d^2\rangle^{\rm EFT} & = &\langle r_{N,0}^2\rangle
  \ +\ 
 {Z_d\over 8\gamma^2}\ +\ {1\over 32 M^2}
\ =\ 4.565~{\rm fm}^2,
\label{eq:EFTcr}
\end{eqnarray}
which is to be compared with the experimental value 
$ \langle r_d^2\rangle = 4.538~{\rm fm}^2$~\cite{Wo94,Bu96,Fr97,EW88}.
The $\nopi$ calculation agrees with data (and
with potential-model calculations too) to within $\sim {1\over 2}\%$.
%%%%%%%%%% Charge FF   %%%%%%%%%%%%%%%
\begin{figure}[!ht]
\hskip 0.75in \psfig{figure=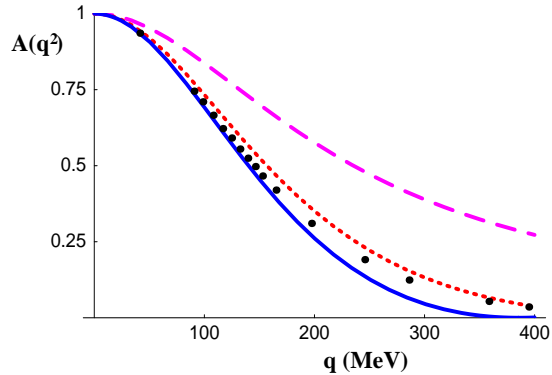,height=2.0in}
\caption{
The form factor $A(q^2)$ 
(dominated by the square of the charge form factor) 
measured in electron-deuteron scattering.
The dashed, dotted and solid curves correspond to the 
LO, NLO and N$^2$LO form factors.
The solid circles correspond to data.}
\label{fig:CFF}
\end{figure}

The magnetic form factor  of the deuteron 
is dominated by contributions from the 
single nucleon magnetic moments.
The Lagrange density describing the magnetic interactions of the nucleons is
\begin{eqnarray}
{\cal L}_{1,B} & = &
{e\over 2 M} N^\dagger
\left( \kappa_0 + \kappa_1 \tau_3 \right) {\bf \sigma} \cdot {\bf B} 
\ N
\ \ \ \ ,
\label{eq:Nucmag}
\end{eqnarray}
where
$\kappa_0 = {1\over 2} (\kappa_p + \kappa_n)$ and
$\kappa_1 = {1\over 2} (\kappa_p - \kappa_n)$
are the isoscalar and isovector nucleon magnetic moments in nuclear magnetons, 
with $\kappa_p=  2.79285~{\rm NM}$
and  $\kappa_n= - 1.91304~{\rm NM}$.
The magnetic field  is conventionally defined 
${\bf B} ={\bf \nabla} \times {\bf A}$.
The magnetic form factor of the deuteron at LO and NLO is 
\begin{eqnarray}
F_M^{(0)} (|{\bf q}|^2) & = & 
2\kappa_0
{4\gamma\over |{\bf q}|} \tan^{-1} \left({ |{\bf q}|\over 4\gamma}\right),
\nonumber\\
F_M^{(1)} (|{\bf q}|^2) & = & 
(Z_d-1) F_M^{(0)} (|{\bf q}|^2)
\ -\ 2 \kappa_0 (Z_d-1)
\ +\ 
L_2\ {\gamma (\mu-\gamma)\over \pi}
\ \ .
\nonumber\\
\label{eq:magFF}
\end{eqnarray}
Note that at LO the result is just the charge form factor multiplied
by the isoscalar anomalous magnetic moment. The first piece of
the NLO result has a similar interpretation, but another piece
also appears at this order. This second contribution represents the 
effect of a two-body
current on $F_M$.
The coefficient, $L_2$, which appears there is
the coefficient of a four-nucleon-one-magnetic-photon
term in the Lagrange density~\cite{Ch99,Ka98C}
\begin{eqnarray}
{\cal L} &=& -e\ \Ltwo\ i\epsilon _{ijk}\left(
N^{T}P_{i}N\right) ^{\dagger }\left( N^{T}P_{j}N\right) {\bf B}_{k}
\  +\  {\rm h.c.}
\label{eq:counter2}
\end{eqnarray}
The value of this coefficient can be found by demanding that the
the deuteron magnetic moment is reproduced at NLO. Since the NLO
result for $\mu_M$ in the Q-expansion is~\cite{Ka98C}
\begin{eqnarray}
  \mu_M & = & {e\over 2 M} \left(\kappa_p\ +\ \kappa_n
  \ +\ \Ltwo\ { 2 M \gamma
    (\mu-\gamma)^2\over\pi}\right), 
\label{eq:deutmag}
\end{eqnarray}
we require that if $L_2$ is evaluated at the scale $\mu=m_\pi$ then
it must have the value~\cite{Ch99,Ka98C} $\Ltwo(m_{\pi })=-0.149 {\rm
fm}^{4}$. This is significantly smaller than the naively estimated
size of $\sim 1$ fm$^{4}$.

%%%%%%%%  Deut Pols %%%%%%%%%%%%%
\subsection{Deuteron Polarizabilities}

The electromagnetic polarizabilities 
of a hadron
measure its deformation
in external electric and magnetic fields 
(for an overview see \cite{Dr97}).
The electric polarizability of the deuteron, $\alpha_{E0}$,
has been investigated thoroughly
with $NN$ potential models~\cite{FP97,LL95,Wi95B,K97}.
To a very high precision one obtains
$\alpha_{E0}=0.6328\pm 0.0017~{\rm fm}^3$
with those techniques.
This is not surprising since the electric polarizability is dominated by the
long-range behavior of the deuteron wavefunction.
Thus, if a model is tuned to reproduce this asymptotic part of the 
deuteron wavefunction then the predicted electric polarizability
should be very close to nature.

Similarly in $\nopi$ $\alpha_{E0}$ can be calculated to high precision.
The contributions from the 
first four orders in perturbation theory give an
electric polarizability of the deuteron which is (including a relativistic
correction)
\begin{eqnarray}
\alpha_{E0}^{(z)}
     & = & 
{\alpha M\over 32\gamma^4}
    \left[ 1 \ \ +\ \  (Z_d-1)
      \ \ +\ \ {2\gamma^2\over 3 M^2}
       \ \ +\ \ {M\gamma^3\over 3\pi} D_P
      \right]
\label{eq:eftpolz}
\end{eqnarray}
where $D_P=\CPzero + 2 \  \CPone + {20\over 3}\  \CPtwo 
\ =\  -1.51~{\rm fm}^3$ is a combination
of P-wave coefficients. The same combination
of P-wave scattering parameters contributes to the E1 amplitude in 
$np\rightarrow d\gamma$.
Higher contributions are estimated to be $\lsim 1\%$.

In fact, in general, insertions 
of higher-dimension operators will also pick up factors of $Z_d$ when 
computed to higher orders and therefore we provocatively write
\begin{eqnarray}
\alpha_{E0}^{(z)}
     & = & 
{\alpha M\over 32\gamma^4}\ Z_d\ 
    \left[ 1\ +\ {2\gamma^2\over 3 M^2} \ +\
    {M\gamma^3\over 3\pi} D_P\ +\ ...
      \right]
\nonumber\\
& = & 0.6325~{\rm fm}^3.
\label{eq:eftpolzZ}
\end{eqnarray}

The ellipses denote higher-order terms, which include the N$^4$LO
contribution from $\siii-\diii$ mixing proportional to $\etasd^2\sim
6.5\times 10^{-4}$.  
We estimate that the uncertainty in the result
shown in Eq.~(\ref{eq:eftpolzZ}) is $\pm 0.002~{\rm fm}^3$, 
which includes the effect of this
S-D mixing.  The first contribution from interactions that are not
constrained by $NN$ phase shifts or gauge invariance occurs at
N$^5$LO, two orders higher than we have calculated, and 
hence contributes an uncertainty of  $\pm 0.0007~{\rm fm}^3$.

%%%%%%%%%  np -> d gamma %%%%%%%%
\subsection{$np\rightarrow d\gamma$}

The capture of low-energy neutrons by protons to form a deuteron is a
reaction which is not only of fundamental interest in nuclear physics,
but also represents a critical input to astrophysical simulations.  
At low energies the total cross-section is dominated by 
M1-isovector capture from $np$ continuum states to the deuteron.  
This is the
energy region relevant for Big-Bang-Nucleosynthesis (BBN), and these
isovector amplitudes have been computed to very high orders in
EFT~\cite{CS99,Ru99}.  Meanwhile, the much smaller isoscalar
amplitudes which contribute to this transition have been computed to
NLO in order to make predictions that are to be compared with imminent
experimental measurements~\cite{Mu00}.

At low energies it is useful to make a multipole decomposition for
the amplitude $np\rightarrow d\gamma $ 
\begin{eqnarray}
\label{eq:silas}
 & &  T \ =\ 
 i e \ X_{M1_{V}}\ \varepsilon ^{abc}\epsilon _{(d)}^{\ast a}\
{\bf k}^{b}\
\epsilon _{(\gamma )}^{\ast c}\ 
U^{T}_{\rm n} \overline{P}^3 \ U_{\rm p} 
\nonumber\\
& + & 
  e\  X_{E1_{V}}\  U^{T}_{\rm n}\ \tau_2\tau_3\ {\bf \sigma }_2\ 
  {\bf\sigma}\cdot\epsilon _{(d)}^*
\ U_{\rm p}\ {\bf P}\cdot\epsilon_{(\gamma)}^*
\nonumber \\
& + & {e X_{M1_{S}} \over\sqrt{2}}  
\ U^{T}_{\rm n} P^a
\left[ 
{\bf k}^a\ \epsilon _{(d)}^{\ast }\cdot \epsilon _{(\gamma )}^{\ast }
-\epsilon
_{(d)}^{\ast }\cdot {\bf k}\ \epsilon _{(\gamma )}^{\ast a }
\right]\  U_{\rm p}
\\
& + & {e X_{E2_{S}}\over\sqrt{2}}
\ U^{T}_{\rm n} P^a
\left[ 
{\bf k}^a\ \epsilon _{(d)}^{\ast }\cdot \epsilon _{(\gamma )}^{\ast }
+\epsilon_{(d)}^{\ast }\cdot {\bf k}\ \epsilon _{(\gamma )}^{\ast a}
-{\frac{2}{n-1}}\epsilon _{(d)}^{\ast a}\ {\bf k}\cdot \epsilon
_{(\gamma )}^{\ast }\right]\  U_{\rm p}
\ \ \ .\nonumber
\end{eqnarray}
Only the lowest partial waves are shown, corresponding to
electric dipole capture of nucleons in a P-wave with amplitude $ X_{E1_{V}}$,
isovector magnetic capture of nucleons in the $\si$ channel with amplitude $
X_{M1_{V}}$,
isoscalar magnetic capture of nucleons in the $\siii$ channel with amplitude $
X_{M1_{S}}$,
and isoscalar electric quadrupole
capture of nucleons in the $\siii$ channel with amplitude $X_{E2_{S}}$.
Higher multipoles are suppressed by additional powers of ${\bf k}$, the
photon momentum. Note that 
$U_{\rm n}$ is the neutron Pauli spinor and $U_{\rm p}$ 
is the proton Pauli spinor, while $|{\bf p}|$ is the magnitude of the 
momentum of each nucleon in the $NN$ center-of-mass frame.
The  photon polarization vector is 
$\epsilon _{(\gamma )}$, and $\epsilon _{(d)}$ is the deuteron polarization
vector. 
For convenience, we define dimensionless variables $\tilde{X}$, as
follows:
\begin{eqnarray}
{ |{\bf P}| M_{N}\over \gamma^{2} + |{\bf P}|^2}
X_{E1_{V}}\  & = &
i {2\over M} \sqrt{\pi\over\gamma^3}\tilde{X}_{E1_{V}}
\quad ,\quad
X_{\pi L_{I}}\  \ =\ 
i {2\over M} \sqrt{\pi\over\gamma^3}\tilde{X}_{\pi L_{I}}
\ \ \ ,
\label{eq:tildedef}
\end{eqnarray}
where $X_{\pi L_{I}}$ are the amplitudes other than $X_{E1_{V}}$.
In terms of the amplitudes given in Eq.~(\ref{eq:silas}) and
Eq.~(\ref{eq:tildedef}) the unpolarized cross-section is 
\begin{eqnarray}
\sigma ={8\pi \alpha \gamma ^{3} \over M_{N}^{5} |{\bf v}|}
\left[ |\tilde{X}_{M1_{V}}|^{2}
  \ +\ |\tilde{X}_{E1_{V}}|^{2}
  \ +\ |\tilde{X}_{M1_{S}}|^{2}
  \ +\ |\tilde{X}_{E2_{S}}|^{2}
\right],
\label{eq:unpol}
\end{eqnarray}
where $\alpha$ is the fine-structure constant, and $|{\bf v}|$ is the
velocity of the incoming neutrons.  Close to threshold
$\tilde{X}_{M1_{V}}$ is the dominant contribution to this cross
section, being larger than the other multipoles by several orders of
magnitude.  Therefore a measurement of $\sigma$ in this kinematic
regime does not constrain $\tilde{X}_{E1_{V}}$, $\tilde{X}_{M1_{S}}$,
or $\tilde{X}_{E2_{S}}$.

The isovector $M1_V$ amplitude up to NLO is
\begin{eqnarray}
 |\tilde{X}_{M1}|^2 & = & 
{\kappa_1^2 \gamma^4 \left( {1\over a_1}-\gamma\right)^2 \over
\left({1\over a_1^2} +|{\bf P}|^2\right) \left(\gamma^2 + |{\bf P}|^2 \right)^2
}
\nonumber\\
& & \left[Z_d
- r_0 { \left( {\gamma\over a_1}+|{\bf P}|^2\right)  |{\bf P}|^2 \over
\left({1\over a_1^2} +|{\bf P}|^2\right) \left( {1\over a_1}-\gamma\right)}
- {L_{np}\over\kappa_1}{M\over 2\pi}
{ \gamma^2 + |{\bf P}|^2 \over {1\over a_1}-\gamma}
\right]
\ \ ,
\label{eq:M1amp}
\end{eqnarray}
where the first term can  be obtained
in ERT~\cite{Au53}. However,  ERT  
cannot reproduce the cross-section for this 
process, giving $10\%$ less than the experimental result for the thermal
neutron capture cross-section~\cite{Co65}
for incident neutrons with $|v|=2200~{\rm m/s}$ of
$\sigma_{\rm th}=(334.2 \pm 0.5) {\rm mb}$.
In the traditional picture this 10\% discrepancy was explained by
Riska and Brown~\cite{RB72} as being due to two-body currents.
These mechanisms contribute significantly to the cross-section for
thermal neutron capture at threshold. Below we will discuss work by
Park, Min, and Rho, which shows that this explanation emerges 
naturally in the EFT with pions~\cite{Pa95,Pa96B}.  

In contrast, in $\nopi$ the
pion is regarded as ``short-range" dynamics and this two-body current
appears as the isovector contact term $L_{np}$. The constant $L_{np}$ is 
an RG-invariant combination of parameters.
$L_{np}$ is related to the couplings that appear in the Lagrangian via
\begin{eqnarray}
L_{np} & = & 
(\mu-\gamma)(\mu-{1\over a_1})
\left[ \Lone -
{\displaystyle{\pi \kappa_1 \over M_{N}}}
\left( 
{\displaystyle{r_{0}^{(\si)}{}^{{}}
    \over \left( \mu -\frac{1}{a^{(\si)}} \right) ^{2}}} +
{\displaystyle{\rho _{d} \over (\mu -\gamma )^{2}}}
\right) \right]
\ \ ,
\nonumber\\
\label{eq:Lnpdef}
\end{eqnarray}
where
the Lagrange density describing four-nucleon-one-magnetic-photon interaction
is
\begin{eqnarray}
{\cal L}
& = & 
e\  \Lone \ (N^T\  P_i \ N)^\dagger (N^T\  \overline{P}^a \ N)\ 
\delta^{a 3}\  {\bf B}_i + {\rm h.c.}
\ \ \ .
\label{eq:Lone}
\end{eqnarray}
The coefficient of this operator, which involves four nucleon fields and a
magnetic photon, is not related to $NN$ scattering
data by gauge invariance. Hence it must
be fitted to data on the interactions of the magnetic photon
with the two-nucleon system. The value for $L_{np}$
is determined by the requirement that we reproduce the cross-section 
$\sigma^{\rm th}$.

Once the value for the $M1_V$ amplitude is fixed at threshold its
energy-dependence is determined by information from the $NN$ system.
But it ceases to be the dominant
contribution once incoming $np$ energies of a few MeV are reached. 
The E1 amplitude becomes increasingly
important as the energy of the emitted photon increases. 
This amplitude has been computed up to N$^3$LO, and the result is:
\begin{eqnarray}
|\tilde{X}_{E1}|^2 & = & 
{|{\bf P}|^2 M^2 \gamma^4 \over \left(\gamma^2+|{\bf P}|^2\right)^4}
\left[ Z_d \ +\ {M\gamma\over 6\pi}
\left({\gamma^2\over 3}+|{\bf P}|^2\right) {\cal D}^{(P)}
\right]
\ \ .
\label{eq:E1amp}
\end{eqnarray}
With this result in hand, and the value for
$L_{np}$ fixed at threshold, the photo-dissociation cross-section
for the inverse process, $\gamma d\rightarrow n p$ can be determined.
The results are shown in Fig.~\ref{fig:nprate}. 

The amplitudes in Eq.~(\ref{eq:M1amp}) and 
Eq.~(\ref{eq:E1amp}) have been computed to 
one higher order by Rupak\cite{Ru99}. This necessitates the inclusion
of relativistic effects and a counterterm for the E1 multipole.
The result is then a calculation of $np\rightarrow d\gamma$
which has an error of $\sim 1\%$.
Since the physics in these EFT calculations is very similar to
the physics of potential-model calculations of this process 
it would seem that these ``traditional" approaches can also
calculate this reaction with comparable precision
if they reproduce $\sigma^{\rm th}$ 
accurately~\cite{Le00}.
Recent
calculations confirm that this is indeed the case.
%%%%%%%%%% np rate   %%%%%%%%%%%%%%%
\begin{figure}[!ht]
\hskip 0.5in \psfig{figure=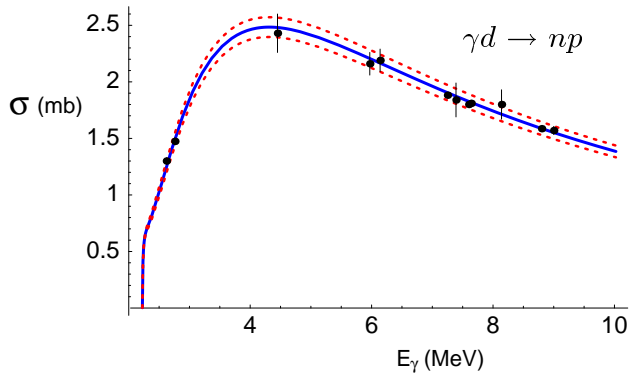,height=2.0in}
\caption{
The photodissociation cross-section for
$\gamma d\rightarrow n p$.
The solid line results from
Eq.~(\protect\ref{eq:unpol}) evaluated with the 
amplitudes Eq.~(\protect\ref{eq:M1amp}) and 
Eq.~(\protect\ref{eq:E1amp})~\protect\cite{CS99}. 
$L_{np}$ is determined by the cross-section
for $np\rightarrow d\gamma$, $\sigma^{\rm th}$.
The dashed lines denote the $\sim 3\%$ theoretical uncertainty
estimated in~Ref.\protect\cite{CS99}.
Rupak has further 
reduced this uncertainty to below $1\%$~\protect\cite{Ru99}.}
\label{fig:nprate}
\end{figure}

We now turn out attention to polarized neutron capture.
For a neutron of polarization $\eta_{\rm n}$
incident upon an unpolarized proton target, 
the cross-sections for the production of right-handed
and left-handed circularly-polarized photons are different.
Defining the asymmetry $A^\gamma_{\eta_{\rm n}} (\theta)$ to be the 
ratio of the  difference and sum of these
two cross-sections, we can express this observable in terms
of the amplitudes $\tilde{X}_{\pi L_{I}}$,
\begin{eqnarray}
  A^\gamma_{\eta_{\rm n}} (\theta) & = &
  \eta_{\rm n} \left[\ 
    \left( P_\gamma (M1)\ +\   P_\gamma (E2) \right)\ \cos\theta
    \ +\   P_\gamma (E1)\ \sin^2\theta
    \right],
\label{eq:gamas}
\end{eqnarray}
where
\begin{eqnarray}
   P_\gamma (\pi L_S) & = &
    {\sqrt{2} {\rm Re}[ \tilde X_{M1_V} \tilde X_{\pi L_S}^*]
     \over  |\tilde X_{M1_V}|^2}
\ ,\ 
   P_\gamma (E1)\ =\  
    { {\rm Re}[ \tilde
      X_{M1_V} \tilde X_{E1_V}^*]\over |\tilde X_{M1_V}|^2}.
\label{eq:pdef}
\end{eqnarray}
Where we have neglected the small $M1_S$, $E2_S$ and $E1_V$
amplitudes in the denominators.

The isoscalar E2 amplitude has been computed to N$^2$LO in 
Ref.~\cite{Ch99B},
\begin{eqnarray}
\tilde X_{E2_s} & = & 
{ \delta\mu_{\cal Q}\ \gamma^2\over4\sqrt{2}} 
-{\etasd\over 10}\left( 1 + {3\over 8}(Z_d-1) -{7\over 32}(Z_d-1)^2\right)
+ {3 \gamma^4\over 40} E_1^{(4)}.
\label{eq:E2mat}
\end{eqnarray}
The deuteron quadrupole moment counterterm appears here, and
once it is fixed there are no free parameters in this expression.
Since $\tilde{X}_{M1_V}$ was
computed to N$^2$LO, this leads to a series for the ratio of
isoscalar E2 to isovector M1 matrix elements which has the numerical
values
\begin{eqnarray}
{\tilde{X}_{E2_{S}} \over \tilde{X}_{M1_{V}}} & = & - \left[\ 1.565\
+\ 0.693\ +\ 0.209 \ \right] \times 10^{-4} \ =\ -2.47\times 10^{-4}.
\label{eq:E2num} \end{eqnarray} 
Note that this series appears to be
converging well, with an expansion parameter of about one third.

The convergence of the isoscalar M1 amplitude is harder to assess.
At NLO this amplitude is:
\begin{eqnarray}
\tilde{X}_{M1_{S}}^{(1)}=
\sqrt{2} \  {\displaystyle{M_{N}\gamma  \over 2\pi }}
\hspace{0.05cm}\Ltwo (\mu -\gamma )^{2}.
\label{eq:l2amp}
\end{eqnarray}
Once again a counterterm that was fitted to a deuteron
static property---in this case the magnetic moment---makes an appearance.
While this is formally the leading contribution,
the smallness of $\Ltwo$ suggests that
the contribution given in Eq.~(\ref{eq:l2amp}) might not dominate over
higher order terms,
and the $M1_{S}$
amplitude might not be
predicted well by $\nopi$ at this order.
Numerically, 
\begin{eqnarray}
{\tilde{X}_{M1_{S}} \over \tilde{X}_{M1_{V}}}=-5.0\times 10^{-4},
\label{eq:M1numS}
\end{eqnarray}
with an uncertainty that we naively estimate to be of $\sim 60\%$.

Using these amplitudes to compute the photon polarizations  $P_\gamma$,
we find
\begin{eqnarray}
P_\gamma (M1)& =& -7.1\times 10^{-4}, \qquad
P_\gamma (E2) \ =\  -3.5\times 10^{-4},
\label{eq:polnums}
\end{eqnarray}
giving a total of $P_\gamma =-1.06\times 10^{-3}$ in the forward
direction.  This is approximately $2/3$ of the experimentally
determined value of~\cite{Ba92} $P_\gamma^{\rm expt} =-(1.5\pm 0.3)
\times 10^{-3}$.  Given the large uncertainty in the calculation of
the $M1_S$ amplitude, and the uncertainty of the measurement, the two
are consistent at this order.  The value~\cite{Ch99} of $P_\gamma
(M1)= -7.1\times 10^{-4}$ agrees with the result obtained using a Reid
soft-core potential~\cite{BK90}.  However, it is significantly smaller
than the ERT value of $P_\gamma (M1)= -9.2\times 10^{-4}$.  However,
given the large uncertainty in the $M1_S$ amplitude of $\nopi$ these
three values are consistent.  They are also consistent with the
calculation of Park {\it et al.}~\cite{Pa00}.  This calculation will
be discussed below, since it is a calculation in an EFT with explicit
pions which uses the counting of Weinberg~\cite{We90,We91}.  The
$\nopi$ value of $P_\gamma(E2)$ quoted in Eq.~(\ref{eq:polnums})
agrees well with that found in Ref.~\cite{Pa00}.  Note that the
zero-range value of $P_\gamma(E2)$: $P_\gamma (E2) = -2.4\times
10^{-4}$, is reproduced at LO in $\nopi$, but it is modified by the
presence of the deuteron quadrupole-moment counterterm. Hence, the
$\nopi$ value of $P_\gamma(E2)$ ends up lying somewhere between this
ERT result, and the result obtained in a Reid soft-core
calculation~\cite{BK90}: $P_\gamma (E2) = -3.7\times 10^{-4}$.

%%%%%%%%%   Neutrinos  %%%%%%%%%%%%%
\subsection{Neutrino-Deuteron Interactions}

Weak interaction processes involving the deuteron are central
to current research efforts in nuclear physics.
In addition to the accelerator based programs that will 
elucidate the 
flavor structure of the nucleon, such as the SAMPLE experiments 
at Bates,
the interactions between neutrinos and the deuteron 
form the core of our efforts to learn  about the neutrino and 
look beyond the standard model of electroweak interactions.
Both in the 
production process, e.g. $pp\rightarrow d e^+\nu_e$, and in detection
at the Sudbury Neutrino Observatory (SNO), 
e.g. $\nu d\rightarrow \nu np$, charged and neutral current 
weak interaction matrix elements 
between the deuteron and continuum states are required.

Over the past few decades
much effort has gone into calculating the
production mechanism $pp\rightarrow d e^+\nu_e$, 
both from  standard non-relativistic quantum mechanics\cite{EB68,BM69}, 
and from  sophisticated  potential model techniques\cite{Sc98}.
Over time, there have
been a series of progressively more sophisticated calculations 
of neutrino-deuteron  scattering, lately
with the goal of providing SNO with high-precision theoretical
predictions of the cross-sections in various 
channels~\cite{Ba88,Ta90,Do92,Ku92,Ku94}.  
The most sophisticated
calculations, 
by Ying, Haxton and Henley (YHH)~\cite{Yi89,Yi92}
and by 
Kubodera and Nosawa (KN)~\cite{Ku92,Ku94},
including the effects of meson exchange currents (MEC's), 
agree with one another
to within $\sim 5\%$~\cite{Ku92,Ku94,Yi89,Yi92,Na00}.

Recently, 
the process $pp\rightarrow d e^+\nu_e$ has been examined with EFT
by Kong and Ravndal~\cite{Ko99A,Ko99B,Ko99C,Ko00}
and by Park, Kubodera, Min and Rho~\cite{Pa98B}.
Elegant expressions and 
numerical values for the weak capture cross-section that are consistent
with previous estimates have been obtained.
The cross-section depends somewhat on the value of 
one four-nucleon-one-weak-gauge-boson interaction, 
with an energy-independent coefficient~\cite{BC00}
$L_{1,A}$, defined by the 
axial currents
\begin{eqnarray}
A_{k}^{S^{(2)}} & = &
-2i\varepsilon _{ijk}L_{2,A}(N^{T}P_{i}N)^{\dagger}
(N^{T} P_{j} N)
\nonumber \\
A_{k}^{a^{(2)}} & = &
L_{1,A}\left( N^{T}P_{k}N\right)^{\dagger }
\left( N^{T} P^aN\right) \ +\  {\rm h.c.}
\ \ \ , 
\label{eq:twoax}
\end{eqnarray}
and the vector currents
\begin{equation}
V_{k}^{S^{(2)}}=2iL_{2}^{s}(N^{T}P_{i}N)^{\dagger }
(\overleftarrow{\nabla }_{i}+\overrightarrow{\nabla }_{i})
(N^{T}P_{k}N)+h.c.
\ \ \ . 
\label{2bodyx}
\end{equation}
Only $L_{1,A}$ contributes at NLO as it is the only two-body operator
connecting the $\si$ and $\siii$ channels.

%%%%%%%%%% cross-section Nu D   %%%%%%%%%%%%%%%
\begin{figure}[!ht]
\psfig{figure=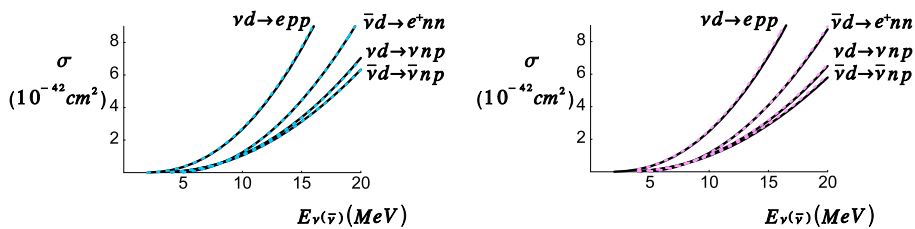,height=1.14in}
\caption{
Inelastic $\protect\nu (\overline{\protect\nu })d$
cross-sections as a function of incident 
$\protect\nu (\overline{\protect\nu})$ energy. 
The solid curves in the left graph are from 
reference~\protect\cite{Na00}, 
while the dashed curves are the EFT results at N$^2$LO, fit with
$L_{1,A}=5.6$~fm$^{3}$. 
The solid curves in the right graph are the results of 
YHH~\protect\cite{Yi89,Yi92}, and the dashed curves are the N$^2$LO 
EFT results~\protect\cite{BC00} fit
with $L_{1,A}=0.94$~fm$^{3}$. 
In both graphs, the dashed curves all lie right on 
top of the solid curves.
}
\label{fig:nuD}
\end{figure}
The neutrino-deuteron 
break-up cross-sections have been determined~\cite{BC00,BC00B}
and it is found that 
$L_{1,A}$ needs to be known at the $10\%$-level
in order to perform a $\sim 1\%$ calculation of the cross-section.
The YHH and KN numerical results can be
recovered with different choices of $L_{1,A}$,
as shown in Fig.~\ref{fig:nuD}, 
confirming 
that the difference between the potential-model calculations 
is short-distance in origin.
Therefore, to predict the break-up cross-section with a precision of better
than $\sim 5\%$, one has two options.
Firstly, compute the $\beta$-decay of tritium, and use this to determine
the EFT counterterm $L_{1,A}$, or equivalently
the MEC's in  potential 
models~\footnote{
With certain assumptions about three-body forces, 
this method of fixing MEC's has been
implemented for $pp\rightarrow d e^+\nu_e$\cite{Sc98}.
}.
Secondly, one can perform an experiment to measure one of the break-up cross
sections to high accuracy, and thereby extract $L_{1,A}$, or the MEC's.
%%%%%%%%%% Ratio  CC/NC   %%%%%%%%%%%%%%%
\begin{figure}[!ht]
\hskip 0.8in \psfig{figure=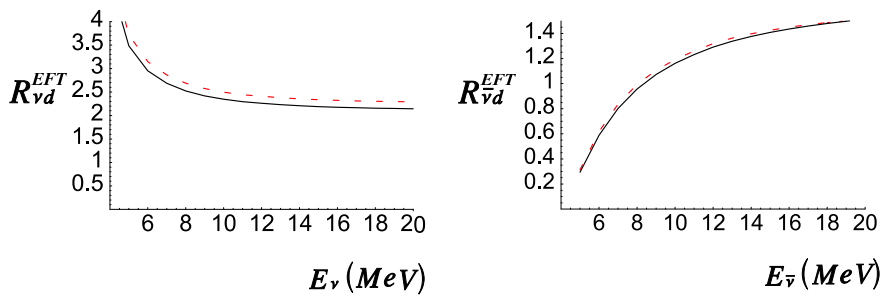,height=1.5in}
\caption{
The ratio of charged current to neutral current 
cross-sections in inelastic
$\protect\nu -d$ and $\overline{\protect\nu }-d$ 
scattering, as functions of incident energy.
Shown are ratios of the EFT N$^2$LO 
results with $L_{1,A}=-20$~fm$^{3}$
(dashed) and $L_{1,A}=40$~fm$^{3}$, demonstrating the insensitivity of
the ratio to the value of $L_{1,A}$~\protect\cite{BC00}.
}
\label{fig:CCNC}
\end{figure}
An important input into determining if neutrinos are changing 
flavor as they move through the interior of the 
Sun and to the Earth is the ratio of charged current to
neutral current cross-sections.
Fig.~\ref{fig:CCNC} shows that this ratio, $R_{\rm EFT}$,
unlike the cross-sections themselves,
is relatively insensitive to the
counterterm $L_{1,A}$ (or equivalently, the MEC's).

\begin{equation}
\sigma(E)=a (E) \ +\ b (E) L_{1,A}
\ \ \ .
\end{equation}
\begin{table}[!ht]
\begin{tabular}{c|ll|ll|ll}
&\multicolumn{2}{c|}
{$\nu_x d\rightarrow n p \nu_x$}
&\multicolumn{2}{c|}
{$\overline\nu_x d\rightarrow n p \overline\nu_x$}
&\multicolumn{2}{c}
{$\nu_e d\rightarrow p p e^-$}
\\
$E_{\nu,\overline{\nu}}$\  (MeV) &$a$ &$b$ &$a$ &$b$ &$a$ &$b$ 
\\
\hline
4 & 0.0287 & 0.00034 & 0.0282 & 0.00034 & 0.146 & 0.0016\cr 
6 & 0.188 & 0.0024 & 0.182 & 0.0024  & 0.574 & 0.0067\cr 
8 & 0.514 & 0.0070 & 0.492 & 0.0069 & 1.34 & 0.017 \cr 
10 & 1.02 & 0.015 & 0.968 & 0.014 & 2.48 & 0.032\cr 
12 & 1.72 & 0.025 & 1.61 & 0.025 & 4.01 & 0.054\cr 
14 & 2.60 & 0.039 & 2.41 & 0.038 & 5.95 & 0.082\cr 
16 & 3.69 & 0.056 & 3.37 & 0.054 & 8.31 & 0.12\cr 
18 & 4.97 & 0.077 & 4.49 & 0.074  & 11.12 & 0.16\cr 
20 & 6.47 & 0.102 & 5.76 & 0.097  & 14.39 & 0.21\cr
\end{tabular}
\caption{
Neutrino-deuteron inelastic cross-sections from 
ref.~\protect\cite{BC00B},
parameterized by 
$\protect\sigma(E)=a (E) + b (E) L_{1,A}$,
where $a (E)$ and $b (E)$ have units of~$10^{-42} {\rm cm}^2$. 
$\nu_X$ denotes a standard model neutrino, $\nu_e$, $\nu_\mu$
or $\nu_\tau$.
}\label{ncccresult}
\end{table}
The cross-sections for two
neutral current and one charged current
neutrino-deuteron break-up channels
are listed in Table~\ref{ncccresult},
parameterized 
by two energy dependent functions, $a (E)$ and $b (E)$,
and the energy independent counterterm $L_{1,A}$, with 
The
magnitude of $L_{1,A}$ is estimated to be $\sim 5~{\rm fm}^3$ 
by NDA and is consistent with the values favored by
potential models.
The uncertainty introduced by
omitting  higher order effects 
is conservatively estimated to be $\sim 3\%$ at threshold and 
$\sim 2\%$ at 20 MeV
(using $L_{1,A}=5.6$ fm$^3$ 
the cross-sections of ref.~\cite{Na00} are reproduced within $\sim 2\%$). 
It can be seen from Table~\ref{ncccresult} that
if $L_{1,A}$ is constrained by a $\sim 1\%$ measurement of the
$\nu_e d\rightarrow e^- p p$ cross-section, then 
the uncertainty in the other channels would also be
$\sim 1\%$.

%%%%%%%%%  Gamma-Deuteron Scattering %%%%%%%%%
\subsection{$\gamma$-Deuteron Compton Scattering}

An area of nuclear physics that 
is still in its infancy is
$\gamma d$ Compton scattering.
It is only recently that experimental measurements have been made
of the differential cross-section for $\gamma d$ scattering, but this has
been limited to just three energies, $E_\gamma = 49,\ 69$ and $95~{\rm MeV}$.
This process is important because it may provide 
the only rigorous way to determine the polarizabilities of the neutron.
The amplitude for $\gamma d$ Compton scattering from an unpolarized
deuteron target is
\begin{eqnarray}
{\cal A} & = & 
-i {e^2\over M} \left[\ 
T_1\  {\bf  \varepsilon}\cdot {\bf \varepsilon}^\prime
\ +\ 
T_2\ {\bf k}^\prime\cdot {\bf  \varepsilon}
{\bf k}\cdot {\bf  \varepsilon}^\prime
\ \right]
 {\bf  \varepsilon}_d\cdot {\bf \varepsilon}_d^\prime
\ \ \ ,
\label{eq:compamp}
\end{eqnarray}
where ${\bf \varepsilon}$ and ${\bf \varepsilon}^\prime$ are the polarization
vectors of the initial and final state photons, and 
${\bf \varepsilon}_d$ and ${\bf \varepsilon}^\prime_d$ 
are the polarization
vectors of the initial and final state deuterons.
It is straighforward to compute the contributions to both form factors,
$T_1$ and $T_2$, order-by-order in the Q-expansion.

%%%%%%%%%% cross-section gamma-Deuteron   %%%%%%%%%%%%%%%
\begin{figure}[!ht]
\psfig{figure=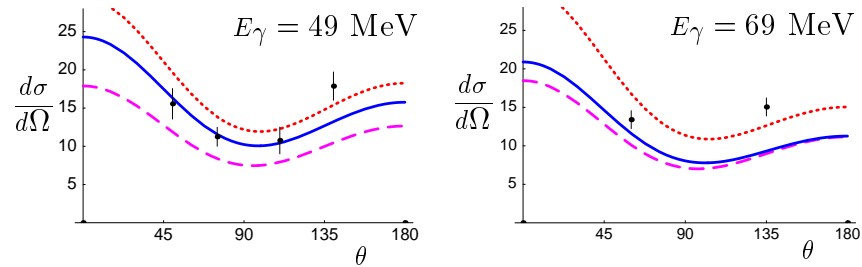,height=1.9in}
\caption{
The differential cross-section for $\gamma d$ Compton scattering
at $E_\gamma=49~{\rm MeV}$ and $69~{\rm MeV}$ in $\nopi$.
The dashed curve is the LO result.
The dotted curve is the N$^3$LO result with the nucleon polarizabilities,
$\alpha$ and $\beta$ set equal to zero.
The solid curve represents the full  N$^3$LO result.
The data is from \protect\cite{Lu94}.
}
\label{fig:gammaD}
\end{figure}
The differential cross-sections
at $E_\gamma=49~{\rm MeV}$ and $69~{\rm MeV}$
are plotted in Fig.~\ref{fig:gammaD}.
At $E_\gamma=49~{\rm MeV}$ the theoretical expression is in very good
agreement with the data at N$^3$LO.
However, at $E_\gamma=69~{\rm MeV}$
there are significant differences between data and the 
N$^3$LO cross-section, particularly at back angles.  
This is no surprise, as the theory should have failed at much lower energies.
It is clear that the electric isoscalar polarizability of the nucleon,
as computed at one-loop in chiral perturbation theory,
makes a significant contribution to the $\gamma d$ scattering cross-section.
There is good reason to believe that precise measurements of the 
cross-sections at low energies will allow for a good determination
of the isoscalar nucleon polarizability and hence the neutron electric
polarizability.

%%%%%%%%%  3-Body Section   NOPI  %%%%%%%%%%%

\section{The Three-Nucleon System}

The scaling of the three-body force is the main issue to be understood
in extending the results of the previous section to the three-nucleon
system. Despite recent progress this is not a fully understood topic
and a complete power counting at all orders does not presently exist.
A number of partial results are known however.  The existence of the
non-trivial fixed point, and the consequent non-perturbative
resummation of the two-body interactions changes the scaling of the
different three-body force terms from their NDA values.  In most
three-body channels this change suppresses the effect of the
three-body forces and makes them appear in higher orders than naively
supposed.  However in the $S-$wave, spin $1/2$ channel (where the
triton and $^3$He are) the opposite happens.  The three-body force in
this channel is enhanced and becomes a LO effect.  The special
physical property shared by no other channel is that, in this channel,
Fermi statistics and/or centrifugal barrier repulsion does not block
all three particles from being at the same point in space.  One can
then expect, on a intuitive basis, that this channel will have a
larger sensitivity to short distance physics. We will see that this
expectation is indeed fulfilled.

When considering three-body systems, in addition to the one- and
two-body Lagrange densities in
Eqs.~(\ref{eq:lagone}) and (\ref{eq:lagtwo}) 
we have also three-body force terms
\begin{equation}
{\cal L}_3 = -D_0 (N^\dagger  N)^3 + \cdots,
\label{eq:lagthree}
\end{equation}
\noindent
where the dots represent terms involving more derivatives. 
 
Using NDA the coefficient $D_{2r}$ of a
three-body force with $2 r$ derivatives should be of order
\begin{equation}
\label{eq:naive}
D_{2r} \sim {4\pi\over M \Lambda^{4+2 r}}
\ \ \ ,
\end{equation}
which would imply that their effect is very suppressed compared to the
two-body interactions. Close to the non-trivial fixed point however,
NDA does not have to hold and $D_{2r}$ can be enhanced by powers of
$\Lambda a $ or $\Lambda/\gamma$.  The natural size of
$D_{2r}(\Lambda)$ is assumed to be set by the size required to appear
in the low energy expansion at the same order as the cutoff dependence
of the sum of two-body diagrams.  Of course, for a particular value of
the cutoff $D_{2r}(\Lambda)$ could be much smaller, but any small
change in the cutoff would restore the ``natural'' value for
$D_{2r}(\Lambda)$.  This same criterion applied to the low energy
expansion around the trivial fixed point is the usual NDA argument
leading to Eq.~(\ref{eq:naive}).  The proximity of the non-trivial
fixed point leads to the necessity of a non-perturbative resummation
of some of the interactions. The ultraviolet behavior of an infinite
sum of diagrams can be very different from the behavior of each
individual diagram and this affects the scaling of the
$D_{2r}(\Lambda)$ coefficients.  In all but one channel,
$D_{2r}(\Lambda)$ will be even smaller than the NDA suggests. In the
$S-$wave, $J=1/2$ channel (where the triton and $^3$He live) though,
the opposite occurs and $D_0$ is a leading order effect.

%%%%%  Spin 3/2 3-body %%%%%%%%%%%%

\subsection{The Spin ${3\over 2}$ Channel}

In the spin $3/2$ channel~\cite{Bedaque:1998mb,BvK97} the spins of all
three nucleons are parallel and Fermi statistics forbids the three
nucleons to occupy the same point in space. The three-body force term
$D_0$ cannot thus contribute in this channel and only three-body terms 
with more derivatives might contribute. Also, only the $^3S_1$ two-body
interactions appear. Making the provisory assumption that the effect
of the three body forces will  be of higher order than our
calculations
we are left with a Lagrangean involving only one- and two-body terms:
 \begin{eqnarray}
\label{eq:l1l2}
{\cal L}_1+{\cal L}_2&=&N^\dagger\left[ i\partial_0 
+  {{\stackrel{\scriptscriptstyle\rightarrow}{\nabla}}^2\over 2 M}
\ +\cdots\right] N
- \CzeroT \left(N^T P^i N\right)^\dagger \left(N^T P^i N\right)
\nonumber\\
&+& 
 {\CtwoT\over 8} 
\left[ (N^T P^i N)^\dagger 
(N^T P^a (\stackrel{\scriptscriptstyle\rightarrow}{\nabla}-
          \stackrel{\scriptscriptstyle\leftarrow} {\nabla})^2 N)
\ +\ {\rm h.c.}\right]+\cdots
\label{eq:donly}
\end{eqnarray}
It is very convenient in three-body calculations to rewrite the
theory described by  Eq.~(\ref{eq:donly}) by introducing a dummy field
$d^i$  with
the quantum numbers of the deuteron (referred to hereafter as the ``deuteron'') 
\begin{eqnarray}
\label{eq:dlag}
{\cal L}_d&=&N^{\dagger}\left[ i\partial_0\ 
+\  {\stackrel{\scriptscriptstyle\rightarrow}{\nabla}^2\over 2 M}\ +\cdots\right] N
+d^{i\dagger} \left[-i \partial_0 
+ {\stackrel{\scriptscriptstyle\rightarrow}{\nabla}^2\over 2 M}\
-\Delta_d\right]d^i\nonumber\\
& + & y_d \left[d^{i \dagger} (N^T P^i N) + \mathrm{h.c.}\right]\ 
+\cdots 
\ \ .
\end{eqnarray}
By performing the gaussian integration over the deuteron field and
treating its kinetic term as a perturbation we recover
Eq.~(\ref{eq:donly}) with time derivatives instead of space derivatives 
in the term proportional to $\CtwoT$ (and  some higher order
terms). This time derivative can be traded by space derivatives by use 
of a field redefinition
\begin{equation}
N\rightarrow N+\frac{y_d^2}{8\ \Delta_d^2}\; P^{i \dagger}
  N^\dagger (N^T P^i N).
\end{equation}
\noindent
The parameters in Eq.~(\ref{eq:donly}) and Eq.~(\ref{eq:dlag}) are
related by
\begin{eqnarray}
\label{eq:ytoC} 
  y_d^2 & = & \frac{(\CzeroT)^2}{M \CtwoT}
\ \ ,\ \ 
\Delta_d\ =\ -\frac{\CzeroT}{M \CtwoT}
\ \ \ .
 \end{eqnarray}
With this choice of parameters Eqs.~(\ref{eq:donly},\ref{eq:dlag})
describe equivalent theories, i.e., they both give rise to the same
S-matrix.  Notice that after the introduction of the deuteron field
there is a large reparametrization invariance in the theory.  We could
for instance, change the coefficient $y_d$ and add an explicit
two-nucleon interaction is such a way that after eliminating the
deuteron field the same Lagrangean was recovered. Thus the Lagrangean
in Eq.~(\ref{eq:dlag}) is {\it not} the most general one involving
nucleons {\it and} deuteron fields. The important point is that after
the elimination of the deuteron field we recover the most general
Lagrangean involving nucleons only.  {}From Eq.~(\ref{eq:ytoC}) and
Eq.~(\ref{eq:cscale}) we arrive at the scaling $y_d^2\sim (4
\pi\Lambda)/M$ and $\Delta_d \sim (4 \pi\Lambda \mu)/M$.  The scaling
above shows that the kinetic term of the $d^i$ field is indeed
sub-leading compared to the term proportional to $\Delta_d$.

At tree level the $d^i$ propagator is given by the constant $i/\Delta_d$. 
\begin{figure}[!htb]
\hskip 0.8in
\psfig{file=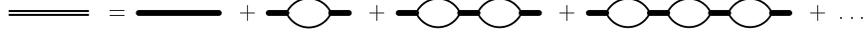,height=0.95in}
 \caption{The leading order
deuteron propagator containing the sum of bubble graphs.}
\label{fig:deuteronpropagator}
\end{figure}
\noindent
The loop corrections shown in Fig.~\ref{fig:deuteronpropagator}
contribute to the same order as the tree level term. 
Summing up these graphs gives
\begin{equation}
\label{eq:dprop}
  i \triangle^{ij}(p) \ =\ 
-{4 \pi i \over M y_d^2}
\   
\frac{\delta^{ij}_d}{\frac{4\pi
  {\Delta_d}}{M y_d^2} -\mu+\sqrt{\frac{\pv^2}{4}-M p_0}}\;\;.
\end{equation}
\noindent
The leading order $d^i$ field propagator is, 
up to a constant, the two-nucleon scattering
amplitude in the $^3S_1$ channel.

We consider now the scattering of a neutron on a deuteron with total
spin $S=3/2$ (the more experimentally accessible proton-deuteron
scattering would include Coulomb effects that would mask the strong
interactions we are interested in here). We  use the field $d^i$ as
the interpolating field for the deuteron. The graphs that 
contribute to this process at leading order are depicted on the top
of the 
Fig.~\ref{fig:quartetfadeev}.
A simple power counting shows that they are all of the same order. The 
tree level diagram, for instance, contains two $dNN$ vertices 
($\sim \Lambda/M$) and
one nucleon propagator ($\sim M/Q^2$), for a total contribution of
order
$\sim \Lambda/Q^2$. The one loop propagator contains two additional
 $dNN$ vertices, two extra nucleon propagators, one $d^i$ field
propagator ($\sim 1/M y_d^2 Q$) and one loop ($\sim Q^5/M$), thus
the one loop graph is also of order $\sim \Lambda/Q^2$. 
All the graphs shown in the first line of Fig.~\ref{fig:quartetfadeev} 
are of the
same order and must be resummed  at LO. This is very 
similar to the two-nucleon scattering case where all graphs involving
only the leading interaction proportional to $\CzeroT$ are leading.
The difference resides on the fact that, while all the graphs for the
LO two-nucleon scattering are easily summed in a geometric
series, the same cannot be done in the three-body sector. The best we
can do is to write an integral equation (shown in the second line of 
Fig.~\ref{fig:quartetfadeev}) whose solution, obtained numerically,
equals the desired infinite sum\footnote{Clearly the validity of 
this integral equation goes beyond all orders in perturbation
theory even though the derivation shown here is  based on diagrams.}.

\begin{figure}[!htb]
\psfig{file=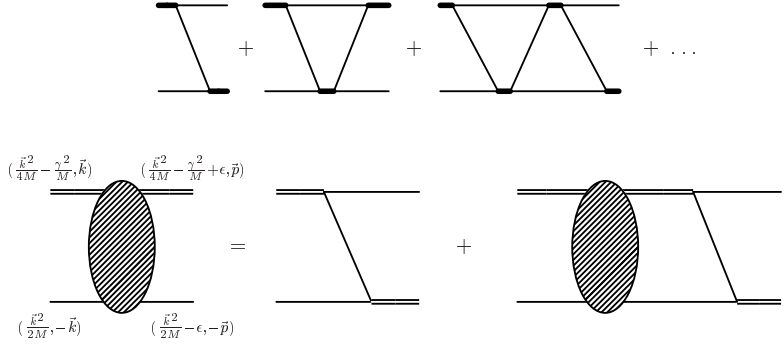,height=2.1in,bbllx=70pt,bblly=560pt,bburx=520pt,bbury=770pt}
\caption{ The first line shows the infinite sum of graphs
giving the LO  neutron-deuteron scattering in the spin $3/2$ 
channels. The second line is the graphical representation of the
Fadeev equation that sums them up.}
\label{fig:quartetfadeev}
\end{figure}

With the kinematic shown in Fig.~\ref{fig:quartetfadeev},
the $S-$wave amplitude satisfies 
\begin{equation}\label{eq:fort}
 t_0(k,p)=-\;\frac{y_d^2 M}{2} {K_0}(k,p)
-\;\frac{1}{\pi}\int\limits_0^\infty dq\; q^2\; t_0(k,q)\;
\frac{1}{-\gamma+\sqrt{\frac{3 q^2}{4}-M E}}\; K_0(q,p)
\end{equation}
\noindent
with
\begin{equation}\label{eq:kernel}
K_0(q,p)=\frac{1}{qp}\;
 \ln\left(\frac{p^2+p q + q^2-M E}{p^2-p q + q^2-M E} \right)
\end{equation}
\noindent
where $E=3 k^2/(4M) - \gamma^2/M$ is the total energy. 
The projection on other partial waves lead to similar equations
\cite{Gabbiani:1999yv}. Notice that when $p=k$,
all external legs are on-shell. This equation is just the Faddeev 
equation for
the case of contact forces derived first in reference
~\cite{Skornyakov} by different methods. 
Eq.~(\ref{eq:fort}) is a one dimensional integral equation (the
variable $k$ appears only as a parameter) and its numerical solution is 
computationally trivial. A little care must be taken with the
singularities of its kernel however.
Although the pole in the real axis due to the
deuteron propagator is regulated by the fact that the energy should be 
thought as having a small positive imaginary part and  the
logarithmic singularity occurring above threshold is integrable, both cause
{\it numerical} instabilities. An efficient way of dealing numerically 
with those instabilities is the
Hetherington--Schick~\cite{HetheringtonSchick,Cahill,Amado} method.

To obtain the neutron-deuteron scattering amplitude we have to multiply the
on-shell amplitude 
$t_0(k,p=k)$ by the wave function renormalization constant,
\begin{equation}
\label{eq:T0}
T_0(k)=\sqrt{Z_0}\; t_0(k,k)\; \sqrt{Z_0}\;\; ,
\end{equation}
\noindent
where
\begin{eqnarray}
 \frac{1}{Z_0} & = &  \ii \frac{\partial}{\partial p_0}\ 
 \frac{1}{\ii\triangle(p)} \Big|_{p_0=-\frac{\gamma^2}{M},\,\pv=0}
\ =\ 
\frac{M^2 y_d^2}{8\pi \gamma}.
\end{eqnarray}
\noindent
Unlike $t_0(k,p)$, the scattering amplitude $T_0(k)$
depends on the parameters $y_d$ and $\Delta_d$ only through the observable
$\gamma=\frac{4\pi \Delta_d}{M y_d^2}+\mu$.

At NLO, we have  contributions coming only from
insertions of the kinetic energy
in the deuteron propagator. 
For book keeping purposes it is convenient to  split the constant 
$\Delta_d$ into $\Delta_{d,-1}+\Delta_{d,0}$ with 
\begin{eqnarray}
  \Delta_{d,-1}&=&-\frac{\CzeroTmone}{M \CtwoT}
\ \ ,\ \ 
  \Delta_{d,0}\ =\ -\frac{\CzeroTzero}{M \CtwoT}=\frac{\gamma_d^2}{M}.
\end{eqnarray}
\noindent $\Delta_{d,0}$ is subleading compared to the leading order
$\Delta_{d,-1}$ operator and is treated in perturbation theory.
With this choice the position of the deuteron pole is not changed by
the NLO corrections to the deuteron propagator 
and $\Delta_{d,-1}$ does not have to be refitted. 

\begin{center}
\begin{figure}[!htb]
\begin{center}
\hskip 0.35in
\psfig{file=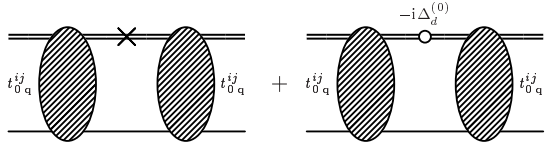,height=1.1in}
\end{center}
 \caption{The NLO contributions to $nd$
      scattering in the quartet channel.}
\label{fig:rangecorrectionquartet}
\end{figure}
\end{center}
The NLO corrections to neutron-deuteron scattering are shown in
Fig.~\ref{fig:rangecorrectionquartet}.
They are simply the range corrections in first order perturbation
theory~\cite{efimovrange,danilovrange,Gabbiani:1999yv}.
The graphs in Fig.~\ref{fig:rangecorrectionquartet} do not mix
different partial waves and give the contribution
\begin{eqnarray}
\label{insertion2}
   \ii t_1(k)&=&\int\limits_0^\infty
   \frac{dq}{2\pi^2}\;q^2\;(\ii t_0(k,q))^2
   \left[ \ii\triangle(\frac{3
   k^2}{4M}-\frac{\gamma^2_d}{M}-\frac{q^2}{2M},\qv)\right]^2\\
   &&\phantom{\int\limits_0^\infty
   \frac{dq}{2\pi^2}}\;\times\;
   \ii{\calI}_d(\frac{3k^2}{4M}-\frac{\gamma^2_d}{M}-\frac{q^2}{2M},\qv)
\nonumber
\;\;,
\end{eqnarray}
\noindent
where the correction to the deuteron propagator is
${\calI}_d(p_0,\pv)
= -\Delta^{(0)}_d - p_0+\frac{\pv^2}{4M}$.
The amplitude in Eq.~(\ref{insertion2})
has to be solved for  numerically as $t_0(k,p)$ is known only
numerically. The diagrams in Fig.~\ref{fig:rangecorrectionquartet}
are ultraviolet finite. The degree of convergence however is not what
one might naively guess, as can be seen by the following argument.
Each one of the diagrams in Fig.~\ref{fig:quartetfadeev} behaves for 
large values of the external momenta $p\gg \gamma$ as $1/p^2$. Their
sum however has a faster decay. To see this consider 
Eq.~(\ref{eq:fort})
for large values of $p\gg k\sim\gamma$. In this case the integral is
dominated by the region of integration $p\sim q$ where the kernel
simplifies and we are left with ~\cite{fadeev,danilovleading}
\begin{equation}
\label{eq:asymptotic}
  t_0(k,p)=
 -\;\frac{2}{\sqrt{3}\pi}\int\limits_0^\infty dq\; t_0(k,q)\;
 \frac{1}{p}\;
 \ln\left(\frac{p^2+p q + q^2}{p^2-p q + q^2} \right)\;\;.
\end{equation}
\noindent
Using the ansatz $t_0(k,p)\sim p^{s-1}$ we find an equation that determines
the power describing the decay of $t_0(k,p)$ for large $p$
\begin{equation}\label{eq:trig}
1+\frac{4}{\sqrt{3}s}\frac{\sin(\pi s/6)}{\cos(\pi s/2)}=0.
\end{equation}
\noindent
The negative solution (the positive ones  lead to divergent
integrals, cannot describe solutions of the integral equation and
are discarded) is $s\simeq -2.166$. The loops in 
Fig.~\ref{fig:rangecorrectionquartet} converge in the ultraviolet as
$\Lambda^{1+2 s}\sim\Lambda^{-3.33}$ as opposed to the naive behavior 
$\Lambda^{-1}$,
valid at each order in a perturbation expansion in the strength of the kernel. 
This fact
has consequences for the estimate of the three-body forces since it
changes their running. As we will see below, in the
triton-$^3$He channel the opposite happens and the three-body force in 
that channel is enhanced compared to NDA
estimates.

To complete the NLO calculation the wave function renormalization
constant $Z$  is found from
\begin{eqnarray}
{1\over Z}&=&{1\over Z_0+Z_1}
\ \simeq\ {1\over Z_0}- {Z_1\over Z_0^2}\ =\ 
{1\over Z_0}-\;\ii\;{\partial\over\partial
p_0}\; \ii{\calI}_d(p_0,\pv)\Big|_{p_0=\frac{\gamma^2_d}{M},\,\pv=0}
\end{eqnarray}
and thus $Z_1=Z_0^2$.
The NLO amplitude  in the quartet channel is
therefore given by
\begin{eqnarray}
T(k)&=&Z\ t(k,k)\ \simeq  (Z_0+Z_1)(t_0(k,k)+t_1(k,k))\nonumber\\ 
&\simeq& T_0(k)+ Z_0 t_1(k,k)+Z_1
t_0(k,k)=T_0(k)+T_1(k)\;\;.
\end{eqnarray}
The phase shift for each partial wave is extracted by expanding both sides of
the  relation 
\begin{equation}\label{kcotg}
  T(k)\simeq T_0(k)+T_1(k)=\frac{3\pi}{M}\;
  \frac{1}{k\cot\delta - ik}
\end{equation}
in $Q$ with $\delta=\delta^{(0)}+\delta^{(1)}+\dots$ and
keeping only linear terms,
\begin{eqnarray}
  \label{eq:deltas}
  \delta^{(0)}&=& \frac{1}{2\ii}\ln\left(1+\frac{2\ii k}{3\pi}
T_0(k)\right)
\ \ ,\ \ 
\delta^{(1)}\ =\ 
  \frac{1}{\ii}\frac{k}{3\pi}\frac{\left(T_0(k)+T_1(k)\right)}
{\left(1+\frac{2\ii k}{3\pi}T_0(k)\right)}\;\;.
\end{eqnarray}

At N$^2$LO there are two kinds of contributions. One comes from 
the same insertions of deuteron kinetic energy and $\Delta^{(0)}$ 
appearing at NLO, but now inserted twice. The other comes from inserting 
the LO operator describing the tensor force appearing in
Eq.~(\ref{eq:sdlagP}). The leading tensor force is generated by the term
\begin{equation}
\label{eq:sdlagP}
{\mathcal L}_{Nd}^{sd}=
y^{sd} d^{i \dagger} N((\stackrel{\scriptscriptstyle\rightarrow}{\nabla}-
      \stackrel{\scriptscriptstyle\leftarrow}{\nabla})^i
(\stackrel{\scriptscriptstyle\rightarrow}{\nabla}-
\stackrel{\scriptscriptstyle\leftarrow}{\nabla})^j 
-\frac{1}{3}\delta^{ij} (\stackrel{\scriptscriptstyle\rightarrow}{\nabla}-
      \stackrel{\scriptscriptstyle\leftarrow}{\nabla})^2)P^j N
\end{equation}
in the Lagrangean Eq.~(\ref{eq:dlag}) 
after the gaussian integration over 
the deuteron fields is performed.
This operators mixes spin and angular momentum and produces a splitting 
between the different three-body amplitudes amplitudes with the same 
spin $S$ and angular momentum $L$
but different values of the total angular momentum $J$. 
This splitting is very important
in determining spin observables like the vector analyzing power 
$A_y$. However the N$^2$LO
calculation of the amplitudes is only a leading calculation of the
splittings between different values of $J$ and cannot be very precise.
Further, the tensor
force appears only linearly and 
after  averaging over the different values of $J$,
its contribution vanishes.
Here we will restrict ourselves to the averaged  phase shifts and, 
consequently, drop the tensor force
contribution at N$^2$LO.
The N$^2$LO calculation (including some higher order contributions that 
can be ignored)
is simply performed by replacing the LO deuteron propagator 
in Eq.~(\ref{eq:dprop}) by
the propagator which includes all effective range effects
\begin{equation}
\label{eq:dproprange}
  i\triangle^{ij}(p) =-\;\frac{4\pi i}{M y_d^2}\;
  \frac{\delta^{ij}}{-\gamma +\frac{1}{2}\rho_d(M
p_0+{\gamma^2}-\frac{\pv^2}{4})
 +\sqrt{\frac{\pv^2}{4}-M p_0}}\;\;.  
\end{equation}
in the integral equation in Eq.~(\ref{eq:fort}).

The calculation for the spin $S=1/2$ channels proceeds in an analogous way to the
one in the spin $3/2$ channels discussed above. The main difference is
that
two-body interactions in the spin singlet channel  also
contribute.
We will introduce another dummy field $T^a$, now with the quantum
numbers of two nucleons in the $^1S_0$ channel
\begin{equation}
{\mathcal L}\rightarrow{\mathcal L} +t^{a  \dagger} \left[(-\ii \partial_0 +
        \frac{\vec{\partial}^2}{4 M})-\Delta_t\right]t^a
   +\;y_t\left[t^{a \dagger} (N^T P^a N) +\mathrm{h.c.}\right]
\end{equation}
with the constants $y_t$ and $\Delta_t$ satisfying relations analogous 
to Eq.(\ref{eq:ytoC}). Again, by integrating over the ``dibaryon'' $t^a$
we recover the $^1S_0$ two-body interactions.
\begin{figure}[!htb]
\begin{center}
\psfig{file=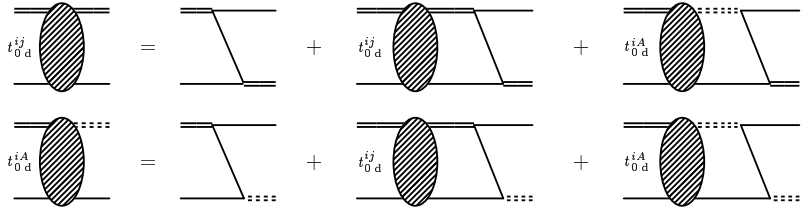,height=1.5in,bbllx=70pt,bblly=650pt,bburx=530pt,bbury=800pt}
\end{center}
\caption{The coupled equations in the spin $1/2$ channels.}
\label{fig:doubletfadeev}
\end{figure}

The spin zero dibaryon field $t^a$  contributes in intermediate states
of neutron-deuteron amplitudes. The analogue of Eq.~(\ref{eq:fort}) is then
a system of two coupled integral equations for the $d+N\rightarrow d+N$ 
($it_d(\vec{k},\vec{p},\epsilon)$)
and $d+N\rightarrow t+N$($it_t(\vec{k},\vec{p},\epsilon)$)
amplitudes. 
\begin{eqnarray}
\label{eq:coupled}
t_d(p,k)&=&
\ {K_0}(p,k)+\frac{2}{\pi}\int_0^\infty\frac{q^2 \;dq}{q^2-k^2}\ 
\ {K_0}(p,q) [t_d(q,k)+3 t_t(q,k)] \\
\label{beq}
t_t(p,k)&=&
3 {K_0}(p,k)+\frac{2}{\pi}\int_0^\infty\frac{q^2 \;dq}{q^2-k^2}\ 
{K_0}(p,q) [3 t_d(q,k)+t_t(q,k)]
\ \ \ .
\nonumber
\end{eqnarray}

\begin{figure}
\begin{minipage}[b]{.49\linewidth}
                 \begin{sideways}                         
                   \psfig{file=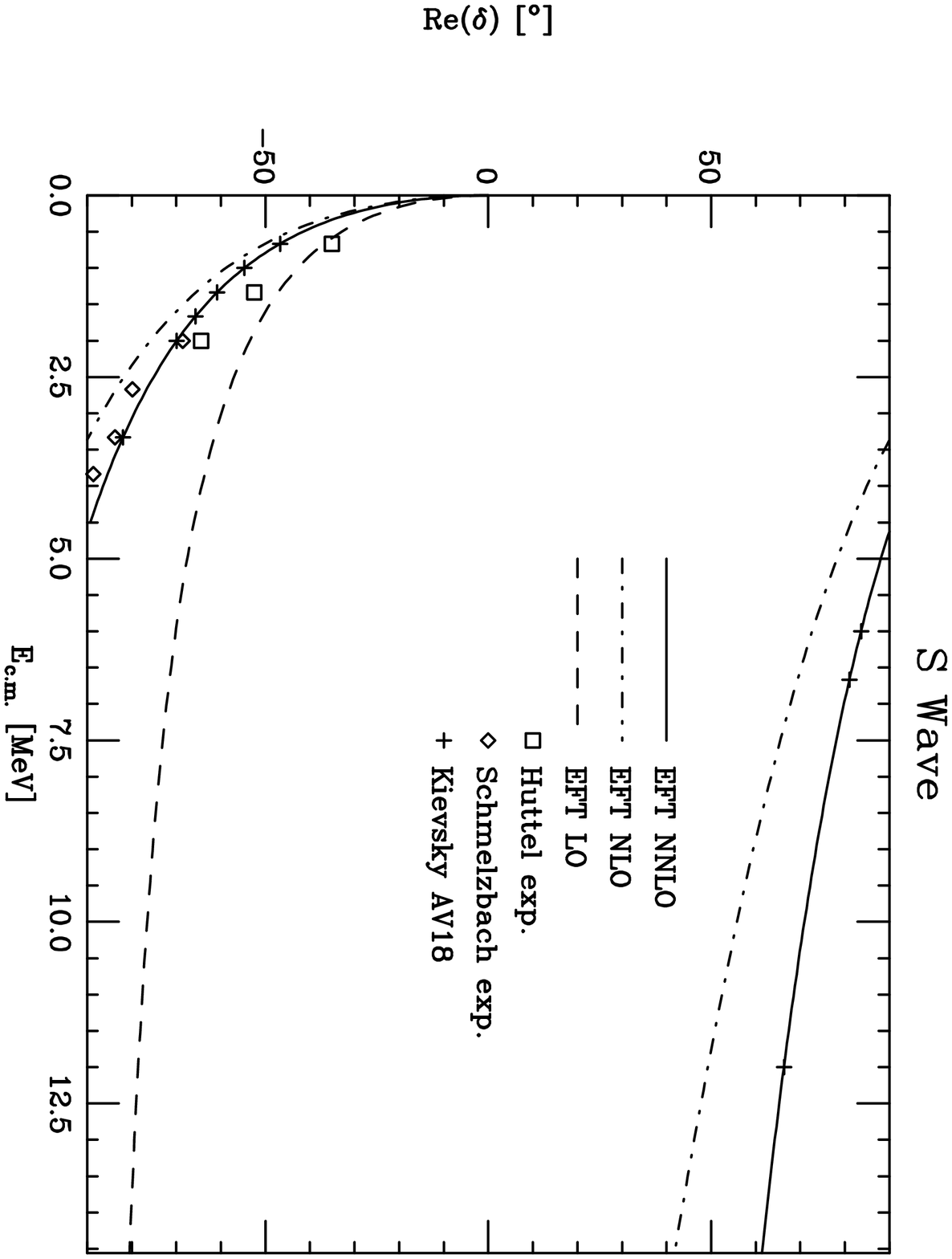,height=2.0in}
                 \end{sideways} 
\end{minipage}\hfill
\begin{minipage}[b]{.49\linewidth}
                  \begin{sideways}                      
                     \psfig{file=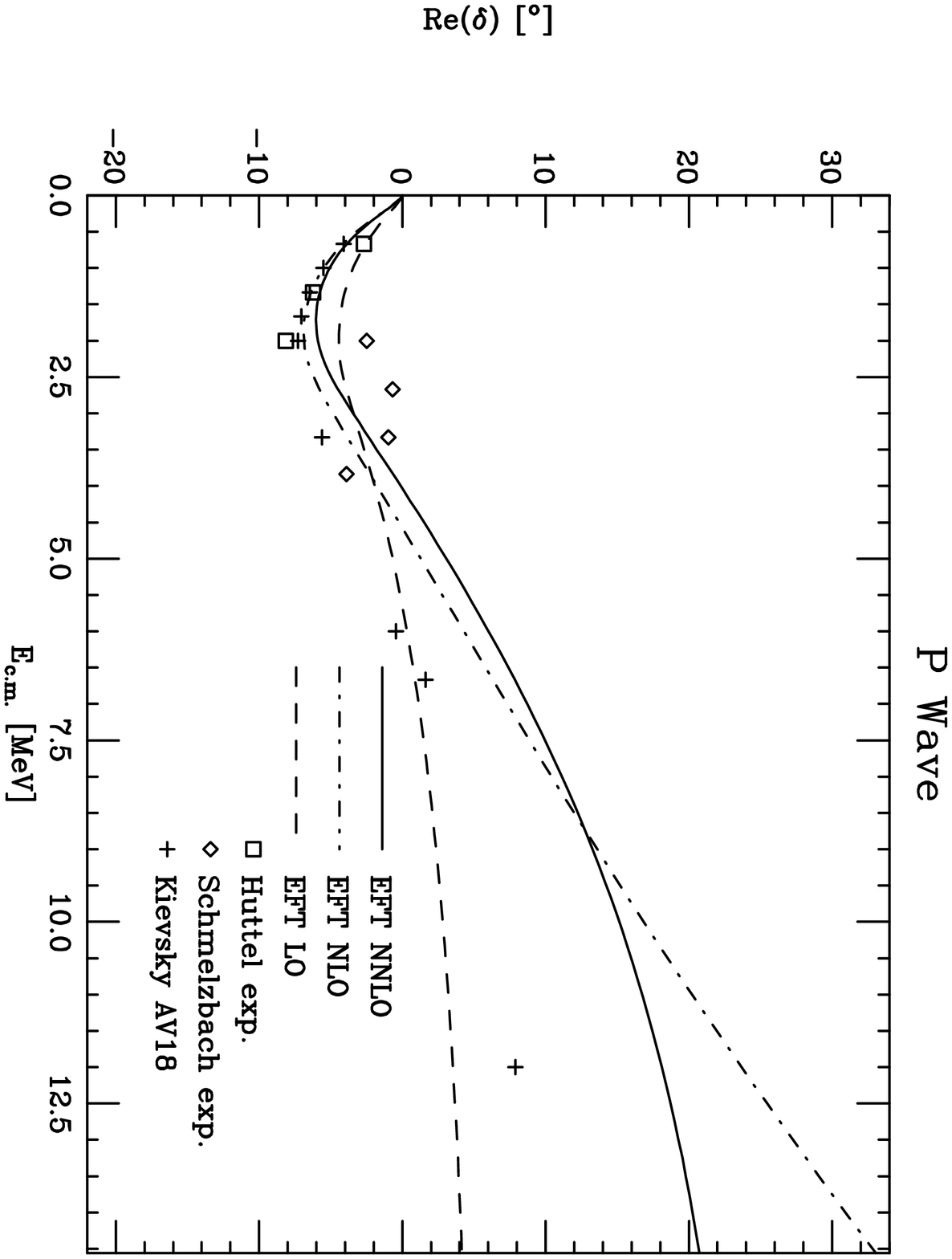,height=2.0in}
                  \end{sideways} 
\end{minipage}\hfill
\begin{minipage}[b]{.49\linewidth}
                 \begin{sideways}                         
                   \psfig{file=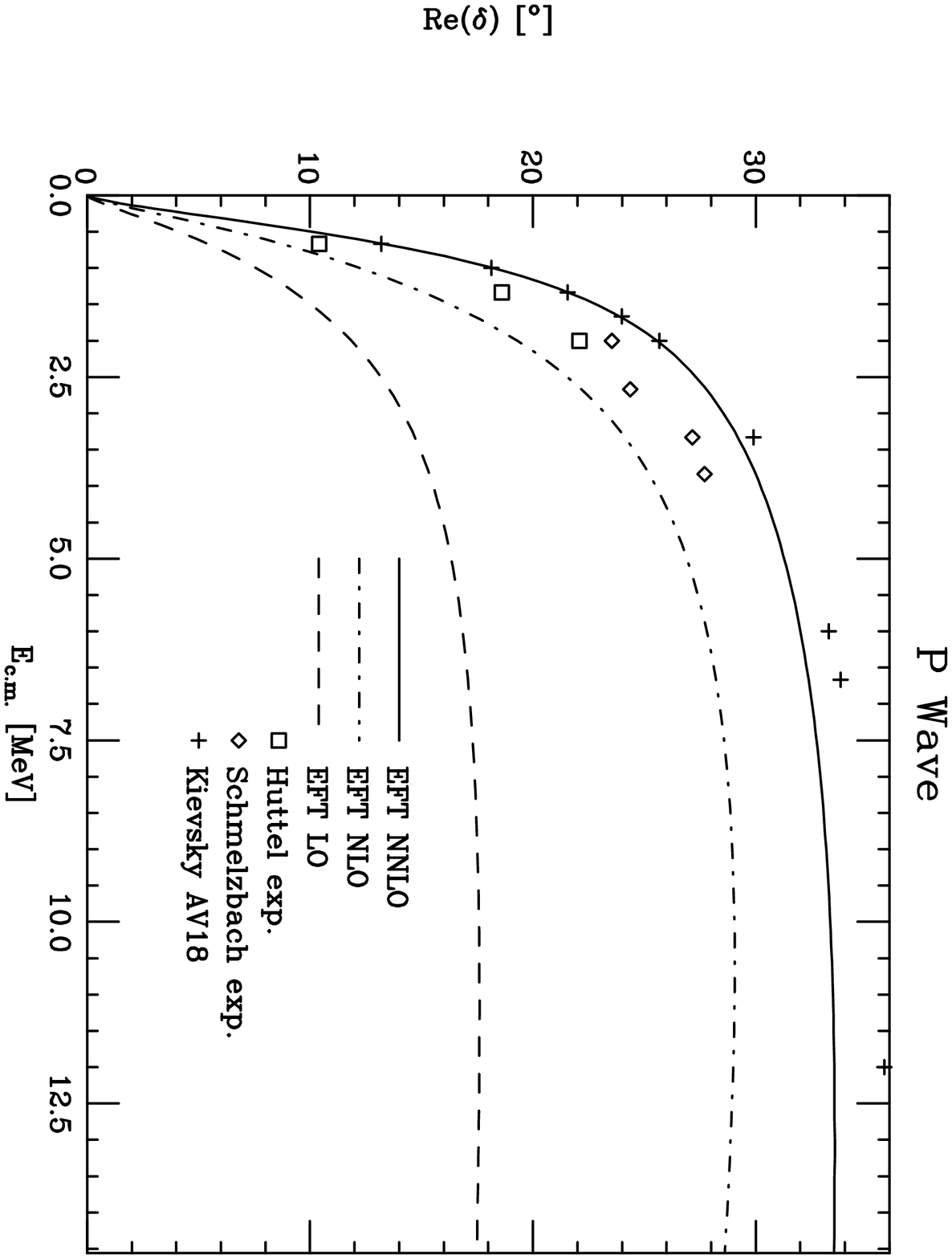,height=2.0in}
                 \end{sideways} 
\end{minipage}\hfill
\begin{minipage}[b]{.49\linewidth}
                  \begin{sideways}                      
                     \psfig{file=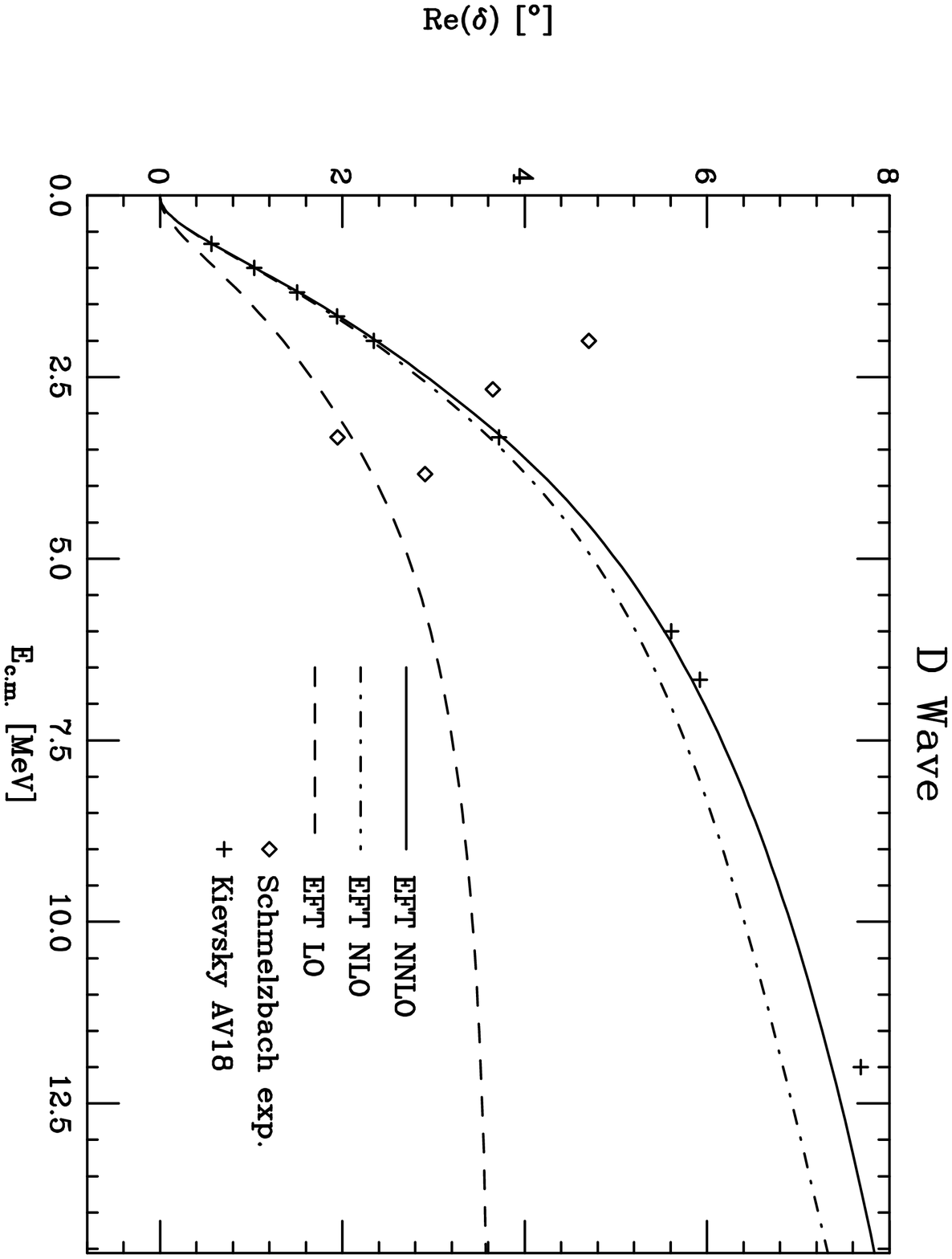,height=2.0in}
                  \end{sideways} 
\end{minipage}\hfill
\begin{minipage}[b]{.49\linewidth}
                 \begin{sideways}                         
                   \psfig{file=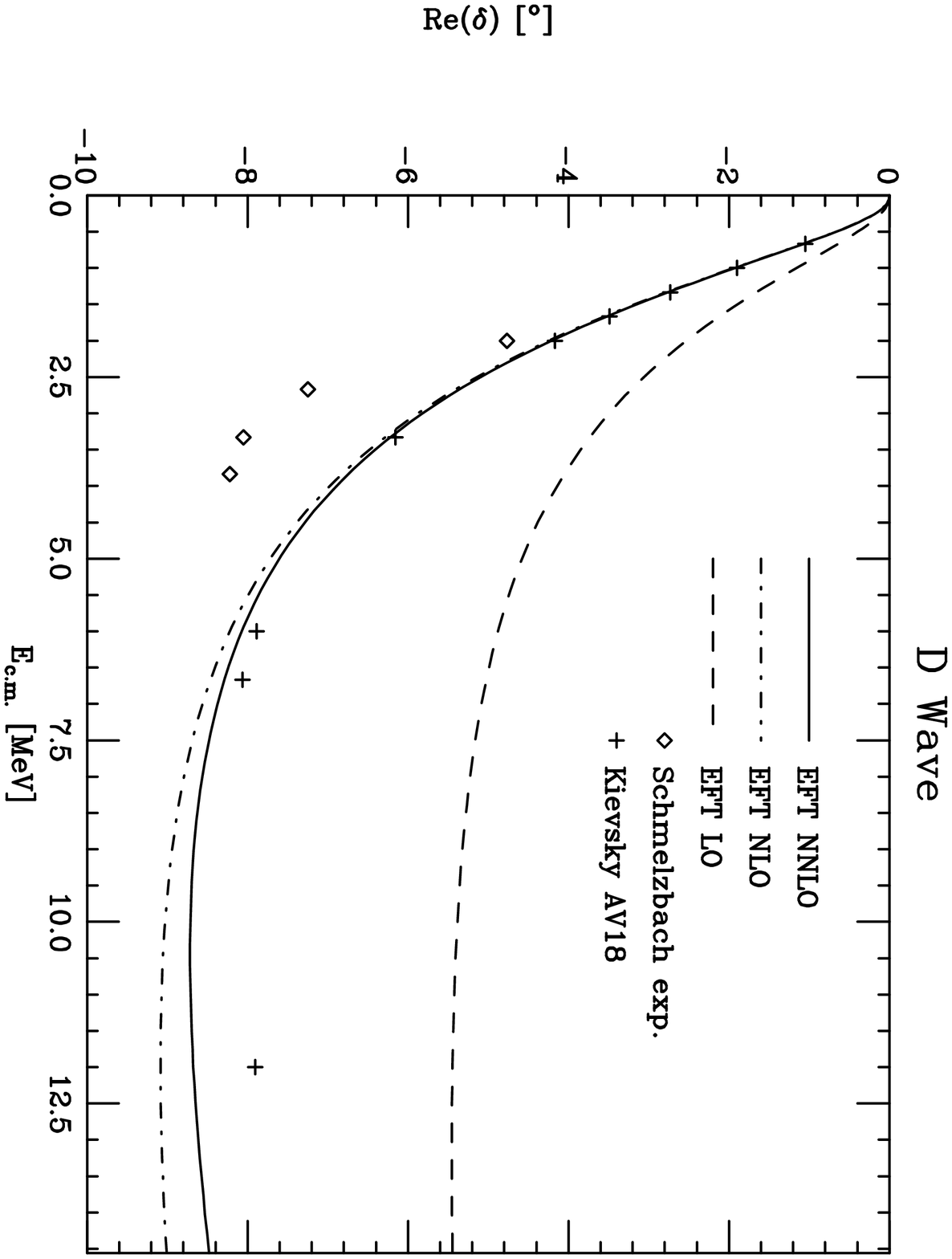,height=2.0in}
                 \end{sideways} 
\end{minipage}\hfill
\begin{minipage}[b]{.49\linewidth}
                  \begin{sideways}                      
                     \psfig{file=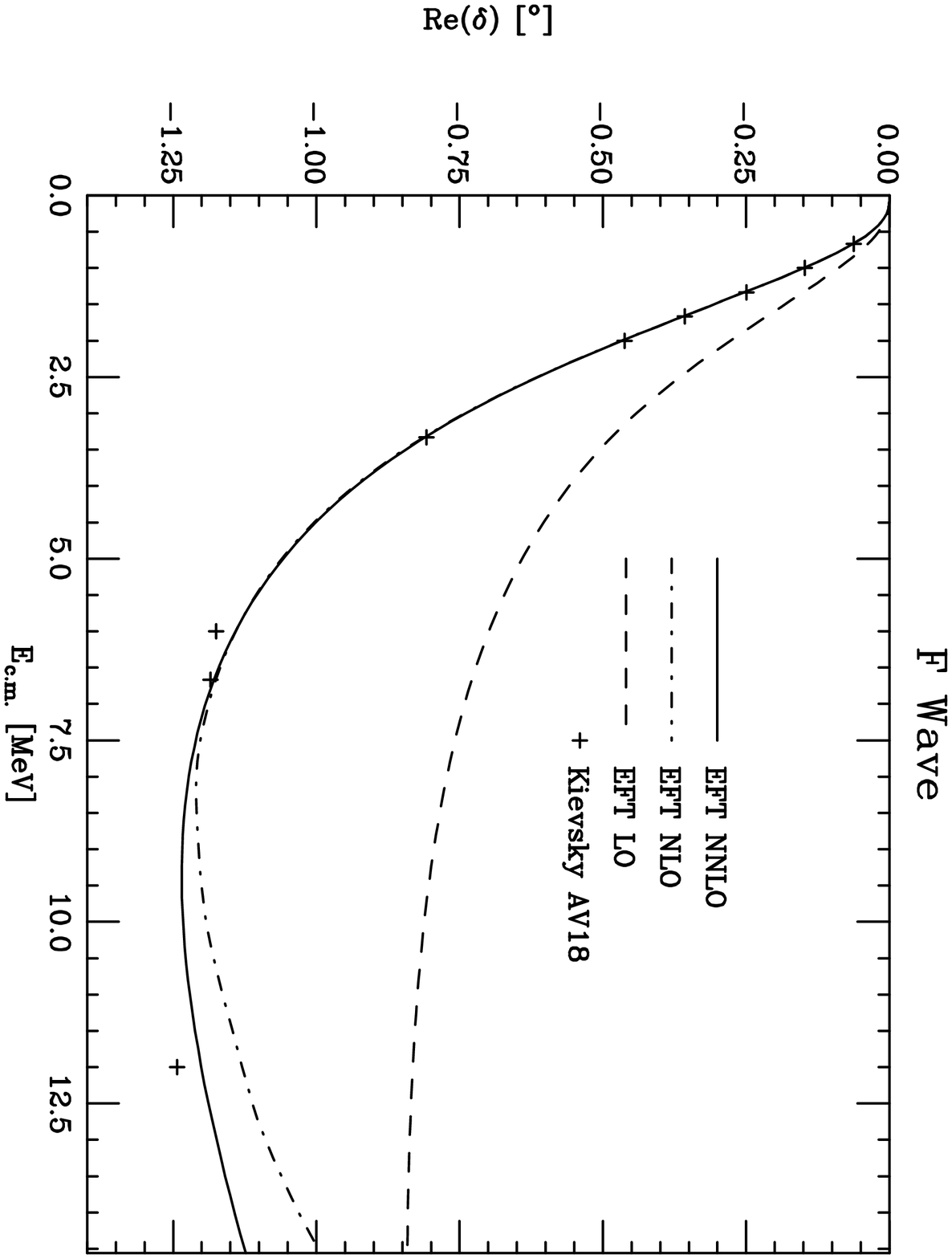,height=2.0in}
                  \end{sideways} 
\end{minipage}\hfill
\caption{Real part of the first  partial waves in
the quartet (left) and doublet (right) channels of $nd$
scattering versus the center-of-mass
energy. The dashed line is the LO, the dot-dashed the NLO and the solid
line the N$^2$LO result. 
}
\label{fig:partialwaves}
\end{figure}

The results of LO, NLO and N$^2$LO calculations
\cite{BvK97,Bedaque:1998mb,Gabbiani:1999yv}
of the phase shifts for the lowest partial waves (averaged over the
different values of $J$) are shown in Fig.~\ref{fig:partialwaves}.
For comparison we include the results of some old phase shift
analyses~\cite{huttel,schmelzbach} (that unfortunately do not contain
error bars) and a recent extensive numerical calculation using modern
potentials~\cite{kievsky,kievskyprivate}. In most cases the N$^2$LO
EFT computation seems to be convergent and accurate at the precision
one would expect ($(\gamma/m_\pi)^3\sim 3\% $).  This theoretical
error is probably an overestimate because the N$^3$LO corrections come
from one insertion of the shape parameter, that is unnaturally small,
three insertions of the range correction, that were already included
in the N$^2$LO calculation as discussed above and one $P-$wave
insertion.  There is a very precise experimental determination
\cite{dilg} of the quartet $S-$wave scattering length
$a_{3/2}^{exp}=6.35\pm 0.02~{\rm fm}$. The N$^2$LO result,
$a_{3/2}^{EFT}=6.33\pm 0.05~{\rm fm}$, compares well with this value
and with calculations using second generation potential models
\cite{friar}.

Since
the scattering lengths $a_d,a_t$, the deuteron binding momentum
$1/\gamma$ and effective range
$r_{0t}$ of the two body system in the $\mathrm{S}$-waves
are all input to the EFT calculation, we
expect that any  potential model which reproduces these numbers 
will  agree
with our results within the estimated error  of an N$^2$LO 
calculation, $\sim 4\,\%$~\cite{friar}.
Higher order calculations in EFT can be easily performed 
and involve only well known two-body interaction parameters.
They too should agree with the latest generation of potential models.
Only at the order where three-body forces first contribute will 
differences between
EFT calculations and realistic potential models arise. In some
observables, like spin observables in
neutron-deuteron scattering,  the contribution from three
body-forces appears at lower orders and it is here where
differences between EFT and potential models will first become apparent.

%%%%%%%  Triton Channel  %%%%%%%%%%%%%%%%%%%%

\subsection{The Triton Channel}
We know turn to the spin $1/2$, $S-$wave channel.
The essential physics point that distinguishes this channel 
from all the others is that, in the triton-$^3$He channel, all three
nucleons can occupy the same point in space while in all the others the
centrifugal barrier and/or Fermi statistics creates an effective
repulsion between the three particles at short distances.
This suggests that the ultraviolet behavior of this channel should be
similar to the case of three spin-zero bosons in a $S-$wave that also 
does not have any short distance repulsion. It turns out that the 
similarities are more
than qualitative and the spin-zero system has all the novel features
we want to illustrate.
We will discuss first the spin-zero boson case to
avoid unessential complications due to the  coupled
equations in the triton-$^3$He channel. After the conceptual issues
are discussed in the bosonic case we will turn to the triton channel.

The integral equation analogous to Eq.~(\ref{eq:fort}) for the three
bosons case is
\begin{eqnarray}\label{eq:bosons}
  t_0(k,p)&=&-\;\frac{y^2 M}{2p k}\ln\left(\frac{p^2+p k + k^2-ME}{p^2-p k
    + k^2-ME}  \right)\\
 &&+\frac{2}{\pi}\int\limits_0^\infty dq\; q^2\; t_0(k,q)\;
 \frac{1}{-\gamma+\sqrt{\frac{3 q^2}{4}-ME}}\;
 {K_0}(q,p),\nonumber
\end{eqnarray}
\noindent The difference from the quartet case  
(Eq.~(\ref{eq:fort})) is that the kernel in  Eq.~(\ref{eq:bosons}) has the
opposite sign and twice the strength. 
This is enough to make the  Eq.~(\ref{eq:bosons}) ill-posed and it can be
shown that its solution is not unique \cite{danilovleading,fadeev}.  
Some physical insight into the 
situation can be obtained by going to coordinate space. 
For that, let us perform the transformation \cite{efimoveffect}
\begin{equation}
\chi(r,\rho)=\int_0^\infty dp\  t_0(k,p) \ 
                     \frac{p \ {\rm sin}(\sqrt{3/4}p\rho)}
                          {-\gamma+\sqrt{\frac{3 p^2}{4}-M E}}\ 
               e^{-r \sqrt{\frac{3 p^2}{4}-M E}}.
\end{equation}
The wave function $\chi(r,\rho)$ satisfies
\begin{equation}
\label{eq:schroedinger}
\left( \frac{\partial^2}{\partial r^2}+\frac{\partial^2}{\partial \rho^2}+
       ME\right)\chi(r,\rho)=0,
\end{equation}
with the boundary conditions
\begin{eqnarray}
&&\chi(r,0)=0
\ \ ,\ \ 
\frac{\partial}{\partial r}\chi(0,\rho) - \gamma\chi(0,\rho)
+\frac{8}{\sqrt{3}}\frac{1}{\rho}
           \chi(\frac{\sqrt{3}}{2}\rho,\frac{1}{2}\rho)=0
\ \ .
\label{eq:boundaryconditions}
\end{eqnarray}
The last boundary condition is equivalent to  Eq.~(\ref{eq:bosons}).
Eqs.~(\ref{eq:schroedinger},\ref{eq:boundaryconditions}) could also be obtained
by considering the three particle problem in the wave-mechanics
formalism and using 
Jacobi coordinates. The variable $r$ corresponds to the distance between 
two of the particles and $\rho$ to the distance between the third particle
and the center of mass of the other two. 
Using polar coordinates $R=\sqrt{r^2+\rho^2},\  \tan\alpha=r/\rho$,
we can see that this complicated boundary condition is simpler for
$R\ll a=1/\gamma$. In this region Eq.~(\ref{eq:boundaryconditions})
becomes
\begin{equation}
\frac{\partial}{\partial\alpha}\chi(R,\alpha=0)+
\frac{8}{\sqrt{3}}\chi(R,\pi/3)=0.
\label{bc2}
\end{equation}
In the region $R\ll a$ we have then a separable problem 
with the solution
\begin{equation}    
\chi(R,\alpha)=\sum_i F_i(R)\  \sin (s_i(\pi/2-\alpha)),
\end{equation}
\noindent where the $s_i$ are the solutions of
\begin{equation}
1-\frac{8}{\sqrt{3} s}
\frac{{\rm sin}(s_i\pi/6)}{{\rm cos}(s_i \pi/2)}=0 .
\end{equation}
Eq.~(\ref{eq:s0}) has two imaginary solutions $\pm is_0$, with
$s_0\simeq 1.006$.
The equation for the radial part $F_0(R)$ is that of a  two dimensional
Schroedinger
equation with an {\it attractive} $1/R^2$ potential
\begin{equation}\label{eq:1r2}
\left(\frac{1}{R}\frac{d}{dR}R\frac{d}{dR}+\frac{s_0^2}{R^2}+ME\right)
    F_0(R)=0.\label{r2}
\end{equation}
As is well known the quantum mechanics problem with an attractive $1/R^2$
potential (or more singular potentials)
is not well defined and its hamiltonian is not bounded from
below. There are arbitrarily deep bound states so the ground state
describes the ``fall to the center''. Of course this conclusion is not 
valid as soon as the bound states start probing distances of the order 
of the cutoff. It indicates though that the spectrum will be dependent
on short distance physics.

Another way of analyzing this system is to consider the equation
describing the asymptotic behavior of the half off-shell amplitude, as we
did in the quartet case in Eq.~(\ref{eq:asymptotic}). In the case of three 
bosons we have
\begin{equation}\label{eq:asymptoticbosons}
p\;  t_0(k,p)=
 \frac{4\lambda}{\sqrt{3}\pi}\int\limits_0^\infty dq\;q\; t_0(k,q)\;
 \ln\left(\frac{p^2+p q + q^2}{p^2-p q + q^2} \right)\;\;,
\end{equation}
\noindent where we introduce a parameter $\lambda$ taking values $\lambda=1$ in
the bosonic case and $\lambda=-1/2$ in the quartet case.
Besides the symmetry under scale transformations
$p\ t_0(k,p)\rightarrow \alpha\ p\ t_0(k,\alpha p)$
Eq.~(\ref{eq:asymptoticbosons}) has a symmetry under inversions
$p\  t_0(k,p)\rightarrow \frac{\alpha}{p}\ t_0(k,\frac{\alpha}{p})$
that will be crucial later.
Using the same ansatz $t_0(k,p)\sim p^{s-1}$ as before we arrive at an
equation determining the asymptotic behavior of $t_0(k,p)$
\begin{equation}
1-\frac{8 \lambda}{\sqrt{3} s}\frac{{\rm sin}s\pi/6}{{\rm cos}(s \pi/2)}=0 
.\label{eq:s0}
\end{equation}
For $\lambda > \lambda_c=\frac{3\sqrt{3}}{4\pi}\simeq 0.4135$
Eq.~(\ref{eq:s0}) has two imaginary solutions $s=\pm i s_0$, with
$s_0\simeq 1.0062$ in the bosonic case $\lambda=1$. This means that for $p\gg
k,\gamma$, $t_0(k,p)$ is a linear combination of $p^{i s_0-1}$ and
$p^{-i s_0-1}$ or
\begin{equation}\label{eq:coslog}
t_0(k,p)=\frac{A}{p} \cos(s_0\log(p)+\delta).
\end{equation}
\noindent The amplitude $A$ is determined by matching the asymptotic
solution in Eq.~(\ref{eq:coslog}) to the full solution in the infrared
region. The phase however remains undetermined. This is another way of 
seeing that there is a one parameter family of solutions to the
equation Eq.~(\ref{eq:bosons}). Of course this undeterminacy is an
artifact of extrapolating  the low energy effective theory to the
ultraviolet region and is eliminated by the introduction of an ultraviolet 
regulator. 

A perturbation theory on the strength of the kernel ($\lambda$) corresponds
to the diagrams on the top line of Fig.~\ref{fig:quartetfadeev}.
The behavior shown in Eq.~(\ref{eq:coslog}) is missed in all orders of
this expansion as can be seen
by expanding Eq.~(\ref{eq:coslog}) in powers of the
kernel strength $\lambda$ ( the dependence on $\lambda$ is implicit in $s_0$)
\begin{equation}
\frac{A}{p} \cos(s_0\log(p)+\delta)=0 + 0 \ \lambda + 0 \ \lambda^2 + \cdots
\end{equation}
\noindent This explains why a perturbative analysis of the ultraviolet behavior
of $t_0(k,p)$ misses the oscillatory behavior in Eq.~(\ref{eq:coslog}).
\begin{figure}[!htb]
\vskip 0.0in
\hskip 0.8in
\psfig{file=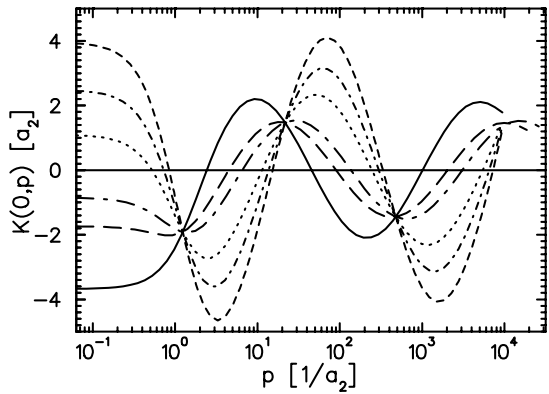,height=2.0in}
 \caption{Half off-shell (K-matrix) amplitudes in
the bosonic case for different values of the momentum cutoff $\Lambda$
in units of the two-body scattering length $a_2$.}
\label{fig:oscillations}
\end{figure}

In Fig.~\ref{fig:oscillations} we can see 
numerical solutions of Eq.~(\ref{eq:bosons}) (or more precisely, the
K-matrix corresponding to the amplitude $t_0(k,p)$  satisfying
Eq.~(\ref{eq:bosons}) with a principal value instead of an $i\epsilon$
prescription) for $k=0$ obtained with different values of the 
momentum cutoff, $\Lambda$. 
The presence of $\Lambda$
breaks the inversion symmetry and selects a phase in
Eq.~(\ref{eq:coslog}),
and further, different values of $\Lambda$
result in very different values of
$t_0(k=0,k=0)$. 
This  $\Lambda$
dependence of a physical quantity indicates something is missing
in our calculation up to now. 
As it appears in
the three-body amplitudes it should be
absorbed by the cutoff dependence of a  three-body force. 
We see now that we were unjustified in dropping the three-body force term from
the integral equation in the case of bosons. To reinstate the three-body
force shown in  Eq.~(\ref{eq:lagthree}) we include the term 
\begin{equation}
{\mathcal L}_{d\ 3}=-\frac{M H y^2}{\Lambda^2}\ d^\dagger d N^\dagger N 
\end{equation}
\noindent in the Lagrangean and choose
$H=D_0 \frac{\Lambda^2\Delta^2}{M y^4}$. 
The integration over the field $d$ generates the three-body force in  Eq.~(\ref{eq:lagthree}).
Graphs containing the three body force 
are now included as shown in 
Fig.~\ref{fig:threebodyforcefadeev}
\begin{figure}[!htb]
\vskip 0.3in
\hskip 0.8in
\psfig{file=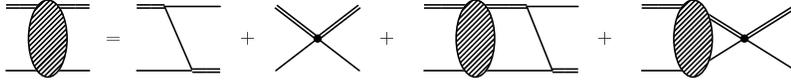,height=0.7in,bbllx=60pt,bblly=690pt,bburx=560pt,bbury=770pt}
\caption{ Equation for the three-body amplitude including the three-body force.}
\label{fig:threebodyforcefadeev}
\end{figure}

The new graphs with a three-body interaction 
change the kernel to
\begin{equation}\label{eq:Hkernel}
{K_0}(q,p)=\frac{1}{qp}\;\left[
 \ln\left(\frac{p^2+p q + q^2-ME}{p^2-p q + q^2-ME} \right)
+\frac{H}{\Lambda^2}\right].
\end{equation}
\noindent It is not currently known how to dimensionally regulate the three-body
integral equation, and from now on we will use a sharp momentum cutoff 
instead
(some progress in this direction was made in the case of a similar 
equation in \cite{Beane:2000ji}).
To be more precise, we will be using two distinct cutoffs: one
regulating the two-body loops in the deuteron propagator that is taken 
to infinity and
another regulating the three-body loops. 
That is why the  kernel of the three-body equation
contains the renormalized form of the deuteron propagator. 
This procedure is justified since keeping an
unrenormalized form for the deuteron propagator amounts to keeping some
sub-leading terms in the kernel that are suppressed by powers of
$q/\Lambda$. These terms are relevant only at the cutoff scale $q\sim
\Lambda$ and can change only non-universal features, like the value of
the bare coupling constants, but not the final physical results.

For $H(\Lambda)\sim 1$ the effect of the three-body force in
Eq.~(\ref{eq:Hkernel})
is important only for $q\sim \Lambda$ thus our analysis of the
asymptotic behavior of the amplitude remains valid in the intermediate 
region $\gamma,k \ll p \ll \Lambda$. A simple argument now shows that
$H(\Lambda)$ can be chosen in such a way as to maintain the 
low-momentum part of the amplitude $t_0(k,p\ll \Lambda)$ and, in
particular, the on-shell point $p=k$, cutoff independent (up to terms
of order $k/\Lambda$). 
For that we define a scale $\gamma \ll \mu\ll\Lambda$ and
rewrite equation Eq.~(\ref{eq:bosons}) as
\begin{eqnarray}
t_0^\Lambda(k,p)&=&
K(k,p) + \frac{2}{\pi}\int_0^\mu dq q^2 t_0^\Lambda(k,q) K_0(q,p)
\nonumber\\
&+&\frac{2}{\pi}\int_\mu^\Lambda dq q^2
t_0^\Lambda(k,q)(K_0(q,p)+\frac{H(\Lambda)}{\Lambda^2} )
+{\mathcal O}(p/\Lambda, \gamma/\Lambda).
\end{eqnarray}
The superscript $\Lambda$  reminds us that $t_0^\Lambda(k,p)$ is the
solution of the integral equation with a particular cutoff $\Lambda$.
Taking another cutoff $\Lambda'$ we can write the equation for
$t_0^{\Lambda'}(k,p)$ as
\begin{eqnarray}
\label{eq:nightmare}
t_0^{\Lambda'}(k,p)&=&K(k,p) 
+ \frac{2}{\pi}\int_0^\mu\ dq q^2 t_0^{\Lambda'}(k,q) K_0(q,p)\nonumber\\
&+& \frac{2}{\pi}\int_\mu^\Lambda\ dq q^2
t_0^{\Lambda'}(k,q)(K_0(q,p)+\frac{H(\Lambda)}{\Lambda^2} )\nonumber\\
&+&\frac{2}{\pi}\int_\Lambda^{\Lambda'}\ dq q^2
t_0^{\Lambda'}(k,q)\left(\frac{H(\Lambda')}{\Lambda{'^2}}+
\frac{1}{q^2}\right)\nonumber\\
&+&\frac{2}{\pi}\int_\mu^\Lambda\ dq q^2
t_0^{\Lambda'}(k,q)\left(\frac{H(\Lambda')}
{\Lambda{'^2}}-\frac{H(\Lambda)}{\Lambda^2}
\right)
\ +\ {\mathcal O}({p\over\Lambda}, {\gamma\over\Lambda}).
\end{eqnarray}
If $H(\Lambda)$  can be
chosen in such a way as to make the terms in the last two lines
of Eq.~(\ref{eq:nightmare}) vanish, the
equations satisfied by $t_0^\Lambda(k,q) $ and $t_0^{\Lambda'}(k,q)$
are the same and, since the equation with a finite cutoff has an
unique solution, the amplitudes are the same. Using the
asymptotic form of the amplitudes
\begin{equation}
t_0^\Lambda(k,\gamma \ll p \ll \Lambda)=t_0^{\Lambda'}(k,\gamma \ll p
\ll \Lambda')
= \frac{A}{p}\cos(s_0 \log(p)/\bar\Lambda),
\end{equation}
\noindent where $\bar\Lambda$ is an arbitrary scale fixing the asymptotic phase
we find an approximate form for $H(\Lambda)$
\begin{equation}
\label{eq:H}
H(\Lambda)\ =\ 
- \ {\sin\left( s_0 \log\left[\Lambda/\overline{\Lambda}\right] 
                 - \arctan\left[1/s_0\right] \right)
\over
\sin\left( s_0 \log\left[\Lambda/\overline{\Lambda}\right] 
                 + \arctan\left[1/s_0\right] \right)
}
\ \ \ 
\end{equation}
\noindent which assures $\Lambda$ independence.
The precise form of $H(\Lambda)$ in Eq.~(\ref{eq:H}) might dependent on
the particular shape of the cutoff, but its qualitative behavior, in
particular the existence of a limit cycle in RG flow, are general.

We can check numerically the argument leading to Eq.~(\ref{eq:H}). For
this we can fix (arbitrarily) the value of the three-body scattering
length and
determine numerically the value of $H(\Lambda)$ needed to reproduce this
same scattering length for any given value of $\Lambda$. This can always
be done since we have one free parameter ($H(\Lambda)$) to fit another
(the three-body scattering length). Using these determined values of
$H(\Lambda)$ we can check that the amplitudes  at finite $k$ are indeed
cutoff independent (up to terms of order $k/\Lambda, \Lambda a$). The
result of this procedure is shown as the dots in
Fig.~\ref{fig:hansplot}. The agreement with Eq.~(\ref{eq:H}) lends
support to the arguments leading to it.
\begin{figure}[!htb]
\vskip 0.0in
\hskip 0.4in
\psfig{file=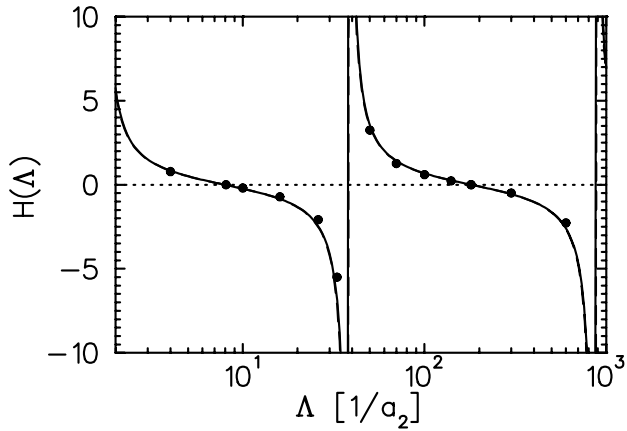,height=2.2in}
 \caption{ Running of the three-body force around a limit cycle. 
The solid line 
shows Eq.~(\ref{eq:H}) and the dots the numerical procedure described in 
the text.}
\label{fig:hansplot}
\end{figure}

The appearance of a new parameter (that can be taken to be the value
$H(\Lambda)$ at some particular value of $\Lambda$ or the scale
$\bar\Lambda$) that is {\it not} determined by two-body physics is the
main result of this analysis.  It implies that different models having
the same two-body scattering length can have quite different
three-body physics. This is shown in a striking manner by plotting the
prediction of a number of different models with very similar low
energy ($p<m_\pi$) two-body physics for two three-body observables
like, for instance, the three-body bound state energy ($B_3$) and the
scattering length of a particle colliding with the bound state of two
others ($a_3$) \cite{phillips}. It is known from the numerical studies
of potential models that those predictions can differ wildly among
different models, but not to scatter all over the $a_3\times B_3$
plane either. They tend to lie along a reasonably well defined line
(Phillips line, see Fig.~\ref{fig:phillips}).  For the point of view
discussed here each point on the Phillips line correspond to a
different value of the three-body parameter $\bar\Lambda$.
\begin{figure}[!htb]
\vskip 0.0in
\hskip 0.6in
\psfig{file=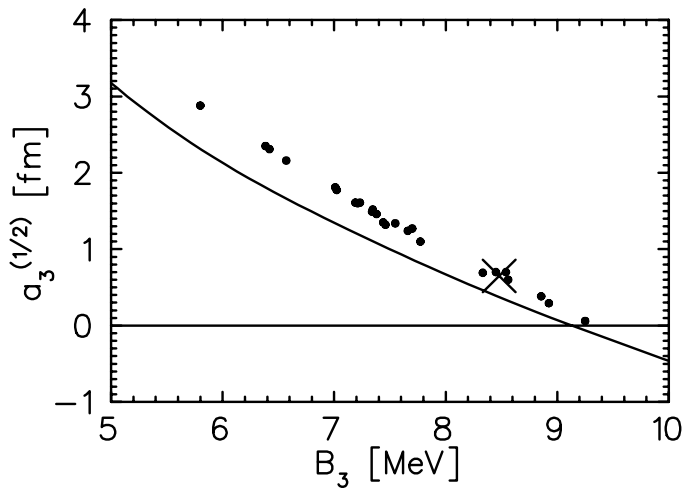,height=2.5in}
\caption{ Results for the doublet $S-$wave
neutron-deuteron scattering length for different models having very similar
two-body scattering lengths and effective ranges, denoted by the dots.
The solid line is the EFT prediction over a range of values for 
$\overline{\Lambda}$. The
cross indicates the experimental value.}
\label{fig:phillips}
\end{figure}

The existence of the Phillips line is usually understood as the result 
of the dependence of the three-body observables on the particular
off-shell extension of the two-body data described by the different
potentials \cite{afnan}. 
This is equivalent to the explanation discussed above since field
redefinitions change both off-shell behavior of amplitudes and
strengths of three-body forces. Looking at those ``off-shell
ambiguities'' as values of short distance counterterms has
the advantage of being a systematic way of mapping them out
and estimating their sizes. In particular it
becomes clear that the ``off-shell ambiguity'' related to the
three-body bound state cannot be fixed by two-body processes involving 
external currents since those are related to different  counterterms.

We now have the understanding necessary to tackle the triton-$^3$He
channel, described by Eq.~(\ref{eq:coupled}) \cite{Bedaque:1999ve}.
There is an interesting limit where   
Eqs.~(\ref{eq:coupled}) decouple. 
The two spin and isospin states of the nucleons can
be put in a $\bf{4}$ representation of 
spin-flavor $SU(4)$. This symmetry is
broken in leading order by the difference between the scattering
lengths in the $^3S_1$ ($\gamma$) and $^1S_0$ ($a$) channels but is  an
approximate symmetry of the system when the typical momenta is $Q\gg
\gamma,1/a$ since
in this case we can take $\gamma\simeq 1/a\simeq 0$. This approximate
symmetry is observed in light nuclei and has interesting consequences 
\cite{wignersu4,Mehen:1999qs}.
In the $SU(4)$ limit the system of equations Eq.~(\ref{eq:coupled}) 
break up into two
equations: one for the sum $t_+(k,p)=t_d(k,p)+t_t(k,p)$ and  
the other for the difference $t_-(k,p)=t_d(k,p)-t_t(k,p)$.  
The three-body force in
Eq.~(\ref{eq:lagthree}) contributes only to the bosonic-like 
equation for $t_-(k,p)$.
It is actually the only three-body force with no derivatives one can
write down, as a simple argument shows.
Fermi statistics
mandates that the three nucleons should be in a completely
antisymmetric representation of $SU(4)$. 
Consequently, the three-body force transforms as
${\bf 4}\otimes\bar{\bf 4}={\bf 1}\oplus{\bf 15}$, but only the 
singlet is separately invariant under the spin and isospin subgroups.
The three-body force in Eq.~(\ref{eq:lagthree}) is generated by the
inclusion of the term
\begin{eqnarray}
{\mathcal L}_{d,\ 3}&=&\frac{2MH(\Lambda)}{\Lambda^2}
   \bigg[ y_d^2 N^\dagger 
(d^i\sigma^i)^\dagger (d^j\sigma^j) N   
+ y_t^2 N^\dagger (t^a \tau^a)^\dagger (t^b \tau^b) N
\,\nonumber\\
&+&\frac{1}{3} y_t y_s \left[ N^\dagger (d^i\sigma^i)^\dagger
(t^a\tau^a) N + h.c. \right] \bigg]
\end{eqnarray}
Upon integrating the field $d^i$ and $t^a$ out we recover
Eq.~(\ref{eq:lagthree}).
\begin{figure}[!htb]
\vskip 0.0in
\hskip 0.5in
\psfig{file=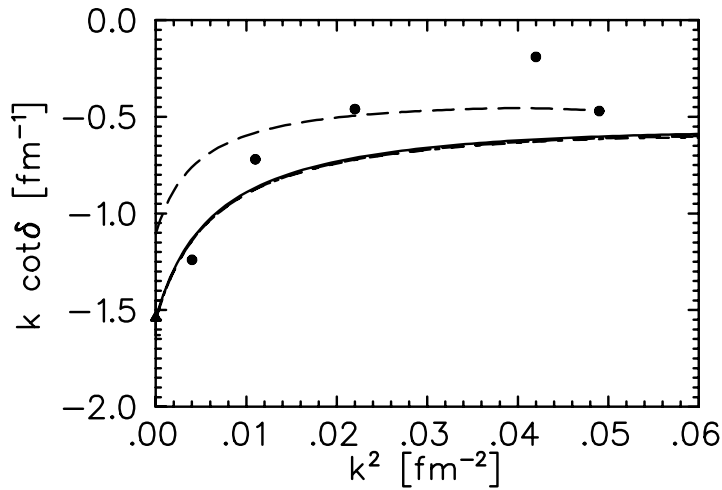,height=2.5in}

\caption{ Doublet $S-$wave neutron-deuteron phase shifts.}
\label{fig:doubletphase}
\end{figure}
Fitting $\bar\Lambda$ to the value of the
experimental
doublet scattering length $a_{1/2}^{exp}$ we arrive at the solid
curve in  Fig.~\ref{fig:doubletphase}. Also shown in 
Fig.~\ref{fig:doubletphase} are the phase shift 
analysis of ~\cite{huttel,schmelzbach} and the experimental value 
of the doublet scattering length.
Using those parameters the prediction for the triton binding energy is 
of $B_3=8.03$ MeV, to be compared with the experimental 
value $B_3^{exp}=8.48$ MeV.
The dashed line was obtained by using
the experimental value of the two-body scattering length, instead of
the deuteron binding energy, to fix $\CzeroT$, while keeping the
three-body force constant. Both procedures differ only by higher
orders terms so their difference provides an estimate of the
higher order corrections.
Comparison with  experiment suggests that the low energy
expansion in the triton
is under control since the errors are within the expected value
for a LO calculation. A more solid statement can
be made only when higher order calculations become available.
A natural question
to ask  then is  whether the three-body force needed to give the
triton its experimental binding energy is ``natural''. For cutoffs of
the order $\Lambda\simeq m_\pi$ , $H(\simeq m_\pi)\simeq 1$ suggesting 
that the relative shallowness of the triton does not require any
further fine tuning besides that in the two-body sector.

Compared to the two-body sector the study of the 
three-nucleon problem with EFT methods is still in its infancy. A
number of conceptual and technical problems have to be understood
before high order calculations can be performed. Among them are the
complete power counting to all orders and the scaling of the
sub-leading three-body forces 
(some recent progress was report in Ref.~\cite{wilson}).

%%%%%%%%%%   Higher Energies   %%%%%%%%%%%%%%%%%%

\section{$NN$ Interactions at Higher Energies}

In order to make progress toward describing multi-nucleon systems,
such as nuclei, an EFT must be constructed that will describe the $NN$
system at momenta $|{\bf p}|\gsim m_\pi$.  It is obvious that one must
include the $\pi$-field explicitly in this kinematic region.  However,
a power-counting must be developed that is both systematic and
convergent.  In what follows we will describe the two different
power-counting schemes that currently exist in this kinematic regime.
Weinberg's power-counting\cite{We90,We91} is based on a momentum and
chiral expansion of the $NN$ potential, in which the momentum
independent, local four-nucleon interaction and one-pion-exchange
(OPE) both enter at leading order.  Higher derivative operators and
the light-quark mass matrix occur at higher orders and are inserted in
perturbation theory.  Calculations that have been performed to date,
indicate that the perturbative expansion appears to be converging in
the two-nucleon sector, as we shall discuss subsequently, however,
there are formal issues that remain to be resolved.  The jury is still
out in the three-nucleon sector, but encouraging results are being
obtained.  In contrast, KSW power-counting provides an explicit
framework in which to systematically expand S-matrix elements as
opposed to an expansion of the (unobservable) potential.  Calculations
are renormalization scale independent order-by-order in the expansion
parameter.  However, a calculation of S-wave $NN$ scattering at
N$^2$LO by Fleming, Mehen and Stewart\cite{Fl00} indicates that, in
fact, treating the pions perturbatively is not converging for momenta
$|{\bf p}|\gsim m_\pi/2$.  They were able to identify the one large
contribution that occurs at N$^2$LO, and found that it persists in the
$m_q\rightarrow 0$ chiral limit.  Hopefully this problem can be
understood and solved in the near future.

This is essentially the current status of the field.  In what follows
both power-countings will be discussed, both their strengths and
currently understood weaknesses.  Emphasis will be placed on the
Weinberg counting due to its present numerical successes and lack of
numerical failures, despite some unresolved formal issues.

\section{Perturbative $\pi$-Exchange and its Problems}

Let us recall how pions couple to nucleons.
The Lagrange density describing the interactions between pions and 
nucleons is
\begin{eqnarray}
{\cal L} & = &  {f_\pi^2\over 4} {\rm Tr} D_\mu \Sigma
D^\mu \Sigma^\dagger
+ 
{ig_A\over 2} N^\dagger {\bf \sigma} \cdot (\xi {\bf D} \xi^\dagger 
- \xi^{\dagger} {\bf
D} \xi)N
\ +\ \cdots
\ \ \ ,
\label{eq:lagOPE}
\end{eqnarray}
where the pion fields are contained
in a special unitary matrix,
\begin{equation}
\Sigma = \xi^2 = \exp {\sqrt{2} i\Pi\over f},\qquad \Pi =
\left(\begin{array}{cc}
\pi^0/\sqrt{2} & \pi^+\\ \pi^- & -\pi^0/\sqrt{2}\end{array} \right)
\ \ \  ,
\label{eq:pimatrix}
\end{equation}
with $f=93\ \MeV$ and
$g_A=1.26$ is the axial coupling constant.
The ellipses represent operators involving more insertions of the light
quark mass matrix, meson fields, and spatial derivatives. 
The leading-order contribution to the $NN$ potential due to the exchange of 
a single ``potential pion''~\footnote{The variation of the pion propagator 
due to the energy transfer between the 
two nucleons is negligible compared to the variation due to the 
three-momentum transfer.} is:
\begin{eqnarray}
V_\pi ({\bf q})
& = & 
-{g_A^2\over 4 f_\pi^2} \ 
{\left({\bf q}\cdot {\bf \sigma}^1\right)
\left({\bf q}\cdot {\bf \sigma}^2\right)
\over |{\bf q}|^2 + m_\pi^2}\ 
\vec{\tau}_1 \cdot \vec{\tau}_2,
\label{eq:OPEpot}
\end{eqnarray}
where ${\bf \sigma}^k$ denotes the spin operator acting on the $k$th
nucleon.  It is clear that this interaction does not have any
dependence upon the renormalization scale $\mu$.  Since the $C_0$
operator scales as $1/(M \mu)$ it would appear that as the system is
evolved into the infrared, the short-range contact interaction will
dominate over one-pion exchange.  From the standpoint of conventional
nuclear physics this is a shocking claim. The tensor force in the
$\siii-\diii$ channel is known to be strong, and this casts doubt on
the applicability of a perturbative treatment of the pion exchanges.
The question that must be addressed is what is the kinematic regime
where this hierarchy is actually present.  The observation that there
is a kinematic regime where pions can be treated in perturbation
theory was first made by Kaplan, Savage and Wise~\cite{Ka98A,Ka98B}
(KSW), and the associated power-counting worked out in these papers is
commonly known as KSW-counting.  KSW-counting has the very attractive
feature that analytic expressions for any scattering amplitude can be
found.

\subsection{$NN$ Scattering with Perturbative Pions}

The $NN$ scattering amplitude has now been computed up to N$^2$LO\cite{Fl00}
in KSW counting. The LO contribution to the amplitude is exactly the
same as in the pionless theory, since, as mentioned above, the pion
scales as $Q^0$, while $C_0$, and the whole bubble chain contributing
to ${\cal A}^{(-1)}$ scale as $Q^{-1}$. Thus, once more:
\begin{equation}
{\cal A}_{-1}  =
-{ C^{(\si)}_0\over  1 + C^{(\si)}_0 {M\over 4\pi}  
\left( \mu + i p \right) 
}\ \ .
\end{equation}

%%%%%%%%%% 1S0 diagrams  %%%%%%%%%%%%%%%
\begin{figure}[!ht]
\hskip 0.5in\psfig{figure=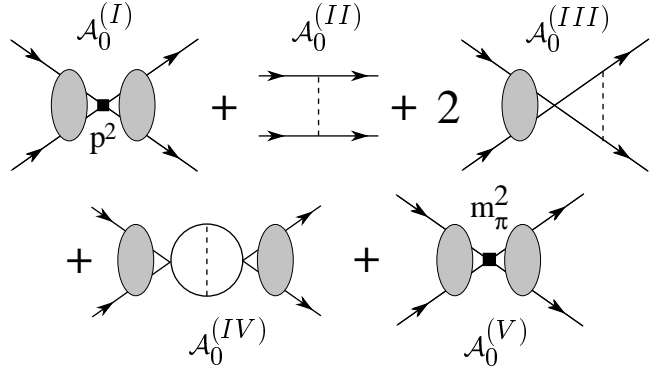,height=2.0in}
\caption{
Feynman diagrams contributing to $NN$ scattering in the 
$\si$ channel.}
\label{fig:S-diags}
\end{figure}

Meanwhile, the Feynman diagrams contributing to the 
NLO amplitude are shown 
in Fig.~\ref{fig:S-diags} and their contribution is
the sum ${\cal A}_0={\cal A}_0^{(I)}+
{\cal A}_0^{(II)}+{\cal A}_0^{(II)}+{\cal A}_0^{(IV)}+{\cal A}_0^{(V)}$.
The local operators at this order involve either two spatial derivatives, 
with coefficient $C_2^{(\si)}$,
or one insertion of the light quark mass matrix, with coefficient 
$D_2^{(\si)}$. Thus, the amplitude is
\begin{eqnarray}
{\cal A}_0^{(I)} \! &=&\! 
-C_2^{(\si)} p^2
\left[ {\CA_{-1}\over C_0^{(\si)}  } \right]^2
\ , \ 
 {\cal A}_0^{(II)}\ = \  
-{g_A^2\over 4f^2} \left(1 -{m_\pi^2\over
4p^2} \ln \left( 1 + {4p^2\over m_\pi^2}\right)\right)
\ ,
\nonumber\\
 {\cal A}_{0}^{(III)} \!&=&\! -{g_A^2\over 2 f^2} 
 \left( {m_\pi M{\cal A}_{-1}\over 4\pi}
\right) \Bigg( {\mu + ip\over m_\pi}
- i {m_\pi\over 2p} X_\pi \Bigg)
\ ,
\nonumber\\
{\cal A}_0^{(IV)} \! &=&\!  -{g_A^2\over 4f^2} 
\left({m_\pi M{\cal A}_{-1}\over 4\pi}\right)^2 
\Bigg(\left({\mu + ip\over m_\pi}\right)^2
- X_\pi - \log\left({m_\pi\over\mu}\right)+ 1\Bigg)
\ ,
\nonumber\\
{\cal A}_0^{(V)} \! &=&\!  - D^{(\si)}_2 m_\pi^2 
\left[ {\CA_{-1}\over C_0^{(\si)}  }\right]^2
\ ,
\nonumber\\
X_\pi \! & = &\!   {1\over 2} 
\ln\left(1+ {4p^2\over m_\pi^2} \right)
- i \tan^{-1} \left({2p\over m_\pi}\right)
\ .
\label{eq:szeroamps}
\end{eqnarray}
The resulting $^1S_0$ phase shifts at LO, NLO~\cite{Ka98B} and 
N$^2$LO~\cite{Fl00} are  shown in Fig.~\ref{fig:S-phase}.
It should be noted that these results are for a particular
fitting procedure and appear to be converging very well. 
Detailed discussions of various fitting
procedures, and whether the pion can truly be treated
in perturbation theory in the ${}^1S_0$ channel,
can be found in Refs.~\cite{Ge98C,CH98A,CH98B,SF98B,KS99,RS99A,RS99B}.
%
%%%%%%%%%% 1S0 phase shift Pis  %%%%%%%%%%%%%%%
\begin{figure}[!ht]
\hskip 1.3in\psfig{figure=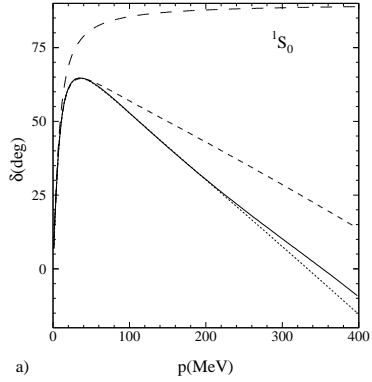,height=2.0in}
\caption{
Fit to the $\si$ phase shift $\delta$.  
The solid line is the Nijmegen fit \protect\cite{St94} to the data. 
The long dashed, short dashed, and dotted lines 
are the LO, NLO, and N$^2$LO results respectively.}
\label{fig:S-phase}
\end{figure}

A similar calculation has been performed in the $\siii-\diii$ coupled 
channels but we will not present
the expressions for the amplitudes here~\cite{Ka98B,Fl00}. 
The resultant function yields the $\siii$
phase shifts shown in Fig.~\ref{fig:St-phase}.
%
%%%%%%%%%% 3s1 and ep1 phase shift in KSW %%%%%%%%%%%%%%%
\begin{figure}[!ht]
\hskip 0.1in\psfig{figure=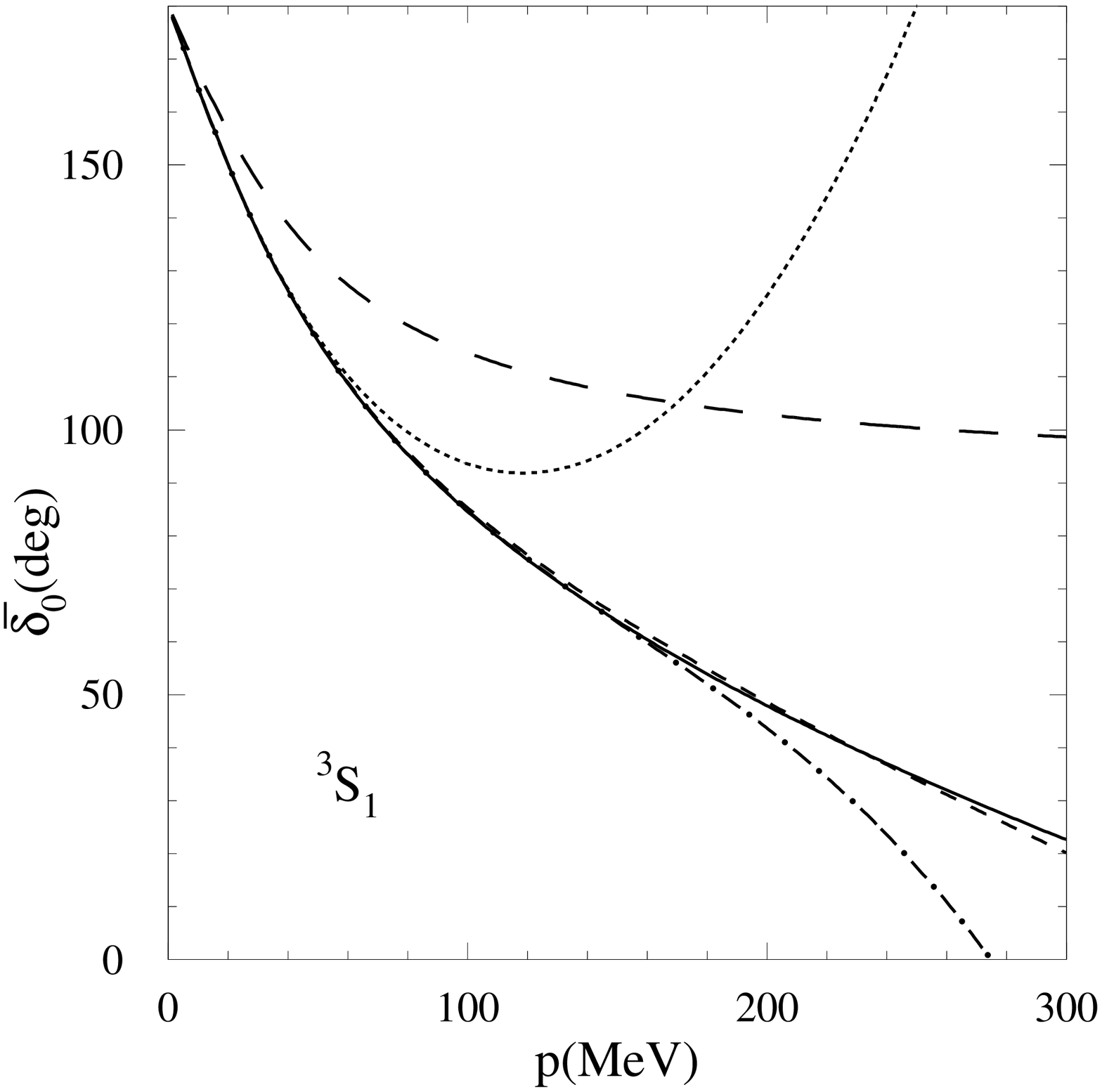,height=2.0in}
\hskip 0.1in\psfig{figure=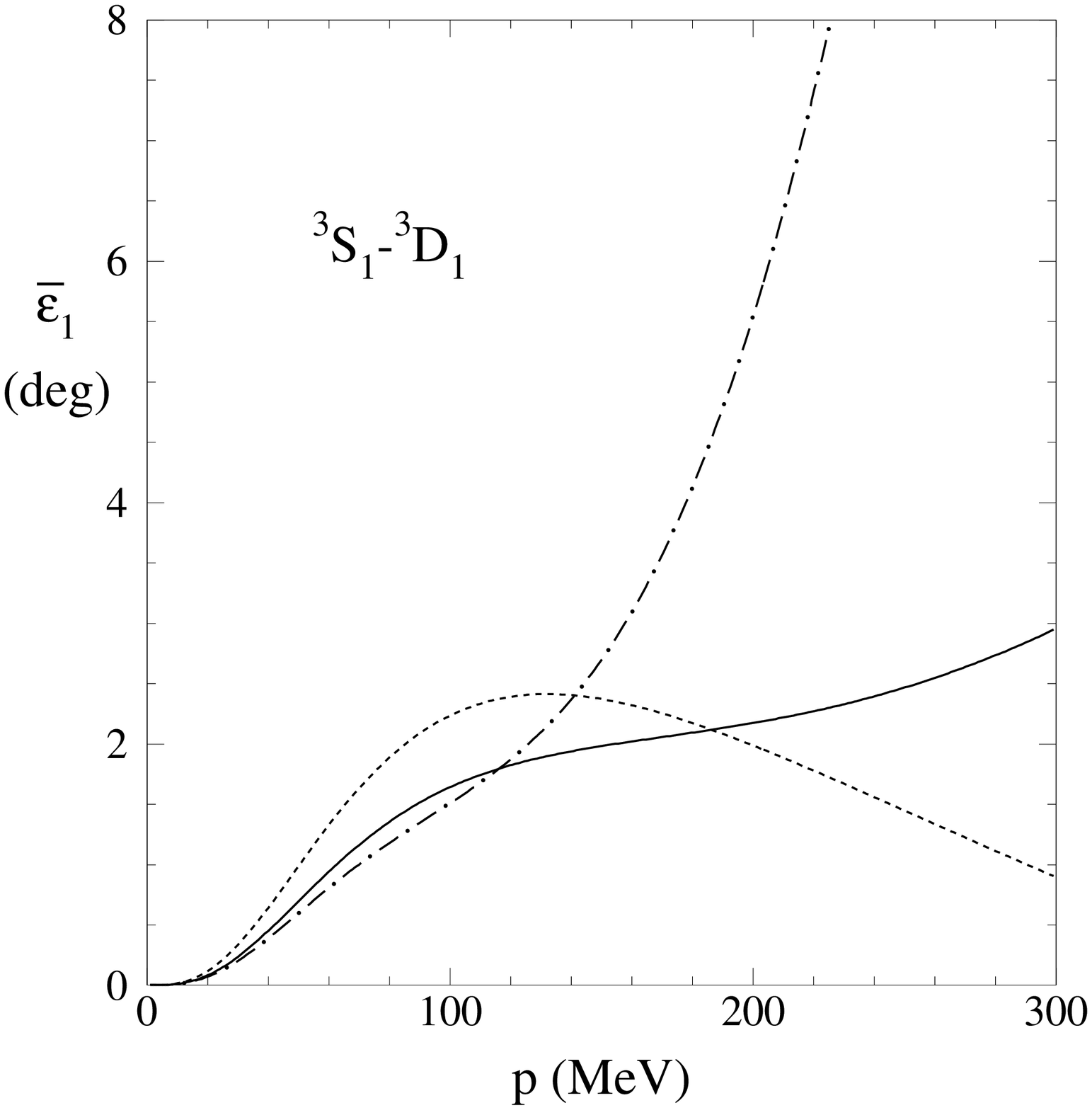,height=2.0in}
\caption{
The $\siii-\siii$ phase shift, $\overline{\delta}_0$
and $\siii-\diii$ mixing parameter, 
$\overline{\varepsilon}_1$ for $NN$ scattering.  
The solid line is the Nijmegen multi-energy 
fit~\protect\cite{St94}, 
the long dashed line is the LO
effective field theory result, 
the short dashed line is the NLO result, and the
dotted line is the N$^2$LO result.  
The dash-dotted line shows the result of
including the parameter $\zeta_5$ 
(defined in Ref.~\protect\cite{Fl00})
which is \emph{higher order} in the power
counting}
\label{fig:St-phase}
\end{figure}
From Fig.~\ref{fig:St-phase} it is clear that the N$^2$LO
phase shift is deviating significantly from the data at 
momenta  $|{\bf p}|\gsim m_\pi/2$.
Given that one expects the perturbative series to converge to the
data at successively higher orders, it would appear that the 
KSW-expansion is in trouble.

This conclusion is reinforced by looking at the $\siii-\diii$ 
mixing parameter, $\overline{\varepsilon}_1$,
defined above. This observable has also
been computed up to N$^2$LO\cite{Fl99A}.
The results at LO, NLO and N$^2$LO are shown 
in Fig.~\ref{fig:St-phase}.

The curves in Fig.~\ref{fig:St-phase} suggest that
KSW power-counting is failing in the $\siii-\diii$ 
coupled channels.
The origin of this failure in the $\siii$-channel
at N$^2$LO has been tracked down to a term
that survives in the chiral limit
\begin{eqnarray}\label{A1lim}
{\cal A}_1^{\siii-\siii} &\simeq&  6\
\left[\ {\cal A}_{-1}^{\siii-\siii}\ \right]^2\ 
{M \over 4 \pi}
 \left({M g_A^2 \over 8 \pi f_\pi^2}\right)^2\ 
 {\pi\over 2}\  |{\bf p}|^3
\label{eq:cause} 
\end{eqnarray} 
where ${\cal A}_{-1}^{\siii-\siii}$ is
the LO $\siii-\siii$ amplitude.  The N$^2$LO amplitude in
Eq.~(\ref{eq:cause}) is large because of the coefficient of $6$ which
is to be compared with the expansion parameter of $\sim 1/3$.  
The fact
that this correction survives in the chiral limit indicates that it
comes from the short distance part of potential pion exchange. 
The problem one encounters in this
coupled channel is really due to the large strength of the tensor
force.  Of course, in the $\siii-\siii$ amplitude this problem only
arises at N$^2$LO, since that is where two-pion-exchange diagrams
first appear. A method to deal with the large non-analytic
terms of Eq.~(\ref{eq:cause}) has yet to be uncovered, although there
are hopes that the solution will come from a better understanding
of the renormalization of singular potentials.

The phenomenology of KSW power-counting in the two-nucleon sector at
relatively low orders has been extensively explored in a number of
processes which are of current experimental interest.  However, none
of the calculations of these processes were carried out to an order
sufficiently high that the strong N$^2$LO correction to ${}^3S_1$
scattering played a role.  Therefore, it may be that these results
represent only low-order terms in a series which ultimately does not
converge. None of the results obtained using KSW power-counting can be
taken too seriously until the convergence issue is resolved. The fact
that the KSW power-counting allows for all calculations to be performed
analytically with dimensional regularization means that it is very
easy to perform gauge invariant calculations for processes involving
photons and other gauge fields.  In particular, $e d$ scattering has
been calculated up to NLO~\cite{Ka98C}, $\gamma d\rightarrow \gamma d$
Compton scattering has been computed (both for unpolarized and
tensor-polarized targets) up to NLO~\cite{Che98A,Che98C}, the
polarizabilities of the deuteron have been computed to high
orders~\cite{Che98B}, parity violation in $np\rightarrow d\gamma$
radiative capture has been explored~\cite{KSSW99}, and also parity
violation in $e d$ scattering due to the deuteron anapole moment has
been investigated~\cite{SS99A,SS99B}. 

Calculations with perturbative pions have also been performed in the
three-body sector~\cite{Bedaque:1999vb}.  
In the $J=3/2$, S-wave channel
there are no unknown
coefficients appearing at NLO, and thus there is no dependence of the
result on any fitting procedure.  The LO calculation involves only
two-body contact interactions.  The NLO corrections come from one
potential pion exchange, taken in perturbation theory. They are of two
kinds: corrections to the deuteron propagator involving only two
nucleons and true three-body diagrams, where the three-nucleons
participate. The diagrams have to be computed numerically since the
half off-shell amplitude $t_0(k,p)$ (solution of Eq.~(\ref{eq:fort}))
is known only numerically. As discussed before, the contribution from
the tensor force vanishes in linear order in the S-wave so there is no
contribution from the tensor part of the one-pion exchange. The
results for the phase shift are shown in Fig.~\ref{fig:quartetksw},
together with the extrapolation from the result obtained with the
theory without pions and phase shift analysis, where it is
available. The results indicate convergence up to very high energies
but are barely distinguishable from the pionless theory. This is
probably a consequence of the high anomalous dimension of the
amplitude $t_0(k,p)$, that decays to zero very rapidly at large $p$.
%
%%%%%%%   Figure  %%%%%%%%%%%
\begin{figure}[t]
\hskip 0.8in\psfig{figure=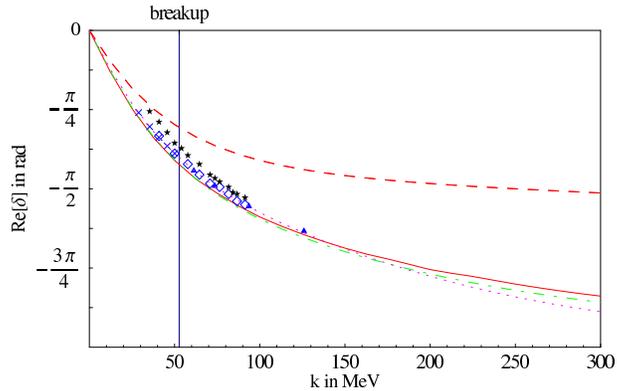,height=2.0in}
\caption{
Neutron-deuteron phase shifts in the S-wave quartet in
LO (dashed), NLO (full), NLO in the pionless theory 
(dot-dashed) and
N$^2$LO in the pionless theory (dotted). Also
shown are the results of PSA of neutron-deuteron
(squares) and proton-deuteron scattering (stars). }
\label{fig:quartetksw}
\end{figure}

\section{Don't Mess with Texas: the Weinberg Program for Physics
of the Two-Nucleon System}

\label{sec-Weinberg}

\subsection{The Proposal}
\label{sec-Weinintro}

One way of dealing with the non-perturbative nature of the pion
exchanges uncovered by higher-order calculations in KSW counting is to
follow Weinberg's original proposal for the $NN$
system~\cite{We90,We91}.  In this proposal one uses baryon $\chi PT$
to compute the $NN$ potential $V_{NN}$ perturbatively. The S-matrix is
then generated by the Schr\"odinger equation.  In this approach,
power-counting is not manifest in the amplitudes for individual
processes, but there is a naive {\it a priori} estimate of the error
in a particular calculation, which must be checked {\it a posteriori}.

Weinberg's separation of all graphs contributing to some process in
the $NN$ system into graphs which are ``two-particle reducible" and
those which are not makes power-counting much simpler.  Since, by
definition, the ``two-particle irreducible" (2PI) graphs do not
contain the low-energy scales $B$ (deuteron binding energy)
and $p^2/M$.  
The 2PI graphs are
precisely those where the energy of all internal legs is of order
$m_\pi$.  It follows immediately that the 2PI graphs have
the power-counting of ordinary $\chi$PT.

The ``Weinberg program" for calculating properties of the $NN$ system
in $\chi$PT then centers around the construction of these 2PI
graphs. Its implementation has three stages.  
Firstly, we must
consider the case where the $NN$ system is not subject to any external
probes, in which case the $NN$ irreducible diagrams just yield the
potential $V$ (e.g. the left-hand graph of
Fig.~\ref{fig-iterOPE}.)  
Constructing an $NN$ potential using the
rules of $\chi$PT will then yield the $NN$ interaction, $V^{(n)}$, up
to a given order $n$ in the chiral power counting.

The second step is to obtain the full $NN$ t-matrix by iterating these
irreducible graphs together with the free two-nucleon Green's
function, $G_0$:
\begin{equation}
G_0(E;{\bf p})={1 \over E - {{\bf p}^2 \over M} + i \eta}.
\end{equation}
($\eta$ is a positive infinitesimal.)  This free Green's function can,
of course, have a small energy denominator that does not obey the
usual $\chi$PT power-counting. Iteration of $V$ with $G_0$ gives
contributions to the t-matrix such as the graph shown on the
right-hand side of Fig.~\ref{fig-iterOPE}. This simply means that we
follow the usual practice in the $NN$ system, and construct the full $NN$
t-matrix, $T$, from the chiral $NN$ interaction $V^{(n)}$, and the
free two-nucleon Green's function $G_0$, via the Lippmann-Schwinger
(or, equivalently, the Schr\"odinger) equation

\begin{equation}
T=V^{(n)} + V^{(n)} G_0 T.
\label{eq:LSE}
\end{equation}

\begin{figure}[t,h,b,p]
\hskip 3.5cm \psfig{figure=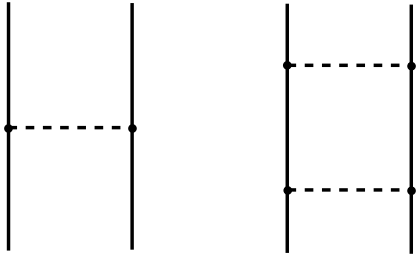,height=3.5cm}
\caption{\label{fig-iterOPE}
One- and two-pion exchange graphs. One-pion exchange is in the
potential and (uncorrelated) two-pion exchange is generated by
solving the Schr\"odinger equation.}
\end{figure}

This point of view is similar to that seen when low-energy effective
actions are derived using Wilsonian RG arguments (see
Refs.~\cite{Ep98C,Bi98B} for more explicit investigations of this
connection). These techniques are widely applied, and have, for
instance, been used to obtain improved actions for lattice gauge
theory.  In all cases the effective action is obtained using a
perturbation theory in the ratio of low-energy scales to the Wilsonian
RG cutoff scale. That Lagrangian is then solved to all orders. The
error of such a calculation is expected to be dominated by the
lowest-dimensional operator omitted from the calculation. This allows
one to make an {\it a priori} estimate of the error, which can then be
verified {\it a posteriori}~\cite{Le97}.  In the same way the Weinberg
program employs $\cpt$ to generate a nonrelativistic,
particle-number-conserving Hamiltonian for the nuclear system.  This
Hamiltonian is valid up to a given order in the chiral expansion and
can be employed in any many-nucleon system at low energy. Once written
down it should be solved to all orders.

With the $NN$ t-matrix in hand implementation of the third step of the
Weinberg program can proceed. In this stage we consider the
interaction of some external probe: a pion, a photon, or some
weakly-interacting particle, with the $NN$ system. Provided this probe
carries momentum of order $m_\pi$ it can have its interaction with the
$NN$ system represented as a sum of irreducible diagrams which forms a
kernel $K_{\rm probe}$ for the process of interest. Again, the
power-counting of $\chi$PT applies to this kernel, since all nucleon
energies appearing in it are, by definition, of order
$m_\pi$. Naturally, this is only true if the external probe energy and
momentum are both of $O(m_\pi)$, but, subject to that limitation,
$\chi$PT allows us to calculate the kernel $K_{\rm probe}$ up to a
given order in the chiral expansion.  The full amplitude for this
reaction is then found by multiplying this kernel $K_{\rm probe}$ by
factors which take account of the initial and final-state interactions
of the $NN$ pair. In other words the amplitude is 

\begin{equation}
{\cal A}=(1 + T G_0) K_{\rm probe} (1 + G_0 T).
\label{eq:probeamp}
\end{equation}
Here $T$ is calculated from Eq.~(\ref{eq:LSE}), using a $V^{(n)}$
which is also computed in $\chi$PT.  For consistency the expansion for
the interaction with the external probe should be taken to the same
chiral order as the order $n$ used in the chiral expansion of the $NN$
potential $V^{(n)}$. Once this is done the result for the amplitude
${\cal A}$ should be the unambiguous prediction of $\chi$PT for the
process of interest up to that order in the chiral expansion.

Up until now our discussion has been completely general, and could
apply to the scattering of particles from the $NN$ bound state, the
deuteron, the break-up of the deuteron by a probe (e.g. $\gamma d
\rightarrow np$) or to the emission of particles from the $NN$ state
(e.g.  proton-proton bremsstrahlung). However, if we are especially
interested in probes of the deuteron we may take the residue of
Eq.~(\ref{eq:probeamp}) at the deuteron pole in both the initial and
final-state.  We then see that the amplitude for the interaction of
the probe with the deuteron is

\begin{equation}
{\cal A}=\langle \psi| K_{\rm probe}|\psi \rangle,
\label{eq:probedeut}
\end{equation}
where the wave function $|\psi \rangle$ is to be calculated
by solving the Schr\"odinger equation for the potential $V^{(n)}$:

\begin{equation}
\left( E - {p^2 \over M} \right) |\psi \rangle=V^{(n)} |\psi \rangle.
\label{eq:Schro}
\end{equation}

This program is a close relative of that pursued by nuclear physicists
in the $NN$ system for many decades. It centers around the same
objects: potentials, wave functions and transition operators, and,
like calculations using $NN$ potential models, it treats the
interaction of the nucleons non-perturbatively.  The main difference
to more traditional potential-model calculations of $NN$-system
properties is the use of $\chi$PT to calculate the potentials and
kernels that are the building blocks of the nuclear-physics
calculation. Below we investigate this connection a little further,
comparing and contrasting the $\chi$PT approach to the physics of the
$NN$ system with physics based on more phenomenological
potentials. Crucially though, the use of $\chi$PT to derive the basic
building blocks that go into potential models allows for
simplification and systematization of the calculations performed using
the successful phenomenological potential models.

\subsection{The Weinberg Program: Power Counting}
\label{sec-problem}

The power counting for contributions to either the
$NN$ potential, $V$, or the kernel for some process, $K_{\rm probe}$, is
just that of $\chi$PT. In other words, we perform
NDA on the kernels of processes, keeping track 
of factors of $p$ and $m_\pi$ in the case of $V$, and $p$, $m_\pi$, and 
$k$---the probe energy/momentum---in the case of $K_{\rm probe}$. 
We generically denote all these
``small" quantities as $P$. 
The expansion is then one in powers
of $P/(\Lambda_\chi,M)$. A diagram which
contributes to $V$ has an order $n$ in this expansion which is
calculated by just multiplying together the powers of $P$ appearing
in the diagram as follows:
\begin{itemize}
\item Each vertex from ${\cal L}^{(n)}$ counts as $P^n$;

\item Each pion line counts as $P^{-2}$;

\item Each loop counts as $P^4$;

\item Each nucleon propagator counts as $P^{-1}$.
\end{itemize}

Note that these last two rules must be modified if we consider
two-nucleon-reducible diagrams as well, since in such diagrams the
nucleon propagator can scale {\it either} as $P^{-1}$ or as
$P^{-2}$, depending on whether the energy present is ``small" or
of order $m_\pi$.

For kernels $K_{\rm probe}$ one additional rule must be added:
\begin{itemize}
\item Two-body contributions to $K_{\rm probe}$ carry an extra
factor of $P^3$.
\end{itemize}
This rule, which means that two-body contributions to
the reaction mechanisms are systematically suppressed
in this low-momentum regime, can be thought of as arising
from the additional ``loop" which occurs when a two-body
reaction mechanism is sandwiched between $NN$ wave functions.
Equivalently, it occurs because there is one less momentum-conserving
delta function in a graph with a two-body contribution to
$K_{\rm probe}$. Either way the rule is again a straightforward
consequence of NDA.

\subsection{Nucleon-Nucleon Potential}
\label{sec-V}

The earliest discussion of constructing a $NN$ potential in the 
context of Weinberg power counting was that
by Weinberg in Refs.~\cite{We90,We91}, 
where $V^{(0)}$ was computed. 
Higher-order calculations were carried out
by Ord\'on\~ez {\it et al.}, where energy-dependent
potentials which represented the results for $V^{(0)}$, $V^{(2)}$, and
$V^{(3)}$ were derived~\cite{Or92,Or94,Or96}. 
The energy-dependence of this potential was not
intrinsic to $\chi PT$ and could be transformed away. 
In
Ref.~\cite{Ep98A} Epelbaum {\it et al.} did this, using unitary
transformations to turn these energy-dependent potentials into
energy-independent ones~\footnote{In Ref.~\cite{FC94} a
similar connection between energy-dependent and energy-independent
forms of the two-pion-exchange $NN$ force was established, thereby
elucidating the difference between old work on two-pion exchange
by Brueckner and Watson~\cite{BW53} and Taketani {\it et al.}~\cite{Ta52}.}.
By construction, this transformation leaves
all S-matrix elements unaffected, and so there is no physical
difference between the potential of Ref.~\cite{Ep98A} and that of
Refs.~\cite{Or92,Or94,Or96}. 
The isospin-violating $NN$ potential can be derived
from $\chi$PT in an analogous fashion~\cite{vK93,vK99}, but we shall
discuss only the charge-symmetric, charge-independent piece of the
$NN$ interaction here. 

The Lagrangian of $\chi PT$ is used to generate an $NN$ potential, $V$
as follows. The interaction $V^{(n)}$ contains all
two-particle-irreducible $NN \rightarrow NN$ graphs that are of chiral
order $n$ or lower.  At leading order ($n=0$), the chiral $NN$
potential has only two pieces~\cite{We90,We91}:
\begin{eqnarray}
  V({\bf q})=-{g_A^2 \over 4 f_\pi^2} \, {{\bf \sigma}_1 \cdot {\bf
      q} \, {\bf \sigma}_2 \cdot {\bf q} \over {\bf q}^2 + m_\pi^2} \, 
  (\vec{\tau}_1 \cdot \vec{\tau}_2) + C_S + C_T {\bf
\sigma}_1 \cdot {\bf \sigma}_2. 
\label{eq:VNN0} 
\end{eqnarray} 
where $C_S$ and $C_T$ are coefficients of 
contact interactions that do not involve spatial gradients.

At any chiral order an analogous separation
into a long- and short-distance $NN$ EFT potential can be made. 
Thus we will follow the authors of Ref.~\cite{Ep99} and write
$V=V_{\pi} + V_{\rm short}$
At low orders in the expansion $V_{\rm short}$ has a
straightforward counting in powers of $P$, with the higher-derivative
operators naively  suppressed by powers of $P/\Lambda$.

Regardless of the scaling of the short-distance part, the
long-distance piece is given by one-pion exchange plus irreducible
two-pion exchange plus three-pion exchange etc. The dimension of
the contributions of these multi-pion-exchange graphs produce in
$V_\pi$ increases by two for each additional pion loop, as can be seen
by simple $f_\pi$ counting.  Consequently, irreducible multiple pion
exchanges are suppressed by powers of $m_\pi/\Lambda$ and
$p/\Lambda$. Having made a chiral expansion for both $V_{\rm short}$
and $V_\pi$, we must truncate both at the same order in the chiral
expansion parameter $(p,m_\pi)/\Lambda$.

The contributions to $V_{\rm short}$ at zeroth and second-order
in $\chi$PT are just~\cite{Ep99}:
\begin{eqnarray}
& & \langle {\bf p}'| V_{\rm short}^{(0)}| {\bf p} \rangle\ =\ 
C_S +  C_T {\bf \sigma}_1 \cdot {\bf \sigma}_2 \nonumber\\
& & \langle {\bf p}'| V_{\rm short}^{(2)}| {\bf p} \rangle\ =\ 
\xi_1 {\bf q}^2 
+ \xi_2 {\bf k}^2 
+ (\xi_3 {\bf q}^2  + \xi_4 {\bf k}^2)
({\bf \sigma}_1 \cdot {\bf \sigma}_2) 
\\
& & \qquad
+ i \xi_5 \frac{1}{2} 
({\bf \sigma}_1 +   {\bf \sigma}_2) \cdot ({\bf q} \times {\bf k}) 
+ \xi_6 ({\bf \sigma}_1 \cdot {\bf q})  ({\bf \sigma}_2 \cdot {\bf q})
+ \xi_7 ({\bf \sigma}_1 \cdot {\bf k})  ({\bf \sigma}_2 \cdot {\bf k})
\nonumber
\label{eq:V2short}
\end{eqnarray}
where ${\bf k}={\bf p}' - {\bf p}$ and ${\bf k}=({\bf p}' + {\bf
p})/2$. 
The translation into the short-range potential that is operative in a
particular partial-wave is then a straightforward matter of taking
matrix elements of these operators. We will not present the explict
formulae here, but two points are immediately apparent:
\begin{enumerate} 
\item This short-distance potential operates only in
S-waves, P-waves, and to mix the $\siii$ and $\diii$ states.

\item In the $\si$ and $\siii$ channels there is a LO
and NLO contribution, with the NLO  piece
being proportional to $(p^2 + p'^2)$ by Hermiticity.

\item In the next five partial waves
(${}^1P_1$, ${}^3P_0$, ${}^3P_1$, ${}^3P_2$,
and $\epsilon_1$) there is a straightforward set of linear
equations which relates the combination of short-distance parameters
that appears in any one partial wave to the parameters appearing in
the potential. 
\end{enumerate}

This combination of facts means that by performing a fit with one
parameter in the partial waves ${}^1P_1$, ${}^3P_0$, ${}^3P_1$,
${}^3P_2$, and $\epsilon_1$, and two parameters in the partial waves
${}^1S_0$ and ${}^3S_1$ we may completely determine the short-distance
potential. 

At NLO the long-distance, pionic part of the
potential is generated by the one-loop diagrams shown in
Fig.~\ref{fig-TPENLO}.  These diagrams are divergent in $\chi$PT and
require renormalization, but the required renormalization may be made
by absorbing the divergences in the constants $\xi_1 \rightarrow
\xi_7$ given above.  Strictly speaking then, the relationship of the
$\xi_1 \rightarrow \xi_7$ which appear in Eq.~(\ref{eq:V2short}) to
the parameters which appear in the Lagrangian involves both the linear
transformation mentioned above and this subtraction of certain forms
arising in the calculation of the graphs shown in
Fig.~\ref{fig-TPENLO}.
\begin{figure}[t,h,b,p]
\hskip 3.5cm \psfig{figure=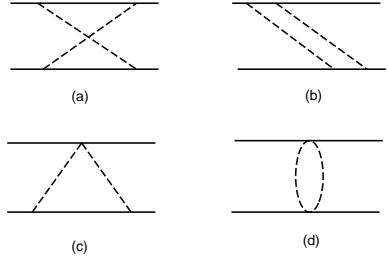,height=3.5cm}
\caption{\label{fig-TPENLO}
The two-pion-exchange graphs which are included in the $NN$
interaction at NLO. }
\end{figure}
After renormalization, the long-distance part of
$V_\pi$ at NLO is~\cite{Or96,Ep99}
\begin{eqnarray}
&& \langle {\bf p}'| V_\pi^{(2)}| {\bf p} \rangle
\ =\ 
- L(|{\bf q}|) {3 g_A^4 \over 64 \pi^2 f_\pi^4} 
\left[({\bf \sigma}_1 \cdot {\bf q})  ({\bf \sigma}_2 \cdot {\bf q})
- {\bf q}^2 ({\bf \sigma}_1 \cdot {\bf \sigma}_2)\right] 
\\
&& - {\vec{\tau}_1 \cdot \vec{\tau}_2 \over 384 \pi^2 f_\pi^4}
L(|{\bf q}|) \left[ 4 m_\pi^2 (5 g_A^4 - 4 g_A^2 - 1) + {\bf q}^2 (23 g_A^4 
- 10 g_A^2 - 1) 
\right.\nonumber\\
& & \left.
\qquad\qquad \qquad\qquad 
+ {48 g_A^4 m_\pi^4 \over 4 m_\pi^2 + {\bf q}^2}\right]
\ ,
\nonumber
\label{eq:Vpi2}
\end{eqnarray}
with
\begin{eqnarray}
L(|{\bf q}|)={1 \over |{\bf q}|} \sqrt{4 m_\pi^2 + {\bf q}^2}
\log \left({\sqrt{4 m_\pi^2 + {\bf q}^2} + |{\bf q}| \over 2 m_\pi}\right).
\end{eqnarray}

At N$^2$LO, $O(P^3/\Lambda^3)$, there are no new short-distance
operators. There are, however, new contributions to $V_\pi$ at this order. 
Where the Weinberg-Tomazawa term entered in the ``triangle"
and football graphs (diagrams (c) and (d) of Fig.~\ref{fig-TPENLO})
and contributed to $V^{(2)}$ we can now make one insertion of an
operator from the pion-nucleon Lagrangian ${\cal L}_{\pi N}^{(2)}$ and
generate contributions to $V_\pi^{(3)}$. After renormalization $V_\pi^{(3)}$
reads
\begin{eqnarray}
& & \langle {\bf p}'|V_\pi^{(3)}|{\bf p}\rangle\ =\ 
-{3g_A^2 \over 16\pi f_\pi^4} 
\left[ -{g_A^2   m_\pi^5 \over 16 M (4m_\pi^2+{\bf q}^2)} 
\right.
\nonumber\\
& & \left.
\qquad\qquad
+ 
\left( 2m_\pi^2(2c_1 -c_3) -{\bf q}^2\,
\left( c_3 + \frac{3g_A^2}{16M} \right) \right) (2m_\pi^2+{\bf q}^2) 
A(|{\bf q}|) \right]
\nonumber\\
&-& {g_A^2 \over 128\pi M f_\pi^4} (\vec{\tau}_1 \cdot \vec{\tau}_2 ) \,
\left[ -\frac{3g_A^2  m_\pi^5}{4m_\pi^2+{\bf q}^2} 
\right.
\nonumber\\
& & \left.
\qquad\qquad\qquad
+ \left( 4m_\pi^2 +
2{\bf q}^2 -g_A^2 (4m_\pi^2 + 3{\bf q}^2) \right)  (2m_\pi^2+{\bf q}^2) 
A(|{\bf q}|) 
\right]
\nonumber\\
&+&  {9g_A^4 \over 512\pi M f_\pi^4} \left( ({\bf \sigma}_1 \cdot {\bf q})
({\bf \sigma}_2 \cdot {\bf q}) -{\bf q}^2 ({\bf \sigma}_1 \cdot{\bf \sigma}_2
)\right) \, (2m_\pi^2+{\bf q}^2) A(|{\bf q}|)  
\nonumber\\
&-& {g_A^2 \over 32\pi f_\pi^4} (\vec \tau_1 \cdot \vec \tau_2 ) \,
\left( ({\bf \sigma}_1 \cdot {\bf q})({\bf \sigma}_2 \cdot {\bf q}) 
-{\bf q}^2 ({\bf \sigma}_1 \cdot{\bf \sigma}_2 )\right) 
\nonumber \\
&& \qquad\qquad
\times
\left[ \left( c_4 + {1 \over 4M} \right) (4m_\pi^2 + {\bf q}^2) 
-{g_A^2 \over 8M} (10m_\pi^2 + 3{\bf q}^2) \right] \, A(|{\bf q}|) 
\nonumber \\
&-& {3g_A^4 \over 64\pi M f_\pi^4} \, i \, ({\bf \sigma}_1 +  
{\bf \sigma}_2 ) \cdot ({\bf p}~' \times {\bf p} ) \, (2m_\pi^2+{\bf q}^2) 
A(|{\bf q}|) 
\nonumber \\
&-& {g_A^2(1-g_A^2) \over 64\pi M f_\pi^4} (\vec \tau_1 \cdot \vec \tau_2 )
\, i \, ({\bf \sigma}_1 +  {\bf \sigma}_2 ) \cdot ({\bf p}~' \times
{\bf p} ) \, (4m_\pi^2+{\bf q}^2) A(|{\bf q}|),
\end{eqnarray}
with
\begin{eqnarray} A(|{\bf q}|)={1 \over 2 |{\bf q}|}
\arctan\left({|{\bf q}| \over 2 m_\pi}\right).
\end{eqnarray} 

The free parameters $c_1$, $c_3$, and $c_4$ are related directly to
data in the $\pi N$ system. The $c$'s have been determined in recent
fits to $\pi N$ scattering data in $\chi$PT~\cite{BM99,FM00}. In
principle then, there are no undetermined parameters in
$V_\pi^{(3)}$. The fit to the $\pi N$ data extrapolated to the region
inside the Mandelstam triangle is used by Epelbaum {\it et al.}  to
fix $c_1$, $c_3$, and $c_4$.  Interestingly, Sugawara and Okubo wrote
down a TPE potential which contained a large part of the correct TPE
presented above as early as 1960~\cite{SO60}.  Stoks and Rijken also
generated all of the two-pion exchange interactions written
above~\cite{SR97}, although their potential also contained many other,
more phenomenological, short-range contributions. Although the fact that
$\chi$PT connects $\pi N$ data to $NN$ data in a model-independent
way makes it similar in philosophy to these, and many other
~\cite{BD71,Co73,La80,RdR96}, works, it was really only 
in Refs.~\cite{Or94,Or96} that a complete
expression for the full chiral TPE was first written down.

This chiral TPE potential was also calculated by Kaiser {\it et al.} as
part of a study we will discuss in more detail below~\cite{Ka97}. Kaiser
{\it et al.} observed that the combination $V^{(2)} + V^{(3)}$ provides strong
scalar-isoscalar attraction.  This attraction is governed by the LECs
$c_1$ and $c_3$. This suggests that if two-pion exchange is calculated
in a fashion consistent with chiral symmetry then there may be no need for
the presence of an explicit $\sigma$-meson in the $NN$ potential.

This potential 
\begin{equation}
V^{[3]}=V^{(0)} \ + \ V_{\rm short}^{(2)} 
\ + \ V_\pi^{(2)} \ + \  V_\pi^{(3)}
\label{eq:chiralpot}
\end{equation}
must now be iterated in the Lippmann-Schwinger equation,
Eq.~(\ref{eq:LSE}). However, it is a poorly-behaved potential from a
quantum mechanical point of view, since it is very singular at short
distances, and the corresponding Hamiltonian is thus unbounded from below. 
This is perhaps not surprising, since we know that the short-distance
physics in $\chi$PT is not the full short-distance physics of QCD. 
At distance scales $\sim 1/\Lambda_\chi$ mechanisms not encoded
in $\chi$PT will begin to play a dynamical role, and will regulate
the behavior of the singular potential $V^{[3]}$. In order to
calculate with the potential Refs.~\cite{Or96,Ep99} introduced
cutoffs on the potential before iteration, defining
a regulated potential
\begin{eqnarray}
\langle {\bf p}'|V_R| {\bf p} \rangle=
f_\Lambda(|{\bf p}'|) \langle {\bf p}'|V| {\bf p} \rangle 
f_\Lambda(|{\bf p}'|),
\end{eqnarray}
where $f_\Lambda(|{\bf p}|)$ is some regulator function that goes
to zero sufficiently fast as $|{\bf p}|$ goes to infinity, 
and obeys $f_\Lambda(0)=1$, e.g.,
\begin{eqnarray}
f_\Lambda(|{\bf p}|)=\exp(-|{\bf p}|^4/\Lambda^4).
\label{eq:evgreg}
\end{eqnarray}
This regulates the potential. Practical implementation of
renormalization is then quite simple. With the regulator mass, $\Lambda$,
fixed, one tunes the short-distance
coefficients~\cite{Le97,Sc97,SF98A,SF98B} so as to reproduce some specified
low-energy observables.

With renormalizable theories, the next step would be to take the
regulator to infinity.  
Here $\Lambda$ should be of the order of the high energy scale $\Lambda_\chi$.
There are two reasons for this.
Firstly, NDA applies when $\Lambda\sim \Lambda_\chi$.
Secondly, in the Weinberg scheme subleading terms are resummed to all orders
and may become large for large $\Lambda$ as the counterterms required to renormalize
these contributions are not present.

The algorithm for generating predictions from the regulated third-order
chiral potential $V_R^{(3)}$ adopted in Ref.~\cite{Ep99} is then:
\begin{enumerate}
\item Choose a regulator mass $\Lambda$.

\item Fit the constants $C_S$, $C_T$, and $\xi_1$--$\xi_7$ to the S
and P-wave phase shifts and $\epsilon_1$ from the Nijmegen 
PSA~\cite{St93} up to $T_{\rm lab}=100$ MeV. 

\item Predict these same phases from $T_{\rm lab}=100$ MeV
to $T_{\rm lab}=300$ MeV, and predict other phase-shifts over
the full energy range.
\end{enumerate}

\begin{figure}[t,h,b,p]
\hskip 0.5cm \psfig{figure=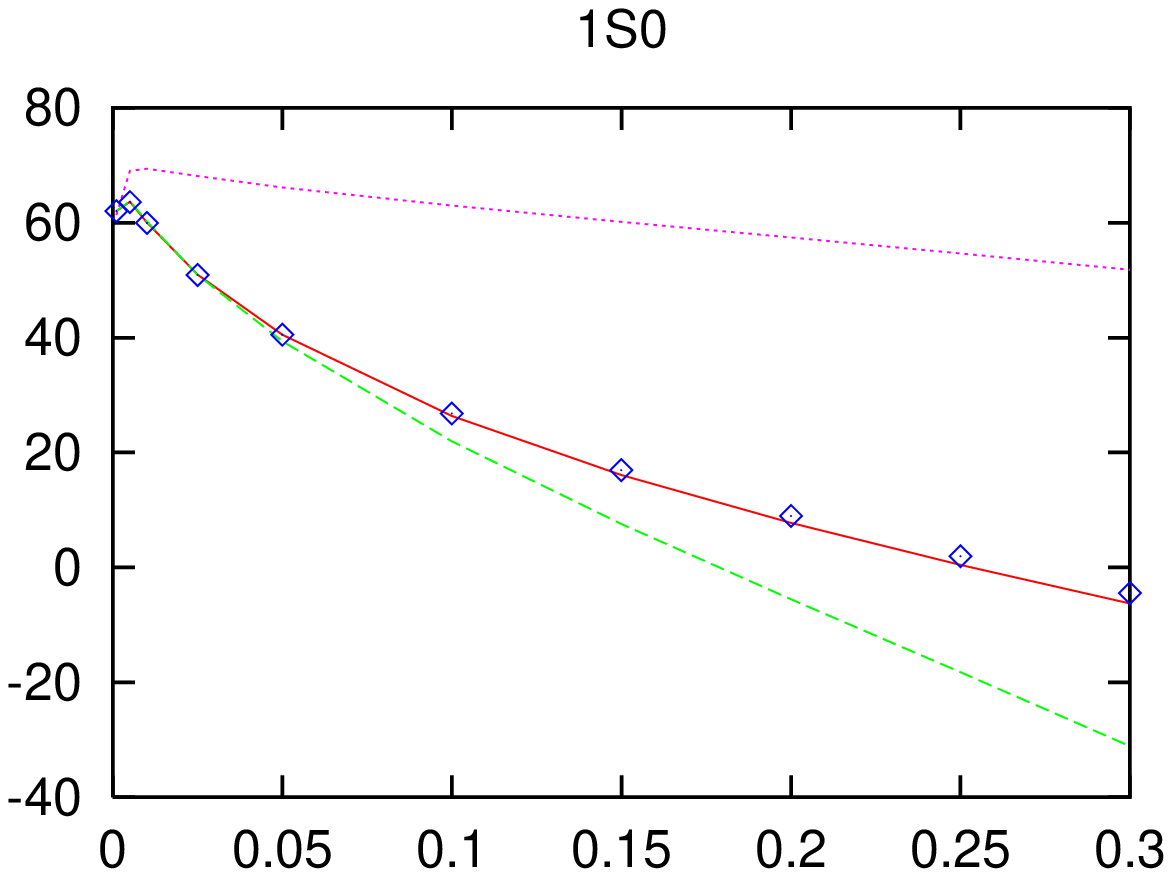,height=3.5cm}
\hskip 1.0cm \psfig{figure=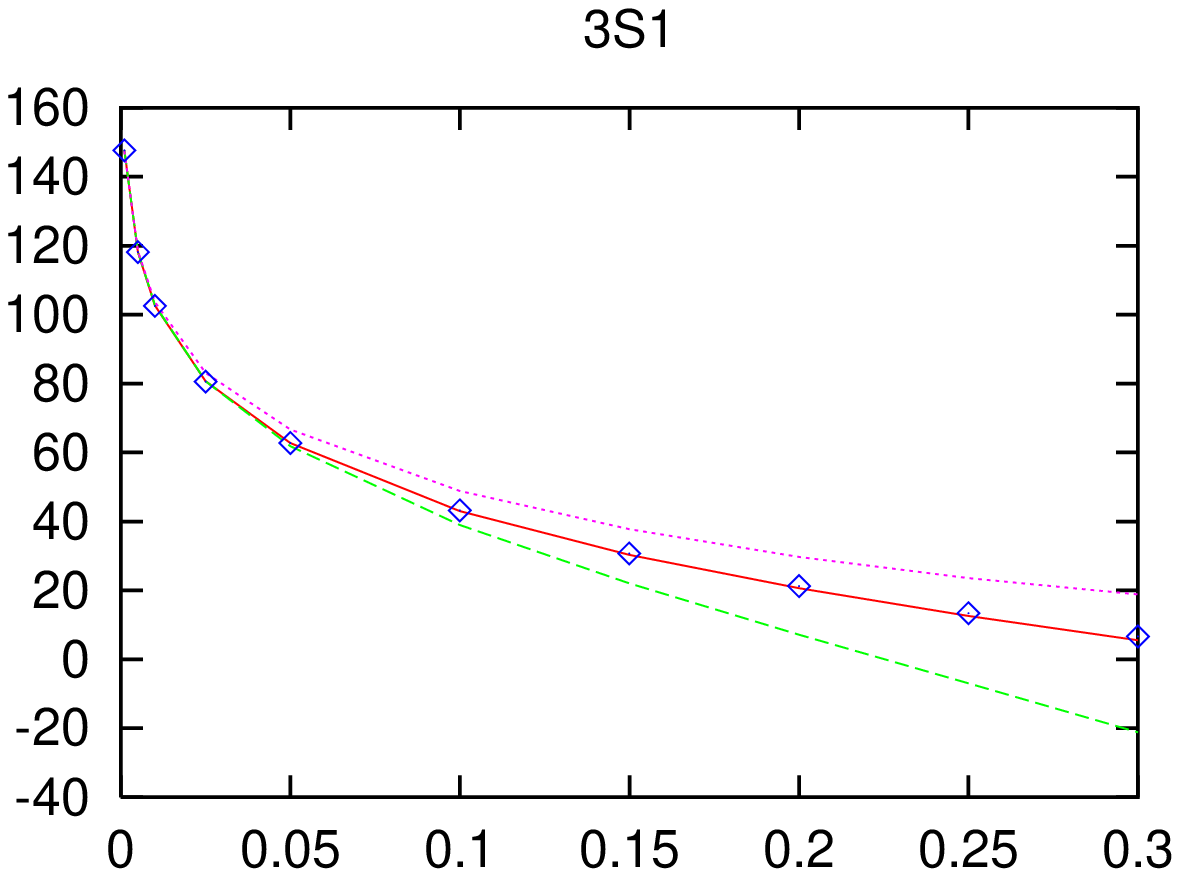,height=3.5cm}
\caption{\label{fig-Swaves}
Predictions for the ${}^1S_0$ and ${}^3S_1$ phase shifts 
from Ref.~\protect\cite{Ep99}. 
The dotted, dashed, and solid curves represent 
the results from the LO, NLO, and N$^2$LO potential.
At NLO and N$^2$LO the two parameters in this channel are fitted to the 
phase shifts below $T_{\rm lab}=100~{\rm MeV}$. 
The squares represent the Nijmegen PSA phases. 
The exponential
regulator was employed with $\Lambda=500$ MeV at NLO, and
$\Lambda=875$ MeV at N$^2$LO.}
\end{figure}
\begin{figure}[h,t,b,p]
\hskip 0.5cm \psfig{figure=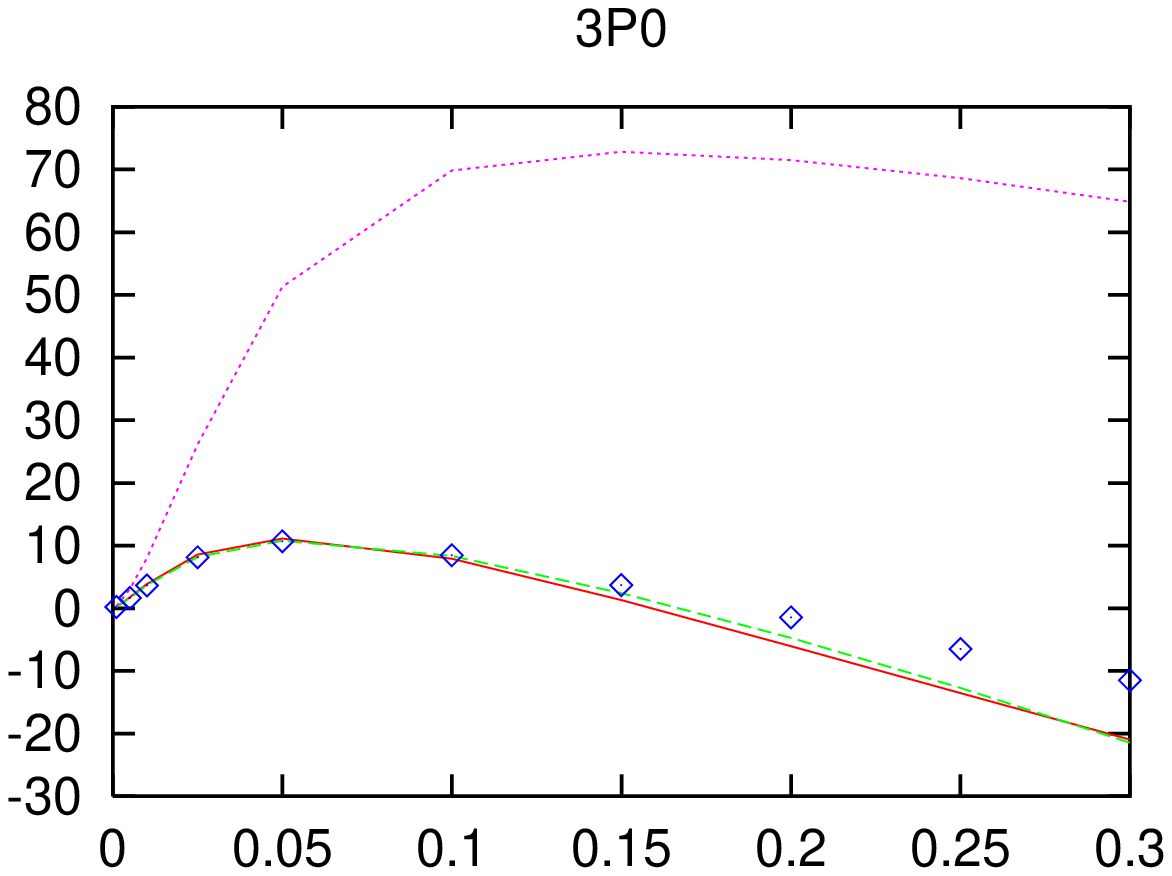,height=3.5cm}
\hskip 1.0cm \psfig{figure=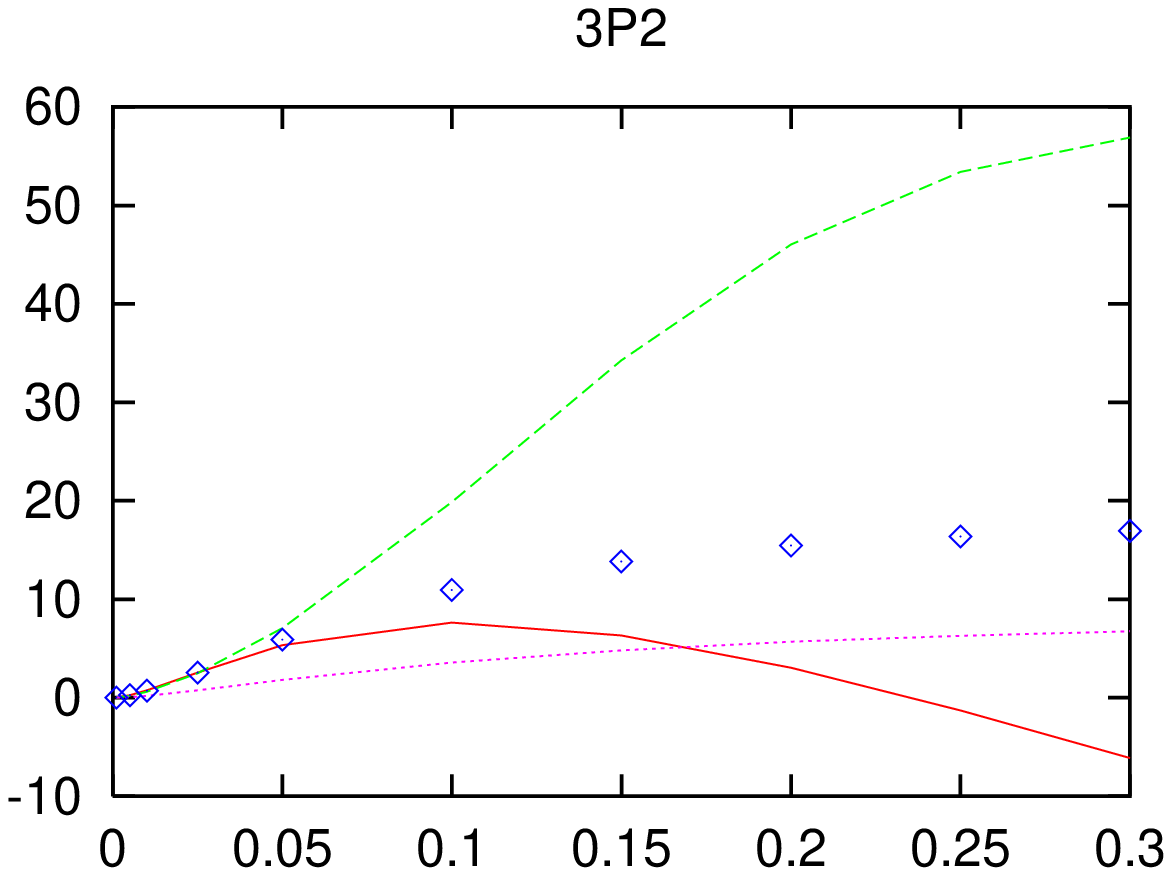,height=3.5cm}
\caption{\label{fig-Pwaves}
The ${}^3P_0$ and ${}^3P_2$ phase shifts 
as calculated in Ref.~\protect\cite{Ep99}. 
In this case there is one parameter at NLO and N$^2$LO, which is
fitted to the phase shifts below $T_{\rm lab}=100~{\rm MeV}$.
Legend as in Fig.~\protect\ref{fig-Swaves}.}
\end{figure}
In general the results for the $S$ and $P$ waves
(where at least one fit parameter was present in each case) 
are quite good, although the behavior of the ${}^3P_2$ phase shift
causes some concern. Some of these results are displayed in
Figs.~\ref{fig-Swaves} and \ref{fig-Pwaves}.
\begin{figure}[h,t,b,p]
\hskip 0.5cm \psfig{figure=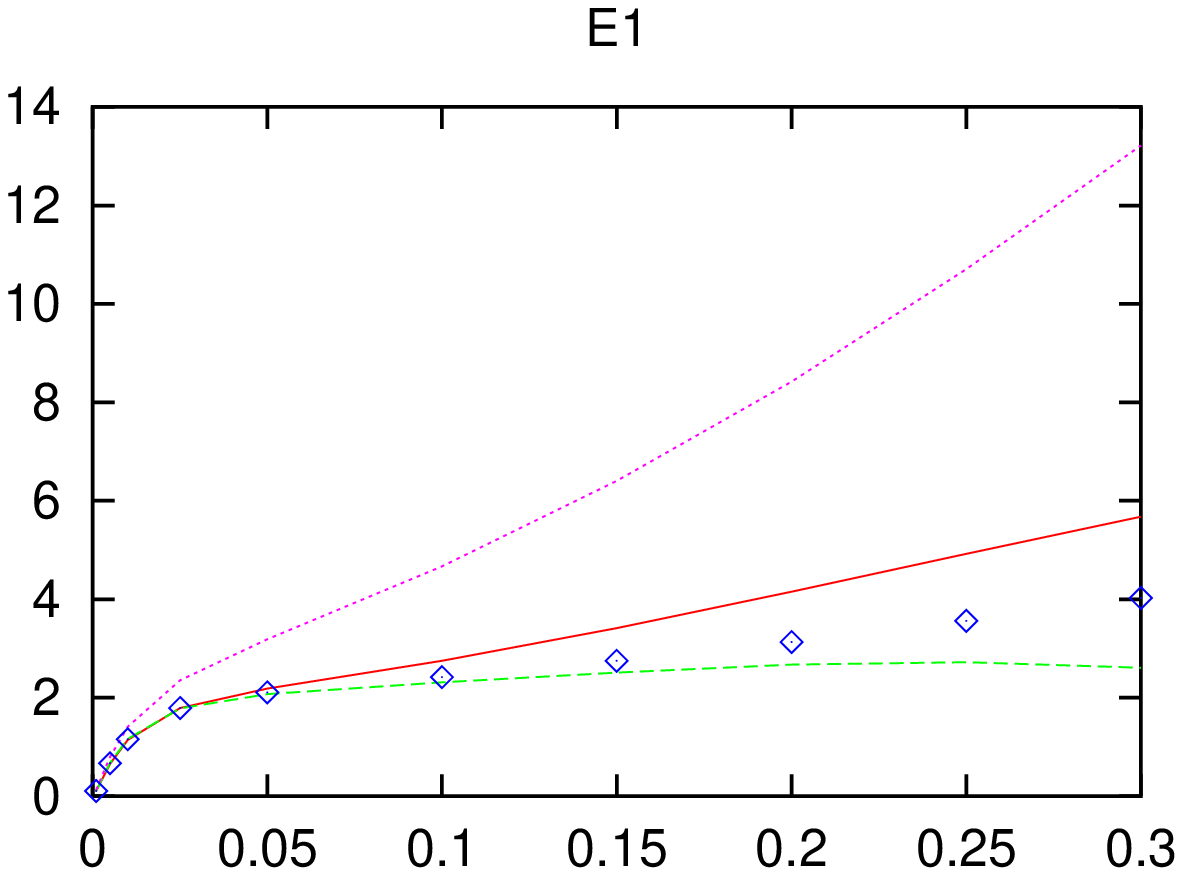,height=3.5cm}
\hskip 1.0cm \psfig{figure=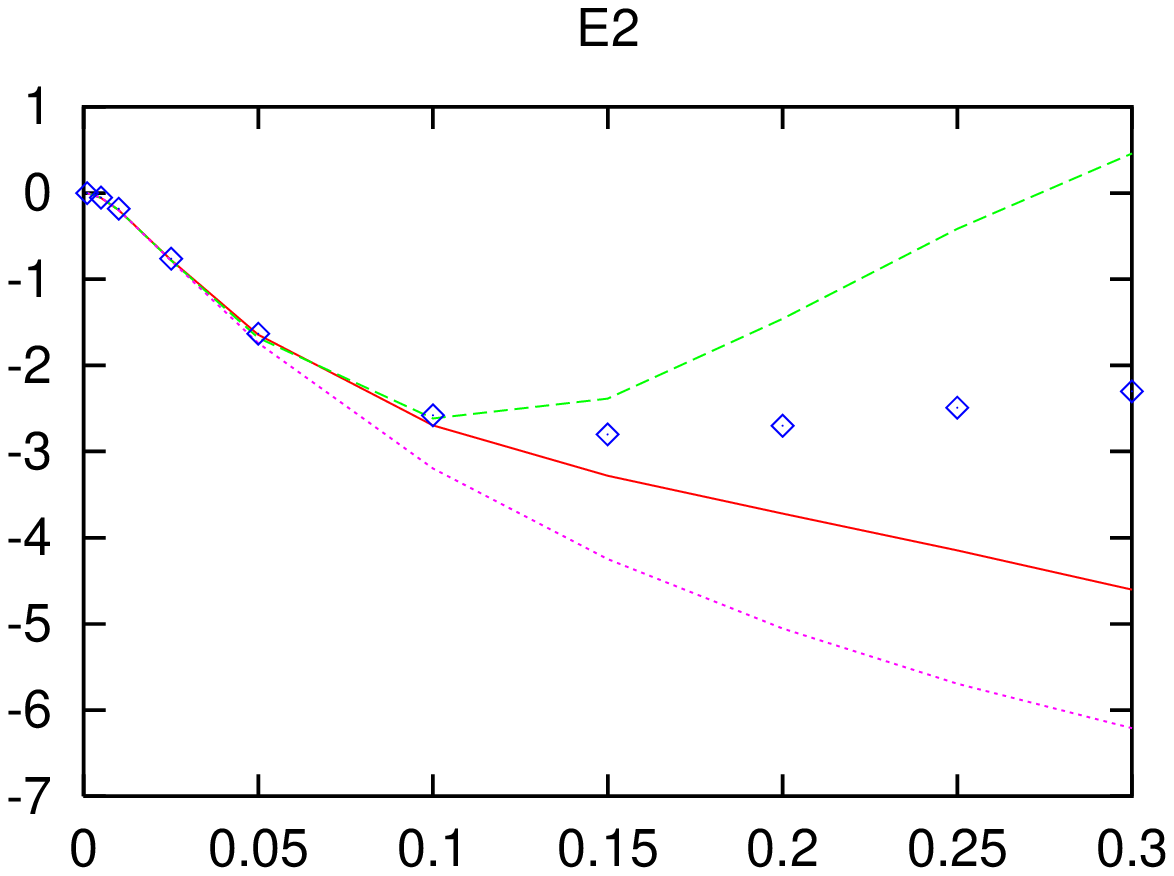,height=3.5cm}
\caption{\label{fig-epsilons}
The mixing parameters $\epsilon_1$ and $\epsilon_2$ from 
Ref.~\protect\cite{Ep99}.  There is one fit parameter
in $\epsilon_1$. The results for $T_{\rm lab} \geq 100$ MeV in
$\epsilon_1$, as well as all of the results for 
$\epsilon_2$ are predictions. Legend as in Fig.~\protect\ref{fig-Swaves}.}
\end{figure}

We stress once again that there are no free parameters to fit in
partial-waves with $L \geq 2$. Thus, the phase shifts calculated there
are parameter-free predictions of this approach. In the $D$-waves, the
results are in quite good agreement with the Nijmegen PSA although
some deviation begins to be seen around $T_{\rm lab}=200$ MeV (see
Fig.~\ref{fig-Dwaves}).  Epelbaum {\it et al.} also note that these
phases are quite sensitive to the value of the cutoff chosen. This
may be because at next order, in $V^{(4)}$, contact interactions enter
which will affect the $D$-wave $NN$ channels. Meanwhile, the mixing
parameter $\epsilon_2$ is not well reproduced above $T_{\rm lab}=100$
MeV, while the results for the triplet $F$-waves are really quite
poor. It seems that for these smaller $NN$ phase shifts a higher-order
calculation is needed for an accurate description. 
\begin{figure}[h,t,b,p]
\hskip 0.5cm \psfig{figure=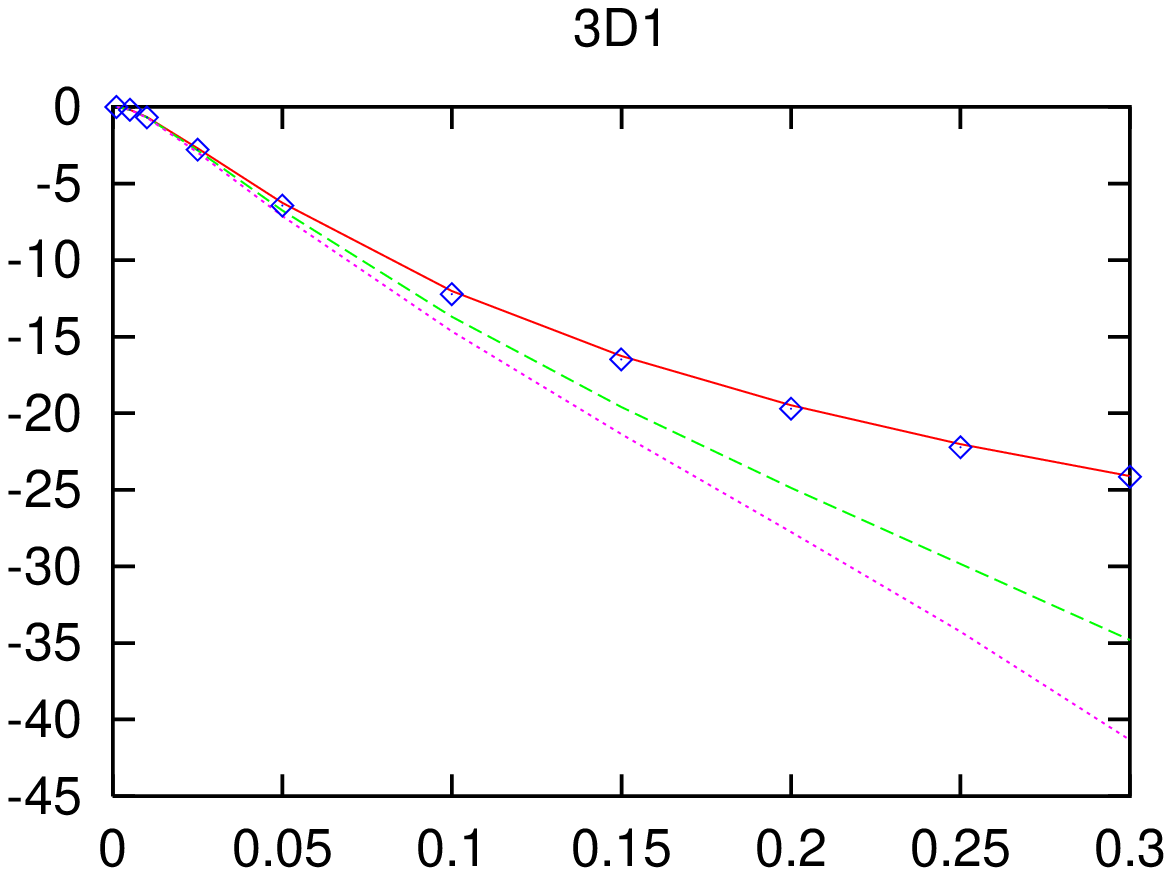,height=3.5cm}
\hskip 1.0cm \psfig{figure=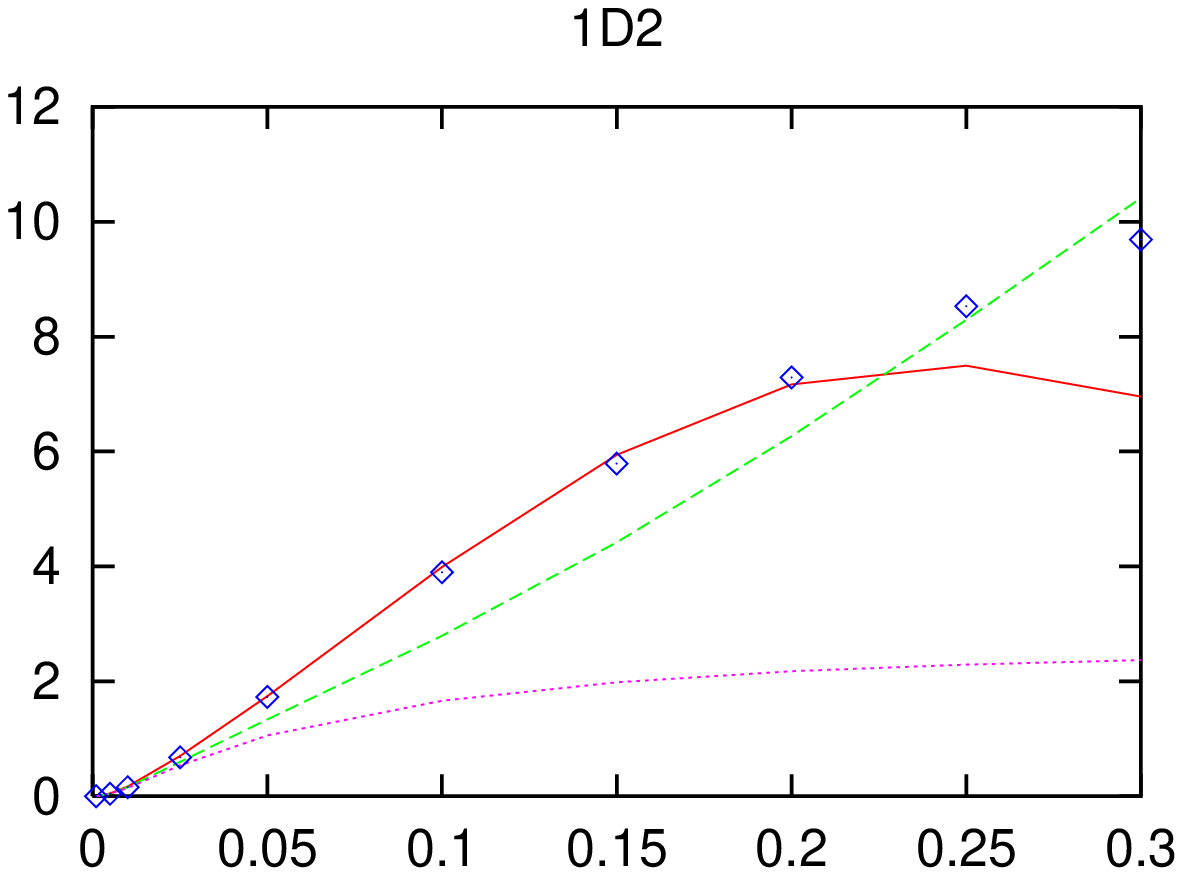,height=3.5cm}
\caption{\label{fig-Dwaves}
The ${}^3D_1$ and ${}^1D_2$ phases from Ref.~\protect\cite{Ep99}. 
In this case there are no free parameters.
Legend as in Fig.~\protect\ref{fig-Swaves}.}
\end{figure}

As the orbital angular momentum is increased the Born approximation
works increasingly well, and there is, in fact, little need to iterate
the potential.  In this way the study of Epelbaum {\it et al.} reduces
to that of Ref.~\cite{Ka97} in the higher partial-waves.  In
that work Kaiser {\it et al.} calculated the following
perturbative $NN$ amplitude
\begin{eqnarray}
T_{\rm pert}=V_\pi^{(0)} \ + \ V_\pi^{(0)} G_0 V_\pi^{(0)} 
\ + \ V_\pi^{(2)} + V_\pi^{(3)}.
\label{eq:Tpert}
\end{eqnarray}
In other words, these authors employed the potentials $V_\pi^{(2)}$ and
$V_\pi^{(3)}$ at tree level, and OPEP in
second Born approximation (i.e. iterated once) as the $NN$ amplitude.
The graphs included are those of Figs.~\ref{fig-iterOPE} and
\ref{fig-TPENLO} together with the triangle and football graphs
with insertions from ${\cal L}_{\pi N}^{(2)}$.  This is an
approximation to the full $NN$ t-matrix obtained by iterating the
chiral $NN$ potential $V^{(3)}$. It works increasingly well as the
angular momentum of the $NN$ state increases, since the centrifugal
barrier ensures that only the long-range, weak, part of the $NN$ force
is probed in these waves. Comparison of the results
of Refs.~\cite{Ka97} and \cite{Ep99} indicates that
iteration begins to be important below $L=4$, but that for $L \geq 4$
the amplitude (\ref{eq:Tpert}) does a good job of reproducing
the phase shifts. 

Indeed, in these peripheral partial waves one-pion exchange alone,
i.e. the interaction $V_\pi^{(0)}$ uniterated, reproduces the
experimental data reasonably well---as has been known for many
years~\cite{Ma87}. For $L=6$ the effects of the parts of $T_{\rm
pert}$ other than $V_\pi^{(0)}$ are really quite difficult to see
below $T_{\rm lab}=200$ MeV. However, for the $G$ and $H$-waves
the inclusion of $V^{(2)}$ and $V^{(3)}$ improves the
agreement with the data.

A calculation of the next-order piece of the chiral $NN$ potential,
$V^{(4)}$ will bring two new ingredients. Firstly, a number of
short-distance operators will arise. The role that these might play
in the ${}^1S_0$ channel has been considered in Ref.~\cite{Hy99}.
In general though, a number of channels will have to be considered:
these operators will have to be
fitted to the $S$, $P$, and $D$ wave phase shifts, as well as to
mixing parameters. Secondly, many two-loop
contributions to the potential---some of which have been calculated in
Refs.~\cite{PR99,Ka99}---will need to be added to the interaction.
The calculation of the full $V^{(4)}$ is a Herculean task.
Nevertheless, we can get some feeling for the size of higher-order
effects by including more freedom in the short-distance part of
the interaction. 

This is essentially the direction taken in a recent calculation by
the Nijmegen group~\cite{Re99}. Rentmeester {\it et al.} calculated 
the piece of the chiral potential $V^{(3)}$ of Eq.~(\ref{eq:chiralpot})
which acts at radii $r > 1/\Lambda_\chi$---i.e. the potential's
long-range tail---and incorporated it into the
Nijmegen PSA machinery. Schematically, they worked with
a potential
\begin{eqnarray}
V(r)=V_{\rm short}(r) \ \theta(R-r) + \left[\sum_{i=0}^3 V_\pi^{(i)} 
+ V_{\rm em}(r) \right] \ \theta(r-R),
\end{eqnarray}
where $V_{\rm em}$ accounts for electromagnetic corrections to the
long-distance part of the potential. As for the short-distance part,
$V_{\rm short}$ is only operative at radii $r<R$, and it depends on
many free parameters. These parameters are tuned so as to reproduce
the experimental phase shifts. Fixing the boundary $R$ at 1.4 fm, and
the constant $c_1=-0.76~{\rm GeV}^{-1}$,
Rentmeester {\it et al.} tuned 23 parameters in $V_{\rm short}$
and the $\pi N$ LECs $c_3$ and $c_4$ to fit 1951 pp scattering
data below $T_{\rm lab}=350$ MeV. 
The resultant minimum $\chi^2$ is very close to the 
theoretical ideal of one per degree of freedom
(see Table~\ref{table-TPEPSA}).

\begin{table}
\caption{Results for the PWA with different long-range
         interactions and $R=1.4$ fm. }
\begin{center}
\begin{tabular}{l|cc}
                              & $\#$Parameters & $\chi^2_{\rm min}$  \\ \hline
 OPE                          &  31  & 2026.2 \\
 OPE + $V^{(2)}$              &  28  & 1984.7 \\ 
 OPE + $V^{(2)}$ + $V^{(3)}$  &  23  & 1934.5 \\ 
\end{tabular}
\end{center}
\label{table-TPEPSA}
\end{table}

This is a strong indication that the chiral TPE is the
intermediate-range part of the $NN$ force. This conclusion is
reinforced by what happens if the chiral TPE is removed from the
interaction. In that case the minimum $\chi^2$ shoots up appreciably,
even if a number of new parameters are introduced in the
short-distance interaction. The final indication that the chiral $NN$
potential is reproducing the physics of the $NN$ system in this
phase-shift analysis is that the values of $c_3$ and $c_4$ obtained
from this fit are $c_3=-4.99 \pm 0.21~{\rm GeV}^{-1}$ and $c_4=5.62
\pm 0.59~{\rm GeV}^{-1}$.  These, together with the value of $c_1$,
agree with the values extracted for these LECs from the $\pi N$ data,
within the significant uncertainties associated with fitting in the
$\pi N$ system.

%%%%%%%%%%%%%%%%%%%%%%%%%%%%%%%%%%%%%%%%%%%%%
\section{External Probes of the Deuteron }
\label{sec-weinbprobesintro}

In scattering processes involving the deuteron EFT has proved more
successful phenomenologically than traditional potential models. Part
of this success is due to the fact that the relation between exchange
currents and potentials is dictated by the rules of field theory and
is therefore unambiguous in EFT.  Perhaps more importantly, EFT
relates scattering processes involving the nucleon alone to nuclear
scattering processes in a controlled manner; all of the phenomenology
of $\cpt$ in the single nucleon sector makes its appearance at some
level in the nuclear sector. Hence, counterterms which are difficult
to measure in experiments with nucleons can sometimes be extracted in
a systematic way from nuclear experiments. This is particularly
important given the absence of suitable neutron targets.  Weinberg
power-counting has led to fruitful computation of many pionic and
photonic probes of the two-nucleon system~\cite{Se98}, including
electron-deuteron scattering, electromagnetic deuteron form factors,
neutral pion photoproduction on the deuteron at
threshold~\cite{Be95,Be97}, neutral pion electroproduction on the
deuteron at threshold~\cite{Br99}, the pion-deuteron scattering
length~\cite{We92,Be97C}, Compton scattering on the
deuteron~\cite{Be99}, as well as $pn$ radiative
capture~\cite{Pa95,Pa96B} and the solar burning process $p p
\rightarrow d e^+ \nu$~\cite{Pa98}.  We will consider each of these
processes in turn.

On the technical side, it should be noted that typical nucleon momenta
inside the deuteron are on the order of $m_\pi$, and consequently, we
expect no convergence problems in the $\cpt$ expansion of any
low-momentum electromagnetic or pionic probe of the deuteron.  We will
use wavefunctions generated using modern nucleon-nucleon potentials as
well as wavefunctions recently computed in $\cpt$. Generally we find
that any wavefunction with the correct binding energy gives equivalent
results to within the theoretical error expected from neglected higher
orders in the chiral expansion.

\subsection{Electron-Deuteron Scattering}
\label{sec-ed}

For simplicity we will work in the Breit frame
in calculating matrix elements of the form,
\begin{equation}
\langle j \, \, {\bf q}/2| J^\mu_{\rm em} | i \, \, {\bf -q}/2\rangle
\end{equation} 
where a virtual photon of
three-momentum ${\bf q}=q \hat{z}$ couples to 
a deuteron electromagnetic current when the deuteron is in
states with specific magnetic quantum numbers $i$ and $j$,
as discussed earlier.
The electromagnetic structure of the
deuteron may be parametrized in terms of the three form factors,
$F_C(q^2)$, $F_M(q^2)$, and $F_Q(q^2)$ 
defined in Eq.~(\ref{eq:emmatdef}).

\subsection{The deuteron current}
In general, the current to which the electron couples in
electron-deuteron scattering may be expressed as:
\begin{equation}
  \langle {\bf P}' \, \, {\bf p}'|J_\mu|{\bf P} \, \, {\bf p} \rangle
  =(2 \pi)^3 j_\mu^{(1)}({\bf p},{\bf q}) \delta ({\bf p}' - {\bf p} - {\bf
    q}/2) + j_\mu^{(2)}({\bf p},{\bf p}';{\bf q}),
\end{equation}
where ${\bf p}$ and ${\bf p}'$ are the initial and final relative
momenta in the deuteron state, and ${\bf q}={\bf P}' - {\bf P}$ is the
three-momentum of the photon exchanged between the electron and the
deuteron. The one- and two-body currents discussed here are depicted in
Fig.~\ref{fig-obctbc}. This is a useful decomposition
because Weinberg power-counting predicts that the two-body
currents are generically suppressed relative to one-body currents.

\begin{figure}[h,t,b,p]
\hskip 1.5cm \psfig{figure=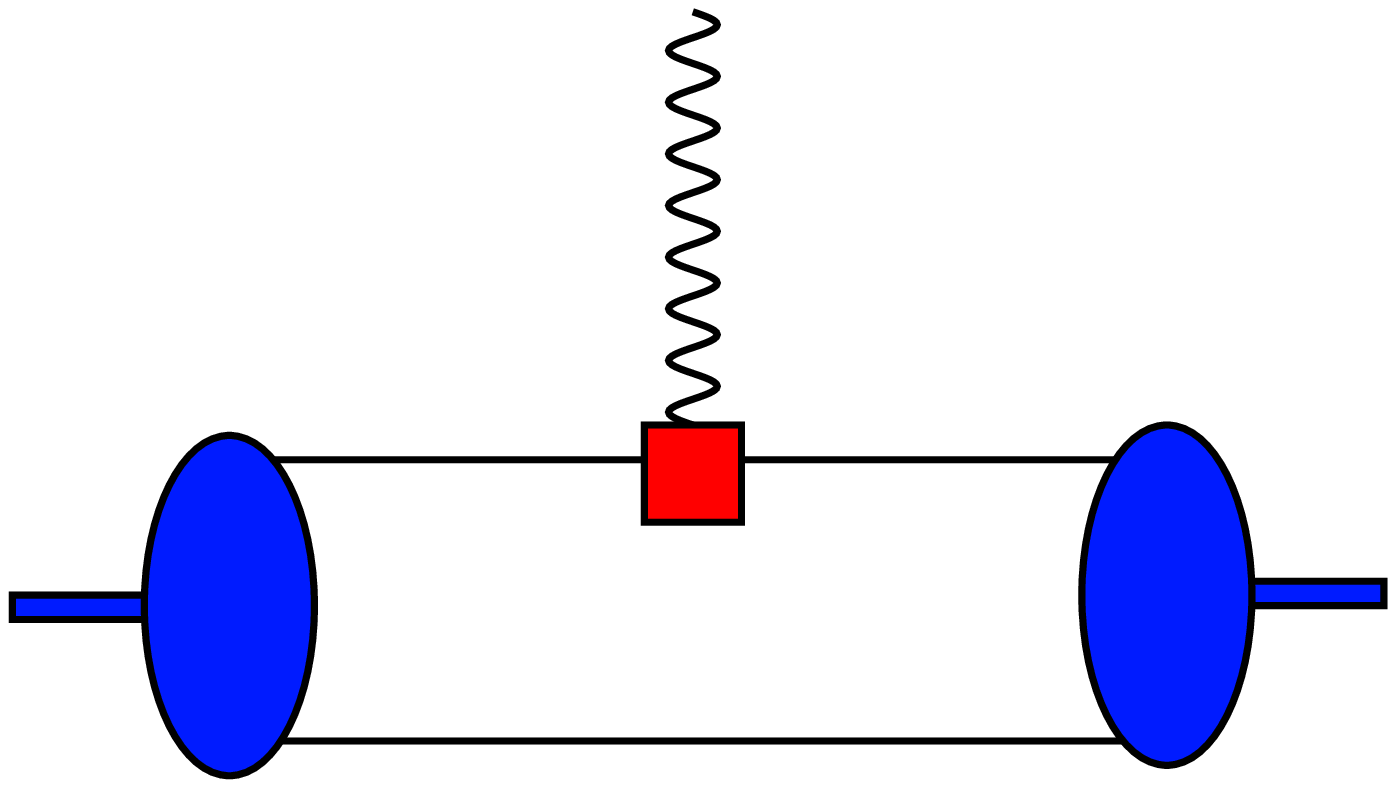,height=2.5cm}
\hskip 1.5cm \psfig{figure=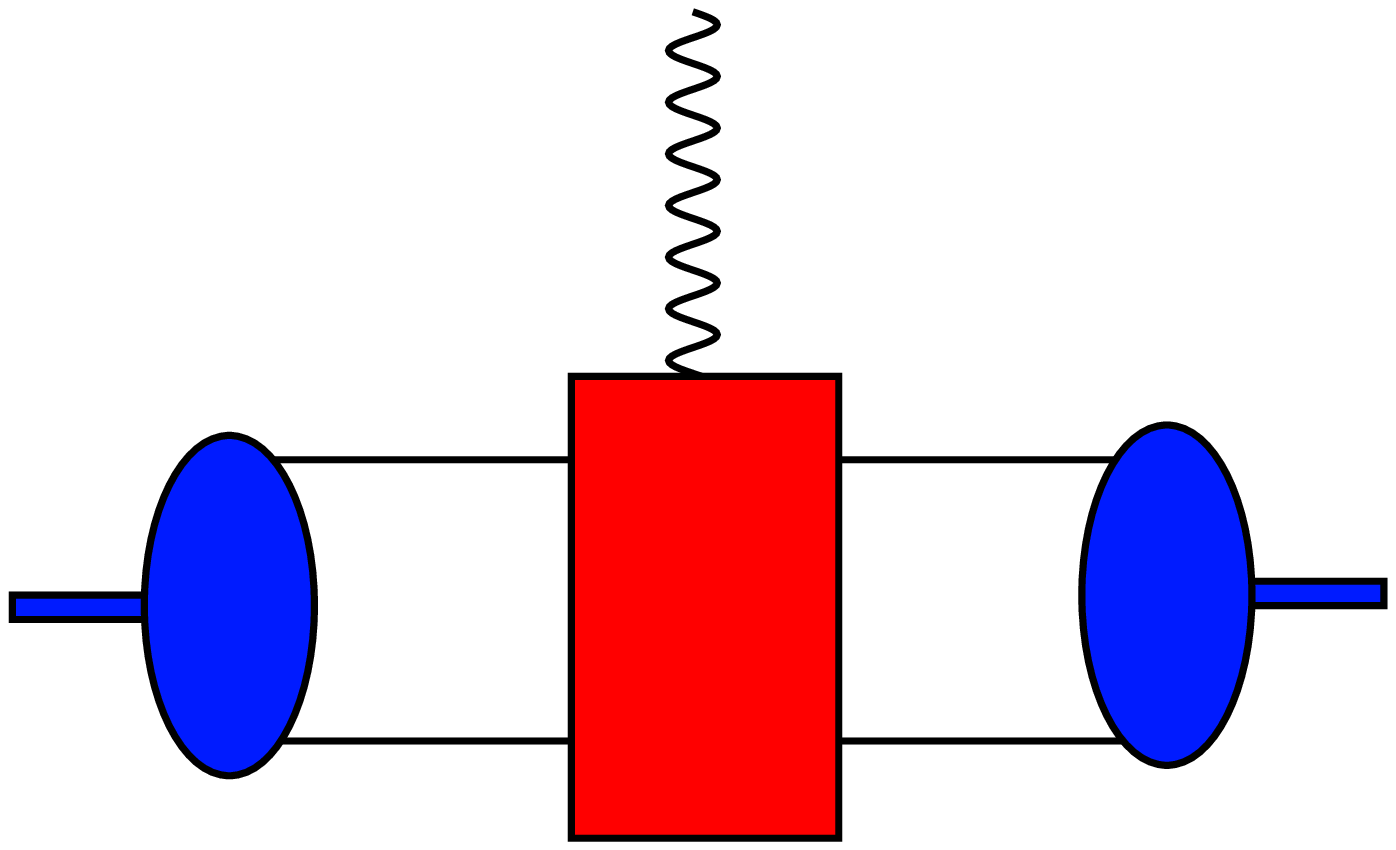,height=2.5cm}
\caption{\label{fig-obctbc} One- and two-body
current contributions to electron-deuteron scattering.
The square and rectangle indicate, respectively a $\gamma NN$ vertex, and 
the two-body $NN \rightarrow NN \gamma$ kernel. 
The blobs are deuteron vertex functions.}
\end{figure}

For on-shell nucleons the most general form that the single-nucleon
current, $j_\mu^{(1)}$, can take, consistent with Lorentz invariance,
time-reversal invariance, and current conservation, is:
\begin{equation}
  j_\mu^{(1)}({\bf p},{\bf q})=e_N \bar{u}({\bf p} + 3 {\bf q}/4)
  \left[F_1(q^2) \gamma_\mu + F_2(q^2) \frac{i}{2 M} \sigma_{\mu \nu}
    Q^\nu \right] u({\bf p} - {\bf q}/4),
\label{eq:job}
\end{equation}
where $e_N$ is the nucleonic charge. For the deuteron case, $F_1$ and
$F_2$ are isoscalar combinations of the proton and neutron Dirac and
Pauli form factors ($i=1,2$)
\begin{eqnarray} 
F_i(q^2)&=&F_i^{(p)}(q^2) +F_i^{(n)}(q^2)
\ \ .
\end{eqnarray}

It is straightforward to calculate these isoscalar Dirac and Pauli
form factors in $\chi$PT~\cite{Be92,Br98B,KM00}. At LO the only
contribution comes from ${\cal L}_{\pi N}^{(1)}$, and corresponds
to an $A_0$ photon coupling to the charge of the nucleon. At 
NLO the current $j_+$ receives contributions from the E1 and M1 vertices
of ${\cal L}_{\pi N}^{(2)}$. Thus the first two orders of the calculation
are trivial, and lead to the currents:
\begin{eqnarray}
  j_0^{(1)}({\bf p},{\bf q})&=& e\ \ ,\ \   
  j_+^{(1)}({\bf  p},{\bf q})\ =\ e \left( {p_+ \over M} + 
             {q \over 2M} (\kappa_p + \kappa_n + 1) \sigma_+ \right)
\ .
\label{eq:jplus}
\end{eqnarray}

Corrections at next order will include relativistic effects, and
the effects of nucleon structure. The former are 
suppressed by ratios of momenta to the nucleon mass, which we denote 
generically by the mnemonic:
$\delta^2\sim p^2/M^2 ,\  p \cdot Q/M^2 ,\  q^2/M^2$.
Corrections at order $\delta^2$ include the usual spin-orbit and
Dirac-Foldy relativistic corrections to the charge
operator~\cite{Fr73}. Meanwhile, the second expansion will be
governed by the parameter $q^2/\Lambda_{\chi}^2$. In the
isoscalar form factors these structure corrections are purely
due to short-distance counterterms, since a photon cannot couple to 
two pions in an isoscalar state.

We now turn to the task of quantifying the effect of two-body
currents on the observables under consideration. The power counting
here was first explained in Ref.~\cite{Pa95} and has been further
worked out for the exchange currents which contribute
to $F_M$ in Ref.~\cite{Pa99}.

The contribution of two two-body pionic mechanisms to the deuteronic
current is depicted in Fig.~\ref{fig-tbcpions}.  It is straightforward
to count the powers of momentum or pion mass in such graphs. 
These current contributions scale {\it at least} as $P^3$~\footnote{Here
  we are counting the charge $e$ as of order $P$ in order to
  simplify matters.}.  Therefore we expect that such meson-exchange
currents will be suppressed by $P^2/\Lambda_\chi^2$ relative to the 
lowest-order $\gamma NN$ interaction, which scales as $P$. 

\begin{figure}[h,t,b,p]
\hskip 1.8cm \psfig{figure=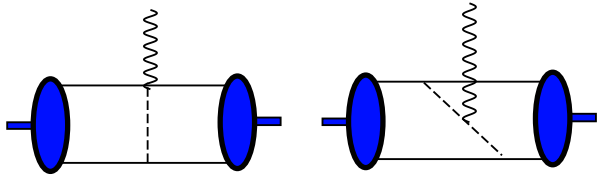,height=2.5cm,width=8.0cm}
\caption{\label{fig-tbcpions} The contribution
        to $e d$ scattering from the LO pionic
        current.  Here all vertices are from the LO chiral
        perturbation theory Lagrangian. The graphs with nucleons one
        and two interchanged also contribute.}
\end{figure}

Since $j_+$ begins only at $O(P^2)$ it might seem that meson-exchange
currents are only suppressed by one power of $P$ in the deuteron form
factor $F_M$, but this is not the case. In fact, although the leading
effect of meson-exchange currents is nominally $O(P^3)$, the
meson-exchange current which occurs at that order is isovector in
character and thus vanishes in the deuteron.  It is well known that
this $O(P^3)$ isovector current is the leading MEC correction to
processes such as electrodisintegration of the deuteron, and $\gamma d
\rightarrow np$ (see, for instance, Ref.~\cite{Ri84,RB72}). In
practice this results in the leading effect from meson-exchange
currents coming at $O(P^4)$, or N$^2$LO in the deuteron form factors.
In such an $O(P^4)$ exchange current the $\gamma \pi$ contact
interaction with the nucleon has a coefficient that is related to
known constants by the requirements of gauge invariance and
reparametrization invariance.  We note that this correction to the
charge operator is phenomenologically important in potential-model
calculations of $F_C$~\cite{CS98,SR91}.

At some order above $O(P^3)$ operators enter whose coefficient is not
constrained by symmetries, but instead is sensitive to short-distance
physics. These have only a small effect in $F_C$ and $F_M$ for $q
\lsim 1~{\rm GeV}$, but they may affect $F_Q$ markedly, since the
natural size of $F_Q$ at $q=0$ is much smaller than that of $F_C$ and
$F_M$. A natural expectation for $\left. F_Q\right|_{q=0} \equiv Q_d$
is:

\begin{equation} Q_d \sim {\eta_{sd} \over \gamma^2}.  \end{equation} Here
$\eta_{sd}$---the asymptotic D-to-S normalization---enters because an
S-wave deuteron cannot emit a quadrupole photon via any one-body
current mechanism. Thus the power-counting for the contribution of
one-body currents to $F_Q$ is exactly the same as for similar
contributions to $F_C$. However, two-body currents can connect the
S-wave pieces of the deuteron initial- and final-state wave functions
one to another and still result in the emission of a quadrupole
photon~\cite{Ch99}. The lowest-order Lagrangian which can generate
this effect is that written down in Eq.~(\ref{eq:counterQ}).  This
two-body operator does not enter in the expansion of the deuteron
current until $O(P^5)$ in Weinberg counting. On the other hand,
crucially it contains no powers of $\eta_{sd}$. Hence, numerically it can
be important at much lower order than naively expected. Counting only
overall powers of $p$, $M$, and $\Lambda$, we see that the
contribution of this operator to $Q_d$ will be of order

\begin{equation}
  \Delta Q_d \, \sim \, {\langle p^3 \rangle \over M \Lambda^2} \, \, 
  {1 \over \Lambda^2}.
\end{equation}
Although this is formally of order $P^5$, numerically it is as important
as N$^2$LO corrections to $Q_d$~\cite{PC99}.

From Ref.~\cite{Or96,Ep99} we have both NLO and N$^2$LO $NN$
potentials, derived from chiral perturbation theory. The Schr\"odinger
equation, Eq.~(\ref{eq:Schro}), can be solved with these potentials to
generate deuteron wave functions. We can then use the systematic
chiral expansion of the deuteron electromagnetic current to calculate
the electromagnetic form factors of the deuteron. These results are
compared with results from a sophisticated potential model, the Nijm93
potential. In both cases we employ the nonrelativistic impulse
approximation with point nucleons, since this represents a calculation
to NLO in the $e d$ scattering kernel.  Before calculating
$F_C$, $F_Q$, and $F_M$ though, we will spend some time discussing 
results for static properties of the deuteron.

\subsection{Static Moments}

The values of the form factors $F_M$ and $F_Q$ at $q^2=0$ are related
to the magnetic and quadrupole moments of the deuteron via the
definitions in Eq.~(\ref{eq:normalization}). 
If no
meson-exchange currents (i.e. no local four-nucleon-one-photon 
operators)
enter the $e d$ scattering kernel,
and there are no relativistic corrections to the current,
then $\mu_d$ and $Q_d$ are given by simple integrals of the deuteron
radial wave functions $u (r)$ (S-wave) and $w (r)$ (D-wave),
\begin{eqnarray}
  \mu_d&=&\kappa_p + \kappa_n - \frac{3}{2}(\kappa_p + \kappa_n - \frac{1}{2}) 
\int_0^\infty dr \, w^2(r);\\
  \mu_{\cal Q} &=&\frac{1}{\sqrt{50}}\int_0^\infty dr \, r^2 \, u(r) w(r) -
  \frac{1}{20} \int_0^\infty dr \, r^2 \, w^2(r).
\end{eqnarray}

Another interesting static property of the deuteron is its 
matter radius, $r_d$. This is dominated by the long-distance
part of the deuteron wave function, since it occurs
in an integral weighted by $r^2$:
\begin{equation}
r_d^2 \ = \ \int_0^\infty \, dr \, r^2 [u^2(r) + w^2(r)].
\end{equation}

Employing the wave functions derived above, along with three
leading potential models~\cite{St94,Wi95,Ma00} leads to the results shown
in Table~\ref{table-static}. The experimental data is included for
comparison only, since we have used a very simple electromagnetic current 
to generate these results.

\begin{table}[h,t,b,p]
\begin{center}
\begin{tabular}{|c|c|c|c|}
\hline
     &   $r_d$  (fm)   & $\mu_d$ ($\mu_N$) & $\mu_{\cal Q}$ (${\rm fm}^2$)  \\ 
\hline \hline
Nijm93        &    1.968        &  0.847       &   0.271     \\ \hline
AV18          &    1.967        &  0.847       &   0.270     \\ \hline
CD-Bonn       &    1.966        &  0.852       &   0.270     \\ \hline
$\chi$PT NLO  &    1.975        &  0.859       &   0.266     \\ \hline
$\chi$PT N$^2$LO &    1.967        &  0.845       &   0.262     \\ \hline
Experiment    &    1.971(6)     &  0.857406(1) &   0.2859(3) \\ \hline
\end{tabular}
\end{center}
\caption{Static moments of the deuteron for the Nijm93, CD-Bonn,
and AV18 potentials, as well as
for the NLO and N$^2$LO $\chi$PT potentials of Ref.~\protect\cite{Ep99}.}
\label{table-static}
\end{table}

The table shows that the deuteron matter radius is the same to three
significant figures in all six calculations. Moreover, the results are
all within the error bar of the value extracted from
hydrogen-deuterium isostope-shift measurements~\cite{Sc93,Mr95}. For
the magnetic moment the situation is not quite so good, but it is
clear that $\mu_d$ is predominantly determined by $\kappa_p$ and
$\kappa_n$. The size of the spread in the theoretical models---and of
their discrepancy with the experimental value~\cite{BC79,ERC83}---is
consistent with the magnitude expected for higher-order contributions
to the deuteron current.

Park {\it et al.}~\cite{Pa98} and Phillips and Cohen~\cite{PC99} have also
performed calculations of the static properties of the deuteron in
EFT-motivated approaches.  Both of these papers
employed one-pion exchange as the long distance part of the $NN$
potential---rather than chiral two-pion exchange---and various
different short-distance regulators. The results are in accord with
the above findings.  Firstly, $r_d$ is extremely well-predicted in the
effective theory, once $A_S$, the asymptotic S-state normalization of
the deuteron wave function, and $B$, the deuteron binding energy, are
reproduced. This is in line with the findings of EFT($\not \! \pi$)
and ERT.  The deuteron radius is largely
determined by the tail of the deuteron wave function. Similarly,
$\mu_d$ is largely insensitive to the short-distance physics, since it
is dominated by $\kappa_p$ and $\kappa_n$.

The large variation in $Q_d$ between the calculations presented in
Table~\ref{table-static} was also seen in the studies of
Ref.~\cite{Pa98} and \cite{PC99}. There is considerable sensitivity to
the short-distance physics in this observable. Presumably, this is why
the five theoretical values all differ markedly from the experimental
number. 
Apparently there are significant contributions to this
quantity from physics beyond those of single nucleon operators.  
One such piece of physics is a two-body current
resulting in the emission of a quadrupole photon~\cite{Ch99}. 
As argued above,
a natural size for the contribution of this operator to $\mu_{\cal Q}$,
will give approximately as large an effect as terms one order in
the expansion beyond what we have considered here.

\subsection{Electromagnetic Deuteron Form Factors}

$F_C$ calculated with the NLO
wave function of $\chi$PT is shown in Fig.~\ref{fig-FC}. 
It is compared
with results from the Nijm93 wave function. 
The agreement is good 
for photon momenta of up to about 500 MeV
for a regulator mass
$\Lambda=600~{\rm MeV}$ in the exponential of Eq.~(\ref{eq:evgreg}).
\begin{figure}[h,t,b,p]
\hskip 3.0cm \psfig{figure=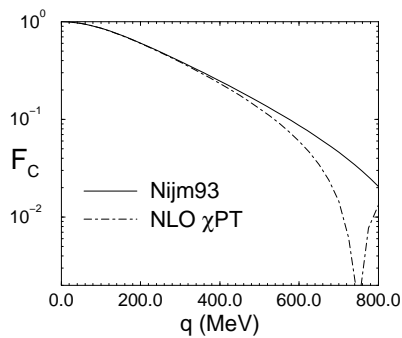,height=4.5cm}
\caption{\label{fig-FC} The charge form factor of the deuteron at NLO. 
The solid line is the result for the
        Nijm93 potential, while the dot-dashed line is a calculation
        with the NLO chiral potential.}
\end{figure}

Results are presented for the deuteron magnetic form factor $F_M$ in
Fig.~\ref{fig-FM}. These plots lead
to conclusions similar to those inferred from the plot of $F_C$.  
This is a consequence of the
dominance of the S-wave pieces of the wave function in determining the
magnetic structure of the deuteron at low $q$.

The situation is somewhat different for the quadrupole form factor
$F_Q$. Since the quadrupole moments obtained in the Nijm93 and NLO
$\chi$PT calculations do not quite agree there is a discrepancy of a
few percent between the two form factors at $q=0$.  On the other hand,
the shape of the two curves is essentially the same out to $q \approx
500$ MeV.
\begin{figure}[h,t,b,p]
\hskip 0.1cm \psfig{figure=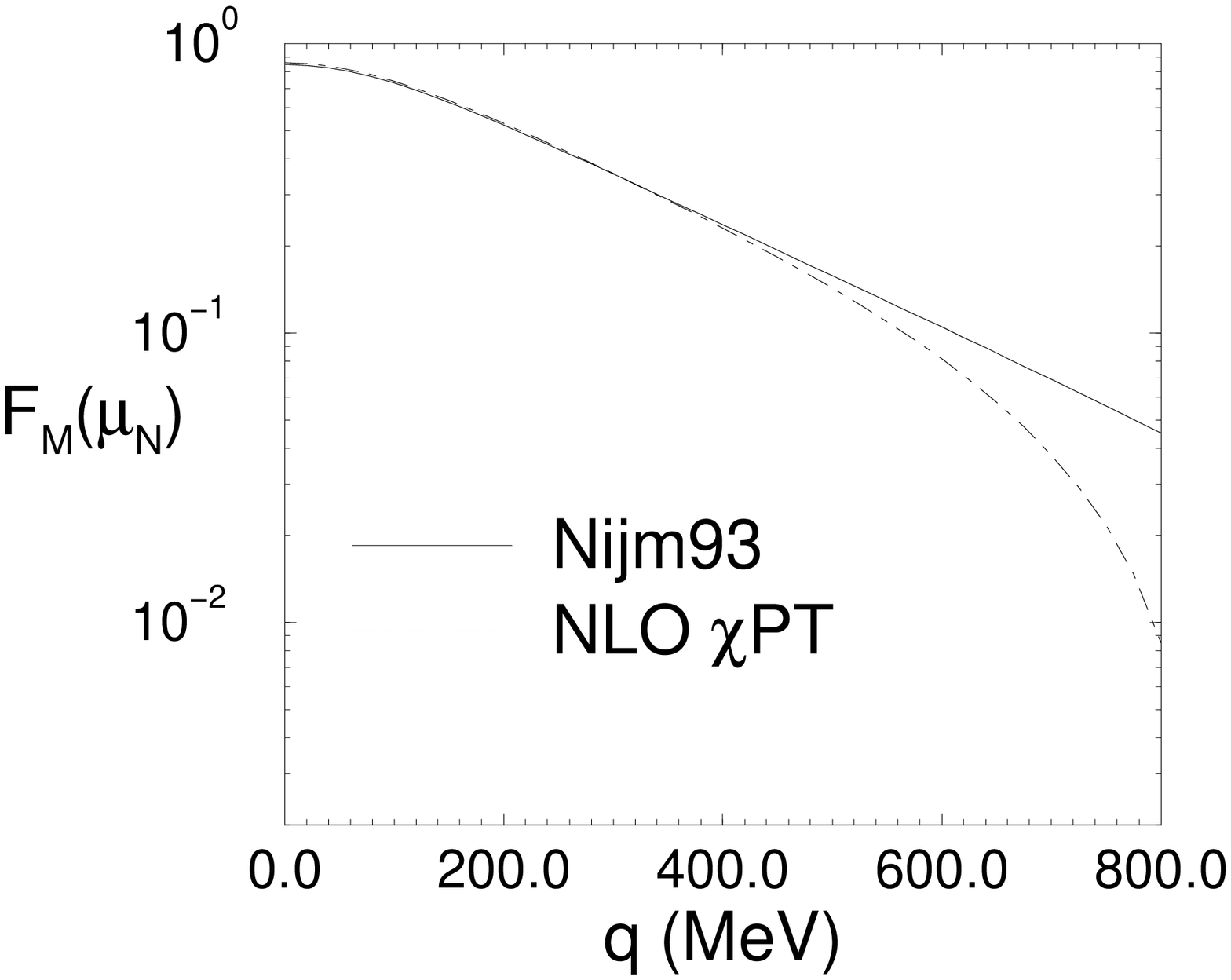,height=4.5cm}
\hskip 0.5cm \psfig{figure=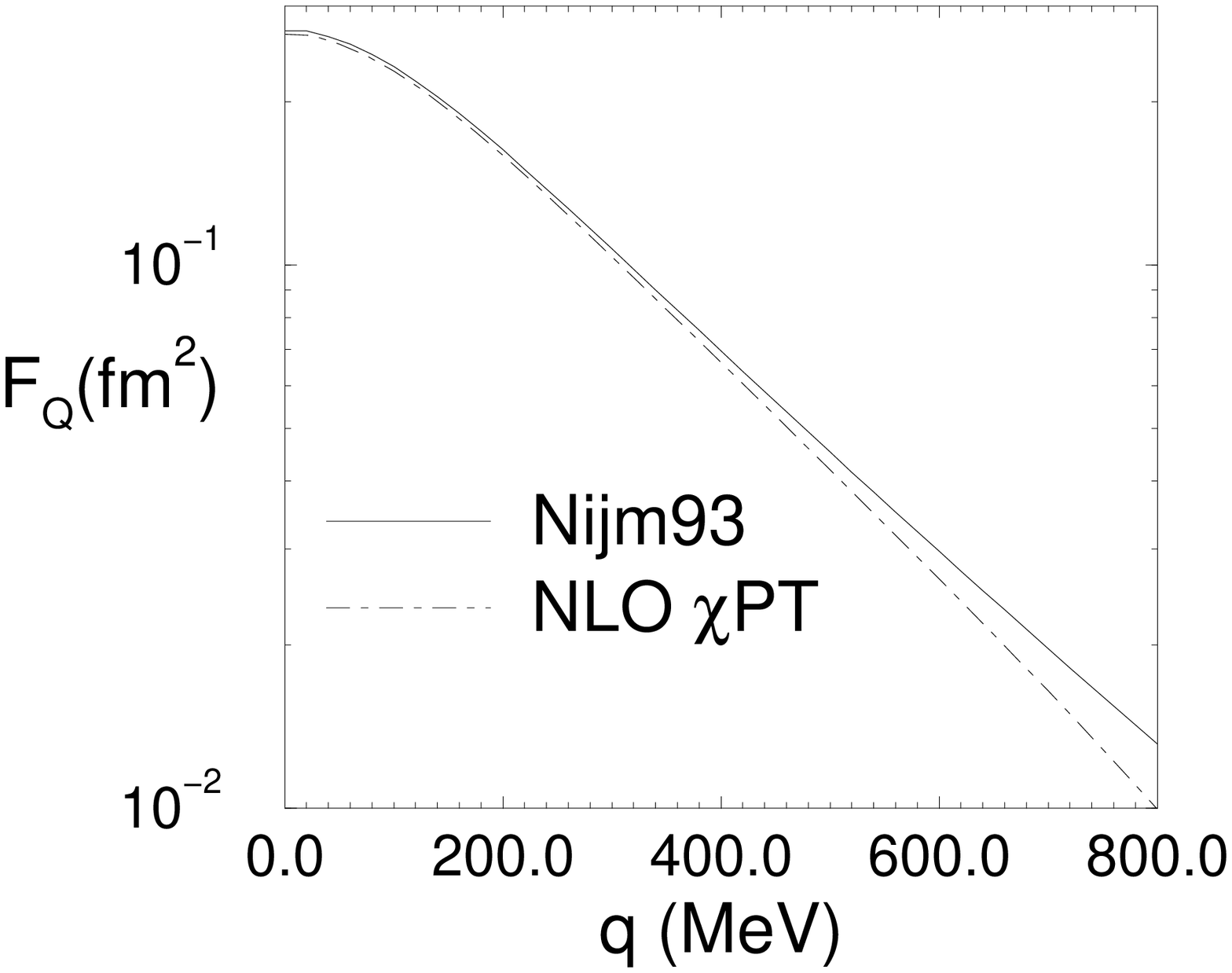,height=4.5cm}
\caption{\label{fig-FM} The magnetic and quadrupole
        form factors of the deuteron for the Nijm93 and NLO
        $\chi$PT wave functions. Legend as in Fig.~\ref{fig-FC}.}
\end{figure}
In fact, the quadrupole moments in both calculations
differ appreciably from the experimental result.
Simple power-counting arguments 
suggest that a two-body counterterm might be important
in $F_Q$. 
To further examinethis problem,
we have calculated the $e d$ tensor-polarization 
observable $T_{20}$ in four different ways. 
We have performed two different types of calculation
for each of the two wave functions: Nijm93 and NLO $\chi$PT:
\begin{itemize}
\item A calculation with single nucleon operators;

\item A calculation where we add---in an admittedly non-systematic
way---a two-body counterterm which shifts the value of the quadrupole moment
in both calculations to the experimental value.
\end{itemize}

\begin{figure}[h,t,b]
\hskip 1.5cm \psfig{figure=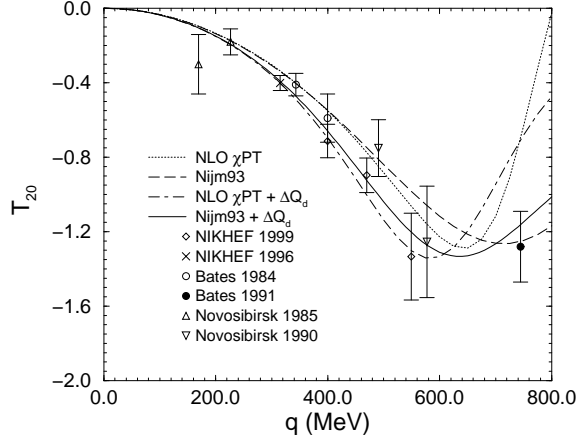,height=6.0cm}
\caption{\label{fig-T20} The $e d$ scattering tensor
        polarization observable $T_{20}$. 
The dashed (dotted) lines correspond 
to insertions of a single nucleon operator
only, using the Nijm93  (NLO $\chi$PT) wave function.
The solid (dot-dashed) line is the Nijm93 (NLO $\chi$PT)
        result when a two-body counterterm is added so that
        $Q_d$ is reproduced. Experimental data is taken from
        Refs.~\protect\cite{Sc84,Dm85,Gi90,Ga94,Fe96,Bo99}.}
\end{figure}

The results of this investigation are presented in 
Fig.~\ref{fig-T20}~\footnote{Thanks to Matt Dorsten for his help
in generating this figure.}. We see that when the same currents
are used the results for $T_{20}$ are the same with both wave functions
out to $q \approx 500$ MeV. This is just a restatement of the results presented
for electromagnetic form factors above. However, we also see that
shifting the value of $Q_d$ from the impulse approximation result
$Q_d=0.27~{\rm fm}^2$ to the experimental value $Q_d=0.29~{\rm fm}^2$,
has a marked effect on $T_{20}$ at momentum
transfers of order a few hundred MeV. Moreover, 
it would seem that once the deuteron quadrupole
moment is reproduced both the potential model and the EFT calculation
describe the data of Refs.~\cite{Sc84,Dm85,Gi90,Ga94,Fe96,Bo99} fairly 
well out to momenta $q \approx 600$ MeV. This suggests that the physics which
determines $T_{20}$ at these momentum transfers is the physics
of one-pion-exchange---which the two potentials here have in common---along
with a deuteron which has the right binding energy {\it and the correct
quadrupole moment}.

%%%%%%%%%%%%%%%%%%%%%%%%%%%%%%%%%%%%%%%%%%%%%
\subsection{$\gamma d \rightarrow \pi^0 d$}
\label{sec-gammadpi0d}

Pion photoproduction on the nucleon near threshold has been studied up
to $O(P^4)$ in $\cpt$ with the delta integrated out in
Ref. \cite{Br96C}.  The differential cross-section for a photon of
momentum $k$ to produce a pion of momentum $q$ at threshold is pure
$S$ wave and given by

\begin{equation}
{{|{\vec k}|}\over{|{\vec q\,}|}}\;{{d\sigma}\over{d\Omega}}
\bigg|_{|{\vec q\,}|=0} 
= |E_{0+}|^2
\end{equation}
where $E_{0+}$ is the electric dipole amplitude.  At LO,
$O(P)$, there are contributions only to charged pion channels arising
from minimal substitution on the pion-nucleon vertex (``Kroll-Ruderman
term'') and from the pion pole. Therefore the  charged channels are 
expected
to be large compared to the neutral channels.  At $O(P^2)$, minimal
substitution on the relativistic correction to the pion-nucleon vertex
contributes to photoproduction on the proton. At $O(P^3)$ there are
tree-level contributions involving the magnetic moment of the
nucleon. Together with the previous orders, this reproduces an old
``low-energy theorem''.  However, at $O(P^3)$ there are also finite
loop corrections where the photon interacts with a virtual pion which
then rescatters on the nucleon.  This large quantum effect was missed
in the old ``low-energy theorem'' and in most models. At $O(P^4)$
there are loops with relativistic corrections at the vertices together
with new counterterms.  Results
\cite{Br96C} for $E_{0+}$ up to $O(P^4)$ from a fit to data
above threshold constrained by resonance saturation of parameters are
shown in Table \ref{T:vkolck:E0+}
\footnote{In this calculation, the dominant isospin breaking effect, 
namely the
charged to neutral pion mass difference 
has been taken into account. A fully consistent calculation 
including all effects from virtual photons and the quark mass differences is
not yet available.
The result for $E_{0+}^{{\pi ^0}n}$ is based on the
assumption that the counterterms entering at $O(P^4)$ are the same
for the proton and the neutron apart from trivial isospin factors.}
.    

\begin{table}[hbt]
%\begin{table}[b!]
% -----------------------------------------------------
% adapted from TeX book, p. 241
\newlength{\digitwidth} \settowidth{\digitwidth}{\rm 0}
\catcode`?=\active \def?{\kern\digitwidth}
% -----------------------------------------------------
\caption{Values for $E_{0+}$ in units of $10^{-3}/m_{\pi^+}$
at various orders in $\cpt$ and in experiments, for the 
different channels.}
\label{T:vkolck:E0+}
\vspace{0.2cm}
\begin{tabular*}{\textwidth}{@{}l@{\extracolsep{\fill}}ccccc}
\hline
$E_{0+}$($10^{-3}/m_{\pi^+}$)& $P$ & $P^2$ & $P^3$ & $P^4$ 
                                                         & Experiment\\
\hline
$\gamma p\rightarrow \pi^+ n$ &    34.0 &    26.4  &    28.9  &    28.2
                                                   &    27.9 $\pm$ 0.5\\
                              &         &          &          &        
                                                   &    28.8 $\pm$ 0.7\\
                              &         &          &          &        
                                                   &    27.6 $\pm$ 0.3\\
$\gamma n\rightarrow \pi^- p$ & $-$34.0 & $-$31.5  & $-$32.9  & $-$32.7 
                                                   & $-$31.4 $\pm$ 1.3\\
                              &         &          &          &
                                                   & $-$32.2 $\pm$ 1.2\\
                              &         &          &          &
                                                   & $-$31.5 $\pm$ 0.8\\
$\gamma p\rightarrow \pi^0 p$ &     0   &  $-$3.58 &     0.96 &  $-$1.16
                                                   &  $-$1.31 $\pm$ 0.08\\
                              &         &          &          &
                                                   &  $-$1.32 $\pm$ 0.08\\
$\gamma n\rightarrow \pi^0 n$ &     0   &     0    &     3.7  &    2.13
                                                   &          ?    \\
\hline
\end{tabular*}
\end{table}

Charged-pion channels are indeed larger than neutral pion channels and
show good convergence and agreement with experimental data.  For
neutral-pion channels convergence is less apparent, but the absolute
values are much smaller.  The $O(P^4)$ result for $\gamma p\rightarrow
\pi^0 p$ is in relatively good agreement with the recent results from
Mainz and Saskatoon.  To the same order there is a prediction for the
near-threshold behavior of the $\gamma n\rightarrow \pi^0 n$ reaction;
the cross-section is considerably larger than that obtained in models
that omit the important one-loop $O(P^3)$ diagrams. In fact, the
neutron cross-section is a factor of four larger than the proton
cross-section, in violation with classical intuition.  Neutral pion
photoproduction on the deuteron is the best candidate for a test of
this peculiar prediction for the neutron.

We now consider pion photoproduction on the deuteron~\cite{Be97}.
Consider a process with two nucleons, a single photon and a single
pion in the initial and final state, respectively. Following the
counting rules outlined above, the contributions are ordered as
follows: ${O(P)}$: At LO, the matrix element is given by tree graphs
with one spectator nucleon, with LO vertices, O(P). This contribution
is nonvanishing only for charged pion photoproduction.  At this order
one finds the Kroll-Ruderman term (Fig. \ref{countmod}a).  ${O(P^2)}$:
The first corrections to the leading terms are the same as above but
with $O(P^2)$ vertices. For example, there is a $1/M$ correction at
this order (Fig. \ref{countmod}b).  ${O(P^3)}$: There are three
classes of corrections at this order: {\it (i)} One-loop graphs with
one spectator nucleon (Fig. \ref{countmod}c).  {\it (ii)} Tree graphs
with one spectator nucleon and with either one $O(P^3)$ vertex
(Fig. \ref{countmod}d) or two $O(P^2)$ vertices.  In the case of
neutral pion photoproduction, the counterterm graphs with one $O(P^3)$
vertex vanish, while graphs with two $O(P^2)$ vertices give rise to
$1/M^2$ corrections proportional to the nucleon magnetic moments.
{\it (iii)} Tree graphs with an interaction between nucleons with only
$O(P)$ vertices. These graphs are of two types: two-nucleon graphs
like the Feynman graph of Fig. \ref{countmod}e and time-ordered graphs
such as those in Fig. \ref{countmod}f.  ${O(P^4)}$: There are many new
contributions at this order, which can be constructed following the
same procedure. Most of these are calculable corrections to the lower
orders, but among the single-scattering diagrams for neutral pion
photoproduction an undetermined counterterm appears.

\begin{figure}[t,h.b,p]
\centerline{\psfig{figure=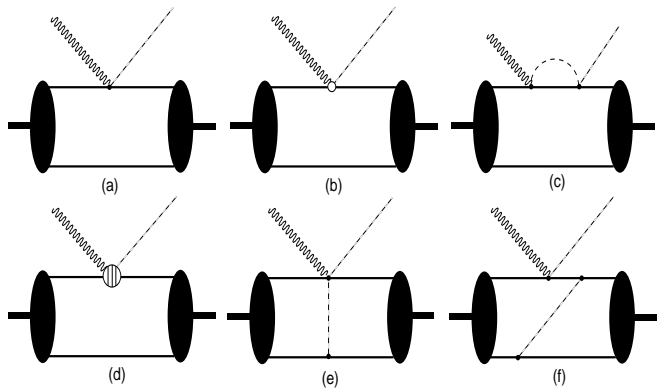,height=2.0in,width=3.4in}}
\caption{Characteristic graphs which contribute at each of the first
three orders in pion photoproduction. 
\label{countmod}}
\end{figure}

In the case of neutral pion photoproduction at threshold a
number of the graphs in Fig. \ref{countmod} will not contribute: those
where the photon line is attached to a pion, and those that go like $S
\cdot q$, where $q$ is the outgoing pion momentum and $S$ is the
covariant nucleon spin.  In particular, all the leading-order graphs
(Fig. \ref{countmod}a) vanish, which immediately suggests that the
cross-section will be smaller than that for charged pion production,
and more sensitive to two-nucleon contributions.  Moreover, the
time-ordered graphs (Fig. \ref{countmod}f) also vanish. It should be
noted that contributions from the delta are minimized at
threshold, and so integrating out the delta should not damage convergence
of the theory too much.  As noted above, to $O(P^4)$ there is one
undetermined parameter, which is fitted to $\gamma p \rightarrow \pi^0
p$ above threshold.

We limit ourselves here to threshold neutral-pion photoproduction on
the deuteron~\cite{Be97}.  As explained in the introduction, this is
of special interest because it can provide information on the
elementary neutron amplitude.

Consider the reaction $\gamma (k) + d(p_1) \to \pi^0 (q) + d(p_2)$ in
the threshold region, $\vec{q} \simeq 0$, where the pion is in an $S$
wave with respect to the center-of-mass (cm) frame.  For real photons
($\vec{\epsilon} \cdot \vec{k} = 0$, with $\vec \epsilon$ the photon
polarization vector), the differential cross-section at threshold
takes the form
\begin{equation}
{{|{\vec k}|}\over{|{\vec q\,}|}}\;{{d\sigma}\over{d\Omega}}
\bigg|_{|{\vec q\,}|=0} =\frac{8}{3}{E_d^2}.
\end{equation}
We now present the results of the chiral expansion of the dipole
amplitude $E_d$ to $O(P^4)$.  The single-scattering contribution is
given by all diagrams where the photon and the pion are absorbed and
emitted, respectively, from one nucleon with the second nucleon acting
as a spectator (the impulse approximation).  Some of these diagrams
are depicted in Fig. \ref{countmod}.  To $O(P^4)$ one finds
\begin{eqnarray} \label{Edss}
{E^{ss}_{d}} &=&
\frac{1+{m_\pi}/{M}} {1+{m_\pi}/{m_d}} \, \biggl\{ \frac{1}{2} \,
({E_{0+}^{{\pi ^0}p}}+{E_{0+}^{{\pi ^0}n}} ) \, \int d^{3}{p} \; 
{{\phi}}^{\ast}(\vec{p}) \, \vec{\epsilon} \cdot \vec{J} \,
{{\phi}}(\vec{p}-\vec{k}/2) \nonumber \\
&-& \frac{k}{M} \, \hat{k} \cdot \int d^{3}{p} \; \hat{p} \,
\frac{1}{2} \, ({P_{1}^{{\pi ^0}p}}+{P_{1}^{{\pi ^0}n}} ) \,
{{\phi}}^{\ast}(\vec{p})\, \vec{\epsilon} \cdot \vec{J} \,
{{\phi}}(\vec{p}-\vec{k}/2) \biggr\},
\end{eqnarray}
evaluated at the threshold value
\begin{equation}
|\vec{k}| =  k_{\rm thr} 
= m_{\pi^0}\left(1 - \frac{m_{\pi^0}^2}{2 m_d}\right) = 130.1 \, {\rm MeV}.
\label{valk}
\end{equation}
Here $m_d$ is the deuteron mass, $\vec J = (\vec{\sigma}_1 +
\vec{\sigma}_2 )/2$, and $\phi$ is the deuteron wavefunction.

Two remarks concerning  Eq.~(\ref{Edss}) are in order:

(1) The elementary $S$-wave pion production amplitudes to $O(P^4)$
\cite{Br96C} are given in Table \ref{T:vkolck:E0+}.  The proton
amplitude that results from a one-parameter fit above threshold
compares well with experiment.  The neutron amplitude is predicted to
be considerably larger than the one obtained from the old ``low-energy
theorem'' that omits the important one-loop $O(P^3)$ single-scattering
diagrams (Fig. \ref{countmod}c).

(2) It is important to differentiate between the $\pi^0 d$ and the
$\pi^0 N$ ($N=p,n$) cm systems.  At threshold in the former, the pion
is produced at rest; it has, however, a small three-momentum in the
latter \cite{KW79,La81}.  Consequently, a single-nucleon $P$-wave
contribution appears which is proportional to the elementary
amplitudes $P_{1}^{{\pi ^0}p}$ and $P_{1}^{{\pi ^0}n}$ that obey the
$P$-wave low-energy theorems \cite{Br96C}

\begin{equation}\label{Plets}
P_{1}^{{\pi ^0}p} = 0.480 \, |\vec{q}\,| \,{\rm GeV}^{-2}, \, \quad
P_{1}^{{\pi ^0}n} = 0.344 \, |\vec{q}\,| \,{\rm GeV}^{-2},
\end{equation}
with

\begin{equation}\label{boost}      
\vec{q} =m_\pi \left[ \left(1 - \frac{m_\pi}{M}\right) \frac{\vec{p}}{M} 
      -\frac{m_\pi}{2M}\left(1 - \frac{5m_\pi}{4M}\right) \hat{k} \right],
\end{equation}
where $\vec{p}$ is the nucleon momentum in the $\pi^0 d$ cm system.
The $P$-wave amplitude for neutral pion production off protons holds
to within 3\%.  Here we neglect the energy dependence of the
elementary $\pi^0 p$ and $\pi^0 n$ $S$- and $P$-wave amplitudes since
the pion energy changes only by 0.4\% for typical average nucleon
momenta in the deuteron.

The two-nucleon contributions  at $O(P^3)$ that survive at
threshold in the Coulomb gauge 
are shown in Fig. \ref{q3newmod}.  All vertices
come from the lowest order Lagrangian.  
\begin{figure}[t,h.b,p]
\centerline{\psfig{figure=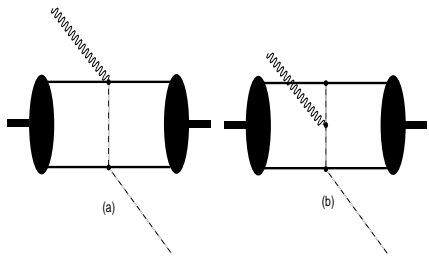,height=1.3in,width=2.2in}}
\caption{
Two-nucleon graphs which contribute to neutral pion photoproduction
at threshold to $O(P^3)$ (in the Coulomb gauge).
All vertices come from ${\cal L}^{(0)}$.
\label{q3newmod}}
\end{figure}

At $O(P^4)$, we have to consider tree graphs with exactly one $O(P^4)$
vertex.  The relevant terms in this Lagrangian fall into two classes,
the first with fixed coefficients due to Lorentz invariance, the
second with four low-energy constants $c_1 , \ldots, c_4$ that are
related to pion-nucleon scattering and two other constants that are
related to the isoscalar and isovector nucleon anomalous magnetic
moments. The pertinent two-nucleon diagrams are shown in
Fig. \ref{fig3mod2}, the blob characterizing an insertion from ${\cal
L}_{\pi N}^{(2)}$. In addition, there are relativistic corrections to
the graphs in Fig.~\ref{q3newmod}.

\begin{figure}[t,h.b,p]
\centerline{\psfig{figure=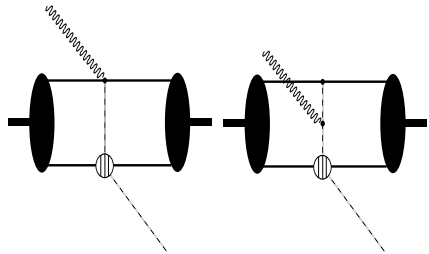,height=1.3in,width=2.2in}}
\vspace{0.75cm}
\caption{
Characteristic two-nucleon graphs contributing at $O(P^4)$ to neutral pion photoproduction.
The hatched circles denote an insertion from ${\cal L}_{\pi N}^{(2)}$.
\label{fig3mod2}}
\end{figure}

One can show that all terms proportional to the low-energy constants
$c_{i}$ vanish at threshold leaving
insertions of
terms $\sim 1/2M$, $\sim g_A/2M$ and $\sim \kappa_{0,1}$. 
Counting powers of momenta, one sees that the contributions from some
graphs at this order are divergent.  This is due to the fact that in
the chiral expansion one has truncated the pion-nucleon form factor
and thus does not suppress the high-momentum components. This
phenomenon also occurs in the calculation of the $NN$ potential in
$\cpt$ \cite{Or92,Or94,Or96}.  As was done there, we introduce an
additional Gaussian cut-off factor of the form $F({\vec q}\,^2) =
\exp\{-{\vec q}\,^2 / \Lambda^2_\pi \}$, where the cut-off
$\Lambda_\pi$ varies between the rho mass and 
$\Lambda_\chi$.  
Notice that the divergences appearing
in other graphs all cancel and thus no further regularization is
needed.

Note that to $O(P^4)$ there are no four-nucleon operators contributing
to the deuteron electric dipole amplitude.  The only possible graphs of this
type that contribute at threshold to this order come from an $NNNN\pi$
vertex, to which a photon is attached either at the outgoing pion leg
or at the vertex itself.  Both vanish for neutral pions as they result
from minimal substitution in a pion derivative.  We therefore conclude
that for threshold neutral pion photoproduction to $O(P^4)$ there is
no contribution from any four-nucleon operator and thus {\it no new,
undetermined parameters appear}.

Using the Argonne V18 \cite{Wi95}, the Reid \cite{Re68}, the Nijmegen
\cite{Na78} and the Paris \cite{Co73,La80} potentials, we find
\begin{equation} 
\label{Edssval}
E^{ss}_{d}=(0.36\pm 0.05) \times 10^{-3}/m_{\pi^+}
\ \ \ .
\end{equation}
The $P$-wave
contribution amounts to a 3\% correction to the $S$ wave. 
The sensitivity of the
single-scattering contribution $E_d^{ss}$ to the elementary neutron
amplitude is given by
\begin{equation}
E_d^{ss} = \left[ 0.36 - 0.38 \cdot (2.13 - E_{0+}^{{\pi ^0}n})
\right] \times 10^{-3}/m_{\pi^+}
\ \ \ .
\label{sensi}
\end{equation}
Consequently, for $E_{0+}^{{\pi
^0}n} = 0$, we have $E_d^{ss} = -0.45$ which is of opposite sign to
the value in Eq.~(\ref{Edssval}) based on the $\cpt$ prediction for
$E_{0+}^{{\pi ^0}n}$ (Table \ref{T:vkolck:E0+}).  If one were to use
the empirical value for the proton amplitude (Table
\ref{T:vkolck:E0+}), the single-scattering contribution would be
somewhat reduced.

We find for the two-nucleon contribution at $O(P^3)$

\begin{equation}
E_d^{tb,3} = (-1.90\, ,\, -1.88 \, , \, -1.85 \, , \, -1.88)
\times 10^{-3}/m_{\pi^+} \,\, ,
\end{equation}
for the Argonne V18, Reid, super soft core \cite{dTS73} and Bonn
\cite{Ma87} potentials, in order.
Further, the result
changes by only a few percent if the chiral potential of
Ref. \cite{Or96} is used~\cite{Be97}.  
From the graphs of $O(P^4)$ we find
\begin{equation}
E_d^{tb,4} = (-0.25\, ,\, -0.23 \, , \, -0.27 )
\times 10^{-3}/m_{\pi^+} \,\, ,
\label{Edtb4}
\end{equation}
for the Argonne V18, Reid and super soft core potentials,
respectively, setting $\Lambda_\pi=1\,$GeV. These amount to
corrections of the order of 15\% to the $O(P^3)$ two-nucleon terms.
If one varies the cut-off $\Lambda_\pi$ from $0.65$ to $1.5\,$GeV, the
$O(P^4)$ two-nucleon contribution using the Argonne V18 potential
varies between $-0.24$ and $-0.29$ (in canonical units), which is a
modest cut-off dependence.

In summary, taking the Argonne V18 potential for definiteness, the
chiral expansion of the electric dipole amplitude $E_d$ is shown in
Table \ref{T:vkolck:Ed}.  
Allowing
$E_{0+}^{\pi^0 p}$ to vary between $-1.0$ and $-1.5$ and for a
fixed value of $E_{0+}^{\pi^0 n} = 2.13$, $E_d^{ss}$ varies between
0.2 and 0.4 (all numbers in canonical units).  The results for the
two-nucleon contributions are stable at $O(P^3)$, whereas we assign a
conservative uncertainty of $\pm 0.1$ to the corrections at next order
due to the cut-off dependence.  We thus obtain the $O(P^4)$ {\it
prediction}~\cite{Be97}

\begin{equation}
E_d = (-1.8 \pm 0.2)  \times 10^{-3}/m_{\pi^+}.
\label{Edtot}
\end{equation}

\begin{table}[t,h.b,p]
\caption{Values for $E_{d}$ in units of $10^{-3}/m_{\pi^+}$
from one-nucleon contributions ($1N$) up to $O(P^4)$,
two-nucleon kernel ($2N$) at $O(P^3)$ 
and at $O(P^3)$ ,
and their sum ($1N+2N$).\label{T:vkolck:Ed}}
\vspace{0.2cm}
\begin{center}
\begin{tabular}{|cccc|}
\hline
$1N$   & \multicolumn{2}{c}{$2N$} &    $1N+2N$        \\
\cline{2-3}
$P+P^2+P^3+P^4$ & $P^3$ & $P^4$   
                                          & $P+P^2+P^3+P^4$  \\
\hline
 0.36  & $-$1.90 & $-$0.25                  & $-$1.79 \\
\hline
\end{tabular}
\end{center}
\end{table}

To see the sensitivity to the elementary neutron amplitude, we set the
latter to zero and find $E_d = -2.6\times 10^{-3}/m_{\pi^+}$ (for the
Argonne V18 potential) which results in a cross-section twice as large
as the one from the $\cpt$ prediction (\ref{Edtot}).  
This is consistent with previous model results
\cite{KW79,La81}.

This is a unique situation: the EFT makes a prediction that differs
significantly from conventional models. Such a test was recently
carried out at Saskatoon
\cite{Be98}.  The experimental results for the pion photoproduction
cross-section near threshold are shown in
Fig.~\ref{piphotodata}, together with the $\cpt$ prediction at
threshold.  Inelastic contributions have been estimated in
Ref.~\cite{Be98} and are smaller than 10\% throughout the range of
energies shown.  At threshold, Ref.~\cite{Be98} extracts
\begin{equation}
E_{d}=(-1.45\pm 0.09)\times 10^{-3}/m_{\pi^+}.
\end{equation}
While agreement with $\cpt$ to order $O(P^4)$ is not better than a
reasonable estimate of higher-order terms, it is clearly superior to
models.  This is compelling evidence of the importance of chiral loops
and of the consistency of nuclear EFT.  A further test will come from
an electroproduction experiment currently under analysis in Mainz.
\begin{figure}[t,h.b,p]
\centerline{\psfig{figure=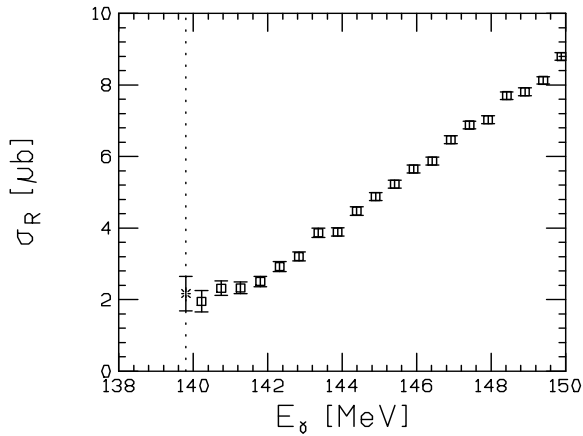,height=2.25in,width=3.in}}
\caption{
Reduced cross-section  $\sigma_R=(k/q)\sigma$ in $\mu$b
for neutral pion photoproduction as function of the photon 
energy in MeV.
Threshold is marked by a dotted line.
Squares are data points from Ref. \protect\cite{Be98}
and the star is the 
$\cpt$ {\it pre}diction of Ref. \protect\cite{Be97}.
Figure courtesy of Ulf Mei{\ss}ner.
\label{piphotodata}}
\end{figure}

%%%%%%%%%%%%%%%%%%%%%%%%%%%%%%%%%%%%%%%%%%%%%
\subsection{$\gamma^* d \rightarrow \pi^0 d$}
\label{sec-gammastardpi0d}

In the case of an off-shell photon the threshold cross-section
takes the form
\begin{equation}
\sigma (k^2)= |E_{d}|^2 + \epsilon_L \,  |L_{d}|^2~,
\end{equation}
with $\epsilon_L = -(k^2/k_0^2)\epsilon$ the longitudinal polarization
for a photon with polarization $\epsilon$. Now in addition to the
transverse multipole $E_{d}$, there is a longitudinal multipole
$L_{d}$ as well. Both of these multipoles are functions of the photon
virtuality, $k^2$.

These multipoles have been computed to $O(P^3)$ in the chiral
expansion. To this order there are graphs with one spectator nucleon
and a contact operator, together with the two graphs of
Fig.~\ref{q3newmod} with an off-shell photon.  The proton electric
dipole amplitude at the photon point is not well described to this
order. However, the $k^2$--dependence of the elementary amplitude for
the proton is dominated by the third order contribution
(see~\cite{Br96} for a discussion).  The ${\cal O}(P^3)$ results for
$E_d$ and $L_d$ are shown in Fig.~\ref{EL}.
\begin{figure}[!ht]
   \centerline{\psfig{figure=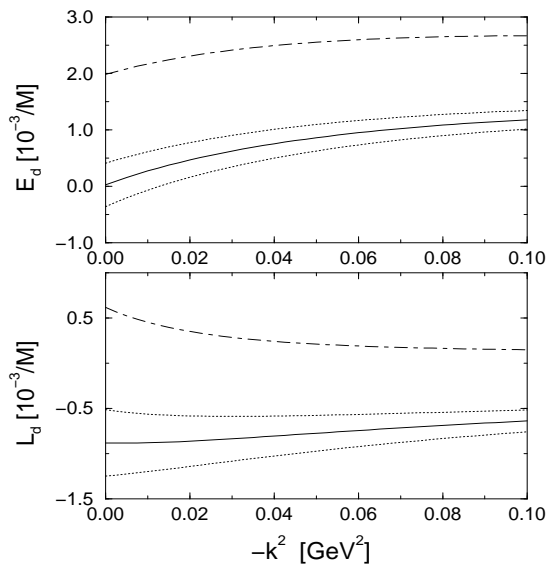,height=3.0in,width=2.8in}}
   \caption{
S--wave multipoles of the deuteron. In the upper and lower panel,
the electric dipole and the longitudinal multipoles are
shown, respectively, by the solid lines. The dot--dashed line
is the single scattering contribution. The dotted lines are obtained
by changing the elementary neutron amplitude by $\pm 1 \times 10^{-3}
/M_{\pi^+}$. 
\label{EL}}
\end{figure}
In both cases, the two-body contribution is large and dominates the
amplitudes. Note that $E_d$ varies more significantly with increasing
$|k^2|$ than $L_d$. As illustrated by the dotted lines in
Fig.~\ref{EL}, which have been obtained by a constant shift of the
$\pi^0 n$ amplitudes by $\pm 1 \times 10^{-3}/M_{\pi^+}$, there is
sensitivity to the elementary scattering off the neutron. To get a
more realistic estimate for the S--wave cross-section, one can adjust
the values of ${X}_{0+}^{\pi^0 p}$ and ${X}_{0+}^{\pi^0 n}$
$(X=\{E,L\}$) at the photon point to the values obtained from the
$O(P^4)$ calculation. More precisely, the electric dipole amplitudes
are taken from ref.\cite{Br96C}, the $L_{0+}^{\pi^0 p}$ from the best
fit obtained in ref.\cite{Br96} and the $L_{0+}^{\pi^0 n}$ using the
resonance saturation estimate with $g_3 = -125.6$ and $X' = -0.23$, as
detailed in~\cite{Br96}.  The so calculated S--wave cross-section is
collected in Table~\ref{elecdtable} for a photon polarization
$\epsilon = 0.67$.  A range is given, which is obtained by adding a
constant shift of $\pm 1
\times 10^{-3}/M_{\pi^+}$ to the elementary $\pi^0 n$ longitudinal
multipole (but keeping the electric dipole amplitude at the same value
as before). This serves to illustrate the sensitivity of the S--wave
cross-section to the so far unmeasured $L_{0+}^{\pi^0 n}$
multipole. For comparison, note that the S--wave cross-section
measured on the proton~\cite{vdB97,Di98} for $-0.10 \le k^2
\le -0.04\,$GeV$^2$ lies between 0.15 and 0.45~$\mu$b. 
For $k^2 \simeq -0.1\,$GeV$^2$, $\sigma \simeq a_0/2$, with $a_0$ the
S--wave cross-section for neutral pion production off the
proton. However, the curvature of $\sigma (k^2)$ is rather different
from the proton case. This is partly due to the interference of the
proton and neutron amplitudes and partly a kinematical effect, since
for a given polarization $\epsilon$ and virtuality $k^2$, $\epsilon_L$
is larger for the proton than for the deuteron. An $O(P^4)$
computation of this process is currently underway.

\begin{table}[htb]
\begin{center}
\footnotesize
\begin{tabular}{|c||c|c|c|c|c|}
%\hline
$-k^2$ [GeV$^2$] & $0.02$  & $0.04$  & $0.06$  & $0.08$ 
                & $0.10$ \\ \hline
$\sigma$  [$\mu$b]       &  0.11   &  0.15  & 0.17  &  0.19 & 0.20
                \\ \hline           
$\sigma$ range [$\mu$b]  & 0.09 -- 0.13   &  0.12 -- 0.18    
                         & 0.14 -- 0.21   &  0.16 -- 0.23   
                         & 0.17 -- 0.24    \\ %\hline           
\end{tabular}
\end{center}
\caption{S--wave cross-section for the scaled single scattering
  amplitudes as explained in the text. The range is obtained by 
  adding a constant shift of $\pm 1\times 10^{-3}/M_{\pi^+}$ to
  the elementary $\pi^0 n$ longitudinal multipole.\label{elecdtable}}
\end{table}

%%%%%%%%%%%%%%%%%%%%%%%%%%%%%%%%%%%%%%%%%%%%%
\subsection{$\pi  d \rightarrow \pi  d$}
\label{sec-pid}

To $O({P^3})$ in $\cpt$ the $\pi$-deuteron scattering length can be written
as~\cite{We92}
\begin{equation}
a_{\pi d}=\frac{(1+\mu)}{(1+\mu /2)}(a_{\pi n} + a_{\pi p})+{a^{(2a)}}+
{a^{(2b,2c)}},
\label{eq:adfirst}
\end{equation}
where $\mu\equiv{m_\pi}/M$ is the ratio of the pion and the nucleon
mass and the $\pi$-N scattering lengths have the decomposition
\begin{figure}[!ht]
   \centerline{\psfig{figure=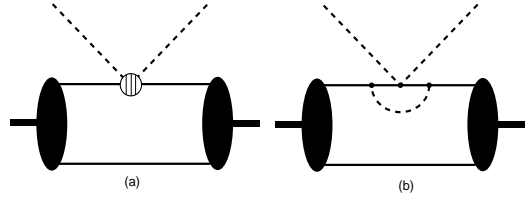,height=1.0in,width=2.7in}}
   \caption{Contributions to
   the $\pi d$ scattering length at ${\cal O}(P^2)$ (a) and 
${\cal O}(P^3)$(b).  
The dots are vertices from ${\cal L}_{\pi N}^{(1)}$ and the
   sliced blob is from ${\cal L}_{\pi N}^{(2)}$.
\label{pidscattLO2mod} }
\end{figure}
$a_{\pi n} + a_{\pi p}=2 a^{+}={a^{(1a)}}+{a^{(1b)}}$
where $a^+$ is the isoscalar S-wave scattering length and the
superscripts refer to the figures. The various diagrammatic
contributions to $a_{\pi d}$ are illustrated in
Fig.~\ref{pidscattLO2mod} and Fig.~\ref{pidscattLOmod}. The leading
contribution, Fig.~\ref{pidscattLO2mod}(a), has a vertex from ${\cal
L}_{\pi N}^{(2)}$ (${P^{2}}$) and is therefore
$O(P^2)$. Fig.~\ref{pidscattLO2mod}(b) has two nucleon propagators
(${P^{-2}}$), one pion propagator (${P^{-2}}$), three vertices from
${\cal L}_{\pi N}^{(1)}$ (${P^{3}}$) and a two-body factor (${P^{3}}$)
and so gives an $O({P^{3}})$ contribution. One can further verify that
the graphs of Fig.~\ref{pidscattLOmod} are $O({P^3})$. Together,
Figs.~\ref{pidscattLO2mod} and
\ref{pidscattLOmod} are all that contribute at $O({P^3})$ at threshold.

The contributions to $a_{\pi d}$ from the graphs of Fig.~\ref{pidscattLO2mod}
are:
\begin{equation}
{a^{(2a)}}= - {{{m_\pi^2}}\over{32{\pi^4}{f_\pi^4}{(1+\mu /2)}}}
\langle{{1}\over{{\vec q}^{\,2}}}\rangle_{\sl wf}
\label{eq:numone}
\end{equation}
\begin{equation}
{a^{(2b,2c)}}={{{g_A^2}{m_\pi^2}}\over
{128{\pi^4}{f_\pi^4}{(1+\mu /2)}}}
\langle{{{\vec q}\cdot{{\vec\sigma}_1}{\vec q}\cdot{{\vec\sigma}_2}}\over
{({\vec q}^{\,2}+{m_\pi^2})^2}}\rangle_{\sl wf}
\label{eq:numtwo}
\end{equation}
where $\langle\vartheta\rangle_{\sl wf}$ indicates that $\vartheta$ is
sandwiched between deuteron wavefunctions.  These matrix elements have
been evaluated using a cornucopia of wavefunctions; results are
displayed in Table~\ref{tab:exp}. Clearly ${a^{(2a)}}$ dominates. This
is the result of the shorter range nature of $a^{(2b,2c)}$.  It is
important to stress that the dominant contribution from these graphs
is quite independent of the wavefunction used. This implies that
$\cpt$, which relies on the dominance of the pion-exchange, is useful
in this context.
\begin{figure}[!ht]
   \centerline{\psfig{figure=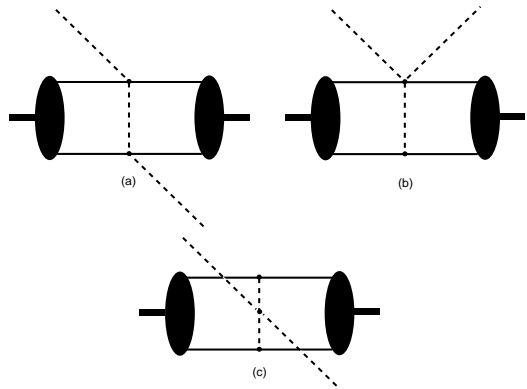,height=2.0in,width=2.7in}}
   \caption{Contributions to the $\pi$-deuteron
   scattering length at ${\cal O}(P^3)$.
\label{pidscattLOmod} }
\end{figure}

To $O({P^3})$ in $\cpt$~\cite{Br93}:
\begin{equation}
4\pi(1+\mu )a^{+}=
{{m_\pi^2}\over{f_\pi^2}}\biggl(-4c_1 +2c_2 -{{g_A^2}\over{4M}}+2c_3 \biggr)
+{{3{g_A^2}{m_\pi^3}}\over{64\pi{f_\pi^4}}},
\end{equation}
where the ${c_i}$ are low-energy constants from ${\cal L}_{\pi
N}^{(2)}$. The sole undetermined parameter entering the
$O({P^3})$ computation of $a_{\pi d}$ is therefore a combination of
$c_1$, $c_2$ and $c_3$:
$\Delta\equiv {-4c_1 +2(c_2 +c_3)}$.
\begin{table}[t]
\caption{Contributions of Fig.~\ref{pidscattLOmod} for various deuteron
wavefunctions in units of $m_\pi^{-1}$.  We use ${f_\pi}=92.4$\,MeV,
${g_A}=1.32$ and ${m_{\pi^+}}=139.6\,$MeV.
\label{tab:exp}}
\vspace{0.2cm}
\begin{center}
\footnotesize
\begin{tabular}{|l|r|r|r|c|}
    \hline ${\sl wf}$ & $a^{(2a)}$ & $a^{(2b,2c)}$ \\ \hline 
    Bonn\protect{~\cite{Ma87}} &
  $-0.02021$ & $-0.0005754$ \\ 
   ANL-V18\protect{~\cite{Wi95}} & $-0.01960$ & $-0.0007919$ \\
   Reid-SC\protect{~\cite{Re68}} & 
  $-0.01941$ & $-0.0008499$ \\ 
   SSC\protect{~\cite{dTS73}} & 
  $-0.01920$ & $-0.0006987$ \\ \hline 
\end{tabular}
\end{center}
\end{table}

There is recent experimental information about both the $\pi$ N and
$\pi d$ scattering lengths~\cite{Ch95,Si95,Si97}.  Since $a^+$
involves constants that are not fixed by chiral symmetry we can use
experimental information about $\pi d$ scattering to predict $a^+$;
the recent Neuchatel-PSI-ETHZ (NPE) pionic deuterium
measurement~\cite{Ch95} gives
\begin{equation}
a_{\pi d}=-0.0259 \pm 0.0011\,{m_\pi^{-1}}.
\label{eq:latestpid}
\end{equation}
For the contributions of Fig.~\ref{pidscattLOmod} we take the average of the
$a^{(2a)}$ and $a^{(2b,2c)}$ values in Table~\ref{tab:exp}:
$a^{(2a+2b+2c)}=-0.0203 \,{m_\pi^{-1}}$.
We then find from Eq.~(\ref{eq:adfirst}):
\begin{equation}
a^{+}=-(2.6 \pm 0.5)\,\cdot\,{10^{-3}}{m_\pi^{-1}}.
\label{eq:apluspred}
\end{equation}
Note that although $a^{(2b,2c)}$ is small, there is a strong
cancellation between $a^{(2a)}$ and $a_{\pi d}$ which leads to a
sensitivity to $a^{(2b,2c)}$. Our value of $a^+$ is not consistent
with the Karlsruhe-Helsinki value~\cite{Ko86},
$a^{+}=-(8.3 \pm 3.8)\,\cdot\,{10^{-3}}{m_\pi^{-1}}$,
or the new NPE value deduced from the strong interaction shifts in
pionic hydrogen and deuterium, which is small and 
positive~\cite{Si95,Si97},
$a^{+}=(0...5)\,\cdot\,{10^{-3}}{m_\pi^{-1}}$.
The result Eq.~(\ref{eq:apluspred}) agrees, however, with the value
obtained in the SM95 partial-wave analysis~\cite{Ar95}, $a^{+}=-3.0\,\cdot\,
{10^{-3}}{m_\pi^{-1}}$.

Given the ambiguous experimental situation regarding $a^{+}$, it seems
most profitable to turn our formula around and use the $\pi d$
scattering data to constrain $\Delta$. We can write
\begin{equation}
\Delta =
\frac{2\pi{f_\pi^2}}{m_\pi^2}(1+\mu /2)
\lbrace a_{\pi d}-({a^{(2a)}}+{a^{(2b,2c)})}\rbrace
+\frac{g_A^2}{4M}\bigl(1-\frac{3M{m_\pi}}{16\pi{f_\pi^2}}\bigr).
\end{equation}
Using Eqs.~(\ref{eq:numone}), (\ref{eq:numtwo}) and (\ref{eq:latestpid})
we find
$\Delta =-(0.08\pm 0.02)\, {\rm GeV}^{-1}$,
where we have taken into account the error in the determination of
$a_{\pi d}$.

In Table~\ref{tab:exp2} we give values of the relevant $c_i$
obtained from a realistic fit to low-energy pion-nucleon scattering
data and subthreshold parameters~\cite{Br97}.  Central values lead to
$\sigma (0)=47.6\,$MeV and $a^+ =-4.7\cdot 10^{-3}{m_\pi^{-1}}$. These
values of the $c_i$ give the conservative determination:
$\Delta =-(0.18\pm 0.75)\, {\rm GeV}^{-1}$.
\begin{table}[!ht]
\caption{Values of the LECs $c_i$ in GeV$^{-1}$ for $i=1,\ldots,3$.  
Also given are the central values (cv) and the ranges for the $c_i$ from
resonance exchange. The $^*$ denotes an input quantity. 
\label{tab:exp2}}
\vspace{0.2cm}
\begin{center}
\footnotesize
\begin{tabular}{|l|r|r|r|c|}
    \hline
    $i$         & $c_i \quad \quad$   &  
                  $ c_i^{\rm Res} \,\,$ cv & 
                  $ c_i^{\rm Res} \,\,$ ranges    \\
    \hline
    1  &  $-0.93 \pm 0.10$  &  $-0.9^*$ & -- \\
    2  &  $3.34  \pm 0.20$  &  $3.9\,\,$ & $2 \ldots 4$ \\    
    3  &  $-5.29 \pm 0.25$  &  $-5.3\,\,$ 
                                     & $-4.5 \ldots -5.3$ \\
    \hline
    $\Delta$ & $-0.18 \pm 0.75$ & $0.8 \,\, $& $-3.0 \ldots +2.6$\\ 
    \hline
  \end{tabular}
\end{center}
\end{table}
Also shown in Table~\ref{tab:exp2} are values of the $c_i$ deduced from
resonance saturation. It is worth mentioning that an independent fit
to pion-nucleon scattering including also low-energy constants
related to dimension three operators finds results consistent with the
fit values of Table~\ref{tab:exp2}~\cite{Mo97}.

%%%%%%%%%%%%%%%%%%%%%%%%%%%%%%%%%%%%%%%%%%%%%
\subsection{$\gamma d \rightarrow \gamma d$}
\label{sec-gammagammad}

Nucleon Compton scattering has been studied in $\cpt$ in
Ref.~\cite{Br92,Br92B}, where the following results for the polarizabilities
were obtained at LO
\begin{eqnarray}
\alpha_p=\alpha_n={{5 e^2 g_A^2}\over{384 \pi^2 f_\pi^2 m_\pi}} &=&12.2 \times
10^{-4} \, {\rm fm}^3; \label{eq:alphaOQ3}\nonumber\\ 
\beta_p=\beta_n={{e^2 g_A^2}\over{768 \pi^2 f_\pi^2 m_\pi}}&=& 1.2 \times 
10^{-4} \, {\rm fm}^3. \label{eq:betaOQ3}
\end{eqnarray}
Recent experimental values for the proton polarizabilities are
\cite{To98}~\footnote{These are the result of a model-dependent
  fit to data from Compton scattering on the proton at several angles
  and at energies ranging from 33 to 309 MeV.}

\begin{eqnarray}
\alpha_p + \beta_p=13.23 \pm 0.86^{+0.20}_{-0.49} \times 10^{-4} \, {\rm fm}^3,
\nonumber\\
\alpha_p - \beta_p=10.11 \pm 1.74^{+1.22}_{-0.86} \times 10^{-4} \, {\rm fm}^3,
\label{polexpt}
\end{eqnarray}
where the first error is a combined statistical and systematic error,
and the second set of errors comes from the theoretical model
employed. These values are in good agreement with the $\cpt$
predictions.

On the other hand, the neutron polarizabilities are difficult to
obtain experimentally due to the absence of suitable neutron targets
and so the corresponding $\cpt$ prediction is not well tested.  One
way to extract neutron polarizabilities is to consider Compton
scattering on nuclear targets.  Consider coherent photon scattering on
the deuteron. The cross-section in the forward direction naively goes
as:

\begin{equation}
\left.{{d \sigma}\over{d \Omega}} \right|_{\theta=0}
\sim (f_{Th} - (\alpha_p + \alpha_n) \omega^2)^2.
\end{equation} 
The sum $\alpha_p + \alpha_n$ may then be accessible via its
interference with the dominant Thomson term for the proton,
$f_{Th}$. This means that with experimental knowledge
of the proton polarizabilities it may be possible to extract those for
the neutron.  Coherent Compton scattering on a deuteron target has
been measured at $E_\gamma=$ 49 and 69 MeV by the Illinois group
\cite{Lu94} and, using tagged photons, in the energy range
$E_\gamma= 84.2-104.5$ MeV at Saskatoon~\cite{Ho00}. Data for
$E_\gamma$ of about 60 MeV is currently being analyzed at Lund.

Clearly the amplitude for Compton scattering on the deuteron involves
mechanisms other than Compton scattering on the individual constituent
nucleons. Hence, extraction of nucleon polarizabilities requires a
theoretical calculation of Compton scattering on the deuteron that is
under control in the sense that it accounts for {\it all} mechanisms
to a given order in $\cpt$.  There exist a few calculations of this
reaction in the framework of conventional potential models
\cite{Wl95,LL98,KM99}.  These calculations yield similar results if
similar input is supplied, but typically mechanisms for nucleon
polarizabilities and two-nucleon contributions are not treated
consistently.  We will see that $\cpt$ provides a framework where this
drawback is eliminated.

\begin{figure}[!ht]
   \centerline{\psfig{figure=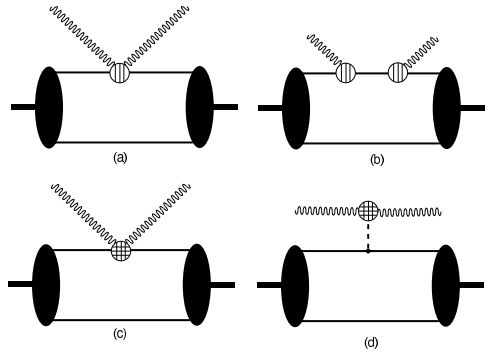,height=1.8in,width=2.5in}}
   \caption{Graphs
   which contribute to Compton scattering on the deuteron at
   ${\cal O}(P^2)$ (a) and ${\cal O}(P^3)$ (b-d).  The
   sliced and diced blobs are from ${\cal L}_{\pi N}^{(3)}$ (c) and
   ${\cal L}_{\pi \gamma}^{(4)}$ (d).  Crossed graphs are not
   shown.
\label{procomptonebodmod} }
\end{figure}

The Compton amplitude we wish to evaluate is (in the $\gamma d$
center-of-mass frame):

\begin{eqnarray}
T^{\gamma d}_{M' \lambda' M \lambda}(\vkayprime,\vkay)&=& \int
\frac{d^3p}{(2 \pi)^3} \, \, \psi_{M'}\left( \vpee + \frac{\vkay -
\vkayprime}{2}\right) \, \, T^{\gamma d \, \, c.m.}_{\gamma N_{\lambda'
\lambda}}(\vkayprime,\vkay) \, \, \psi_M(\vpee)\nonumber\\ 
&+& \int \frac{d^3p \, \, d^3p'}{(2 \pi)^6} \, \, \psi_{M'}(\vpeeprime) \, \,
T^{2N}_{\gamma NN_{\lambda' \lambda}}(\vkayprime,\vkay) \, \, \psi_M(\vpee)
\label{eq:gammad}
\end{eqnarray}
where $M$ ($M'$) is the initial (final) deuteron spin state, and
$\lambda$ ($\lambda'$) is the initial (final) photon polarization
state, and $\vkay$ ($\vkayprime$) the initial (final) photon
three-momentum, which are constrained to
$|\vkay|=|\vkayprime|=\omega$.  The amplitude $T^{\gamma d \, \,
c.m.}_{\gamma N}$ represents the graphs of
Fig.~\ref{procomptonebodmod} and Fig.~\ref{combined}b where the photon
interacts with only one nucleon.  Of course this amplitude must be
evaluated in the $\gamma d$ center-of-mass frame. The amplitude
$T^{2N}_{\gamma NN}$ represents the graphs of Fig.~\ref{combined}b
where there is an exchanged pion between the two nucleons.

The LO contribution to Compton scattering on the deuteron is
shown in Fig.~\ref{procomptonebodmod}(a).  This graph involves a
vertex from ${\cal L}_{\pi \gamma}^{(2)}$ and so is $O({P^{2}})$.
This contribution is simply the Thomson term for scattering on the
proton.  At next order, $O({P^{3}})$, there are several more graphs
with a spectator nucleon Fig.~\ref{procomptonebodmod}(b,c,d) and as
well as graphs involving an exchanged pion with leading order vertices
(Fig.~\ref{combined}a) and one loop graphs with a spectator
nucleon (Fig.~\ref{combined}b). The full amplitudes are given in
Ref.~\cite{Be99}.

\begin{figure}[t,h.b,p]
   \centerline{\psfig{figure=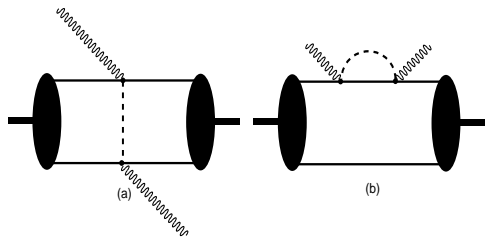,height=1.2in,width=2.5in}}
   \caption{Graphs which contribute to Compton scattering 
on the deuteron at ${\cal O}(P^3)$. 
Crossed graphs are not shown.}
\label{combined} 
\end{figure}

For the wave function $\psi$ we use the energy-independent Bonn OBEPQ
wave function parameterization which is found in Ref.~\cite{Ma87} as
well as the chiral wavefunction computed to N$^2$LO
in Ref.~\cite{Ep99}.  The photon-deuteron $T$-matrix
in Eq.~(\ref{eq:gammad}) is then calculated and the laboratory differential
cross-section evaluated directly from it:

\begin{equation}
  {{d \sigma}\over{d \Omega_L}}={{1}\over{16 \pi^2}}
  \left(\frac{E_\gamma'}{E_\gamma}\right)^2 \frac{1}{6} \sum_{M'
    \lambda' M \lambda} |T^{\gamma d}_{M' \lambda' M \lambda}|^2,
\end{equation}
where $E_\gamma$ is the initial photon energy in the laboratory frame
and $E_\gamma'$ is the final photon energy in the laboratory frame.
Fig.~\ref{plot.49} shows the
results at $E_\gamma = $ 49, 69, and 95 MeV. 
For comparison we have included the
calculation at $O(P^2)$, where the second contribution in
Eq.~(\ref{eq:gammad}) is zero, and the $\gamma N$ $T$-matrix in the
single-scattering contribution is given by the Thomson term on a
single nucleon. It is remarkable that to $O(P^3)$ no unknown
counterterms appear.  All contributions to the kernel are fixed in
terms of known pion and nucleon parameters such as $m_\pi$, $g_A$,
$M$, and $f_\pi$.  Thus, to this order $\cpt$ makes {\it predictions}
for Compton scattering.

The curves show that the correction from the $O(P^3)$ terms gets
larger as $\omega$ is increased, as was to be expected. Indeed, while
at lower energies corrections are relatively small, in the 95 MeV
results the correction to the differential cross-section from the
$O(P^3)$ terms is of order 50\%, although the contribution of these
terms to the {\it amplitude} is of roughly the size one would expect
from the power-counting: about 25\%.  Nevertheless, it is clear, even
from these results, that this calculation must be performed to
$O(P^4)$ before conclusions can be drawn about polarizabilities from
data at photon energies of order $m_\pi$. This is in accord with
similar convergence properties for the analogous calculation for
threshold pion photoproduction on the deuteron~\cite{Be97}.
\begin{figure}[!ht]
\hskip 0.1in\psfig{figure=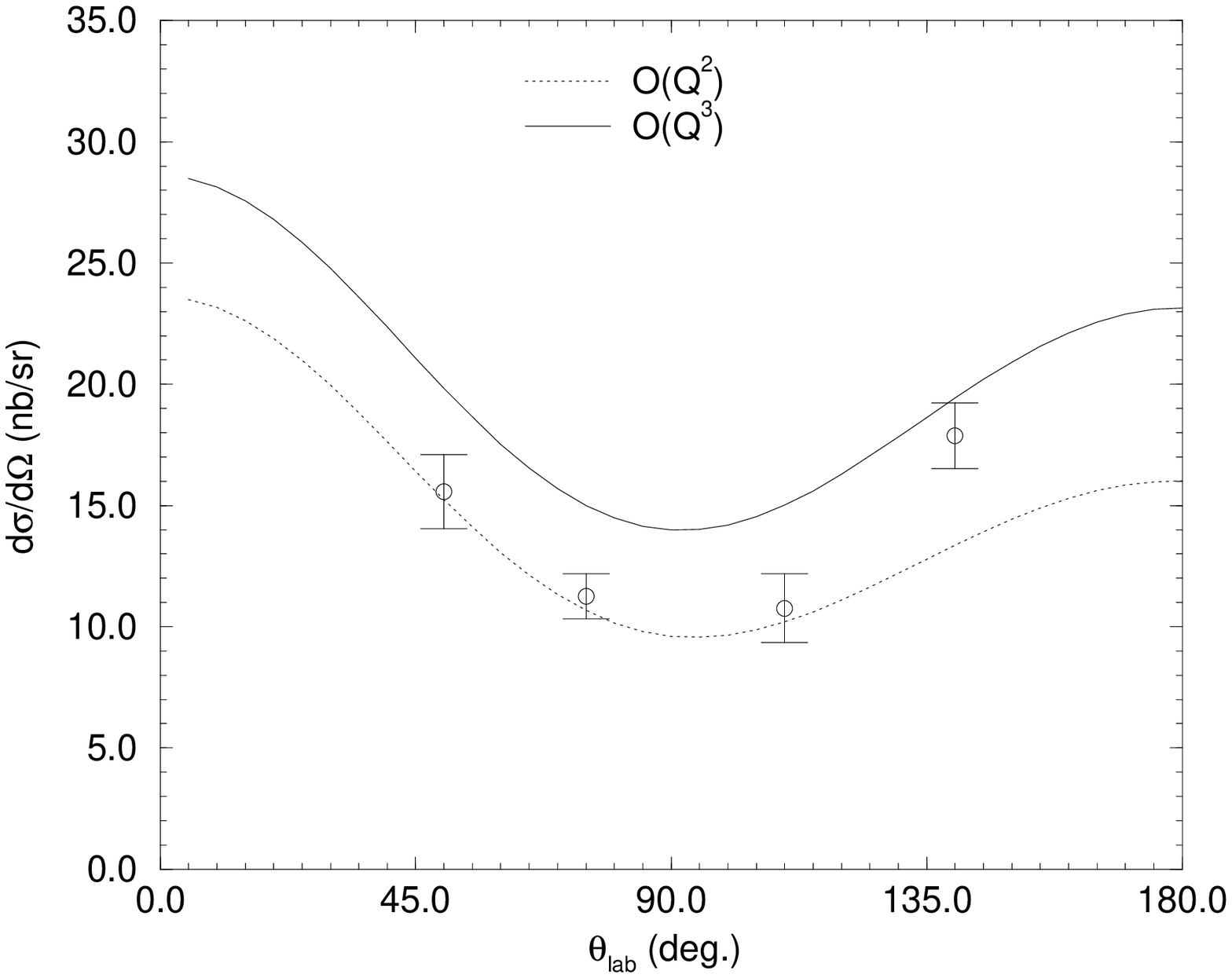,height=1.2in,width=1.5in}
\hskip 0.1in\psfig{figure=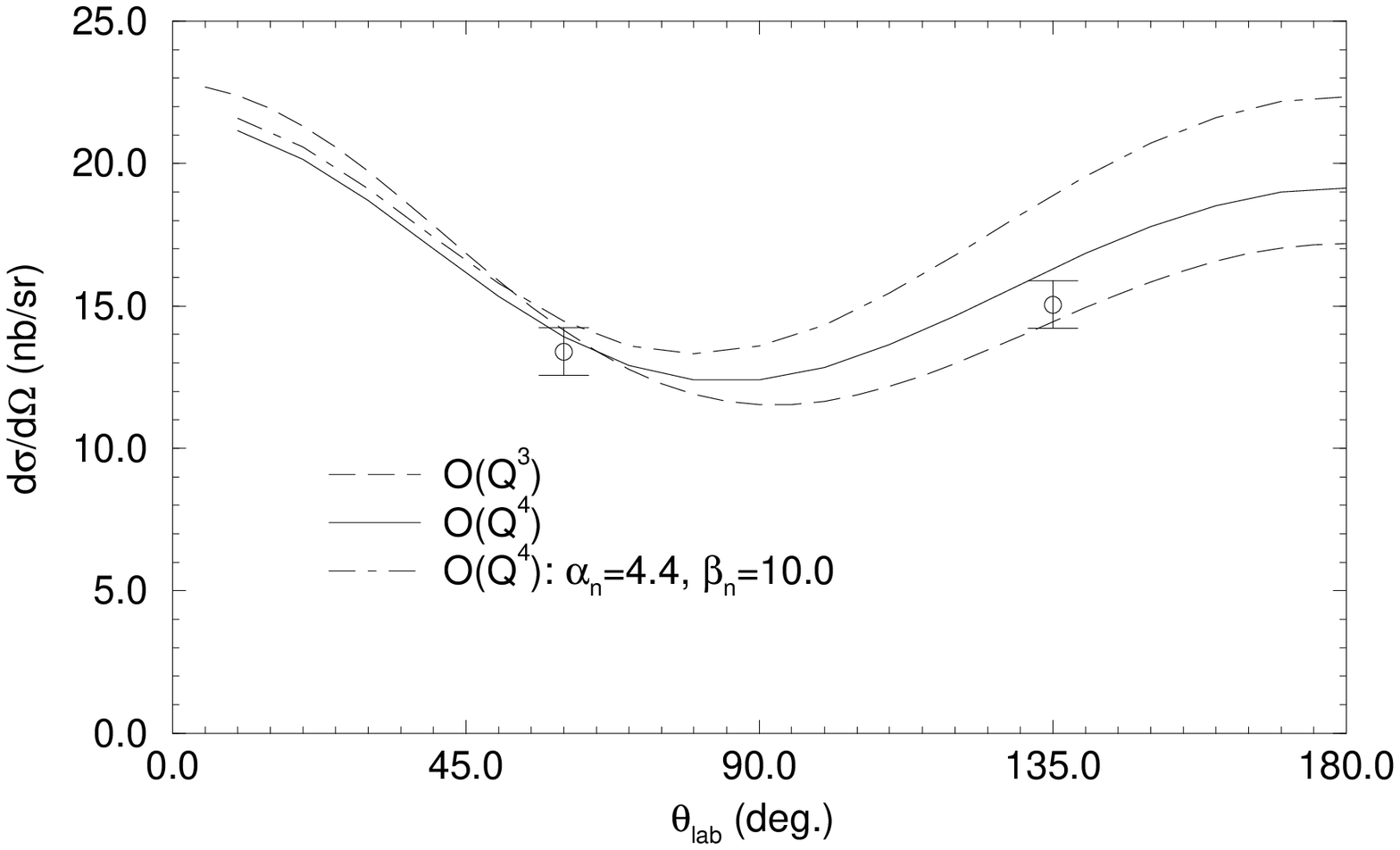,height=1.1in,width=1.4in}
\hskip 0.1in\psfig{figure=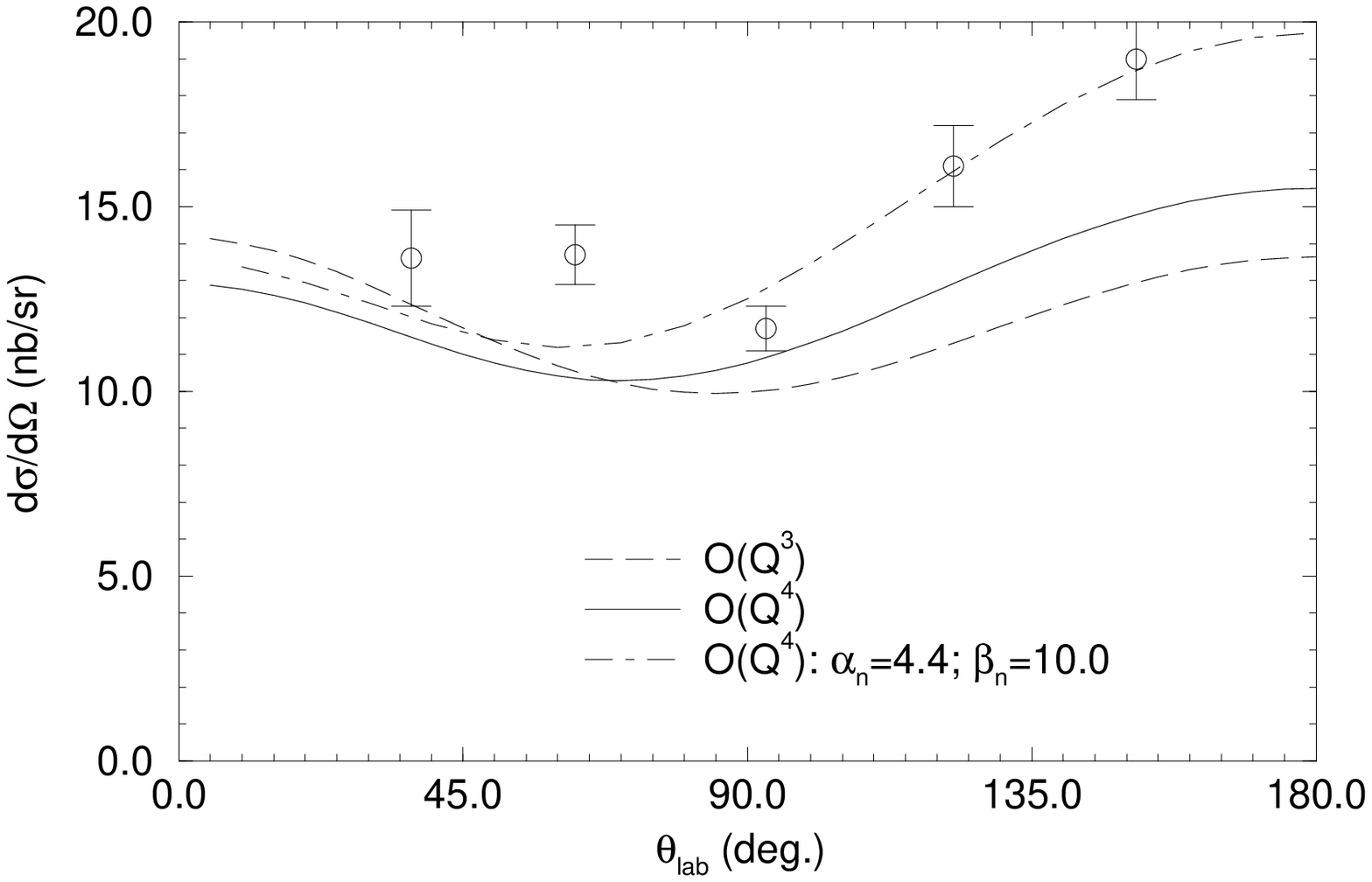,height=1.1in,width=1.4in}
   \caption{Results of the
   $O(P^2)$ (dotted line) and $O(P^3)$ (solid line) calculations
   for 
$E_\gamma=49~{\rm MeV}$, $69~{\rm MeV}$ and $95~{\rm MeV}$
respectively from left to right.
\label{plot.49}}
\end{figure}

We have also shown the six Illinois data points at 49 and 69
MeV~\cite{Lu94} and the Saskatoon data at 95
MeV~\cite{Ho99}. Statistical and systematic errors have been added in
quadrature. It is quite remarkable how well the $O(P^2)$ calculation
reproduces the 49 MeV data. However, it is clear that the agreement at
forward angles is somewhat fortuitous, as there are significant
$O(P^3)$ corrections.  Meanwhile, the agreement of the $O(P^3)$
calculation with the 69 MeV data is very good, although only limited
conclusions can be drawn, given that there are only two data points,
each with large error bars.

These results are qualitatively not very different from other existing
calculations.  At 49 and 69 MeV the $O(P^3)$ results are very close to
those in Ref. \cite{Wl95} and a few nb/sr higher, especially at
back angles, than those of Refs. \cite{LL98,KM99}.  
At 95 MeV the $O(P^3)$ result is close to
that of Ref. \cite{LL98}, higher by several nb/sr at back angles
than Ref. \cite{KM99}, and several nb/sr lower than the calculation
with no polarizabilities of Ref. \cite{Wl95}~\footnote{At this
energy Ref. \cite{Wl95} only presents results with
$\alpha_p+\alpha_n=\beta_p+\beta_n=0$, which in turn are considerably
less forward peaked than the corresponding calculation of
Ref. \cite{LL98}.}. 
However the EFT calculation is the only one to
incorporate the full single-nucleon amplitude instead of its
polarizability approximation. 
The
tendency to higher relative cross-sections in the backward directions
is at least in part due to this feature.

Although nominally the domain of validity of the Weinberg formulation
extends well beyond the threshold for pion production, the
power-counting fails at low energies well before the Thompson limit is
reached.  Consider the $O(P^4)$ contribution shown in
Fig. \ref{procomptinfra2mod}.  We can use this graph to illustrate the
transition to the very-low energy regime $\sim m_\pi^2/M$.  It is
easy to see that this graph becomes comparable to the order $P^3$
graph of Fig.~\ref{combined}b when
$|\vpee |^2\sim \omega M$.
Here $\vpee$ is a typical nucleon momentum inside the deuteron and
$\omega$ is the photon energy.  Since our power-counting is predicated
on the assumption that all momenta are of order $m_\pi$, we find that
power-counting is valid in the region
$m_\pi^2/M\ll P \ll \lsc$ .
Therefore, in the region $\omega \sim B$ the Weinberg power-counting
is not valid, since the external probe momentum flowing through the
nucleon lines is of order $P^2/M$, rather than order $P$. It is in
this region that the Compton low-energy theorems are
derived. Therefore our power-counting will not recover those
low-energy theorems. Of course the upper bound on the validity of the
effective theory should increase if the $\Delta$-resonance is included
as a fundamental degree of freedom~\cite{JM91}.
\begin{figure}[!ht]
   \centerline{\psfig{figure=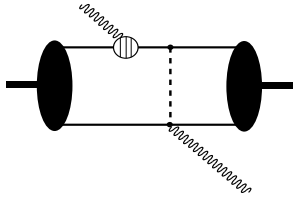,height=1.0in,width=1.5in}}
   \caption{Interaction which
   contributes to Compton scattering on the deuteron at
   ${\cal O}(P^4)$. 
The sliced blob represents a $1/M$ correction vertex from
   ${\cal L}_{\pi N}^{(2)}$.
\label{procomptinfra2mod}}
\end{figure}

There are $O(P^4)$ counterterms which shift the polarizabilities. We
use this freedom to fit the SAL data using the incomplete $O(P^4)$
calculation.  A reasonable fit at backward angles can be achieved with
$\alpha_n =4.4$ and $\beta_n =10$, which are in startling disagreement
with the $O(P^3)$ $\cpt$ expectations. That a fit can be achieved with
these values of the polarizabilities is not surprising; continually
increasing $\beta_p+\beta_n$ at approximately constant
$\alpha_p+\alpha_n$ decreases the cross-section at forward angles and
increases it at back angles. In fact, it seems that if
$\beta_p+\beta_n$ is sufficiently large then the character of the
cross-section at 95 MeV can change completely from forward peaked to
backward peaked. We have also plotted the cross-section at 69 MeV with
$\alpha_n =4.4$ and $\beta_n =10$. This curve is then excessively
forward peaked and misses the Illinois data. This situation poses an
interesting theoretical puzzle. Of course, a full $O(P^4)$ calculation
in $\cpt$ is necessary if the neutron polarizability is to be
extracted from the Saskatoon data in a systematic way.

%%%%%%%%  npd gamma  %%%%%%%
\subsection{$np \rightarrow d \gamma$}
\label{sec-npdgamma}

The chiral filter hypothesis is an idea based on current 
algebra~\cite{CR71,Ku78}. The hypothesis is that when a long-wavelength
photon or axial probe interacts with the nucleus the dominant
meson-exchange current contribution comes from the interaction
of the probe with the soft pions. The arguments given above 
for $e d$ scattering suggest that, as first
argued by Rho~\cite{Rh91}, this hypothesis
can be organized into a systematic statement of the relative
importance of various contributions to the $NN$ system current,
using Weinberg power counting.

This can be clearly seen in the calculation of threshold
$n p \rightarrow d \gamma$. In Ref.~\cite{Pa95,Pa96B} Park
{\it et al.} computed the M1 isovector matrix element
for this reaction up to $O(P^5)$ in Weinberg counting.
At $O(P^2)$ the LO  contribution to the M1 transition
enters. This is just the isovector counterpart of the one-body operator
given above in Eq.~(\ref{eq:jplus}):
\begin{equation}
{\bf \mu}_{\rm 1B}={e \over 2 M} \left[(1 + \kappa_p) {\bf \sigma}_p 
+ \kappa_n {\bf \sigma}_n\right].
\end{equation}
At $O(P^3)$ the two two-body currents depicted in Fig.~\ref{fig-tbcpions}
enter (although, of course, now a continuum $np$ state is the
final-state of the system). We will denote their contribution to the operator
in question by ${\bf \mu}_{2 {\rm tree}}$. Park {\it et al.}
derive an analytic expression for this operator
in co-ordinate space:
\begin{eqnarray}
{\bf \mu}_{2 {\rm tree}}&=&e {g_A m_\pi^2 \over 8 f_\pi^2}
\left(\vec{\tau}_1 \times \vec {\tau}_2\right) 
\left\{ \left({\bf \sigma}_1 \times
{\bf \sigma}_2\right) \left[{2 \over 3} (1 + m_\pi r) - 1\right] \right.
\nonumber\\
&& \qquad \left. - \left[ \hat{r} \ \hat{r} \cdot \left({\bf \sigma}_1 \times
{\bf \sigma}_2\right) - {1 \over 3}  \left({\bf \sigma}_1 \times
{\bf \sigma}_2\right)\right] \left[1 + m_\pi r\right]\right\}
{e^{-m_\pi r} \over m_\pi r}.\nonumber\\
\end{eqnarray}
This exchange current was derived many years ago by Chemtob and
Rho~\cite{CR71}, but it now has a definite place in the expansion
of the two-nucleon current in powers of the small parameter $P$.

The correction at next order, $O(P^4)$, vanishes, and so the next
term in the series occurs at $O(P^5)$. At this order three types of term
enter.  
\begin{enumerate} 
\item Loop corrections to the $\pi NN$ and
$\gamma\pi NN$ vertices. Thus, the only corrections comes
from the counterterms which renormalize the divergences that appear in
the loop graphs contributing to the $\gamma \pi NN$ vertex. Park
{\it et al.} chose to fix these counterterms using resonance
saturation. These low-energy constants could also
be obtained by fitting to data in the single-nucleon sector. 
The topological structure of these graphs is the same as the
 left-hand graph in 
Fig.~\ref{fig-2BCOP5},and
these contributions are denoted by ${\bf \mu}_{2 {\rm vertex}}$.

\item There are also a number of one-loop, two-body exchange currents
(not shown) denoted by  
${\bf \mu}_{2 {\rm loop}}$. 

\item There is the right-hand graph in Fig.~\ref{fig-2BCOP5}. 
This is a contact interaction arising from the Lagrangian 
(\ref{eq:Lone}). Park {\it et al.} argue that this contribution
is suppressed in their ``hard-core cutoff scheme" (HCCS)~\cite{Pa95,Pa96B}. 
However, if this
process was reconsidered in the light of their more recent work, where
a ``modified hard-core cutoff scheme" (MHCCS) was employed, then the effect
of this graph would be non-zero.
\end{enumerate}

Without this last contribution the magnetic moment operator
is just the sum of all of these contributions, 
${\bf \mu}={\bf \mu}_{\rm 1 B} + {\bf \mu}_{2 {\rm tree}}
+ {\bf \mu}_{2 {\rm vertex}} + {\bf \mu}_{2 {\rm loop}}$
each of which
is calculable in terms of $g_A$, $f_\pi$, $\kappa_N$, or
low-energy constants from the single-nucleon sector. 
This operator must now be sandwiched between deuteron and
$np$ scattering state wave functions. For this purpose
Park {\it et al.} employed the AV18 wave functions. The contribution
from the one-body operator ${\bf \mu}_{1 B}$ gave a cross-section for
threshold neutron capture,
$\sigma_{\rm 1B}=305.6~{\rm mb}$,
very close to the result from ERT~\cite{Au53}. This simply means that
the threshold capture cross-section is dominated by the physics
of the $NN$ wave function tail, which is the same in ERT and
the AV18 potential. 

The evaluation of the two-body magnetic-moment operators
is a little more subtle. We write the contribution
of any given operator ${\bf \mu}$ to the M1 matrix element
as:
\begin{equation}
\delta_X={\langle d |{\bf \mu}_X|n p\rangle \over 
\langle d |{\bf \mu}_{\rm 1B}|n p \rangle}.
\end{equation}
Park {\it et al.} cut-off the contribution
from the operators ${\bf \mu}_X$ at $r=r_c \sim 1/\Lambda$, since the physics
of this short-distance region will not be well-represented
in $\chi$PT. Multiplying the operators written above by $\theta(r-r_c)$ and
employing the AV18 wave functions they can then evaluate the ratios
$\delta$ as functions of $r_c$. The dependence on $r_c$ is rather weak
for a range of cutoffs from $r_c=0$ to $r_c=0.7~{\rm fm}$.
Summarizing the results at $r_c=0.6$ fm~\cite{Pa96B}
$\delta_{2 {\rm tree}}=2.592 \%$,
$\delta_{2 {\rm vertex}}=1.477 \%$, and 
$\delta_{2 {\rm loop}}=0.362 \%$.
This leads to a theoretical prediction:
$\sigma=334 \pm 3~{\rm mb}$,
in excellent agreement with the experimental cross
section $\sigma^{\rm th}$. 
\begin{figure}[!ht]
\hskip 2.4cm \psfig{figure=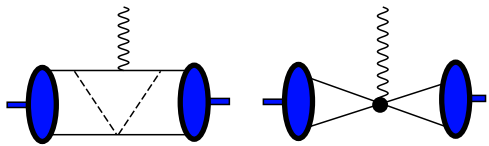,height=2.0cm}
\caption{
Two two-body currents which
contribute to $E2_S$ and $M1_S$ at $O(P^5)$.}
\label{fig-2BCOP5} 
\end{figure}

As in the case of $\nopi$ this approach can also be used to evaluate
the suppressed amplitudes $M1_S$ and $E2_S$~\cite{Pa00}.  The results
obtained agree with those discussed previously, but differ somewhat in
the error estimates.

\subsection{$pp \rightarrow d e^+ \nu_e$}

\label{sec-solar}

Similar computations were performed in Ref.~\cite{Pa98} 
for the solar proton burning process $p p \rightarrow d e^+ \nu_e$.
The dominant piece comes from the one-body
axial current. Park {\it et al.} show that this contribution
can be evaluated with low-energy experimental data, $\rho_d$,
and $\eta_{sd}$, as long as one is careful to include
Coulomb corrections in the $pp$ wave function. This gives a result:
${\cal M}_{1B}^{\rm exp}=4.858 \pm 0.006~{\rm fm}$,
where we have included the errors from the experimental data, added linearly.
In contrast direct evaluation of the matrix element with the AV18
wave function yields:
${\cal M}_{1B}^{\rm AV18}=4.859~{\rm fm}$.
Neither of these numbers include a ``theory-dependent" part of the
matrix-element which comes from the interference between the Coulomb
and strong nuclear $pp$ interactions. Using AV18 wave functions
this additional piece of ${\cal M}_{1B}$ is computed to be:
${\cal M}_{1B}^{C+N}=0.011~{\rm fm}$.
The contribution of vacuum polarization in the $pp$ final state is also small,
of order $0.5\%$ of ${\cal M}_{1B}$. Effects of two-photon exchange are 
smaller still.
In fact, all of these numbers are robust under the interchange of
the AV18 wave function with an EFT-motivated wave function, provided
that in the EFT wave function the short-distance interaction is
tuned to reproduce the scattering length and effective range of the
$pp$ system.

The strength
of these short-distance pieces is unconstrained by chiral
symmetry, although we expect it to be ``natural". As in Ref.~\cite{Pa96B},
in Ref.~\cite{Pa98} it was argued that this counterterm contribution
to $pp \rightarrow d e^+ \nu_e$ will be suppressed
by the presence of the hard core in the $NN$ interaction.
It would be interesting to see how the use of the MHCCS 
would modify the conclusions
of Ref.~\cite{Pa98}. 
However, as in $\nopi$, this would
require some way to determine the counterterm contribution. 
This would need
to be done either by measuring a low-energy neutrino cross-section
in the $NN$ system, or by calculating the rate for the $\beta$-decay
of tritium, including the counterterm, and then fixing the counterterm
to reproduce the extremely well-known experimental number,
as discussed previously.

In Ref.~\cite{Pa98} only the axial currents of ``pion range" were 
evaluated. These gave a contribution of between 2 and 5\% to the
overall matrix element, depending on what cutoff radius was
employed. It would seem that this strong cutoff dependence
might be due to the omission of the corresponding counterterm
from the calculation. Regardless, the number resulting
from this $\chi$PT calculation is still accurate at the level of
a few per cent. Park {\it et al.} quote results for the astrophysical
S-factor~\cite{BM69}:
\begin{equation}
S_{pp}(E)=\sigma(E) E e^{2 \pi \eta},
\end{equation}
where $\sigma$ is the cross-section for  $pp \rightarrow d e^+ \nu_e$,
and $\eta=M \alpha/(2 p)$. The value obtained in Ref.~\cite{Pa98} is:
\begin{equation}
S_{pp}(0)=(4.05 \times 10^{-25}) 
(1 \pm 0.0015 \pm 0.0048)^2~{\rm MeV}~{\rm barn},
\end{equation}
where the first error is due to the uncertainty in the one-body matrix
element and the second is Park~{\it et al.}'s estimate of the error on
the two-body matrix element.

In contrast the canonical value of Kamionkowski and Bahcall~\cite{KB94},
calculated using a variety of $NN$ potential models, is:
\begin{equation}
S_{pp}(0)=(3.89 \times 10^{-25}) (1 \pm 0.011)~{\rm MeV}~{\rm barn}.
\end{equation}
The more sophisticated potential-model calculation of Ref.~\cite{Sc98}
gives a much smaller error bar
\begin{equation}
S_{pp}(0)=(3.88 \times 10^{-25}) (1 \pm 0.003)~{\rm MeV}~{\rm barn},
\end{equation}
where the error comes both from a combination of small effects neglected
in the calculation of the $NN$ wave functions and from 
the observed sensitivity to the choice of modern 
$NN$ potential.

%%%%%%  Pion Prod  %%%%%%%%
\subsection{Pion Production}

\label{sec-piprod}

The pion production reactions:
$p + p \rightarrow p + p + \pi^0$, 
$N + N \rightarrow d + \pi$, and 
$N + N \rightarrow p + n + \pi$
would seem to be a natural candidate for calculation in $\chi$PT.
However, the situation here is complicated by the appearance of a new
scale. Since the energy of the incoming nucleons in any of these
processes is of order $m_\pi$ the typical nucleon momentum in the
initial state is~\cite{Co96} $\sqrt{M m_\pi}$.
This renders the $\chi$PT expansion for these processes poorly
convergent, since the expansion is now one in $\sqrt{m_\pi/M} \sim 0.4$.
This makes it difficult to reproduce the experimental data, which is
in fact of very high quality, in  both the case of neutral pion~\cite{Co96} 
and charged pion~\cite{daR00} production.
There have been many attempts to treat
the S-wave production of pions to high-enough order that some level of
convergence is seen~\cite{Pa96,Sa97,Ha97,Dm99,Ge99,An00}, but, at this stage
it seems that the series is just not well-behaved.

One solution to this is to resum certain mechanisms in the $\chi$PT 
expansion in some non-systematic way, as discussed in Ref.~\cite{vK96,Br99B}. 
Alternatively, one can turn to the production of P-wave pions. There
one might hope that the series is better behaved, since the
LO  contribution is large~\cite{Ha00}. The results from
early studies of the production of P-wave pions in $\chi$PT are
encouraging, but the expansion parameter remains large. The advantage
of employing $\chi$PT to study the production of P-wave pions is that, if
the series is well-behaved enough, the experimental data
might be able to be employed to extract information on contact 
operators involving four nucleons and a single pion. 
It is exactly this type of operator which enters in the novel 
three-nucleon forces discussed in Ref.~\cite{Hu99}. The study
of P-wave pion production therefore provides us with an opportunity
to use $\chi$PT to systematically connect the three-nucleon
system with pion-production in the two-nucleon system.

%%%%%%%  Paulos three body stuff  %%%%%%

\subsection{Three-Body Forces}

In the Weinberg program the order in which the three-body forces
start contributing depends on whether or not one is including the
isobars explicitly. Let us first discuss the case where the
$\Delta$'s are present \cite{vK94}. The three-body forces
then first appear at order  ${\mathcal O}(P^2)$ and are given by
tree diagrams involving $\Delta$'s as intermediate states and
zero, one and two pion exchanges. At order  ${\mathcal O}(P^3)$
three classes of contributions appear. The first involves two-pion
exchanges and a one-nucleon-two-pion vertex. The three independent
one-nucleon-two-pion vertices can be fixed by the study of
pion-nucleon scattering. The second class involves one pion
exchange and a two-nucleon-one-pion vertex. The two independent
coefficients of the two-nucleon-one-pion vertices can be
determined by analyzing pion-deuteron scattering and/or pion
production in nucleon-nucleon scattering. One particular linear
combination of these two coefficients was determined by
considering $p-$wave pion production in $N+N\rightarrow
N+N+\pi$~\cite{birajerryhanhart}. It is also known that if those
coefficients have natural size they can appreciably affect the
vector-analyzing power $A_y$ in proton-deuteron scattering that is
not very well described by phenomenological two-body forces only
\cite{birathree2}. Finally the last category includes contact
three-body pieces that can only  be determined by fitting to
three-body observables. In the version of the theory where the
isobars are integrated out the three-body forces have no
contribution at order $P^2$ and start at order $P^3$. The
contributions have the same form as the case with $\Delta$'s, but
with different coefficients.

Only recently the first calculations of three-body observables in
the Weinberg scheme were performed. 
A calculation at the order $P^2$ without explicit $\Delta$'s shows
two main features~\cite{Ep00C}. 
The binding energy of the triton is found to be
close to the experimental value but shows a cutoff dependence that
seems to be too large for a NLO calculation. The values obtained
were $B_3=8.284$ for $\Lambda=540$ MeV and $B_3=7.546$ for
$\Lambda=600$, 
but with a $10\%$ variation in the binding energy for a
$10\%$ variation in the cutoff. It is presently
unclear whether cutoff dependence is somewhat accidental or
indicates that three-body forces that could cancel this dependence
appear at lower order than predicted by  
Weinberg's power-counting, as shown to be the case in the
pionless theory. The second feature is that spin-observables are
well described by the  NLO calculation, including the
vector-analyzing power $A_y$.

%%%%%%%%%   Weinbergs Problems  %%%%%%%%%%%%
\subsection{Formal Problems with Weinberg Power Counting}

As we have already discussed,
at LO in Weinbergs power-counting there are contributions to 
$V_0^{(2)}$ from both the local four-nucleon operators, 
$  C_{S,T}$ and  from the exchange of a single potential pion,
giving a momentum space potential of 
\begin{eqnarray}
V_0^{(2)}({\bf p},{\bf p^\prime}) = C \ - \ \({g_A^2\over
2f_\pi^2}\){({\bf q} \cdot\sigma_1   {\bf q} \cdot\sigma_2)(\tau_1\cdot\tau_2)
\over ({\bf q}^2+m_\pi^2)\ }
\ \ \ \ ,
\label{eq:ope}
\end{eqnarray}
where $C$ denotes the  combination of $C_{S,T}$ appropriate for a
given spin-isospin channel.
The LO amplitude results from solving the Schr\"odinger
equation with this potential.  This can be justified in the 
EFT only if $M\sim P^{-1}$, otherwise diagrams at different
orders in Weinbergs expansion are numerically of the same size.
At two-loops in the ladder sum there is a  
logarithmic divergence in the graph shown in Fig.~\ref{wein_fig3}a
that  must be regulated.
\begin{figure}[t]
\hskip 0.75in\psfig{figure=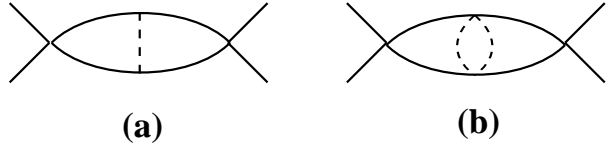,height=0.9in}
\caption{Graphs  with logarithmic divergences at LO in 
Weinberg's power counting.
The solid lines are nucleons and the dashed lines are pions.
\label{wein_fig3}}
\end{figure}
In dimensional regularization the divergent part of this graph is 
\be
-{1\over \epsilon} {g_A^2 m_\pi^2 M^2\over 128\pi^2 f_\pi^2} C^2
\ \ \  ,
\label{eq:pole}
\ee
which requires a counterterm with a single insertion of the light 
quark mass matrix, for instance
\begin{equation}
-{1\over 2} C_S^{(m)} 
Tr \left[m_q (\Sigma + \Sigma^\dagger)\right] 
\left(N^\dagger N\right)^2
-{1\over 2} C_T^{(m)} 
Tr \left[m_q (\Sigma + \Sigma^\dagger)\right] 
\left( N^\dagger \sigma N\right)^2
\ +\ ...
\ \ \ .
\end{equation}
However, the coefficients of these operators must scale like $M^2$,
and since $ m_\pi^2 M^2 \sim P^0$ these formally higher order
operators in Weinberg's power-counting are required at LO to absorb
divergences in the time-ordered products of the LO potential,
$V_0^{(2)}$.  Ignoring the multi-pion vertices arising from these
operators, they can be re-absorbed into the LO operators with
coefficients $C_{S,T}$.  We are then in the situation where there is
no chiral expansion, multiple insertions of the light quark mass
matrix are not suppressed compared to LO interactions.

The Schr\"odinger equation sums diagrams that are of different order
in $P$. Paradoxically, it is consistent with Weinberg's power counting
{\bf not} to use the Schr\"odinger equation. In that case the near
threshold bound states and the associated behavior of the scattering
amplitudes are not recovered.  It was shown in Ref.~\cite{Ka96} that
taking $M\sim P^{-1}$ corrects this problem.  However, a more
intricate scaling might in fact be in operation.

Resolving these issues is currently the most important line of
investigation in multi-nucleon EFT, since only then will there be an
acceptable formal basis for the EFT at scales $|{\bf k}|\gsim m_\pi$.

\section{Gains over the Traditional Approach}
\label{sec-NNpots}

Traditionally, low-energy phenomena in light nuclei have been
described using the Schr\"odinger equation with a model for the $NN$
potential. In practice, all sensible potentials include the
one-pion exchange~ potential (OPEP) as their long-range tail.  The
inclusion of OPEP is required on theoretical grounds, as it uniquely
describes the longest-range part of the strong interaction. This can
also be seen experimentally, since OPEP is essential in the
description of the higher partial waves in $NN$ 
scattering~\cite{Ma87,Ka97}. 
On the other hand, apart from their common one-pion 
exchange tails, different
$NN$ potential models look rather dissimilar.  Some include a set of
single (or, in some cases, double) boson-exchanges to generate the
short-distance piece of the interaction, as in the
Nijmegen~\cite{St94,Na78} or the original Bonn
potentials~\cite{Ma87,Ma89}.  Others have a more phenomenological
short-distance structure, e.g. the Reid potential~\cite{St94,Re68},
the Paris potential~\cite{Co73,La80}, or the Argonne
potentials~\cite{Wi95,Wi84}~\footnote{The CD-Bonn
  potential~\cite{Ma96,Ma00} nominally involves single-meson exchanges at
  short distances, but the mass of the scalar-isoscalar meson
  varies from partial wave to partial wave.  Therefore in classifying
  OBE potentials as we have done here it is not entirely clear which
  class to place it in.}. In all cases the short-distance potential
must be rich enough to describe the existing data.  Once a particular
form is chosen the Schr\"odinger equation is solved and the free
parameters are fit to the $NN$ scattering data and certain deuteron
bound-state properties. This is now an extremely sophisticated
enterprise. Potentials exist which fit the $np$ and $pp$ scattering
data up to laboratory energies of 350 MeV with a $\chi^2$ per degree
of freedom close to one~\cite{St94,Wi95,Ma00}.

The treatment of short-distance physics employed in many of these
potential models can be motivated from an EFT point of view.
Potentials ranging from the modern AV18~\cite{Wi95} to the venerable
Reid potential~\cite{Re68} are based on the idea that the details of
the short-distance physics in the $NN$ problem are not important. It
{\it is} important to introduce a structure for the short-distance
operators that parameterizes our ignorance about that part of the
interaction, but the detailed form of that parameterization is not
crucial. This is perhaps not surprising: physically it is just not
possible to resolve the details of short-distance physics using
low-momentum observables. Therefore attempts to get the ``right''
short-distance $NN$ potential using low-momentum data are futile.
Potential models have predictive power precisely {\it because} the
details of the short-distance physics do not matter greatly.

It should therefore be clear that the goal of a $\chi$PT $NN$
potential is not to learn about the short-distance structure of the
$NN$ interaction. The chiral Lagrangian quite deliberately makes no
dynamical assumptions about dynamics at scales of order
$\Lambda_\chi$, and so, by construction, the $\chi$PT $NN$ potential
contains only gross information about the physics which takes place at
this scale. There is, however, the chance to learn something about the
pion physics which is the provenance of $\chi$PT. 

Of course, ``traditional" potential models are not just employed
to fit $NN$ scattering data. They also do a very good job in
describing a wide array of low-energy data obtained using electroweak
probes. To calculate the response of the nucleus to these probes,
single-nucleon currents and potential-model wave functions are often
employed.  In some cases such a construction does not lead to a
conserved electromagnetic current, and so additional ``model-independent'' 
two-body currents are added to restore current
conservation (see, for instance~\cite{Ri84}).  Conversely, transverse
currents are not constrained by current conservation and so are always
``model-dependent''. The usual practice is to model these
currents via something akin to the resonance saturation hypothesis
used to estimate counterterms in the single-nucleon sector of
$\chi$PT. The results for these transverse currents
reflect dynamical assumptions
about the important physical mechanisms, e.g. that pionic,
$\rho$-meson, $\rho \pi \gamma$, $\omega \pi \gamma$, $\Delta \pi
\gamma$, and $\Delta \rho \gamma$ meson-exchange currents dominate the
transverse currents. 

Employing $\chi$PT to calculate these current operators leads to two
gains over these more standard approaches. Firstly, as in the case of
the $NN$ potential, it {\it simplifies} the picture of the dynamics
down to its essential features: the ``long-distance" part of the
current operator comes from the physics of pions and nucleons
interacting with the probe, while the short-distance part must be
parameterized in some way.  Secondly, and perhaps of more practical
importance, the calculation of current operators in $\chi$PT {\it
systematizes} the treatment of the current operators in the $NN$
system.  In particular, $\chi$PT constrains the number and form of
current operators allowable at a given order and mandates the type of
pion-loop and pion-exchange mechanisms which should be included in the
calculation. This means that, as in all EFT calculations, errors can
be estimated in a reliable way. Such error estimation is possible in
potential-model calculations (see, for instance~\cite{Sc98}), but it
is significantly simpler in $\chi$PT.

%%%%%%%%% Nuclei %%%%%%%%%%%%%%%
\section{The Nuclear Many-Body Problem as an EFT}

In order for the EFTs that we have discussed in the previous pages
of this chapter to have any impact whatsoever on nuclear structure,
a way of solving the many-body problem must be developed, that is consistent
with the power counting.
As we do not know exactly what the final EFT will look like in the momentum
region relevant to nuclei, we now turn our attention to the problem of solving 
the nuclear many-body problem for the hamiltonian given in Eq.~(\ref{eq:HUrbana}).
The hope is that the EFT program will allow enable us to match onto
Eq.~(\ref{eq:HUrbana}) and provide the tools to systematically calculate
corrections to it.
Equally important is that the many-body problem has to be solved anyway at 
some point! 
Although this problem has consumed nuclear physicists for several decades, the tools
are still not available to make reliable calculations in 
bound systems with more than $\sim 10$ nucleons.  
Furthermore the techniques employed in systems of 6-10 nucleons --
Green's function Monte Carlo (GFMC) and variational methods -- are 
sufficiently technical and challenging numerically that very
few theorists have access to them.  In still heavier
systems the techniques of choice, such as the shell model, are
phenomenological, providing no basis for systematic corrections 
or evaluating errors.  For these reasons
there is some interest in the possibility that the ideas of EFTs
may point the way to new many-body techniques that are both
simpler and more systematic and controlled.
Although the first such steps in this direction have been taken
only recently, here we summarized some recent work (that of 
Haxton and collaborators\cite{Ha99A,Ha99B}) 
(for others investigations of multi-nucleon dynamics see e.g.
Ref.~\cite{Lutz00,Fu99})
which suggests that
conventional nuclear models do have controlled effective theory
counterparts in which modern techniques, such as the renormalization
group, can be employed to simplify the classical nuclear physics
problem of point nucleons interacting through a static potential.
A related and very interesting development that we will not
discuss involves similar progress in systematically treating
low-density nuclear systems\cite{Ham00,Kr00}.  
The use of dimensional regularization in this scenario leads to great
simplifications in various calculations.

The canonical microscopic model for nuclear structure -- the shell
model (SM) -- is motivated in many textbooks by Brueckner's 
treatment of nuclear matter.  The language is at least superficially
reminiscent of an EFT.  The starting point is a collection of nonrelativistic
nucleons interacting through a potential, e.g., of the Nijmegen or
Argonne/Urbana type within an infinite Hilbert space
\begin{equation}
H = {1 \over 2} \sum_{i,j=1}^A (T_{ij} + V_{ij}),
\end{equation}
where $T_{ij}$ is the relative nonrelativistic kinetic energy
operator and $V_{ij}$ the nucleon-nucleon potential.  The related 
SM Hamiltonian acts in a restricted space and 
employs a softer ``effective" potential,
\begin{equation}
H_{SM} = {1 \over 2} \sum_{i,j=1}^A (T_{ij} + V^{eff}_{ij}).
\end{equation}
Motivating $H_{SM}$ is the notion that the determination of
$V^{eff}$ might be simpler than solving the original A-body problem:
the foundation of Brueckner theory is that high-momentum contributions
to the wave function might be integrated out in a rapidly converging
series in $\rho_{nuclear}$ or, equivalently, in the number of 
nucleons in high-momentum states interacting at one time outside
the SM space.  This assumption -- sometimes termed the hole-line
expansion -- clearly has some connections with our discussion 
of the power-counting for three-body forces, and thus may
find some justification through EFTs.

As will become apparent later, a typical SM space will 
contain explicitly $\sim$ 60\% of the
wave function that resides at long-wavelengths, thus guaranteeing at 
least a qualitative description of A-body correlations important to 
soft collective modes.  Implicitly the high-momentum components 
are swept into a rather poorly defined ``effective interaction,"
often determined empirically.  The strength
of the SM resides in the first of these two aspects: the technology developed
for direct diagonalizations in large SM spaces is quite remarkable, including 
recent progress in Lanczos-based methods \cite{Ca98},
in treatments of light nuclei involving many shells \cite{Na98},
and in Monte Carlo sampling \cite{ko97,Ho96}.  
Its 
weakness is the numerous uncontrolled approximations that 
become apparent when one tries to view the shell model as a 
faithful effective theory (ET).  
However, we show below that
the same numerical strides that have advanced shell model
diagonalizations now allow us to remove these uncontrolled
approximations.  
The resulting ET has many
differences with the shell model and many similarities to
the EFTs discussed previously.

Among the SM uncontrolled approximations are the following:\\
1) Even in LO, where only the pairwise interaction 
of high-momentum nucleons is included in $H^{eff}$, the 
functional form of the resulting effective interaction is 
not as simple as assumed in the SM,
\begin{equation}
\langle | H^{eff} | \rangle_{SM} 
\equiv \langle \alpha \beta | H^{eff} | \gamma \delta \rangle
\end{equation}
where the Greek symbols label single-particle shell-model states.
As nucleons in high-momentum states can scatter pairwise 
repeatedly, this expression can at best be valid in some LO
sense: beyond LO three-, four-, and higher-body body operators
must be successively induced.
Furthermore the energy denominators describing this
scattering depend on the total many-body energy\cite{Zh95}, not simply
the single-particle states participating in the scattering
(apart from special cases, like a LO calculation restricted to a
single shell).  Thus the SM simplification of $H^{eff}$ is ad hoc. \\
2) SM wave functions are orthogonal and normed to unity.
In ET the effective wave functions are naturally defined as the
restrictions of the true wave functions $|\Psi_i \rangle$ to the
model space
\begin{equation}
|\Psi_i \rangle \begin{array}[t]{c} \longrightarrow \\
ET \end{array} \langle SM | \Psi_i 
\rangle  | SM \rangle \equiv |\Psi_i^{eff} \rangle.
\end{equation}
Thus the norms are less than unity and orthogonality, which holds
for the full wave functions, is lost for their restricted
counterparts. \\
3) Shell model interactions frequently depend on fictitious 
parameters such as ``starting energies," introduced to 
adjust the energy denominators in the high-momentum ladder
diagrams.  Answers depend on these parameters, as well as
on others (e.g., the oscillator strength if the SM is
defined in terms of harmonic oscillator Slater determinants,
and the number of included oscillator quanta) that define
the division between the SM and excluded high-momentum spaces.
Clearly the predictions of a correct ET must be independent
of all such physically irrelevant parameters. \\
4) Perhaps most serious, the important issue of effective 
operators is almost never addressed in a meaningful way.  In
many cases practitioners adopted a phenomenological $H^{eff}$
which, while successful in producing spectra, provides no
diagrammatic basis for calculating effective operators or
wave function normalizations.  Even in cases where 
$H^{eff}$ is derived from some underlying NN interaction, 
the practice is generally to then employ bare operators.
In some well-studied cases, such as allowed $\beta$ decay in
the $1p$ and $2s1d$ shells, it is then recognized that a 
phenomenological renormalization (e.g., $g_A \rightarrow$ 1)
of operators greatly improves agreement with experiment.
But the origin of this renormalization and its evolution with
momentum transfer $q$ are left unclear.  The situation is 
very unsatisfactory and undercuts the shell model as a 
predictive tool.

%%%%%%  Block-Horowitz  %%%%%%%%%
\subsection{Self-consistent Bloch-Horowitz Solutions}
  
We consider the canonical nuclear structure problem of 
nonrelativistic point nucleons interacting through a realistic
NN interaction, such as the Nijmegen potential discussed previously,
or the similar Argonne V18 \cite{Wi95} 
or Reid93 \cite{Re68}
potentials.  The question is whether the uncontrolled approximations
in the shell model can be removed, leaving a more complicated
but still tractable effective theory.  The approach involves three
major steps:\\

\noindent
$\bullet$ Formulating a treatment of effective interactions 
and operators that exploits the basic assumption in Brueckner
theory --- that interactions at high momenta can be integrated
out in a cluster expansion (essentially an expansion in 
$\rho a^3$, where $\rho$ is the nuclear density and $a$ an
interaction range) --- but is otherwise exact.  The convergence
of the expansion could then be tested numerically and should
depend on the operator under study and the momentum transfer.
Alternatively (and preferably) one might be able to make
connections to EFTs that would justify the expansion.\\

\noindent
$\bullet$ To find numerical tricks for implementing this
formulation, demonstrating their validity in cases (e.g,
A=2,3,4) where the expansion can be carried to all orders,
so that the answers should then agree with Faddeev and
other exact methods.\\

\noindent
$\bullet$ To imbed the formulation in a heavier nucleus, 
where the cluster expansion can be carried out only partially.\\

There is some reason for optimism that if the first two goals
can be achieved, the third might yield very accurate results: 
the Argonne group 
cluster variational Monte Carlo effort on $^{16}$O appeared to
yield nearly exact results when clusters up to A = 5 were
included.  
Below we describe progress toward the first two goals,
contrasting an exact ET of the deuteron and ${}^3$He with the
shell model to illustrate the shortcomings of the later.

The approach is sketched in Fig.~\ref{eq:cluster}.  
The Hilbert space is
divided into a long-wavelength ``SM'' space, defined
by some energy scale $\Lambda_{SM}$, and a high-momentum 
space that extends to some scale $\Lambda_\infty$.  Potential 
models effectively contain ultraviolet cutoffs embodied in their
hard but finite short-range cores.  Thus if $\Lambda_\infty$ 
is set to $\sim$ 3 GeV/$\hbar \omega \sim 140$ for the potentials 
mentioned above, ground state energies will be accurate to 
better than 1 keV.  That is, for sufficiently large $\Lambda_\infty$
\begin{equation}
H^{eff}_{(i)}(\Lambda_{SM},\Lambda_{\infty}) \begin{array}[t]{c}
\longrightarrow \\ \Lambda_\infty \mathrm{~large} \end{array}
H^{eff}_{(i)}(\Lambda_{SM}).
\end{equation}
All correlations within the ``SM'' space are included, but the
high-momentum correlations in the excluded space are limited 
to n-body, where n is the cluster size.  Thus the lowest order
effective interaction is
\begin{equation}
H^{eff}_{(n=2)} \equiv H^{eff}_{(n=2,0)}
\end{equation}
It corresponds to embedding the A-body ladder diagram of 
Fig.~\ref{eq:cluster}(b)
between SM states: A-2 of the nucleons are spectators, with
the remaining pair scattering via a two-body ladder.
The notation (n=2,0) states that the
two-body cluster has no explicit dependence on the nuclear 
density, varying as $\rho^0$.
The n=3 A-body ladder of Fig.~\ref{eq:cluster}(c) is similarly
\begin{equation}
H^{eff}_{(n=3)} \equiv H^{eff}_{(n=2,1)} + H^{eff}_{(n=3,0)}
\end{equation}
where $H^{eff}(n=2,1)$ is the two-body part of the three-body
ladder
\begin{equation}
\langle \alpha \beta | H^{eff}_{(n=2,1)} | \alpha' \beta' \rangle =
\sum_{\gamma \leq k_F} \langle \alpha \beta \gamma | H^{eff}_{(n=3)}
| \alpha' \beta' \gamma - \alpha' \gamma \beta' + ... \rangle,
\end{equation}
where $k_F$ denotes the Fermi level. 
This decomposition -- which is done only to emphasize the content
of the cluster expansion -- illustrates that $H^{eff}_{(n=3)}$
contains $H^{eff}_{(n=2,0)}$ as well as a correction to 
the two-body interaction
that depends linearly on the density and is obtained by identifying one
ingoing SM single-particle leg of the three-body ladder with one outgoing leg,
summed over all occupied states.  It also contains a true
three-body piece $H^{eff}_{(n=3,0)}$, where three SM ingoing 
single-particle states 
connect to three distinct outgoing states 
after undergoing a series of scatterings outside the SM
space.  The point is simple pedagogy: treatments of successively
larger clusters in the high-momentum space correct the lowest-order
two-body $H^{eff}_{(n=2,0)}$ by adding terms proportional to the
$\rho$, $\rho^2$, etc., in the spirit of Brueckner theory.  It
also adds true three-body terms, true four-body terms, etc.
Thus an expansion through four-body clusters yields
$H^{eff}_{(n=2,2)}$, the two-body interaction corrected 
through order $\rho^2$,
$H^{eff}_{(n=3,1)}$, the three-body interaction through 
order $\rho$, and $H^{eff}_{(n=4,0)}$,
a density-independent true four-body interaction.
We emphasize that the convergence of such an expansion is still
a matter of conjecture, despite its physical plausibility and
some justification from EFT power counting.  The procedure has
not yet been given a chance to succeed because the individual
steps -- certainly those beyond the two-body level -- have not
been carried out.  We now discuss procedures that might 
make such an expansion feasible.
\begin{figure}[!ht]
\hskip 0.1in\psfig{bbllx=-3.0cm,bblly=3.0cm,bburx=14cm,bbury=22cm,figure=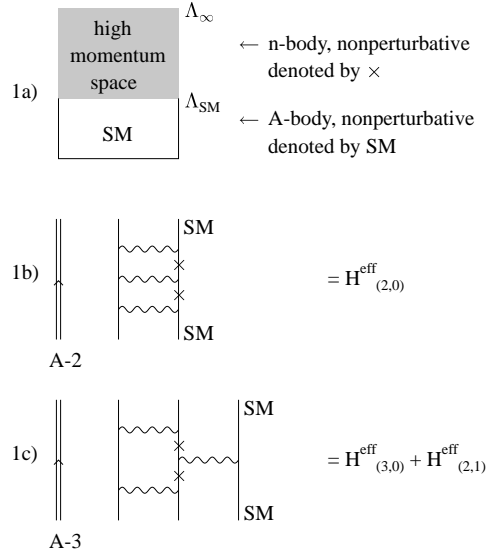,height=3.0in}
\caption{Cluster expansion of the effective interaction.}
\label{eq:cluster}
\end{figure}
  
The calculation begins with a definition of the ``SM" space.  The
goals of handling bound states and of generating an effective 
interaction that is translationally invariant leaves one sensible
choice, many-body states constructed from harmonic oscillator
Slater determinants.  To exploit the relative/center-of-mass separability
of harmonic oscillator Slater determinants, one must separate the
SM and high-momentum spaces so that all configurations satisfying
\begin{eqnarray}
E \leq \Lambda_{SM} \hbar \omega
\end{eqnarray}
are retained in the former.  For example, a SM calculation of 
$^{16}$O with
$\Lambda_{SM} = 4 + \Lambda_0$, where $\Lambda_0$ is the number of
quanta in the $^{16}$O closed shell, would include all
$4 \hbar \omega$ configurations, e.g.,
$0p0h$, $2p2h$, and $4p4h$ excitations of nucleons from the $1p$ 
shell into the
$2s1d$ shell, $1p1h$ excitations of a $1s$ shell nucleon into 
the $3s2d1g$ shell, etc.  One can define the projection operator
onto the high-momentum space by
\begin{equation}
Q_{SM} = Q(\Lambda_{SM},b).
\end{equation}
where $b$ is the oscillator parameter.  Thus the included or ``SM"
space is defined by two parameters, $\Lambda_{SM}$ and $b$.  
The preservation of 
translational invariance is also important numerically, as it
reduces the two-body ladder to an effective one-body problem, etc.

The resulting Bloch-Horowitz (BH) equation \cite{Bl58} is then
\begin{eqnarray}
H^{eff} &=& H + H {1 \over E - Q_{SM}H} Q_{SM} H \nonumber \\
H^{eff} |\Psi_{SM} \rangle &=& E |\Psi_{SM} \rangle~~~|\Psi_{SM} \rangle
 = (1-Q_{SM}) |\Psi \rangle
\label{eq:BlHo}
\end{eqnarray}
where $|\Psi \rangle$ is the exact wave function and $H |\Psi \rangle
= E |\Psi \rangle$.  The difficulty posed by this equation is the
appearance of the unknown energy eigenvalue in the equation for
$H^{eff}$.  Thus this system must be solved self-consistently.
Note that there is no explicit reference to the harmonic oscillator
in this equation: it enters only implicitly through $Q_{SM}$ in
distinguishing the long-wavelength ``SM" space from the remainder
of the Hilbert space.
We emphasize that a proper solution of the BH equation must yield
results (energies and operator matrix elements) that are 
independent of $Q_{SM}$.

There is an extensive literature on this and similar equations,
often involving a division of $H$ into an unperturbed $H_0$ and
a perturbation $H_1 = H-H_0$ \cite{Ba73,Ku90}.  There are well-known
pathologies with this division involving the effects of near-by
intruder states on the perturbation expansion \cite{Sh72,Sh73}.  
Here we explore another approach that is nonperturbative and 
involves, in effect, a computer summation of diagrams.  The
method is based on the Lanczos algorithm and offers a remarkably
simple solution to the issue of self-consistency.

In the Lanczos algorithm a basis for representing a Hamiltonian
is formed recursively in such a way that the resulting 
Hamiltonian is tridiagonal.  Given a Hermitian operator $H$ and
and an initial normalized vector $|v_1\rangle$, the successive steps are
\begin{eqnarray}
H|v_1\rangle &=& \alpha_1 |v_1\rangle + \beta_1 |v_2\rangle \nonumber \\
H|v_2\rangle &=& \beta_1 |v_1\rangle + \alpha_2 |v_2\rangle
+ \beta_2 |v_3\rangle \nonumber \\
H|v_3\rangle &=& ~~~~~~~~~~~\beta_2 |v_2\rangle + \alpha_3 |v_3\rangle
+ \beta_3 |v_4\rangle~~etc.
\label{eq:LanHam} 
\end{eqnarray}
so that the $H$ takes the form
\begin{equation}
H \rightarrow \left( \begin{array}{cccc} \alpha_1 & \beta_1
& 0 & \cdots \\ \beta_1 & \alpha_2 & \beta_2 & \cdots \\
0 & \beta_2 & \alpha_3 & \cdots \\ \vdots & \vdots & \vdots & 
\end{array} \right) \left( \begin{array}{c} |v_1\rangle \\
|v_2\rangle \\ |v_3\rangle \\ \vdots \end{array} \right)
\label{eq:Hamgoto}
\end{equation}
The remarkable property of this algorithm has to do with 
truncating the process in Eq.~(\ref{eq:LanHam})
after $n$ steps, where $n$ can be much smaller than the dimension
of the Hilbert space.  The resulting truncated matrix in 
Eq.~(\ref{eq:Hamgoto})
then contains the information needed to reconstruct the exact
2$n$-1 lowest moments of H over the eigenspectrum.  As extremum
eigenvalues are crucial to higher moments, one common application
of the Lanczos algorithm is in determining such eigenvalues
and their associated eigenfunctions.  Another is to begin with
the vector $|v_1\rangle = \hat{O}|g.s.\rangle$ and then use the
algorithm to calculate the moments of the response of the ground
state $|g.s.\rangle$ to the operator $\hat{O}$.  A small number
of moments, e.g., $\sim$ 100, often is sufficient to construct
a response function with a numerical resolution comparable to that achieved
experimentally.

A third application \cite{Ha74} is in constructing fully
interacting Green's functions.  One finds
\begin{equation}
{1 \over E-H} |v_1\rangle = g_1(E) |v_1\rangle + g_2(E) |v_2\rangle
+ \cdots
\label{eq:lgf}
\end{equation}
where the $g_i(E)$ are continued fractions that depend on 
${\alpha_i,\beta_i}$ and where E appears only as a parameter.
For example,
\begin{equation}
g_1(E) = {1 \over {E - \alpha_1 - {\beta_1^2 \over
E - \alpha_2 - {\beta_2^2 \over E - \alpha_3 - \beta_3^2} \cdots}}}
\end{equation}
  
It follows that the BH equation can be solved
self-consistently with only a single solution of the effective
interactions problem, even in cases where multiple bound
states are needed.  The procedure is: \\
  
\noindent
$\bullet$ For each relative-coordinate vector in the SM space
$|\gamma\rangle$, form the excluded-space vector
$|v_1\rangle \equiv Q_{SM}H|\gamma\rangle$ and the corresponding
Lanczos matrix for the operator $Q_{SM}H$.  Retaining the
resulting coefficients ${\alpha_i,\beta_i}$ for later use,
construct the Green's function for some initial guess for
$E$ and then the dot product with $\langle \gamma' | H$ to find
$\langle \gamma' | H^{eff}(E) | \gamma \rangle$. \\
  
\noindent
$\bullet$ Perform the ``SM" calculation to find the
desired eigenvalue $E'$ which, in general, will be different
from the guess $E$.  Using the stored ${\alpha_i,\beta_i}$,
recalculate the Green's function for $E'$ and $H^{eff}(E')$
then redo the ``SM" calculation.  Repeat until convergence,
i.e., until the input $E'$ in the Green's function equals 
the desired output ``SM" eigenvalue.\\
  
\noindent
$\bullet$ Then proceed to the next desired bound state, e.g., the
first excited state, and repeat the above step.  Note that it is not
necessary to repeat the $H^{eff}$ calculation.  The eigenvalue
taken from the ``SM" calculation is, of course, that of the
first excited state.  The procedure then generates distinct
$H^{eff}(E')$s for each desired state.\\

The attractiveness of this approach is that the effective 
interactions part of the procedure, which is relatively time consuming
as it requires one to perform a large-basis Lanczos calculation
for each relative-coordinate starting vector in the ``SM" space,
is performed only once.  The diagonalization in
the model space is generally much faster: modern workstations 
can handle even large-dimension shell model calculations
(sparse matrices of $d \sim 10^6$) quickly ($\sim$ 30 minutes).
In practice we found that self-consistency is achieved 
easily: six to eight cycles is typical.  (More cycles
are required for states with small binding energies.)
Thus it is quite practical to derive the exact $H^{eff}(E)$s
for a series of bound states.

In our view the Lanczos approach to the effective interactions
problem appears to be remarkably simpler than the standard 
procedures of the field, particular in view of the need for
extensions to multi-nucleon ladders.  The traditional approach
divides the effective interactions problem into an energy-dependent
piece (often called the Q-box) and an energy-dependent one
(represented by folded diagrams).  Frequently the energy 
dependence is removed by making a unitary transformation to
a nonHermitian effective Hamiltonian.  The Lanczos procedure
is also much more in the spirit of EFT: the full Hamiltonian 
in the high-momentum space is never constructed.  Rather, it is
replaced by a much smaller Lanczos matrix which contains 
exactly the most relevant long-wavelength information of the
full matrix, the 2$n$-1 lowest moments.  Thus the procedure 
can be viewed as a numerical ET in which this
information is recursively extracted.
  
Now we discuss the results of applying this procedure to the
simplest nuclei, d and ${}^3$He, carrying the above process to
completion (two-body and three-body ladders, respectively).
The motivation is two-fold: demonstrate the numerical procedures
we described above, and provide for the first time exact effective
theory results that can be compared to those of the shell model.

The harmonic oscillator mode expansion must be sufficient 
to represent both the long-distance tails of bound states and
the short-distance ``hard core" scattering predicted by
realistic NN potentials.  (The Argonne V18 ${}^1S_0$ potentials
are shown in Fig.~\ref{fig:wick2}.)  
Inclusion of high-momentum states through $\Lambda_\infty \sim 50$
yields a deuteron binding energy accurate to $\sim 60 {\rm keV}$;
extending this to $\sim 140$ produces a result accurate to
one keV.
Fig.~\ref{fig:wick3} shows the rate of convergence as a function of $b$ and 
$\Lambda_\infty$.
The convergence for the binding of ${}^3$He (not shown) is 
slightly more rapid, a feature related to the deeper binding,
as will become clear later.
\begin{figure}[!ht]
\hskip 0.5in\psfig{bbllx=0.0cm,bblly=9.0cm,bburx=18cm,bbury=22.5cm,figure=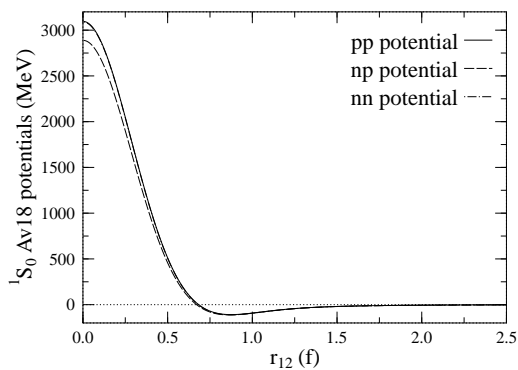,height=2.0in}
\caption{The Argonne V18 ${}^1S_0$ potentials.}
\label{fig:wick2}
\end{figure}

The binding energies and operator matrix elements for simple
systems like ${}^3$He can, of course, be calculated
by other methods, e.g., Faddeev techniques or GFMC.
We thus want to stress that the point of the 
following discussion is to do analogous calculations in the
context of an effective theory, so that we begin to see the
shortcomings of conventional techniques like the shell model
as well as possibilities for overcoming those shortcomings.
A first test of the techniques outlined above is to solve the
BH equation for some SM-like space to then see
if the resulting self-consistent energy is, indeed, the 
correct value.  For model spaces of 2, 4, 6, and 8$\hbar \omega$
in the case of the deuteron we obtained a binding energy of
-2.224 MeV (using $\sqrt{2} b=1.6f$ and $\Lambda_\infty = 140$).  
The exact result is -2.2246 MeV.  Our ${}^3$He result for
same SM spaces and
$\Lambda_\infty = 60$ is -6.87 MeV, in agreement with the
corresponding GFMC result of -6.87 $\pm$ 0.03 MeV.
\begin{figure}[!ht]
\hskip 0.5in\psfig{bbllx=0.cm,bblly=9.0cm,bburx=18cm,bbury=22.5cm,figure=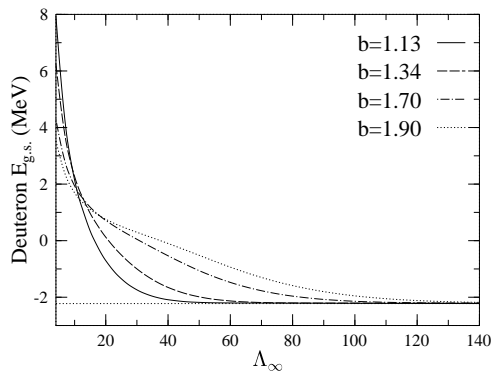,height=2.0in}
\caption{Deuteron ground state energy convergence as a function of
$\Lambda_\infty$ for several choices of the oscillator parameter $b$.
This b is defined as in the independent-particle SM: the corresponding
b for relative motion is $\sqrt{2}$ times the values shown.
It is clear that if an inappropriate size is chosen for the basis states
(e.g., b=1.9), the rate of convergence can be greatly slowed.}
\label{fig:wick3}
\end{figure}

More interesting is the evolution of the wave functions, 
shown in Tables~\ref{table:one} and ~\ref{table:two}.  
The tables were generated as above: for some large value of
$\Lambda_\infty$ SM wave functions were generated for various
choices of $\Lambda_{SM}$ by interating the SM 
calculation until the self-consistent energy is fully converged.
The deuteron and ${}^3$He calculations involve two- and
three-body ladder sums in the excluded space, yielding 
sets of two- and three-body ``SM" matrix elements of $H^{eff}$
for the model spaces.  The deuteron calculation is rather
trivial; for $\Lambda_\infty \sim 50$ the ${}^3$He BH 
calculation involves a dense matrix of dimension $\sim 10^4$,
still rather modest by current SM standards.  The matrix is
dense because we work in relative Jacobi coordinates, rather
than an m-scheme, utilizing standard Talmi-Brody-Moshinshy 
methods \cite{Mo69}.  (See Ref. \cite{So98} for details.)
The results in Table~\ref{table:two} 
were obtained with approximately 100 Lanczos
iterations: it is apparent that the convergence is then quite good.  
The wave functions must be 
normalized according to Eq.~(\ref{eq:BlHo}): 
this involves calculating
unity as an effective operator.  We will return to this point
below.
\begin{table}
\caption{ET results for the deuteron ground state wave function 
calculated with the Argonne V18 potential~\protect\cite{Ha99A,Ha99B}.  
The columns on the
right correspond to different choices of the ET model space,
the analog of a SM space.  The rows correspond to the resulting amplitudes
for the designated, selected configurations $\mid n, {}^SL_J \rangle$.  
The quantities
within the parentheses are the square of the norm of the 
effective wave function, e.g., the probability that the deuteron
resides in the corresponding ``SM'' space.}
\label{table:one}
\vspace{0.2cm}
\begin{center}
\begin{tabular}{|r|r|r|r|r|r|r|}
\hline
\hline
 & \multicolumn{6}{c|}{amplitude} \\ \cline{2-7}
basis state & 0$\hbar \omega$  & 2$\hbar \omega$ &
4$\hbar \omega$ & 6$\hbar \omega$ & 8 $\hbar \omega$ & exact \\ \cline{2-7}
 & (65.9\%) & (79.5\%) & (86.1\%) & (91.3\%) & (93.0\%) & (100\%) \\ \hline
 $\mid$ 1, ${}^3S_1 \rangle$ & 0.81155 & 0.81154 & 0.81155 & 0.81155 & 
0.81152 & 0.81155 \\ \hline
 $\mid$ 2, ${}^3S_1 \rangle$ & 0.00000 & -0.31483 & -0.31483 & -0.31483 & 
-0.31482 & -0.31483 \\ \hline
 $\mid$ 1, ${}^3D_1 \rangle$ & 0.00000 & 0.19524 & 0.19524 & 0.19524 & 
0.19523 & 0.19524 \\ \hline
 $\mid$ 3, ${}^3S_1 \rangle$ & 0.00000 & 0.00000 & 0.24945 & 0.24945 & 
0.24944 & 0.24945 \\ \hline
 $\mid$ 4, ${}^3S_1 \rangle$ & 0.00000 & 0.00000 & 0.00000 & -0.20851 & 
-0.20850 & -0.20851 \\ \hline
 $\mid$ 5, ${}^3S_1 \rangle$ & 0.00000 & 0.00000 & 0.00000 & 0.00000 & 
0.12596 & 0.12596 \\ \hline\hline
\end{tabular}
\end{center}
\end{table}

The tables show the lovely evolution of the wave function in ET,
an evolution quite unlike that of typical shell model calculations.
The wave functions obtained in different model spaces agree 
over overlapping parts of their Hilbert spaces.  Thus as one
proceeds through 2$\hbar \omega$, 4$\hbar \omega$, 6$\hbar \omega$,
... calculations, the ET wave function evolves only by adding 
new components in the expanded space.  The normalization of the
wave function grows accordingly.  Thus, for ${}^3$He,
the 0$\hbar \omega$ ET calculation contains 0.311 of the full
wave function in the effective space; the 0+2+4$\hbar \omega$
result is 0.700.  

This evolution will not arise in the standard SM because the wave
function normalization is set to unity regardless of the model
space.  It will also not arise for a second reason, illustrated in
Table~\ref{table:three}.  The matrix elements of $H^{eff}$ are crucially 
dependent on the model space: the listed results for ${}^3$He
show that a typical matrix $\langle \alpha | H^{eff} | \beta \rangle$
changes very rapidly under modest expansions of the model space,
e.g., from 2$\hbar \omega$ to 4$\hbar \omega$.  Yet it is 
common practice in the shell model to expand calculations
by simply adding to an existing SM Hamiltonian new interactions
that will mix in additional shells.  
We suspect the behavior found for ${}^3$He is generic in ET
calculations: it arises because a substantial fraction of the
wave function lies near but outside the model space 
(e.g., see Table~\ref{table:two}).  An
expansion of the model space changes the energy denominators 
for coupling to some of these configurations, and moves 
other nearby configurations from the excluded space to the model
space.  Naively, relative changes in effective interaction matrix
elements of unity are expected.
\begin{table}
\caption{As in Table~\ref{table:one}, only for ${}^3$He.  
The basis states are
now designated somewhat schematically as $\mid N, \alpha \rangle$,
where $N$ is the total number of oscillator quanta and 
$\alpha$ is an index representing all other quantum numbers.}
\label{table:two}
\vspace{0.2cm}
\begin{center}
\begin{tabular}{|r|r|r|r|r|r|r|}
\hline
\hline
 & \multicolumn{6}{c|}{amplitude} \\ \cline{2-7}
state & 0$\hbar \omega$ & 2$\hbar \omega$ &
4$\hbar \omega$ & 6$\hbar \omega$ & 8 $\hbar \omega$ & exact \\ \cline{2-7}
 & (31.1\%) & (57.4\%) & (70.0\%) & (79.8\%) & (85.5\%) & (100\%) \\ \hline
 $\mid 0, 1 \rangle$ & 0.55791 & 0.55791 & 0.55791 & 0.55795 & 0.55791 & 
0.55793 \\ \hline
 $\mid 2, 1 \rangle$ & 0.00000 & 0.04631 & 0.04613 & 0.04618 & 0.04622 & 
0.04631 \\ \hline
 $\mid 2, 2 \rangle$ & 0.00000 & -0.48255 & -0.48237 & -0.48243 & -0.48243 & 
-0.48257 \\ \hline
 $\mid 2, 3 \rangle$ & 0.00000 & 0.00729 & 0.00731 & 0.00730 & 0.00729 & 
0.00729 \\ \hline
 $\mid 2, 4 \rangle$ & 0.00000 & 0.16707 & 0.16698 & 0.16706 & 0.16706 & 
0.16708 \\ \hline
 $\mid 2, 5 \rangle$ & 0.00000 & 0.00566 & 0.00564 & 0.00565 & 0.00565 & 
0.00566 \\ \hline
 $\mid 2, 6 \rangle$ & 0.00000 & -0.00017 & -0.00017 & -0.00017 & -0.00017 & 
-0.00017 \\ \hline
 $\mid 4, 1 \rangle$ & 0.00000 & 0.00000 & -0.02040 & -0.02042 & -0.02043 & 
-0.02047 \\ \hline
 $\mid 4, 2 \rangle$ & 0.00000 & 0.00000 & 0.11267 & 0.11274 & 0.11275 & 
0.11289 \\ \hline
 $\mid 4, 3 \rangle$ & 0.00000 & 0.00000 & -0.04191 & -0.04199 & -0.04208 & 
-0.04228 \\ \hline
 $\mid 4, 4 \rangle$ & 0.00000 & 0.00000 & 0.28967 & 0.28978 & 0.28978 & 
0.29001 \\ \hline
 $\mid 4, 5 \rangle$ & 0.00000 & 0.00000 & 0.01059 & 0.01059 & 0.01059 & 
0.01059 \\ \hline
 $\mid 4, 6 \rangle$ & 0.00000 & 0.00000 & -0.00213 & -0.00212 & -0.00211 & 
-0.00210 \\ \hline
 $\mid 4, 7 \rangle$ & 0.00000 & 0.00000 & 0.00998 & 0.01000 & 0.01000 & 
0.01000 \\ \hline
 $\mid 4, 8 \rangle$ & 0.00000 & 0.00000 & -0.11319 & -0.11327 & -0.11330 & 
-0.11335 \\ \hline
 $\mid 4, 9 \rangle$ & 0.00000 & 0.00000 & 0.08446 & 0.08447 & 0.08446 & 
0.08448 \\ \hline
 $\mid 4, 10 \rangle$ & 0.00000 & 0.00000 & -0.08613 & -0.08626 & -0.08632 & 
-0.08638 \\ \hline
 $\mid 4, 11 \rangle$ & 0.00000 & 0.00000 & -0.00210 & -0.00211 & -0.00211 & 
-0.00211 \\ \hline
 $\mid 4, 12 \rangle$ & 0.00000 & 0.00000 & -0.00252 & -0.00254 & -0.00256 & 
-0.00257 \\ \hline
 $\mid 4, 13 \rangle$ & 0.00000 & 0.00000 & -0.00020 & -0.00020 & -0.00020 & 
-0.00020 \\ \hline
 $\mid 4, 14 \rangle$ & 0.00000 & 0.00000 & -0.00010 & -0.00010 & -0.00010 & 
-0.00010 \\ \hline
 $\mid 4, 15 \rangle$ & 0.00000 & 0.00000 & -0.00012 & -0.00013 & -0.00012 & 
-0.00012 \\ \hline \hline
\end{tabular}
\end{center}
\end{table}
\begin{table}
\caption{Selected BH 3-body effective interaction matrix 
elements~\protect\cite{Ha99A,Ha99B} for ${}^3$He,
in MeV, illustrating the strong dependence on the ``SM" space.  The Argonne
V18 potential was used.}
\label{table:three}
\vspace{0.2cm}
\begin{center}
\begin{tabular}{|r|r|r|r|r|}
\hline
\hline
  & 2$\hbar \omega$ &
4$\hbar \omega$ & 6$\hbar \omega$ & 8 $\hbar \omega$  \\ \hline
$\langle 0, 1 \mid H^{eff} \mid 2, 1 \rangle$ & -4.874 & -3.165 & -0.449 
& 1.279 \\ \hline
$\langle 0, 1 \mid H^{eff} \mid 2, 5 \rangle$ & -0.897 & -1.590 & -1.893 
& -2.208 \\ \hline
$\langle 2, 1 \mid H^{eff} \mid 2, 2 \rangle$ & 6.548 & -2.534 & -4.144 
& -5.060 \\ \hline \hline
\end{tabular}
\end{center}
\end{table}

Now we turn to the question of operators.  The standard procedure
in the SM is to calculate nuclear form factors with 
bare operators, or perhaps with bare operators renormalized 
according to effective charges determined phenomenologically
at $q^2$ = 0, using SM wave functions normed to 1.
As we now have a series of exact effective interactions
corresponding to different model spaces, we can test
the validity of this approach.  The results for the elastic
magnetic form factors for the deuteron and $^3$He are shown in
Fig.~\ref{fig:wickmags}.  
\begin{figure}[!ht]
\hskip -0.1in\psfig{bbllx=0.cm,bblly=9.0cm,bburx=18cm,bbury=22.5cm,figure=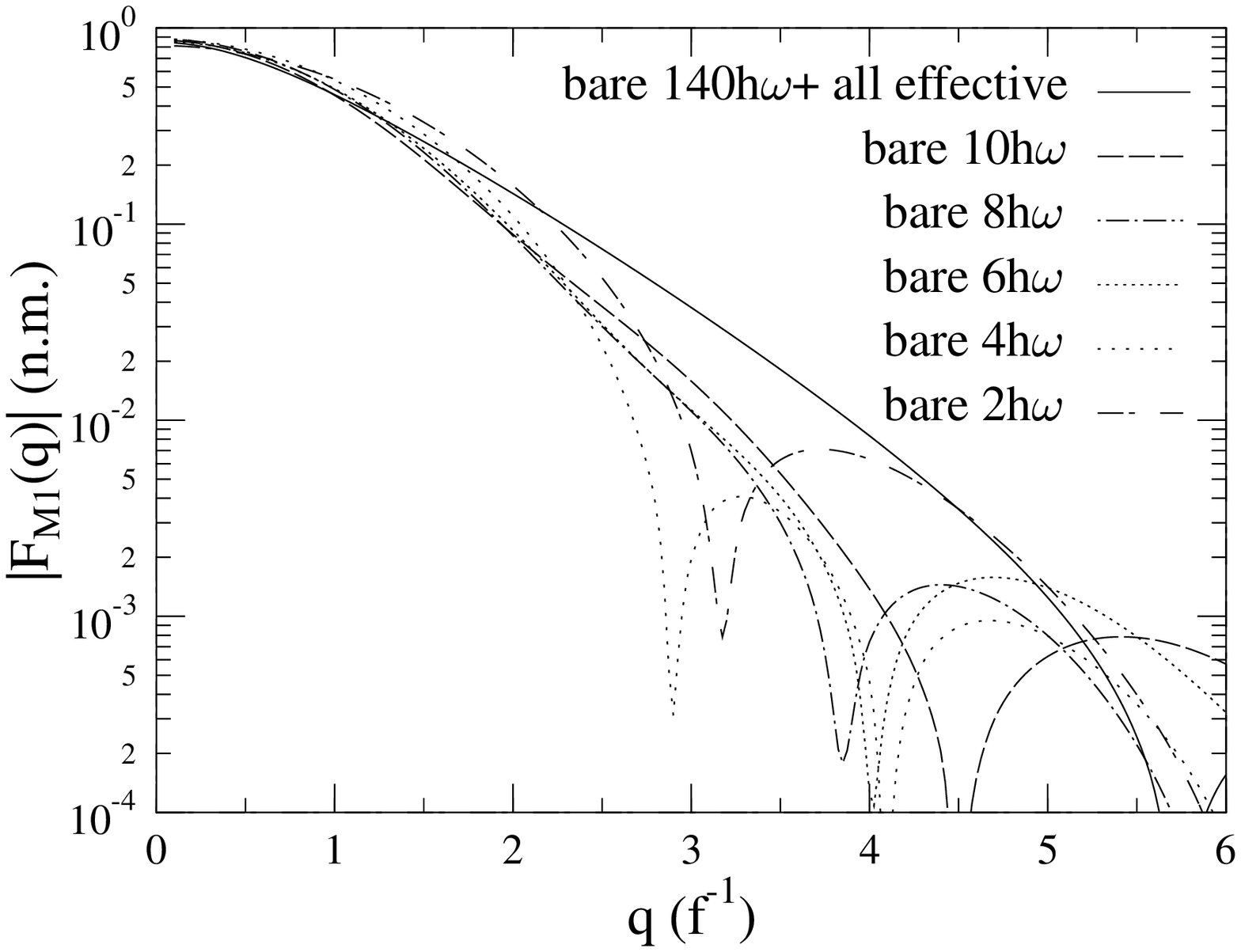,height=1.7in}
\hskip 0.01in\psfig{bbllx=0.cm,bblly=9.0cm,bburx=18cm,bbury=22.5cm,figure=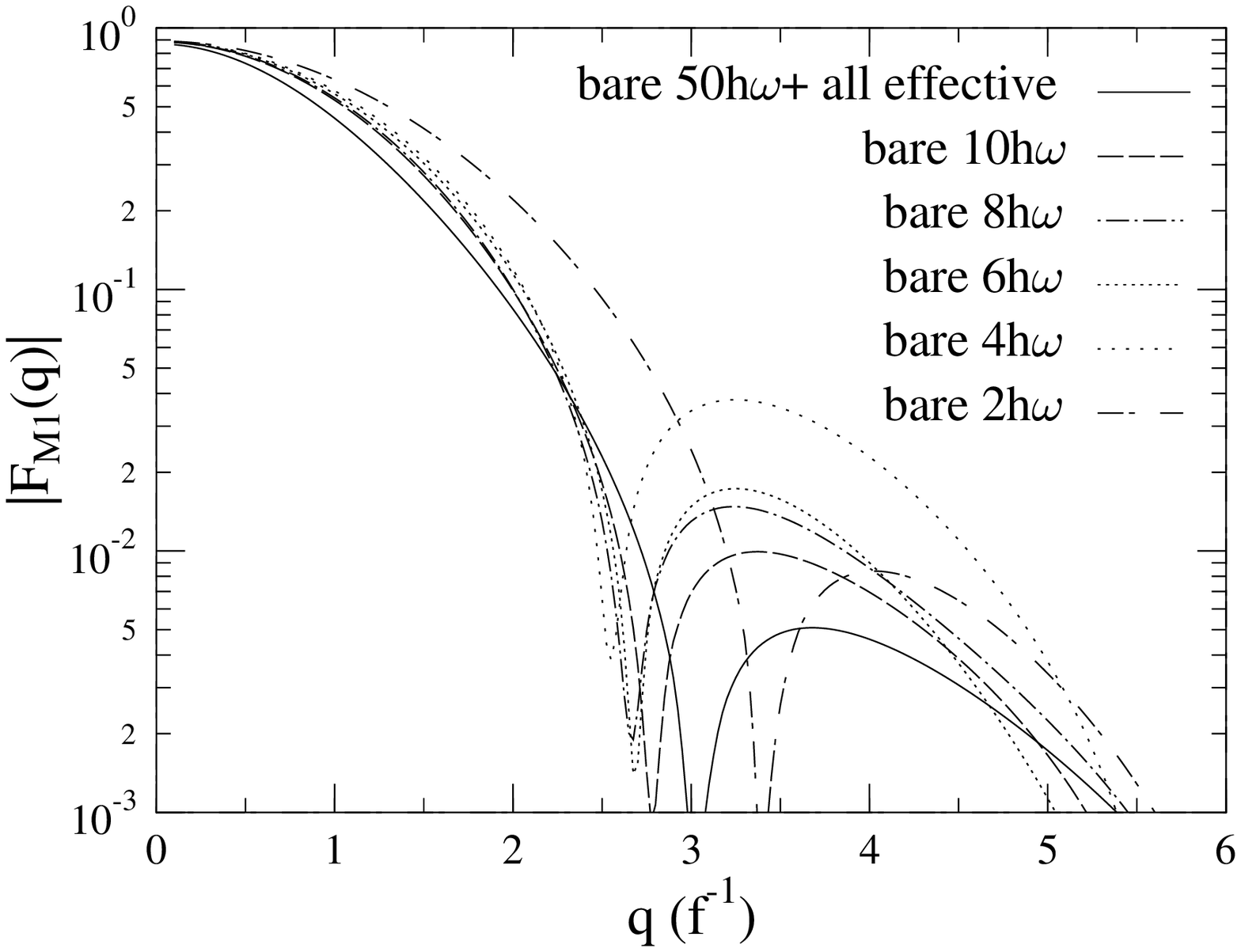,height=1.7in}
\caption{The magnetic elastic form factor for the deuteron (left panel)
and $^3$He (right panel)
calculated with the exact $H^{eff}$, SM wave functions normalized
to unity, and a bare operator are compared to the exact result
(solid line)~\protect\cite{Ha99A,Ha99B}.  
When effective operators and the proper wave
function normalizations are used, all results become identical
to the solid line.}
\label{fig:wickmags}
\end{figure}
One sees in each case that by the time one 
reaches a momentum transfer $q \sim 2.5/f$, random numbers are
being generated: bare operators used in conjunction with exact
effective wave functions generate results that differ by an
order of magnitude, depending on the choice of the model space.
This is not surprising, of course.  If one considers the 
operation
\begin{equation}
e^{i \vec{q} \cdot \vec{r}} \sigma \tau_3 | g.s. \rangle 
\end{equation}
at momentum transfers $\gsim 2k_F$, where $k_F$ is the Fermi 
momentum, most of the resulting amplitude should reside outside
the long-wavelength model space, in any simple view of the 
nucleus.  That is, the strength resides entirely in the effective
contributions to the operator.  If these components are ignored,
the results have to be in error.

Clearly the effective interaction and effective operator have to
be treated consistently and on the same footing.  If $\hat{O}$
is the bare operator, one finds
\begin{equation}
\langle \Psi_f | \hat{O} | \Psi_i \rangle \equiv \langle \Psi_f^{eff}
 | \hat{O}^{eff} | \Psi_i^{eff} \rangle 
\end{equation}
where
\begin{equation} 
\hat{O}^{eff} = (1 + HQ_{SM} {1 \over E_f - HQ_{SM}}) \hat{O}
(1 + {1 \over E_i - Q_{SM}H}Q_{SM}H)
\end{equation}
and where the effective wave function normalization of 
$|\Psi_i^{eff} \rangle$ and $|\Psi_f^{eff} \rangle$, mentioned
earlier, must be determined using the effective operator $\hat{1}$,
e.g.,
\begin{equation}
1 = \langle \Psi_i | \Psi_i \rangle = \langle \Psi_i^{eff} |
(1 + HQ_{SM}{1 \over E_i - HQ_{SM}})(1 + {1 \over E_i - Q_{SM}H}
Q_{SM}H) | \Psi_i^{eff} \rangle
\end{equation}
These expressions can be evaluated with the Lanczos Green's function
methods described earlier.  When this is done, all of the effective calculations,
regardless of the choice of the model space, yield the same result,
given by the solid lines in Fig.~\ref{fig:wickmags}.

We would argue, based on this example, that many persistent
problems in nuclear physics --- ranging from the renormalization
of $g_A$ in $\beta$ decay to the systematic differences
between measured and calculated M1 electromagnetic form factors ---
very likely are due to naive treatments of operators, treatments
that fail to satisfy the basic rules of ETs.  It
should be apparent from the above example that no amount of work
on $H^{eff}$ will help with this problem.  What is necessary is
a diagrammatic basis for generating $H^{eff}$ that can be 
applied in exactly the same way to evolving $\hat{O}^{eff}$.
{}From this perspective, phenomenological derivations of $H^{eff}$
by fitting binding energies and other static properties of 
nuclei are not terribly helpful, unless one intends to 
simultaneously find phenomenological renormalizations for 
each desired operator in each $q^2$ range of interest.

The point of this discussion has been largely pedagogical.
To our knowledge the examples given above are the only ones in
the classical nuclear physics literature in which a ``model
space" calculation has been formulated in a way that satisfies the 
basic rules of ET.  The results ---
particularly the invariance of energies and operator matrix
elements under changes in how we define $Q$ --- are obvious 
from the perspective of ET.  The fact that standard techniques
like the SM violate so many of the rules of ETs 
is cause for optimism: much can be done to improve the rigor of
such nuclear physics tools.  As we have demonstrated here, new 
numerical methods can be developed to handle the full ET problem 
at the cost of only modest additional effort.

%%%%%%  Renormalization Group %%%%%%%%%%
\subsection{Crossing the Many-Body Border}

The effort to derive SM effective interactions directly from 
realistic NN interactions was largely abandoned in the early 
1970's when it became apparent that the high-momentum scattering
problem was nonperturbive.  A series of efforts involving 
perturbative expansions in either the potential $V$ or the 
corresponding $G$-matrix (the sum of the two-body ladder diagrams)
yielded discouraging results, with higher-order terms often
dominating over lower-order ones\cite{Bar99}.  In particular
Shucan and Weidenmuller showed that any overlap in the spectra
of included and excluded, high-momentum states would lead to
a nonperturbative series\cite{Sh72,Sh73}.  Thus, the uncontrolled
approximations in the SM approach grew out of necessity: 
it became a largely phenomenological tool in part because 
early efforts to provide a more rigorous basis for the model
ended in failure.

Given our two SM-like ET calculations, the deuteron and ${}^3$He,
it is interesting to return to this old problem of the 
nonperturbative behavior of the nonrelativistic many-body
problem, using these examples as laboratories.   
Today's challenge, in fact, is more modest than that of the
1970's: modern computers coupled with new algorithms, like the
Lanczos solution of the BH equation introduced 
above, allow us to solve rather complex nonperturbative problems.
To extend the calculations above to ${}^4$He and heavier nuclei
one has to achieve only modest increases in efficiency,
such as a substantial lowering of $\Lambda_\infty$.  
Below we summarize some very recent work by Haxton and Luu\cite{Luu00} that
suggests that the goals of the 1970's --- a perturbative 
treatment of the nonrelativistic nuclear physics problem ---
may not be out of the question.  Through renormalization group 
and other techniques, the most severely nonperturbative contributions 
to the potential can be systematically extracted and treated.  The approach is
quite similar to that of EFTs.  We conclude our discussion
of the many-body problem by sketching this effort to cross the
border between EFTs and conventional nuclear models.

The deuteron calculations were performed for $\Lambda_\infty \sim$
140, a value where the hard core of the AV18 potential is 
resolved sufficiently well that the resulting binding energy is
good to 1 keV.  Results are variational in this parameter: as
$\Lambda_\infty$ is increased, the correct binding energy is 
approached from above.  The evolution was explored in Ref.~\cite{Ha99A}
and found to vary as
\begin{equation}
E_{gs} \sim -2.175 \mathrm{MeV} + 3.178 \mathrm{MeV} 
e^{-1.055 (\Lambda_\infty/50)^2}
\end{equation}
to very good accuracy, a functional dependence that already suggests
some nonperturbative behavior.

The BH calculation determines the full set of matrix
elements $\langle \alpha | H^{eff} | \beta \rangle$ for the 
SM space.  These individual matrix elements carry information
about the convergence not easily seen in $E_{gs}$.
Their behavior in perturbation theory can be examined by expanding
the BH propagator
\begin{equation}
{1 \over E-QH} = {1 \over E-H_0} + {1 \over E-H_0} Q(V-V_0) {1 \over E-H_0} + \cdots,
\end{equation}
evaluating the series term by term.  Here $H_0$ and $V_0$ are the
harmonic oscillator Hamiltonian and potential.  
We thus lower $\Lambda_\infty$ and test whether states above 
$\Lambda_\infty$ can be treated perturbatively.

Once a scale of $\Lambda_\infty \sim 70$ is reached, striking 
differences appear in the rate of convergence of different matrix
elements.  All are greatly improved in first and second order
perturbation theory.  Those matrix elements well away from the 
``boundary" at $\Lambda_\infty$ converge quickly to their 
correct values in this way.  However, those matrix elements
$\langle \alpha | H^{eff} | \beta \rangle$ where either
$| \alpha \rangle$ or $| \beta \rangle$ resides in the last
shell (with $N=\Lambda_\infty$) converge only very slowly
to the correct values, after the initial improvement in low-order
perturbation theory.  Typical $\sim 10^3$ orders of perturbation
theory are required to produce the correct value.
This clearly suggests that slow convergence is
associated with the relative kinetic energy operator $QT$ contribution
to $QH$, as the only transitions to states outside the Hilbert space
generated by this operator have $\Delta n$ = 1.
At large $r$ the strength of this transition becomes quite
large, $-Q V_0(r) | \alpha \rangle$, reflecting the
unphysical asymptotic behavior of harmonic oscillator wave
functions.  This amplitude 
propagates nonperturbatively as $(V-V_0)/H_0 \sim 1$.  Eventually
enough high-momentum harmonic oscillator wave functions are coupled
together to produce the softer asymptotic fall-off characteristic
of the correct bound-state wave function.

As discussed in Ref.~\cite{Luu00}, the required resummation is 
guided by the observation that the true potential, $V(r)$, falls off
properly at large $r$.  Thus a reorganization of the BH
equation in which the propagator is always sandwiched between
$V(r)$ should remove the unwanted propagation.  This leads to the
following recasting of the BH equation:
\begin{eqnarray}
\langle \alpha | H^{eff} | \beta \rangle &=& \langle \alpha | T | \beta \rangle +
(\langle \hat{\alpha} | - \langle \alpha |) E-T (| \hat{\beta} \rangle - | \beta \rangle) \nonumber \\
&& + \langle \hat{\alpha} | V + V {1 \over E-QH} QV | \hat{\beta} \rangle
\end{eqnarray}
where
\begin{equation}
| \hat{\alpha} \rangle = {E \over E-QT} | \alpha \rangle
\label{eq:kegf}
\end{equation}
If $\alpha$ and $\beta$ are not in the last shell, $| \hat{\alpha} \rangle = | \alpha \rangle$
and $| \hat{\beta} \rangle = | \beta \rangle$, so that the above 
rewriting of the BH equation reduces to the original form.
Otherwise, a modified wave function generated by Eq.~(\ref{eq:kegf})
must be used.

This discussion is a necessary diversion from the principal topic of
this section: this source of nonperturbative behavior dominates
one that is of more interest (and more difficult to remove),
hard core scattering.  It should be clear the the lowering of the
scale $\Lambda_\infty$ at the cost of an evaluation of Eq.~(\ref{eq:kegf})
represents a tremendous savings: in place of a dense matrix $QH$
whose elements have to evaluated numerically, we have a sparse
matrix $QT$ whose matrix elements are known analytically.  The kinetic
energy Green's function is thus trivial to evaluate via
Eq.~(\ref{eq:lgf}).  For example, in the case of the deuteron, $QT$
is tridiagonal in the harmonic oscillator basis.  Thus the Green's
function can be expressed analytically in the form of Eq.~(\ref{eq:lgf}).

Once the replacement $| \alpha \rangle \rightarrow | \hat{\alpha} \rangle$
is made, $\Lambda_\infty$ can be lowered to $\sim 40$ while maintaining
1 keV accuracy, if one works to third order in perturbation theory.
But with further lowering, errors arise in the perturbative
expansion that persists even if calculations are carried to very
high order.  At $\Lambda_\infty =$ 30, third order perturbation theory 
reduces $\sim$ 10\% errors in $\langle n'=1 l'=0 | H^{eff} | n=1 l=0 \rangle$
to $\sim$ 0.2\%.  But an error in excess of 0.1\% --- corresponding to
50 keV --- persists after 10
additional orders of perturbation.  Numerically one can verify that 
the nonperturbative tail is generated by the scattering at very
small $r$.  As expected, the nonperturbative contributions to $s-d$ matrix elements are 
much smaller than those for $\langle n'=1 l'=0 | H^{eff} | n=1 l=0 \rangle$
and other $s-s$ matrix elements.

We sketch an elegant solution to this problem described in Ref.~\cite{Luu00}.
The most singular short-ranged contributions are removed from the 
potential analytically through a renormalization group (RG) procedure.
The method starts with the well-known Talmi integral expansion of
harmonic oscillator radial matrix elements and involves a systematic
removal of the most singular of these by introducing local,
separable operators familiar from EFT treatments.  The 
coefficients of these operators are then run by the RG equations.
The result establishes a very close connection between 
potential descriptions and EFT treatments, and also allows a clean
separation of long-distance contributions to the potential that
can be treated perturbatively.

As our goal here is to illustrate the procedure, we work to
LO.  (The NLO and N$^2$LO results are given in Ref.~\cite{Luu00}.)
One begins with the Talmi integral expansion for radial harmonic oscillator
matrix elements
\begin{eqnarray}
\langle n' l' | V_{sr}(r) | n l \rangle &=& \int^\infty_0
R_{n'l'}(r) V_{sr}(r) R_{nl}(r) r^2 dr \nonumber \\
&=& \sum_p B(n',l';n,l;p) I_p(b)
\label{eq:bcoef}
\end{eqnarray}
where
\begin{equation}
I_p(b) = {2 \over \Gamma(p+3/2)} \int^\infty_0 e^{-r^2/b^2} V_{sr}(r) {r^{2p+2} dr \over b^{2p+3}}.
\label{eq:talmi}
\end{equation}
The $B(n',l';n,l;p)$ are known analytically\cite{BM67}.
Here $V_{sr}(r)$ denotes the short-range contributions to the potential
(we return to this point below).  The integrals $I_p$ can be viewed
as a systematic expansion of the nonperturbative hard core in terms
of the parameter $(r_c/b)^2$, where $r_c$ is a distance associated
with the size of the hard core that remains unresolved at scale $\Lambda_\infty$.
Thus the LO term in the expansion is
\begin{equation}
I_{p=0}^{(0)} = {2 \over \Gamma(3/2)} {1 \over b^3} \int_0^\infty V_{sr}(r) r^2 dr
\label{eq:ip0}
\end{equation}
while in NLO one includes
\begin{eqnarray}
I_{p=0}^{(1)} &=& {2 \over \Gamma(3/2)} {1 \over b^3} \int_0^\infty V_{sr}(r) r^2 (1 - {r^2 \over b^2}) dr \nonumber \\
I_{p=1}^{(0)} &=& {2 \over \Gamma(5/2)}  {1 \over b^5} \int_0^\infty V_{sr}(r) r^4 dr .
\label{eq:ip1}
\end{eqnarray}
In N$^2$LO one obtains
\begin{eqnarray}
I_{p=0}^{(2)} &=& {2 \over \Gamma(3/2)} {1 \over b^3} \int_0^\infty V_{sr}(r) r^2 (1 - {r^2 \over b^2} + {r^4 \over 2b^4}) dr \nonumber \\
I_{p=1}^{(1)} &=& {2 \over \Gamma(5/2)}  {1 \over b^5} \int_0^\infty V_{sr}(r) r^4 (1 - {r^2 \over b^2}) dr \nonumber \\
I_{p=2}^{(0)} &=& {2 \over \Gamma(7/2)}  {1 \over b^7} \int_0^\infty V_{sr}(r) r^6 dr .
\label{eq:ip2}
\end{eqnarray}
and so on.

Now the radial component of the $s$-wave potential $V_0 \delta({\bf r})$
is
\begin{equation}
V_\delta^{(0)}(r) = V_0 {1 \over 4 \pi r^2} \delta(r).
\end{equation}
Substituting this into Eq.~(\ref{eq:bcoef}) yields
\begin{equation}
\langle n' l'=0 | V_\delta^{(0)}(r) | n l=0 \rangle = B(n',0;n,0;0) I_0^\delta(b)
\end{equation}
where
\begin{equation}
I_0^\delta(b) = {2 \over \Gamma(3/2)} {1 \over 4 \pi b^3} V_0
\end{equation}
It follows that the contact interaction will produce the exact 
lowest-order Talmi integral contribution (Eq.~(\ref{eq:ip0})) to $V_{sr}$ provided
\begin{equation}
V_0 = 4 \pi \int_0^\infty V_{sr}(r) r^2 dr .
\end{equation}

Using a normalization that will be convenient later, we can then
rewrite $V_{sr}({\bf r})$ as
\begin{equation}
V_{sr}({\bf r}) = V_{sr}^{(1)}({\bf r}) + 
b^3 {\pi^2 \over 2} \hbar \omega a_0^{ss}(\Lambda_\infty) \delta({\bf r})
\end{equation}
where $V_{sr}^{(1)}$ is a new potential whose leading-order behavior
is of order $(r_c/b)^2$ relative to $V_{sr}$ (determined by subtracting
from the original Talmi integral expression for $V_{sr}$
the contribution from Eq.~(\ref{eq:ip0})).
The coefficient $a_0^{ss}(\Lambda_\infty)$ is a dimensionless 
coupling
(Note that throughout this discussion, we have suppressed the 
spin and isospin of $V_{sr}({\bf r})$.  Thus $V_{sr}(r)$ represents
the radial function obtained after spin and isospin matrix elements
have been taken.  For example, the components of the AV18 
potential contributing to $s-s$ transitions is
\begin{equation}
V_{sr}({\bf r}) = V_1(r) + V_2(r) \vec{\tau}_1 \cdot \vec{\tau}_2 +
V_3(r) \vec{\sigma}_1 \cdot \vec{\sigma}_2 +
V_4(r) \vec{\sigma}_1 \cdot \vec{\sigma}_2 \vec{\tau}_1 \cdot \vec{\tau}_2
\end{equation}
so that
\begin{equation}
V_{sr}(r) = V_1(r)-3V_2(r)+V_3(r)-3V_4(r)
\end{equation}
for the deuteron ($l=0,s=1,t=0$).)

The above procedure is, in fact, general.  The NLO 
contribution (Eq.~(\ref{eq:ip1})) can be removed from $s-s$ wave deuteron matrix elements
of $V_{sr}({\bf r})$ by introducing a second, higher-order contact operator
\begin{equation}
a_2^{ss}(\Lambda_\infty) {1 \over 2} (\overleftarrow{\nabla}^2 \delta({\bf r}) +
\delta({\bf r}) \overrightarrow{\nabla}^2)
\end{equation}
where
\begin{equation}
a_2^{ss}(\Lambda_\infty) = {8 \over 3 \pi b^5 \hbar \omega}
\int_0^{\infty} V_{sr}(r) r^4 dr,
\end{equation}
while the N$^2$LO contributions can be removed by
\begin{equation}
a_4^{ss}(\Lambda_\infty) (\overleftarrow{\nabla}^2 \delta({\bf r}) \overrightarrow{\nabla}^2 +
{3 \over 10} (\overleftarrow{\nabla}^4 \delta({\bf r}) + \delta({\bf r}) \overrightarrow{\nabla}^4))
\end{equation}
where
\begin{equation}
a_4^{ss}(\Lambda_\infty) = {2 \over 9 \pi b^7 \hbar \omega}
\int_0^{\infty} V_{sr}(r) r^6 dr.
\end{equation}
Likewise there are coeficients of additional contact operators
(denoted $a_2^{sd}$, $a_4^{sd}$, and $a_4^{dd}$ in Ref.~\cite{Luu00})
that remove the Talmi integral contributions to $s-d$ and $d-d$
matrix elements to N$^2$LO. 

Up to this point we have simply rewritten the ``bare" potential ---
the potential that acts in the space defined by $\Lambda_\infty$ ---
in an entirely equivalent form, exploiting the Talmi integral expansion
and the fact that the unique translationally-invariant two-body
contact operators generate those Talmi integrals.  Now we consider,
in LO, the effects on $a_0^{ss}$ of integrating out the single 
shell at $\Lambda_\infty$, thereby mapping the original problem 
into an effective one with $\Lambda_\infty \rightarrow \Lambda_\infty - 2$.
We require that some low-energy matrix element
\begin{equation}
\langle n' 0 | V_{sr}({\bf r}) | n 0 \rangle
\end{equation}
remain invariant, so that $a_0^{ss}(\Lambda)$ must run.
The relevant equation for $a_0^{ss}$ then is the single-shell limit
of the BH equation, evaluated in LO (which eliminates $sd$ and $dd$
contributions to transitions and the Green's function).  The 
result is \cite{Luu00}
\begin{equation}
a_0^{ss}(\Lambda-2) = a_0^{ss}(\Lambda) + {\Gamma({\Lambda+3 \over 2})
\over ({\Lambda \over 2})!} {a_0^{ss}(\Lambda)^2 \over
E_0 - (\Lambda + {3 \over 2}) - a_0^{ss}(\Lambda) {\Gamma({\Lambda+3
\over 2}) \over ({\Lambda \over 2})!}}
\label{eq:rgeq}
\end{equation}
where $E_0 = E/\hbar \omega$ is a dimensionless energy.  ($E$ is the
BH energy being determined selfconsistently.)
This leading-order RG equation is {\it independent} of the matrix
element chosen to evaluate the running: it is an operator equation,
independent of the running scheme.  Thus, to this order, the most singular Talmi
integral can be subtracted exactly from all matrix elements in the
low-momentum space.
This difference equation thus tells us how to run $a_0^{ss}(\Lambda_\infty)$
to any desired new scale $\Lambda$.

While the RG equation is a difference equation, the connections with
EFT become even clearer by introducing the natural definition of
the derivative
\begin{equation}
{d a_0^{ss}(\Lambda) \over d \Lambda} \equiv {a_0^{ss}(\Lambda) - a_0^{ss}(\Lambda-2) \over 2}.
\end{equation}
As will be apparent from results given below, the terms in the
denominator of Eq.~(\ref{eq:rgeq}) that compete with $\Lambda$ are
one power lower in $\Lambda$.  Thus for large $\Lambda$ the RG
equation can be written in a more conventional form
\begin{equation}
\Lambda {d a_0^{ss}(\Lambda) \over d \Lambda} \sim {1 \over 2}
\sqrt{{\Lambda \over 2}} a_0^{ss}(\Lambda)^2
\end{equation}
which has the solution
\begin{equation}
a_0^{ss}(\Lambda) \sim - \sqrt{{2 \over \Lambda}}.
\end{equation}
As $\Lambda$ is a dimensionless nonrelativistic energy, this is 
equivalent to running as $1/p$, where $p$ is the momentum, a 
result familiar from EFT.  Similarly, $a_2^{ss}$ runs as 
$\Lambda^{-3/2}$ and $a_4^{ss}$ as $\Lambda^{-5/2}$.

This procedure is followed through order $(r/b)^4$ in Ref.~\cite{Luu00},
which involves a series of coupled RG equations for $a_0^{ss}$,
$a_2^{ss}$, $a_4^{ss}$, $a_2^{sd}$, $a_4^{sd}$, and $a_4^{dd}$.
Beyond lowest order the running is no longer scheme independent;
the fit in Ref.~\cite{Luu00} is to the longest wavelength properties
of the low-energy system, the set of matrix elements coupling the
$n=1$ and $n=2$ oscillator shells.  Thus the analogy with standard
EFT approaches is very close.  In effect, the primary difference
is that in EFTs the strengths of the local operators are determined
by directly fitting data.  Here a phenomenological potential ---
itself fit to data --- determines the initial value of the 
local operators that are then run by the RG equation.  It is also
apparent that the detailed short-range form of the potential
is irrelevant to shell-model-inspired effective theories:
the needed quantities are Talmi integrals over the short-range
part of the potential.  Any change in the detailed radial behavior
of the potential that preserves the value of the Talmi integrals
will also leave the low-energy theory unchanged.

The numerical results in Ref.~\cite{Luu00} are quite promising:
in N$^2$LO, the hard core (below 0.7f) nonperturbative contribution 
to the scattering can be stepped down to $\Lambda \sim 10$ ---
that is, to nearly a typical SM scale --- with a loss of accuracy
of $\sim$ 7 keV.  The portion of the core one can treat accurately
increases as the order of the RG calculation is increased.
As the potential is known, one can --- if one is willing to invest 
sufficient effort --- carry this procedure out to an arbitrary
number of orders.  The numerical results 
are given in Fig.~\ref{fig:wickrg}.  One sees that the accuracy
with which one can run to small $\Lambda$ continues to improve
according to the work invested (LO, NLO, and N$^2$LO) in the RG
equations.
\begin{figure}[!ht]
\hskip 1.0in\psfig{figure=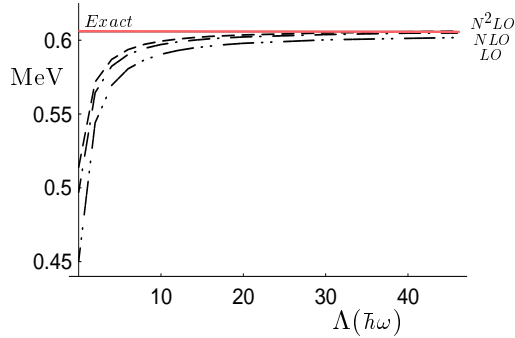,height=1.8in}
\caption{The results of integrating out the hard-core contribution
($r_c$ = 0.7f) to the AV18 potential by the RG method, in LO, NLO, and 
N$^2$LO.
These orders correspond to the removal of successively less
less singular contributions of the leading-order Talmi integrals,
and thus can be viewed as an expansion in $(r_c/b)^2$.  One
can lower the scale of the low-energy theory from $\Lambda_\infty = 50$
to 10 analytically with the loss of $\sim$ 7 keV in the accuracy
of the deuteron ground-state energy in N$^2$LO.}
\label{fig:wickrg}
\end{figure}

The detailed use of this Talmi integral subtraction and related 
RG procedures will not be discussed here, though it is treated in
Ref.~\cite{Luu00}.  The essential idea is to exploit the 
RG equations to treat the nonperturbative scattering due to the
hard core, then use perturbation theory to handle the remaining
softer parts of the potential.  Thus $V_{sr}(r)$ is defined in
terms of the Talmi integrals for $V(r) \theta(r_c-r)$, where $r_c$
is the radius of the unresolved hard core. 
The marriage of perturbation theory and RG techniques can be done
in a number of ways: many aspects of this question remain to be
explored.  However it is clear these two tools are complementary.
If one starts at very high $\Lambda_\infty$, the part of the hard
core that cannot be treated by perturbation theory 
is limit to a very small region near $r \sim 0$.
However to minimize the effects of omitted higher order terms
(i.e., Talmi integrals) in the RG equation for matrix elements
with large values of $n$, $r_c$ must be made quite small.  As $\Lambda_\infty$
is lowered, $r_c$ can be extended without loss of accuracy, 
but perturbation theory --- an expansion in 1/$\Lambda$ ---
becomes less effective.  

The extensions to three- and four-body clusters have not yet been
carried out, though nothing in the procedures is specific to the
deuteron.  The RG equations will, of course, involve  
the full set of general three- and four-body translationally 
invariant contact operators that can be constructed, just as 
EFT treatments of the three-body problem were required to
introduce three-body contact terms.  Thus the RG equations
become more complicated, but not conceptually more difficult.

%%%%%%  Conclusions %%%%%%%%%%%%%%
\section{Conclusion}

In this chapter we have given a somewhat detailed overview of the
progress that has been made in applying the ideas of EFT to nuclear
physics.  This will ultimately provide the bridge between QCD and the
rigorous study of multi-nucleon processes.  The field is still
somewhat embryonic, but as we have shown, impressive (and rigorous)
results have already been obtained.  There is much more to come!

\section{Acknowledgments}

We are grateful to our very learned colleagues who are responsible for
much of our lack of understanding of this field. 
We would like to thank Harald Grie\ss hammer, Tom Lu, and 
all our collaborators.
This work is
supported in part by the U.S. Dept. of Energy under Grants
No. DE-FG03-97ER$0$14 (NT@UW-00-021) (S.B., D.P. and M.S.) and
DOE-ER-41132-101 (P.B. and W.H.).

\end{document}